\documentclass[english,showpacs,twocolumn,superscriptaddress,showpacs,showkeys]{revtex4-1}
\usepackage{graphicx}
\usepackage{chngcntr}
\usepackage[export]{adjustbox}
\usepackage{rotating}
\usepackage{dcolumn}
\usepackage{bm}
\usepackage[usenames,dvipsnames]{color}
\usepackage{amsmath}
\usepackage{amssymb}
\usepackage{subcaption}
\usepackage[utf8]{inputenc}
\usepackage[colorlinks=true,linkcolor=blue,citecolor=green,urlcolor=magenta]{hyperref} 
\usepackage[nameinlink]{cleveref} 
\setcitestyle{super}
\graphicspath{{./Files/}{./Hydro_Results/}{./2D_Results/}{./3D_Results/}{./Recurrence/}}

\begin{document}


\title{Pseudo-spectral solver versus grid-based solver: A quantitative accuracy test using \textit{GMHD3D} and \textit{PLUTO4.4}}

\author{Shishir Biswas} 
\email{shishir.biswas@ipr.res.in}
\email{shishirbeafriend@gmail.com}
\affiliation{Institute for Plasma Research, Bhat, Gandhinagar, Gujarat 382428, India}
\affiliation{Homi Bhabha National Institute, Training School Complex, Anushaktinagar, Mumbai 400094, India}


\author {Rajaraman Ganesh}
\email{ganesh@ipr.res.in}
\affiliation{Institute for Plasma Research, Bhat, Gandhinagar, Gujarat 382428, India}
\affiliation{Homi Bhabha National Institute, Training School Complex, Anushaktinagar, Mumbai 400094, India}

%

%
\date{\today}
\begin{abstract}
We provide a thorough comparison of the GMHD3D code and the PLUTO4.4 code for both two- and three-dimensional hydrodynamic and magnetohydrodynamic problems. The open-source finite-volume solver PLUTO4.4 and the in-house developed pseudo-spectral \textcolor{black}{multi-GPU} solver GMHD3D both can be used to model the dynamics and turbulent motions of astrophysical plasmas. Although GMHD3D and PLUTO4.4 utilize different implementations, it is found that simulation results for hydrodynamic and magnetohydrodynamic problems, such as the rate of instability growth, 3-dimensional turbulent dynamics, oscillation of kinetic \& magnetic energy, and recurrence dynamics, are remarkably similar. However, it is shown that the pseudo spectral solver GMHD3D is significantly more \textcolor{black}{superior} than the grid based solver PLUTO4.4 for certain category of physics problems.
\end{abstract} 

\maketitle

\section{Introduction}
Plasma is a collection of charged particles that is frequently represented as a fluid on which electromagnetic body forces operate. It has been observed that a spatially averaged model termed the ``fluid model'' is particularly efficient in predicting the behavior of the plasma when a large number of charged particles is present. Therefore, Maxwell's equations coupled with the equations of hydrodynamics become the primary governing equations for the motion of the charged-fluid element in the presence of an self-generated electric and magnetic field. The subject that studies the self-consistent evolution of such a magnetized plasma fluid is known as MagnetoHydroDynamics (MHD). In order to better understand the behavior of astrophysical plasmas like those found in the Sun or other young stars, the theories of HydroDynamics (HD) \cite{Kraichnan:1965, Brachet:1988, Shishir_POF:2022} and MagnetoHydroDynamics (MHD) \cite{biskamp:2003, beresnyak_mhd:2019, Schekochihin:2022} are frequently employed to investigate HD turbulence and magnetized plasma turbulence, respectively.  The theory of MHD is useful for a wide variety of purposes, including the analysis of three-dimensional magnetized plasma turbulence, which is essential for understanding the fundamental behavior of astro-plasmas present in the Sun and other young stars \cite{choudhuri_book:1998}, the physics of magnetic reconnection, and the careful operation of complex fusion reactors like Stellarators and Tokamaks \cite{Tanna:2019}.


Plasmas are inherently turbulent, whether they are found in burning stars or fusion devices. Because of nonlinear interactions across different scales of length, energy can cascade through different modes in a fully formed turbulent plasma medium. It has long been a problem in fluid dynamics to characterize the nature of the cascade of kinetic energy for a given initial spectrum. Almost all plasma physics problems - from relativistic jets \cite{jets_dipanjan:2020} to angular momentum transport in accretion disks \cite{MRI_Balbus:1998} - require an in-depth understanding of the turbulence in the plasma on a variety of length scales. Understanding plasma turbulence is essential for controlling the disruption \cite{ghosh_disruption:2021} of plasma in such experimental devices, thereby enhancing plasma confinement for fusion plasmas and enabling the prediction of extreme events in astrophysical objects and stellar matter. Additionally, the magnetic field-lines combined with such a plasma flow provide mechanism for the transfer of energy and dynamics within the plasma. Due to the strong coupling, the dynamics of a completely developed turbulent plasma present in stellar objects need to be treated appropriately. Understanding and accurately analyzing the defining characteristics of this plasma turbulence is essential for forecasting events in our nearest star, our ``Sun'', or in the fusion reactors operating in the many laboratories.


An important challenge in astrophysical plasmas is the creation of multi-scale magnetic fields, which occurs in the Sun, newborn stars, accretion disks, and other astronomical entities. The ``Dynamo Theory'' of Parker \cite{parker:1955} is one of the first explanations for the generation of such magnetic multi-scale fields. Such large or intermediate-scale magnetic field is generated at the expense of the plasma kinetic energy, which primarily governs the dynamics of the charged fluid (plasma) via a time-dependent Lorentz force (back-reaction) term added to the Navier-Stokes equation, thereby self-consistently influencing the dynamics of the fluid flow. Consequently, the turbulence in an MHD plasma is fundamentally different from hydrodynamic turbulence. Search for a rapid growth of magnetic fields in astrophysical objects remains one of the most fascinating areas of study \cite{Schekochihin_dynamo:2004,Schekochihin_dynamo:2007,squire_dynamo:2015,kunz_dynamo:2018,kunz_squire_schekochihin:2020,amitava_dynamo:2021,Biswas:2023,Biswas_Shear_dynamo:2023}.


In general, one needs to solve the set of coupled partial differential equations in order to deal with the complex astrophysical MHD phenomena outlined above. It is extremely challenging to solve the set of coupled nonlinear MHD equations analytically. For this reason, high-performance numerical solvers are required to accurately represent the physics problems occurring in plasmas. In order to simulate the MHD systems on a wide scale, including astrophysical entities and laboratory scenarios, it is necessary to develop highly scalable codes.


For the purpose of simulating plasma flows in astrophysics and in the lab, a variety of numerical MHD solvers have been developed. Popular examples of such codes include ZEUS \cite{ZEUS:1992}, BATS-R-US \cite{BATSRUS:1999}, FLASH \cite{FLASH:2000,FLASH:2008}, PLUTO \cite{PLUTO_Mignone:2007}, ATHENA \cite{ATHENA:2008}, NIRVANA \cite{NIRVANA:2008}, M3DC1 \cite{M3DC1:2008}, GHOST \cite{GHOST, MININNI:2011}, BIFORST \cite{BIFROST:2011}, PEGASUS \cite{PEGASUS:2014}, ENZO \cite{ENZO:2014}, CAFEQ \cite{CAFE:2015}, GKEYLL \cite{GKEYLL:2017,GKEYLL:2020}, HMHD \cite{HMHD_PPPL}, CANS+ \cite{CANS+:2019}, DEDALUS \cite{DEDALUS:2020}, PENCIL \cite{PENCIL:2021}, CLT \cite{CLT:2021} to mention some.


\textcolor{black}{At the Institute for Plasma Research [IPR], INDIA, we have recently upgraded an already existing three-dimensional compressible single GPU MHD solver \cite{rupak_thesis:2019,Rupak_IEEE:2018} to multi-node, multi-card GPU architecture [GMHD3D] \cite{GTC} using OpenAcc \& MPI and achieved considerable speed increases across 32 P100 GPU cards  \cite{ANTYA}.} The continuity, momentum, and energy equations for fluid and magnetic variables, with a thermodynamic closure for pressure, are solved using the GMHD3D solver using a pseudo-spectral technique. The solver currently employs OpenMPI/4.0.1 for its multi-node communication and the AccFFT library \cite{Accfftw:2016} for FFT operations. PyEVTK \cite{VTK}, a data converter (ASCII to BINARY) written in Python, is designed to dump data in VTK binary format for the sake of visualization. After dumping the data file to binary an open source visualization softwares like, VisIt 3.1.2 \cite{visit} and Paraview \cite{paraview} are used for visualization.


In this study, we provide a comprehensive comparison between the aforementioned in-house pseudo-spectral MHD solver (GMHD3D) \cite{GTC} and an open source grid based MHD solver PLUTO4.4 \cite{PLUTO_Mignone:2007} for some specific physics problems. \textcolor{black}{A number of earlier works \cite{KOOIJ:2018, Capuano:2023} have reported comparative analyses between several numerical codes.} The primary goal of \textcolor{black}{current} investigation is to validate the precision of the recently developed GPU solver and to compare the superiority of a pseudo-spectral solver to that of a grid-based solver atleast for certain class of physics problems. 

The organization of the paper is as follows. In Sec. II we present about the dynamic equations. About our numerical solver and simulation details of the solver are described in Sec. III.  Section IV is dedicated to the simulation results that we obtained from both the codes. Finally the summary and conclusions are listed in Sec. V.


\section{Governing Equations}\label{Equations}
The governing equations for the single fluid MHD plasma are as follows,
\begin{eqnarray}
&& \label{density} \frac{\partial \rho}{\partial t} + \vec{\nabla} \cdot \left(\rho \vec{u}\right) = 0\\
&& \frac{\partial (\rho \vec{u})}{\partial t} + \vec{\nabla} \cdot \left[ \rho \vec{u} \otimes \vec{u} + \left(P + \frac{B^2}{2}\right){\bf{I}} - \vec{B}\otimes\vec{B} \right]\nonumber \\
&& \label{velocity} ~~~~~~~~~ = \frac{1}{R_e} \nabla^2 \vec{u} + f\\
&& \label{EOS} P = C_s^2 \rho \\
&& \label{Bfield} \frac{\partial \vec{B}}{\partial t} + \vec{\nabla} \cdot \left( \vec{u} \otimes \vec{B} - \vec{B} \otimes \vec{u}\right) = \frac{1}{R_m} \nabla^2 \vec{B}
\end{eqnarray}
for the above said system of equations, $\rho$, $\vec{u}$, $P$ and $B$ represent the density, velocity, kinetic pressure and magnetic fields respectively.  $f$ is the external driver available in the system. All quantities are appropriately normalised as discussed below. \textcolor{black}{GMHD3D suite also provides a choice between utilizing the energy equation or the equation of state. For all cases studied here, we have used equation of state (see Eq. \ref{EOS}).}

We define Alfven speed as, $V_A=\frac{u_0}{M_A}$, here $M_A$ is the Alfven Mach number of the plasma flow and $u_0$ is a typical velocity scale. Sound speed of the fluid is defined as $C_s = \frac{u_0}{M_s}$, where $M_s$ is the sonic Mach number of the fluid flow \textcolor{black}{and the dynamic sound speed $C_s$ contains the inherent information regarding the temperature of the system}. The initial magnetic field present in the plasma is calculated from relation $B_0 = V_A\sqrt{\rho_0}$, $\rho_0$ is the initial density of the flow. The time is normalized to Alfven times as $t = t_0*t'$, $t_0 = \frac{L}{V_A}$ and length to a typical characteristic length scale L.

The dimensionless numbers are defined as, $R_e = \frac{u_0L}{\mu}$, $R_m = \frac{u_0L}{\eta}$ , here $R_e$ and $R_m$ are the kinetic Reynolds number and magnetic Reynolds number, $\mu$ \& $\eta$ are the kinematic viscosity and magnetic diffusivity. Magnetic Prandtl number is also be defined as, $P_M = \frac{R_m}{R_e}$. The symbol '$\otimes$' represents the dyadic between two vector quantities.

\label{equations}

All magnetic variables, including Equation \ref{Bfield}, are disabled for addressing the physics of hydrodynamic systems.
\section{About numerical solver \& Simulation Details}
Before comparing the open-source grid-based MHD solver PLUTO4.4 with the in-house developed pseudo-spectral MHD solver GMHD3D, we provide additional details on both numerical solvers.   
\subsection{Details of \textit{GMHD3D} suite}

\begin{figure}
	\includegraphics[scale=0.50]{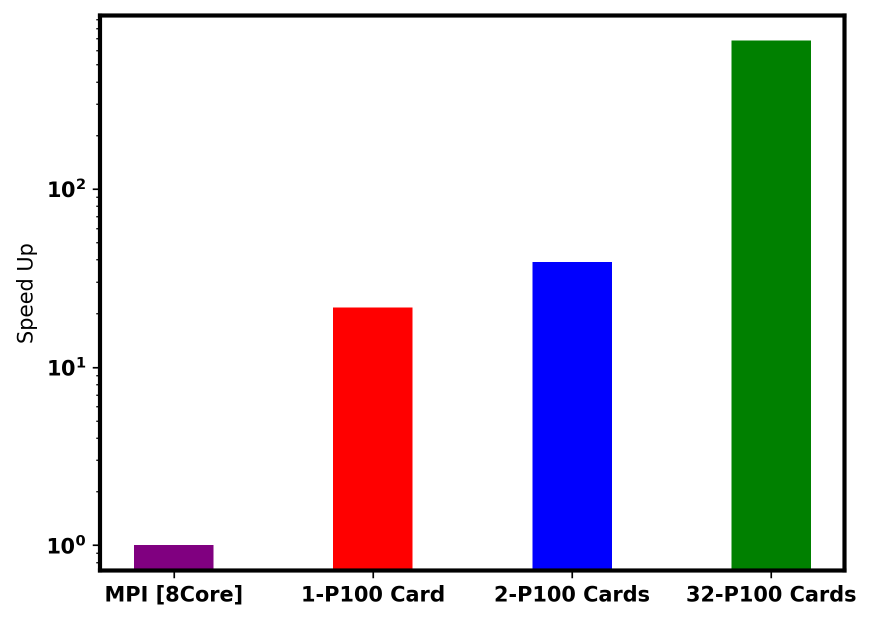}
	\caption{\textcolor{black}{Speed Up of GMHD3D code. A 675.5x speedup is obtained across 32 P100 GPU cards in comparison to the MPI version, and a 32x speedup in comparison to a single GPU version. Vertical axis set in log for plotting purpose only.}}
	\label{Bar Plot}
\end{figure}

\textcolor{black}{In order to study the plasma dynamics governed by MHD equations described above, we have recently upgraded an already existing well bench-marked single GPU MHD solver, developed in house at Institute For Plasma Research to multi-node, multi-card (multi-GPU) architecture. After multi-GPU upgrade, we obtain a 675.5x speedup across 32 P100 GPU cards in comparison to the MPI version, and a 32x speedup in comparison to the single-GPU version (See Table \ref{wall_time} \& Fig. \ref{Bar Plot}) \cite{GTC}}. This GPU based magnetohydrodynamic solver (\textit{GMHD3D}) is capable of handling very large grid sizes.

\begin{table}
	\centering
	\begin{tabular}{ |c||c| }
		\hline
		Architecture & Wall time (in Hrs)\\
		\hline
		8 Core (Intel Xeon 6148) & 60.80 \\
		1 NVIDIA P100 Card & 2.88 \\
		2 NVIDIA P100 Cards & 1.45 \\
		32 NVIDIA P100 Cards & 0.09 \\
		\hline
	\end{tabular}
	\caption{\textcolor{black}{Time taken for hundred iteration at $512^3$ grid resolution using GMHD3D.}}
	\label{wall_time}
\end{table} 

 As outlined before, \textit{GMHD3D} is  a multi-node, multi-card, three dimensional (3D), weakly compressible, pseudo-spectral, visco-resistive solver \cite{GTC}. This suite (GMHD3D) includes both 2-dimensional and 3-dimensional HydroDynamic (HD) and MagnetoHydrodynamic (MHD) solvers. It uses pseudo-spectral technique to simulate the dynamics of 3D magnetohydrodynamic plasma in a cartesian box with periodic boundary condition. By this technique one  calculates the spatial derivative to evaluate non-linear term in governing equations with a standard $\frac{2}{3}$ de-aliasing rule \cite{dealiasing:1971}. OpenACC FFT library (AccFFT library \cite{Accfftw:2016}) is used to perform Fourier transform and Adams-bashforth time solver, for time integration.

 For 3D iso-surface visualization, an open source Python based data converter to VTK (Visualization Tool kit) by ``PyEVTK'' \cite{VTK} is developed, which converts ASCII data to VTK binary format. After dumping the state data files to VTK, an open source visualization softwares, VisIt 3.1.2 \cite{visit} and Paraview \cite{paraview} is used to visualize the data. The further details of GMHD3D suite are given in Table \ref{Input GMHD3D}.

 \begin{table*}
 	\centering
 	\begin{tabular}{ |c||c| }
 		\hline
 		Equations & Compressible Navier-Stokes + Maxwell's Equations\\
 		Dimension & 2D \& 3D\\
 		Physics Modules & 2D Hydrodynamics, 3D Hydrodynamics, 2D MHD, 3D MHD\\
 		Spatial Derivative Solver  & Pseudo-Spectral\\
 		Time integration  & Adams-Bashforth, Runge-Kutta 4\\
 		Architecture & Single GPU, Multiple GPU\\
 		Parallelization &  OpenACC, OpenMPI\\
 		 Libraries  &  cuFFT, AccFFT\\
 		Precision  &  Double\\
 		Language  &  Fortran 95\\
 		Visualization  &  Gnuplot, Python, VisIt, Paraview\\
 		\hline
 	\end{tabular}
 	\caption{Features of GMHD3D suite.}
 	\label{Input GMHD3D}
 \end{table*}

As we mentioned above, we have upgraded a well bench-marked single GPU MHD solver to multi-GPU architecture [GMHD3D] \cite{GTC}, we only crosschecked the upgraded solver accuracy with the existing one and observe that results match upto machine precision. Few more benchmark details can be found in some earlier works \cite{GTC,Shishir_POF:2022,Biswas:2023}.

\subsection{Details of \textit{PLUTO4.4} code}

PLUTO4.4 is a multi-physics, multi-algorithm, high resolution code that can solve hyper sonic flows in one, two, and three spatial dimensions \cite{PLUTO_Mignone:2007}. In order to solve the system of non-linear equations, a finite volume/finite difference approach is employed. PLUTO4.4 is parallelized with the help of the MPI Library via global domain decomposition.


Different Reconstruction algorithms are available in PLUTO4.4; however, we employ PARABOLIC reconstruction, which employs the piece wise parabolic method (PPM) to determine the spatial order of integration. The stencil requires either 3 or 5 zones and was implemented by Migone et al. \cite{Mignone_Reconstraction:2014}. We consider the RK3 scheme for time stepping because it is compatible with PARABOLIC reconstruction \cite{PLUTO_Mignone:2007}.


Since the governing equation requires an Isothermal equation of state, we set the EOS module in PLUTO4.4 to be in the ISOTHERMAL state \cite{PLUTO_Mignone:2007}. For the isothermal equation of state, Migone et al. demonstrated that the hlld solver provides the highest precision of any Riemann solver \cite{Mignone_EOS:2007}. For this reason, we employ hlld Riemann solver for PLUTO4.4 throughout our study. Table \ref{Input PLUTO} contains a complete listing of all input modules for PLUTO4.4.

\begin{table*}
	\centering
	\begin{tabular}{ |c||c| }
		\hline
		Reconstruction & PARABOLIC\\
		Time Steeping & RK3\\
		EOS & ISOTHERMAL\\
		Div. B Control & CONSTRAINED TRANSPORT\\
		Resistivity & EXPLICIT\\
		Viscosity & EXPLICIT\\
		Limiter & MC LIM\\
		\hline
	\end{tabular}
	\caption{Initial input modules for PLUTO4.4.}
	\label{Input PLUTO}
\end{table*}

\subsection{Cost metric comparison}
\textcolor{black}{
All simulations using the PLUTO4.4 and GMHD3D codes were executed on the 1 PetaFlop ANTYA \cite{ANTYA}  supercomputer located at the Institute for Plasma Research in India. PLUTO4.4 utilizes a dual configuration of 20 CPU cores, namely the Intel Xeon 6148 model, operating at a clock speed of 2.4 GHz. The system is equipped with a total of 384 GB DDR4 RAM.  For simulations utilizing GMHD3D code we have used GPU nodes of ANTYA cluster with similar specification along with two NVIDIA tesla P100 GPU cards in a single node with 16 GB RAM each.}

\begin{figure*}
 	\centering
 	\begin{subfigure}[b]{0.32\textwidth}
 		\centering
 		\includegraphics[width=\textwidth]{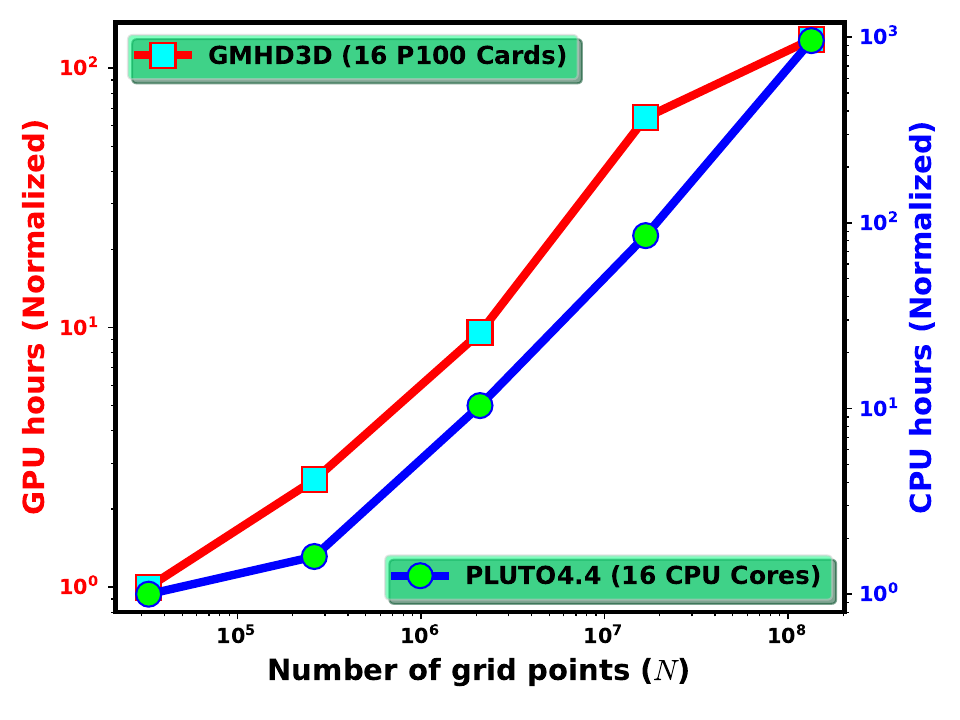}
 		\caption{}
 			\label{cost a}
 	\end{subfigure}
 	\hfill
 	\begin{subfigure}[b]{0.32\textwidth}
 		\centering
 		\includegraphics[width=\textwidth]{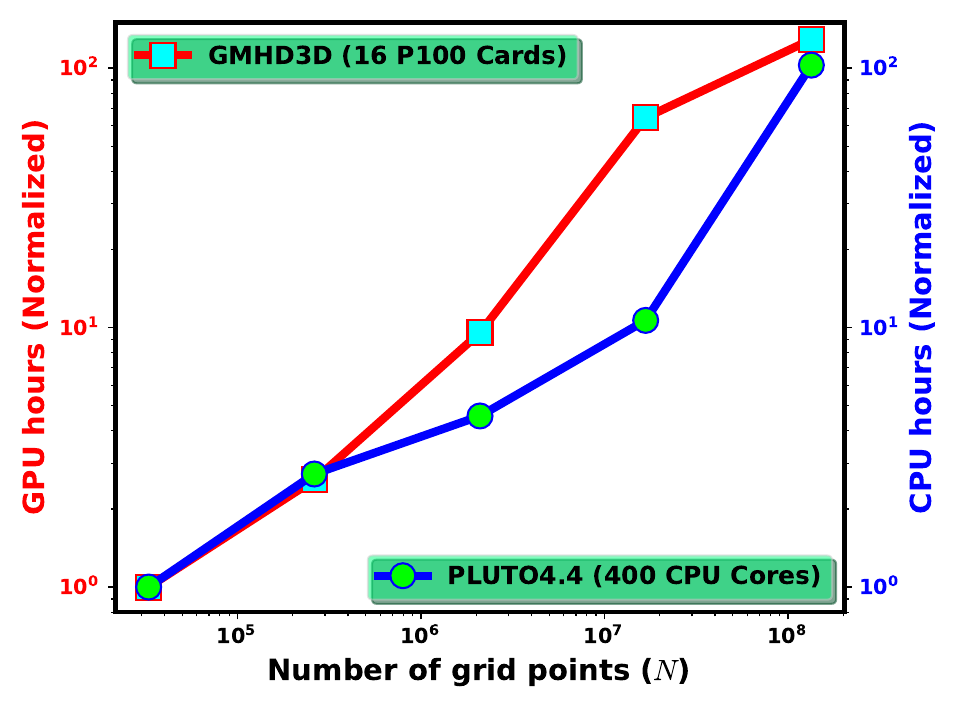}
 		\caption{}
 			\label{cost b}
 	\end{subfigure}
 	\hfill
 	\begin{subfigure}[b]{0.32\textwidth}
 		\centering
 		\includegraphics[width=\textwidth]{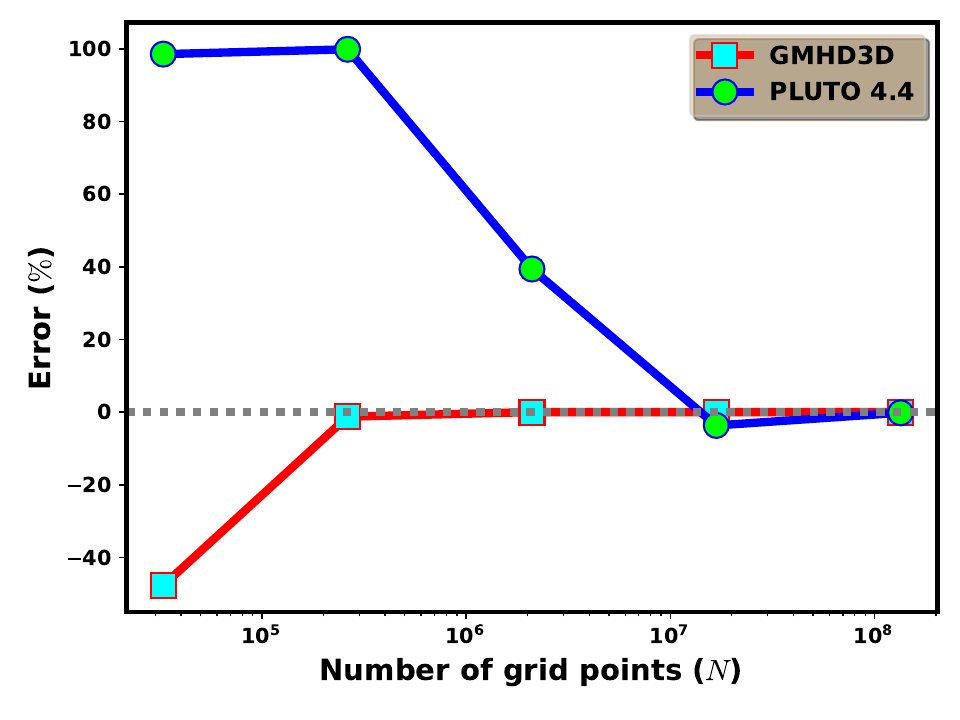}
 		\caption{}
 			\label{Accureacy}
 	\end{subfigure}
 	\caption{\textcolor{black}{Cost metric comparison in terms of GPU/CPU hours (normalized) versus number of grid points for GMHD3D and PLUTO4.4. For both algorithms, the time stepping is kept constant, and the wall time is calculated for a total of $10^4$ iterations. For this investigation, we have utilized a total of 16 GPU cards (P100) for GMHD3D and (a) 16 CPU cores \& (b) 400 CPU cores (Intel Xeon 6148) for PLUTO4.4.} \textcolor{black}{(c) Comparison of cost vs accuracy  as a function of the number of grid points ($N$) for GMHD3D and PLUTO4.4. The accuracy is measured by the value of the error percentage ($Error (\%)$) as explained in the text.}}
 	\label{Efficiency New}
 \end{figure*}

\textcolor{black}{
 To conduct a cost metric comparison between the two solvers, a series of simulation runs have been performed, varying the number of resources (CPUs and GPUs) and grid points. We have plotted the normalized computational costs for the GPUs and CPUs in relation to the grid resolutions (refer to Fig. \ref{cost a} \& \ref{cost b}). From Fig. \ref{cost a} \& \ref{cost b} it is readily understood that the computational expenses increases linearly for both CPUs and GPUs as grid points increases. It can be observed from Fig. \ref{cost a} \& \ref{cost b} that the normalized computational time of 16 GPUs is nearly similar to the computational time of 400 CPUs in cases when the computational workload is significant.}

\textcolor{black}{
It is widely acknowledged that CPUs exhibit a higher power consumption, as compared to CPUs, GPUs which have a lower power consumption because of their shared memory architecture. Therefore, the maintenance of 400 CPUs would result in higher computational expenses, including electrical power consumption, cooling, and rack space, as compared to the maintenance of 16 GPU cards. This, in turn, signifies the cost-effectiveness of the GPU solver.}

\textcolor{black}{
It is also important to note that, the GPUs performances would also depend on the architectures. The computational efficiency of P100 cards, for instance, is lower than that of V100 and A100 cards. Consequently, additional investigation is required to determine the efficacy of the GMHD3D solver on the A100 architecture. Furthermore, comparing two GPU solvers is more meaningful than comparing the computational expenses of a CPU solver and a GPU solver, and to the best of our knowledge, the GPU version of the PLUTO4.4 code is currently being developed. Hence, such comparisons are what we plan to include in future communications.}

\textcolor{black}{The calculation of the accuracy (value of error $\%$) is determined by the following formula: 
\begin{equation}
Error (\%) =  100*\frac{W_1 - W_2}{W_1}
\end{equation}  
where, $W_1$ represents the original expected value, while $W_2$ represents the value observed in the numerical simulation. In order to determine the accuracy, expressed as a percentage of error, we have considered an individual test problem that was utilized for the purpose of conducting a cost comparison analysis. The difference in peak values are calculated for various grid points using both codes. We have plotted the value of percentage of error ($Error (\%)$) for the GPU solver (GMHD3D) and CPU solver (PLUTO 4.4) in relation to the grid resolutions (see Fig. \ref{Accureacy}). It is evident from Fig. \ref{Accureacy} that the GPU solver (GMHD3D) converged rapidly to errors of the order of less than $1\%$, whereas PLUTO 4.4 requires a more precise grid (higher grid resolution) in order to converge. Fig. \ref{Accureacy} demonstrates that the GMHD3D code fulfills the accuracy criteria at the lowest cost, whereas the PLUTO 4.4 code is the most expensive. This indicates the cost-effectiveness and accuracy of the GPU solver compared to the CPU solver being discussed.}




\section{Numerical Tests} 
It has been evident from the preceding discussion that we intend to provide a detailed comparison between two alternative framework solvers, one of which is an in-house developed pseudo-spectral solver and the other being an open source grid based solver. We have considered some well-known test problems in two- and three-dimensional hydrodynamics and magnetohydrodynamics to accomplish this. \textcolor{black}{For example,
\begin{itemize}
	\item 2-dimensional Kelvin-Helmholtz instability (Details are described in \ref{2D KH}).
	\item Dynamics of 3-dimensional Taylor-Green (TG) vortex (Details are described in \ref{3D TG}).
	\item Coherent  nonlinear oscillations using 2D Orszag-Tang (OT) Flow (Details are described in \ref{2D OT}).
	\item Coherent  nonlinear oscillations using 2D Cats Eye (CE) Flow (Details are described in \ref{2D CE flow}).
	\item Coherent  nonlinear oscillations using 3-dimensional astrophysical Flows (Details are described in \ref{3D flows}).
	\item Coherent  nonlinear oscillations for driven Flows (Details are described in Appendix \ref{Appen B}).
	\item Recurrence dynamics in 3D MHD plasma (Details are described in \ref{Recurrence}).
\end{itemize}
}

 Parameter information for each individual test problem is provided in their respective subsections.
\subsection{Test 1 [Hydrodynamics]: 2-dimensional Kelvin-Helmholtz instability}\label{2D KH}

\begin{figure*}
	\centering
	\begin{subfigure}{0.49\textwidth}
		\centering
		\includegraphics[scale=0.55]{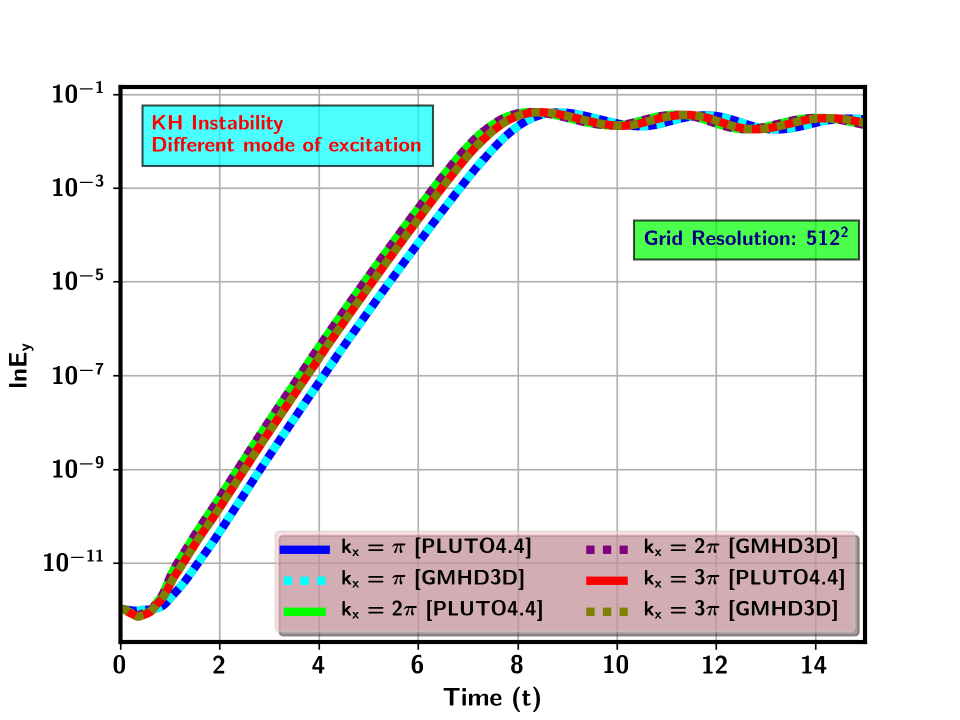}
		\caption{}
		\label{mode Energy}
	\end{subfigure}
	\begin{subfigure}{0.49\textwidth}
		\centering
		\includegraphics[scale=0.55]{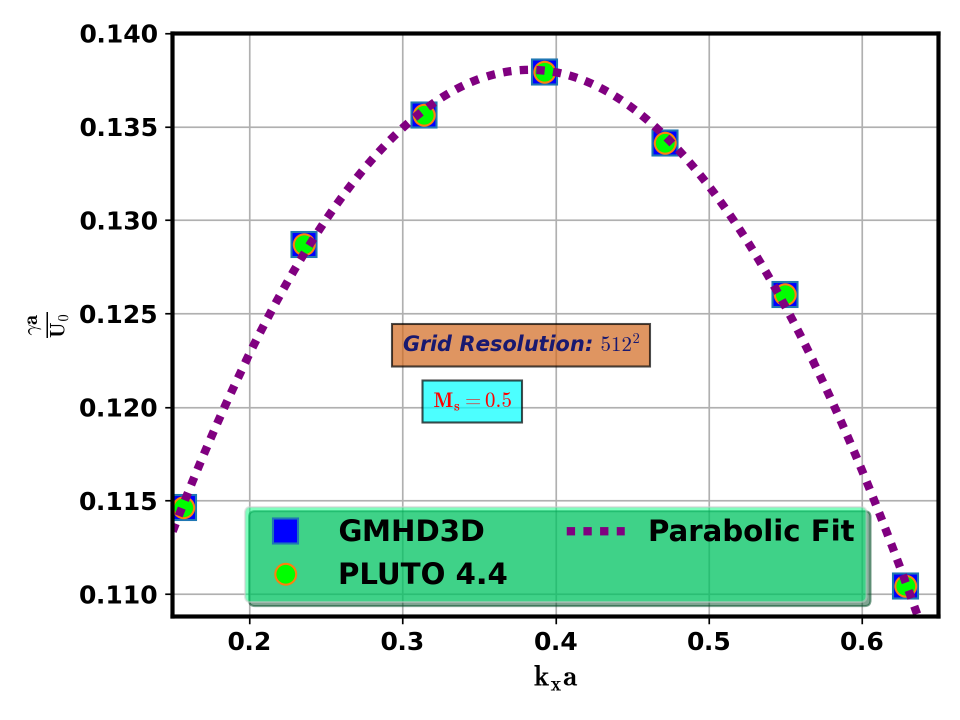}
		\caption{}
		\label{Mode vs Growthrate}
	\end{subfigure}
	\label{KH mode of perturbation}
	\caption{(a) Comparison of kinetic energy from both GMHD3D and PLUTO4.4 code at different mode number of perturbation in the direction perpendicular to the flow direction is evaluated with time. (b) The growth rate ($\frac{\gamma a}{U_0}$) of KH instability at different normalized mode of perturbation ($k_x a$) with sonic Mach number $M_s = 0.5$, where $a$ is the shear width. The calculated growth rates from both the codes are exactly identical and it is fitted perfectly by an inverted parabola which is identical to Keppens et al \cite{keppens_KH:1999}. \textcolor{black}{Simulation Details: Grid resolution $N = 512^2$, Time stepping $dt = 10^{-4}$.}}
\end{figure*}

Using the GMHD3D solver and the PLUTO4.4 solver, we have investigated the 2-dimensional Kelvin-Helmholtz (KH) instability for hydrodynamic systems. We assume a simulation box with dimensions $L_x = 1$ and $L_y = 2$, and that the initial pressure ($p_0$) and initial density ($\rho_0$) are each to be unity. We apply a shear velocity along the $x$-direction of the form, 
\begin{equation}
	u_x = U_0 \left[\tanh \left(\frac{y-\frac{L_y}{3}}{a}\right) - \tanh \left(\frac{y-\frac{2L_y}{3}}{a}\right) -1 \right]
\end{equation}
where $U_0 = 0.645$ is the shear flow strength and $a = 0.05$ is the shear width. We introduce a sinusoidal perturbation in the direction perpendicular to the initial flow velocity, of the form,
\begin{equation}
	\begin{aligned}
		u_y = u_{y0} \sin(k_x x) \exp \left[-\frac{\left(y-\frac{L_y}{3}\right)^2}{\sigma^2} \right]\\
		+ u_{y0} \sin(k_x x) \exp \left[-\frac{\left(y-\frac{2L_y}{3}\right)^2}{\sigma^2} \right]
	\end{aligned}
\end{equation}
with $\sigma = 4a$, to this initial configuration. Here, $u_{y0} = 10^{-4}$ represents the amplitude of the velocity perturbation, and $k_x$ represents the mode of the velocity perturbation. For our system, the Sonic Mach number is defined as $M_s = \frac{U_0}{C_s}$, where $C_s$ is the sound speed. We have investigated KH instability in the compressible limit using these parameter spaces. While maintaining the Sonic Mach number constant (i.e. $M_s = 0.5$), we run simulations of the KH instability with both the solver (GMHD3D \& PLUTO4.4) at different modes of perturbation ($k_x$) [See Fig. \ref{mode Energy}]. Also, by tracking the $y$-direction kinetic energy ($E_y = \int \frac{1}{2}\rho v_y^2 dx dy$), we are able to calculate the growth rate ($\gamma$) of the KH instability in both codes [See Fig. \ref{mode Energy}]. The relationship between the KH growth rate ($\gamma$) and the perturbation wave number ($k_x$) is also calculated. As can be seen in Fig. \ref{Mode vs Growthrate}, a perfect inverted parabola fits the predicted growth rates for varied mode numbers from both the GMHD3D and PLUTO4.4 codes. From Fig. \ref{Mode vs Growthrate}, we can further conclude that the KH instability is stabilized for both small and large wave numbers, with the maximum growth rate occurring at $k_x a = 0.4$. This finding agrees well with that of Keppens et al \cite{keppens_KH:1999}.

\begin{figure*}
	\centering
	\begin{subfigure}{0.49\textwidth}
		\centering
		\includegraphics[scale=0.55]{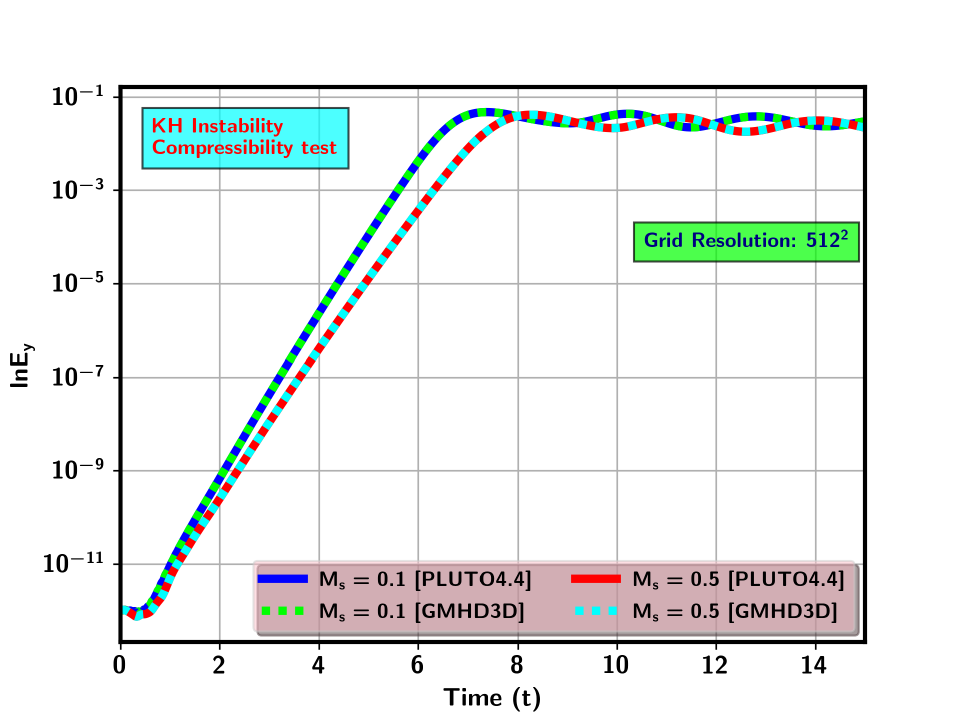}
		\caption{}
		\label{mode Energy Different MS}
	\end{subfigure}
	\begin{subfigure}{0.49\textwidth}
		\centering
		\includegraphics[scale=0.55]{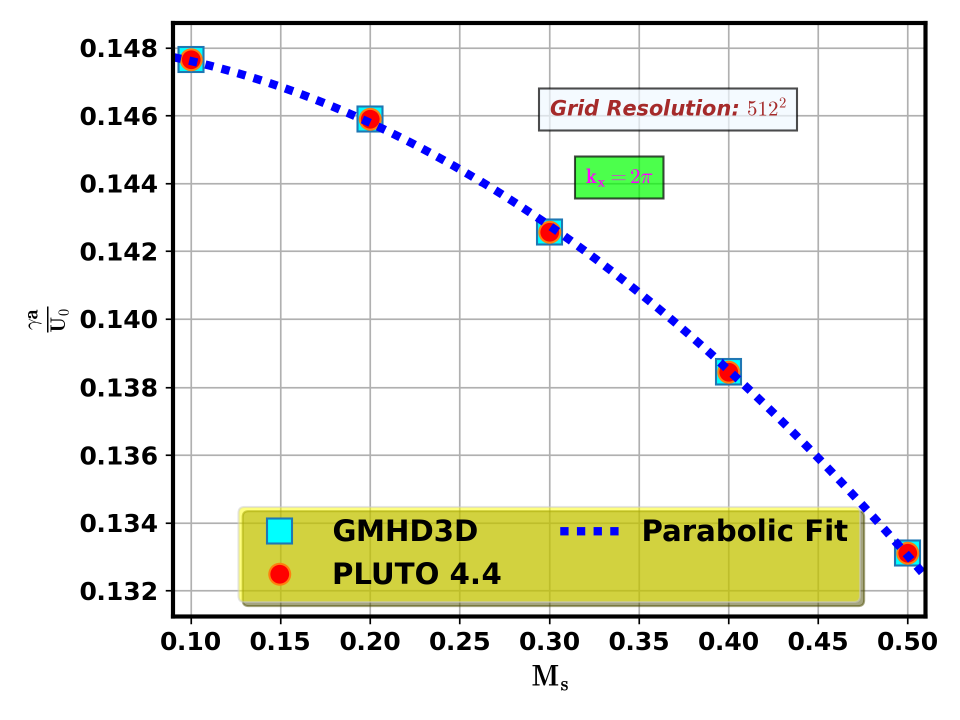}
		\caption{}
		\label{Mode vs Growthrate different Ms}
	\end{subfigure}
	\caption{(a) Comparison of kinetic energy from both GMHD3D and PLUTO4.4 code at different sonic Mach number ($M_s$) in the direction perpendicular to the flow direction is evaluated with time. (b) The growth rate ($\frac{\gamma a}{U_0}$) of KH instability at different sonic Mach number ($M_s$) with $k_x = 2\pi$. The calculated growth rates from both the codes are exactly identical and it is fitted perfectly by an inverted parabola which is identical to Keppens et al \cite{keppens_KH:1999}. \textcolor{black}{Simulation Details: Grid resolution $N = 512^2$, Time stepping $dt = 10^{-4}$.}}
\end{figure*}

Next, by holding the mode number of excitation at $k_x = 2\pi$, we investigate the impact of compressibility on KH instability [See Fig. \ref{mode Energy Different MS}]. Similar growth rates are obtained from both codes for various Sonic Mach numbers [See Fig. \ref{Mode vs Growthrate different Ms}]. Also, the KH growth rates as a function of $M_s$ exhibit an inverted parabolic nature [See Fig. \ref{Mode vs Growthrate different Ms}], as determined by Keppens et al \cite{keppens_KH:1999}. 

\begin{figure*}
	\centering
	\begin{turn}{90} 
		\normalsize{\textbf{\textcolor{blue}{GMHD3D}}}
	\end{turn}
	\begin{subfigure}{0.24\textwidth}
		\centering
		\includegraphics[scale=0.07550]{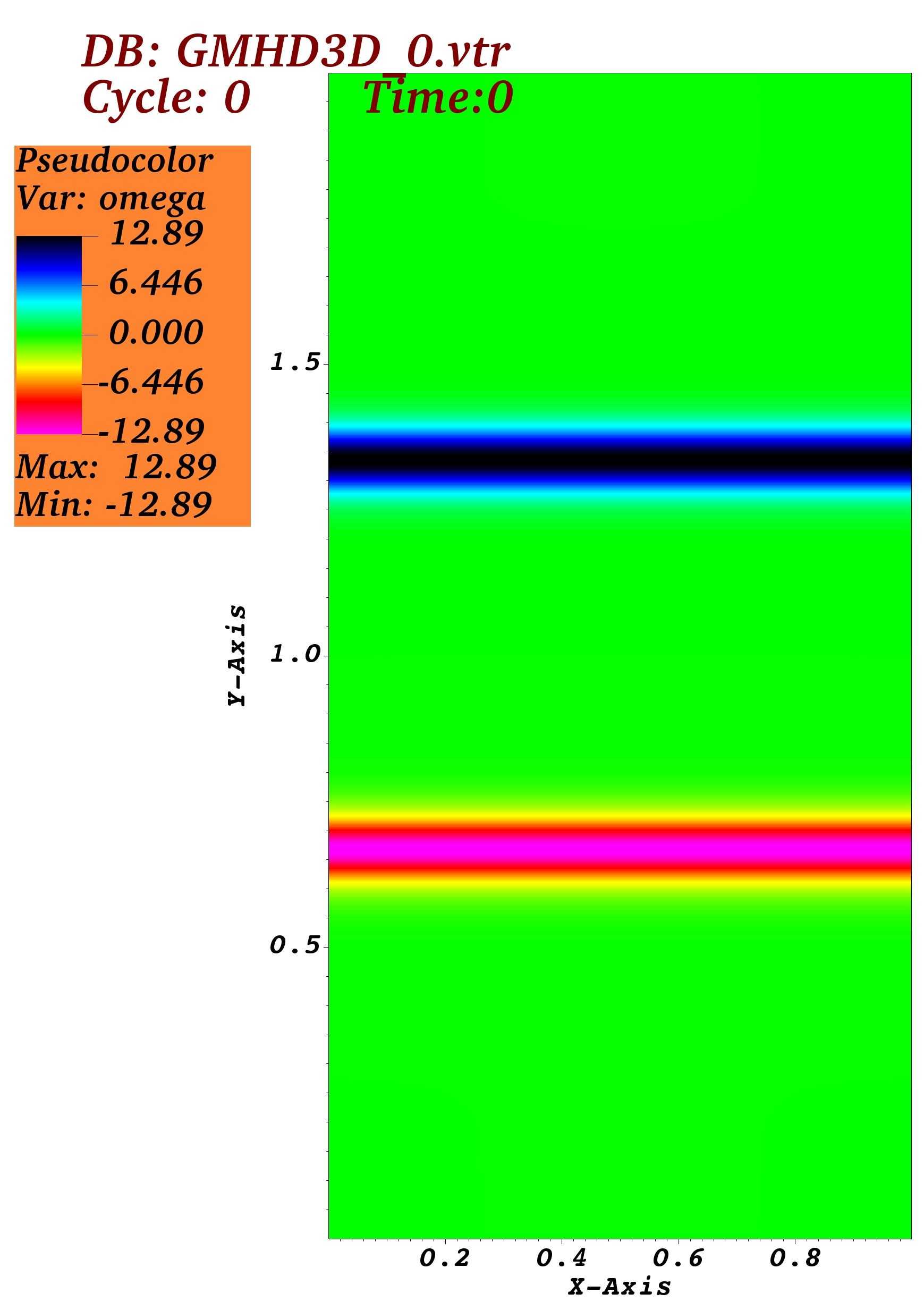}
		\caption{}
	\end{subfigure}
	\begin{subfigure}{0.24\textwidth}
		\centering
		\includegraphics[scale=0.07550]{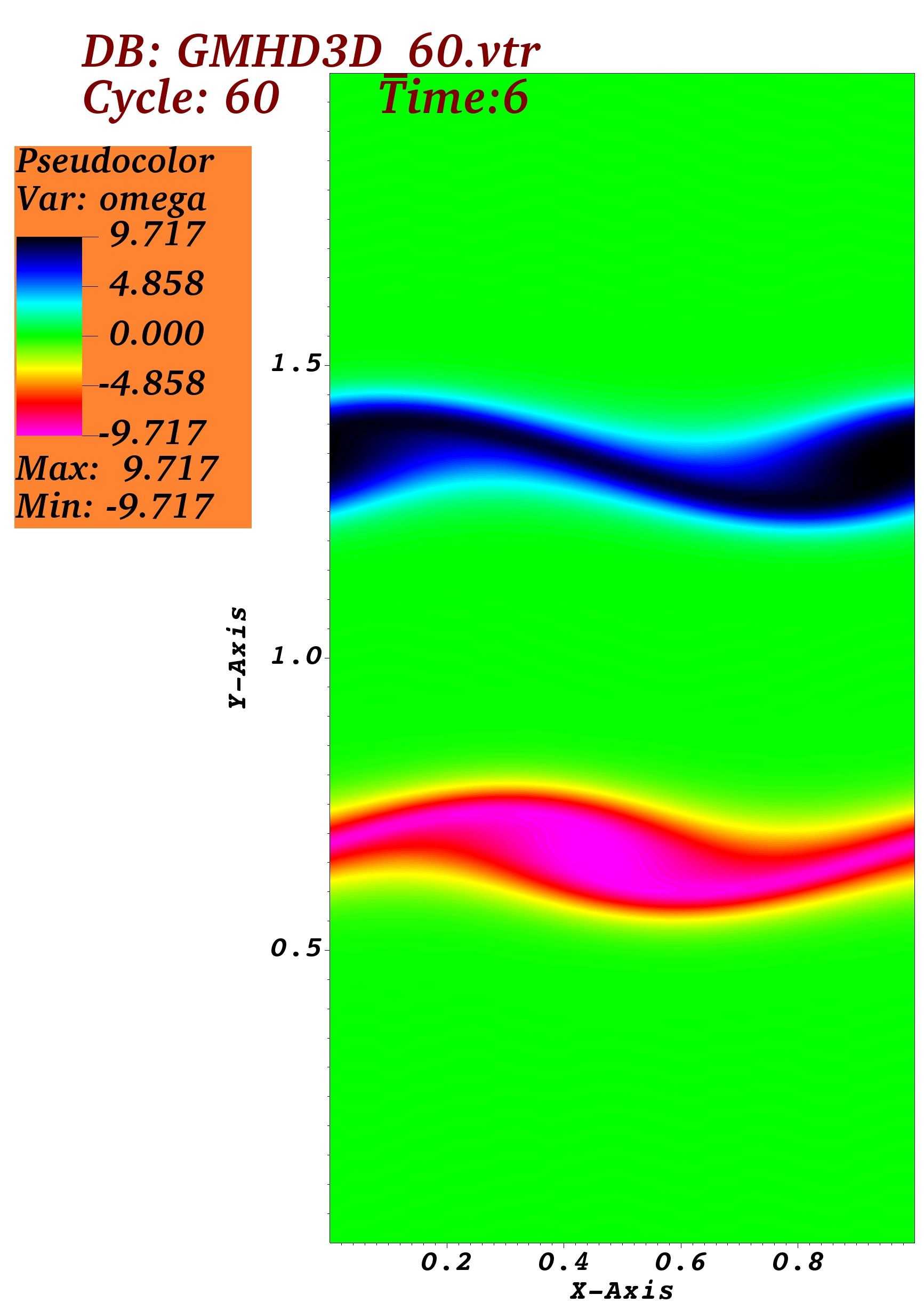}
		\caption{}
	\end{subfigure}
	\begin{subfigure}{0.24\textwidth}
		\centering
		\includegraphics[scale=0.07550]{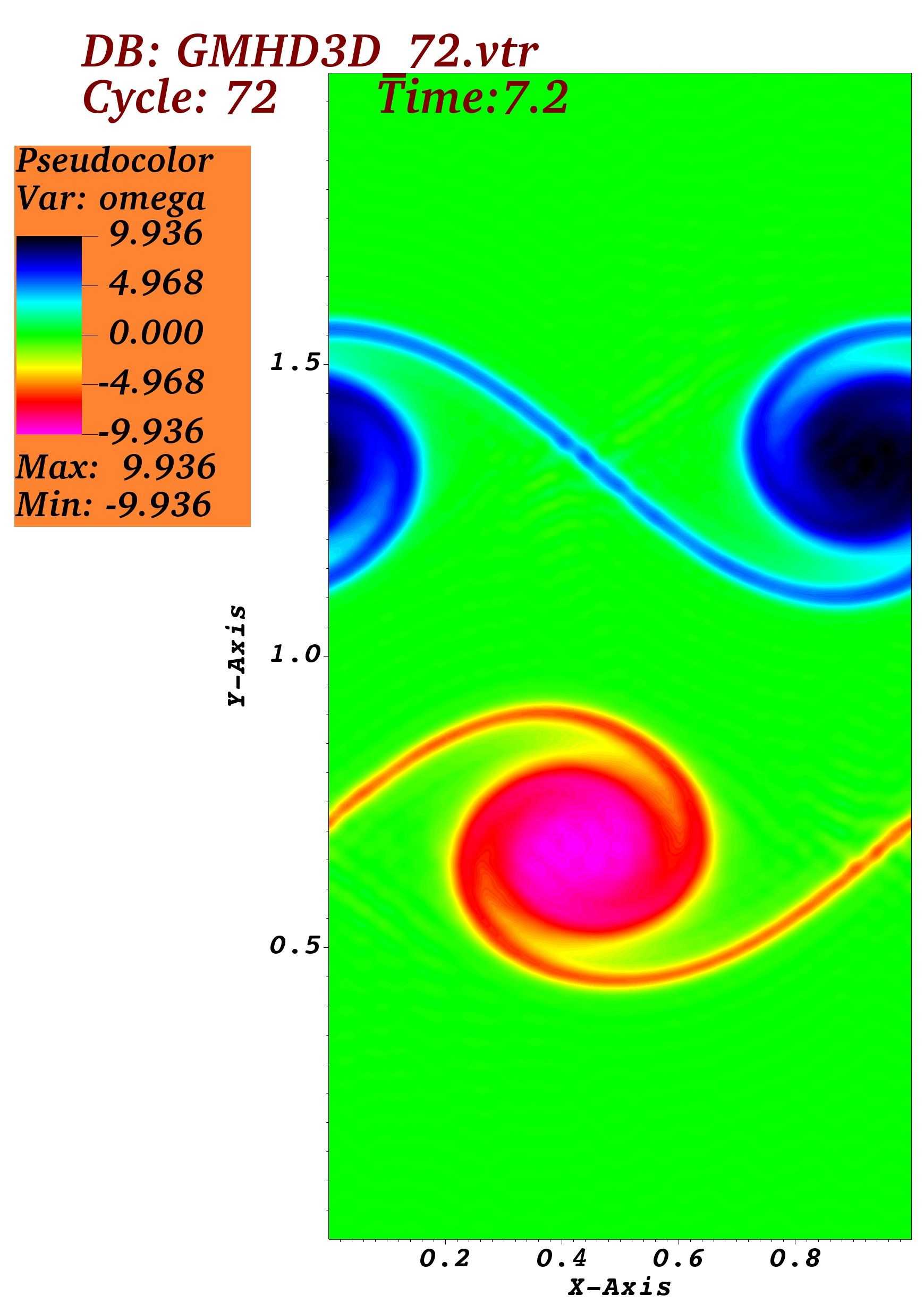}
		\caption{}
	\end{subfigure}
	\begin{subfigure}{0.24\textwidth}
		\centering
		\includegraphics[scale=0.07550]{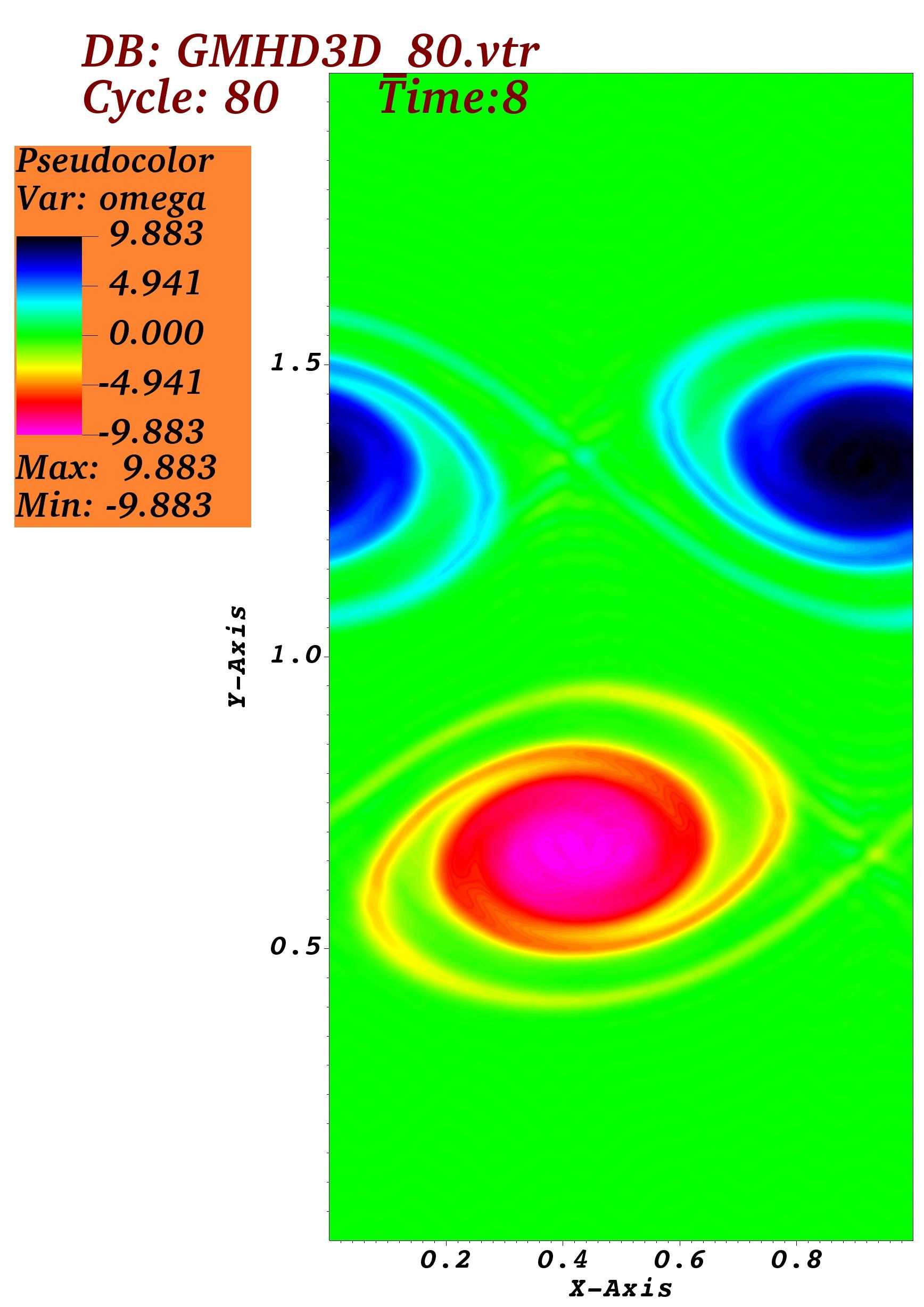}
		\caption{}
	\end{subfigure}
	\begin{turn}{90} 
		\normalsize{\textbf{\textcolor{blue}{PLUTO4.4}}}
	\end{turn}
	\begin{subfigure}{0.24\textwidth}
		\centering
		\includegraphics[scale=0.07550]{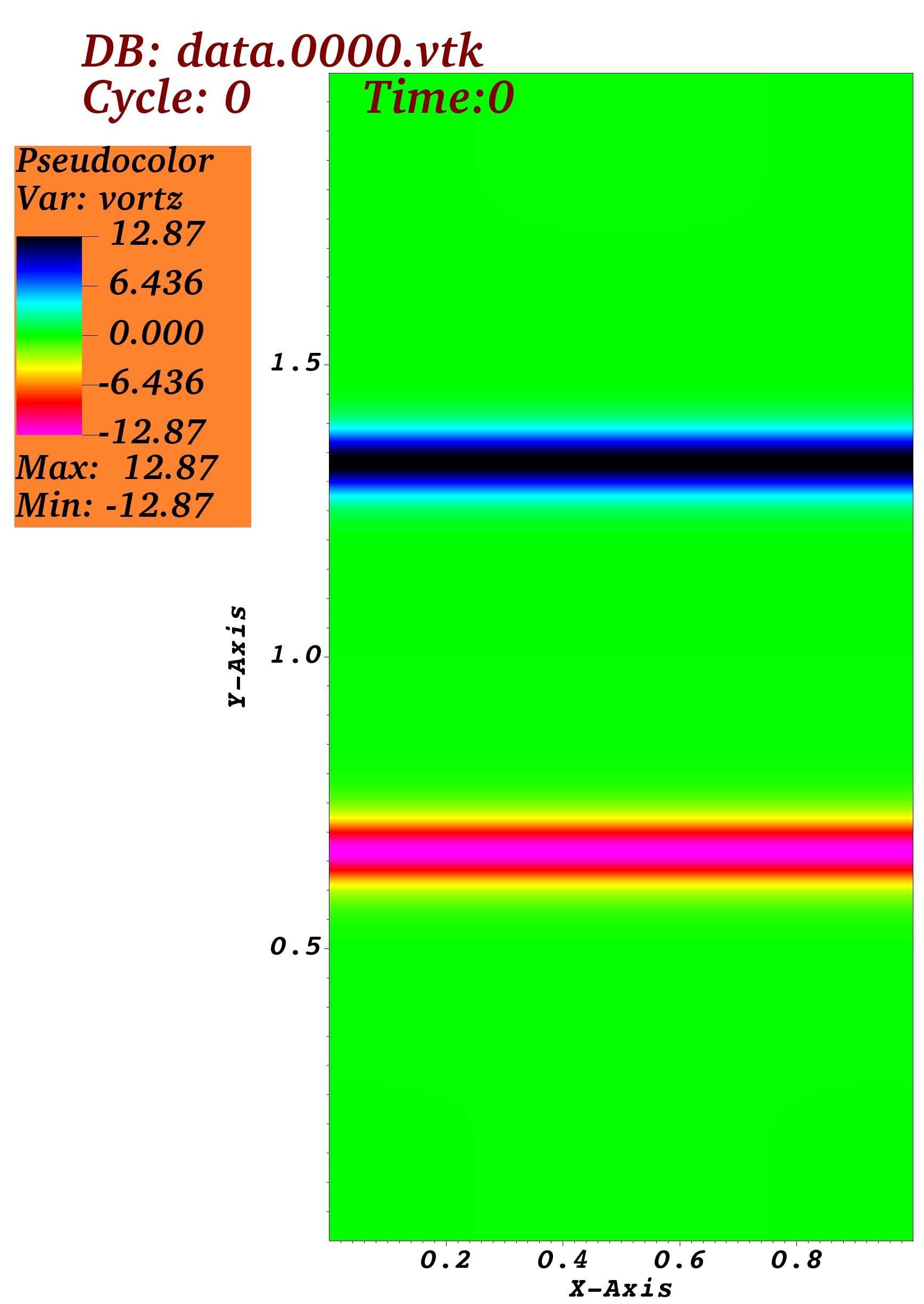}
		\caption{}
	\end{subfigure}
	\begin{subfigure}{0.24\textwidth}
		\centering
		\includegraphics[scale=0.07550]{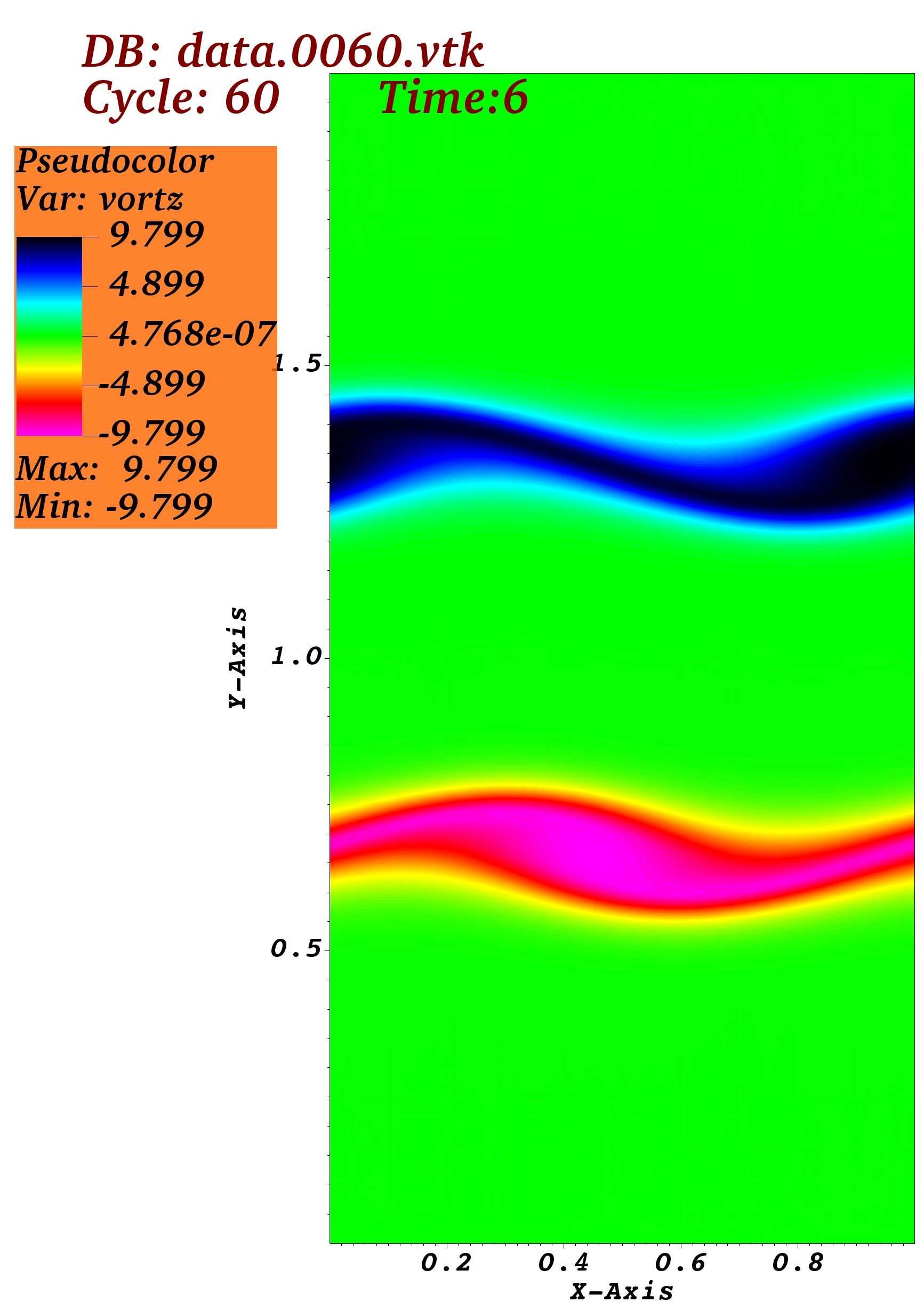}
		\caption{}
	\end{subfigure}
	\begin{subfigure}{0.24\textwidth}
		\centering
		\includegraphics[scale=0.07550]{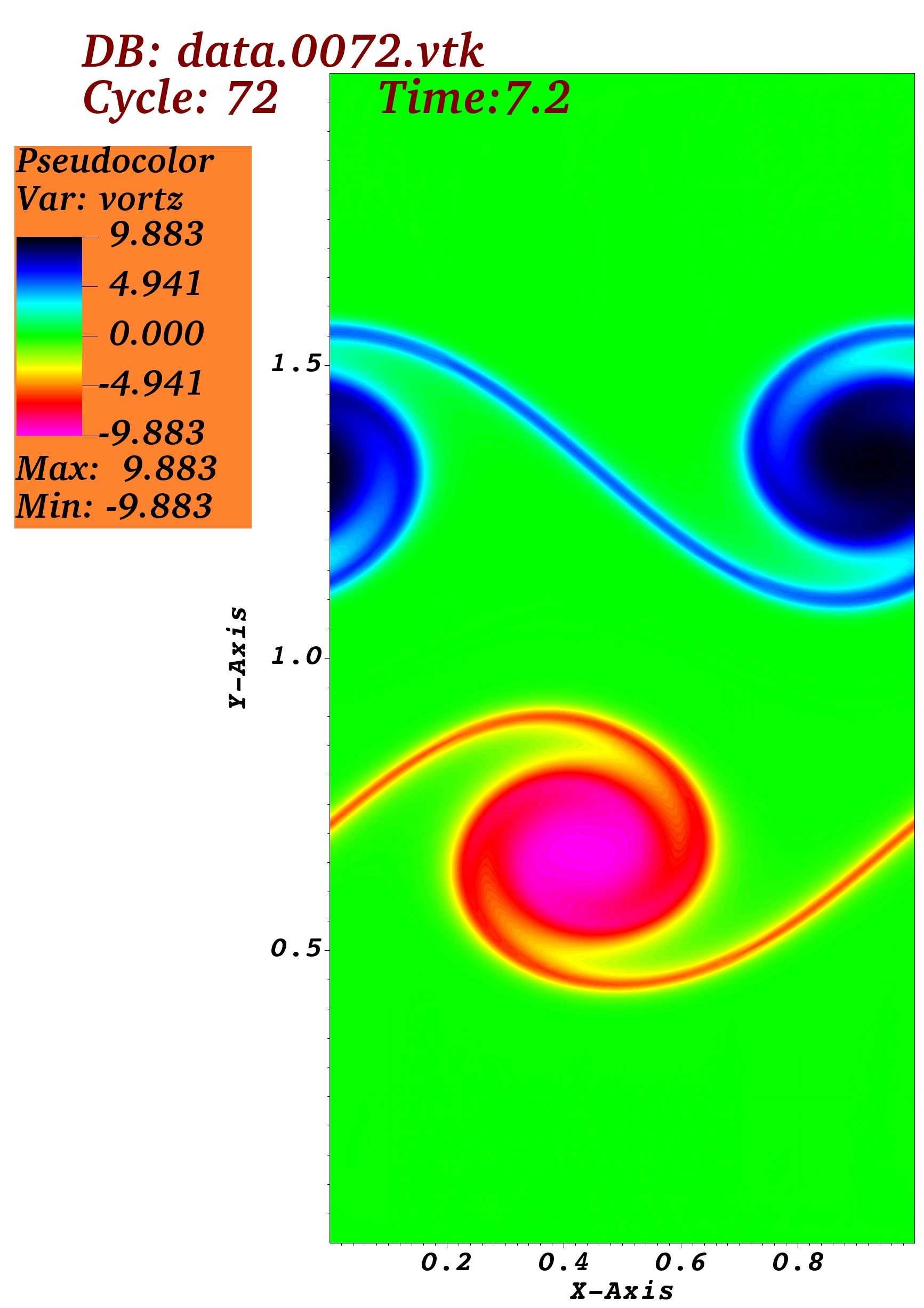}
		\caption{}
	\end{subfigure}
	\begin{subfigure}{0.24\textwidth}
		\centering
		\includegraphics[scale=0.07550]{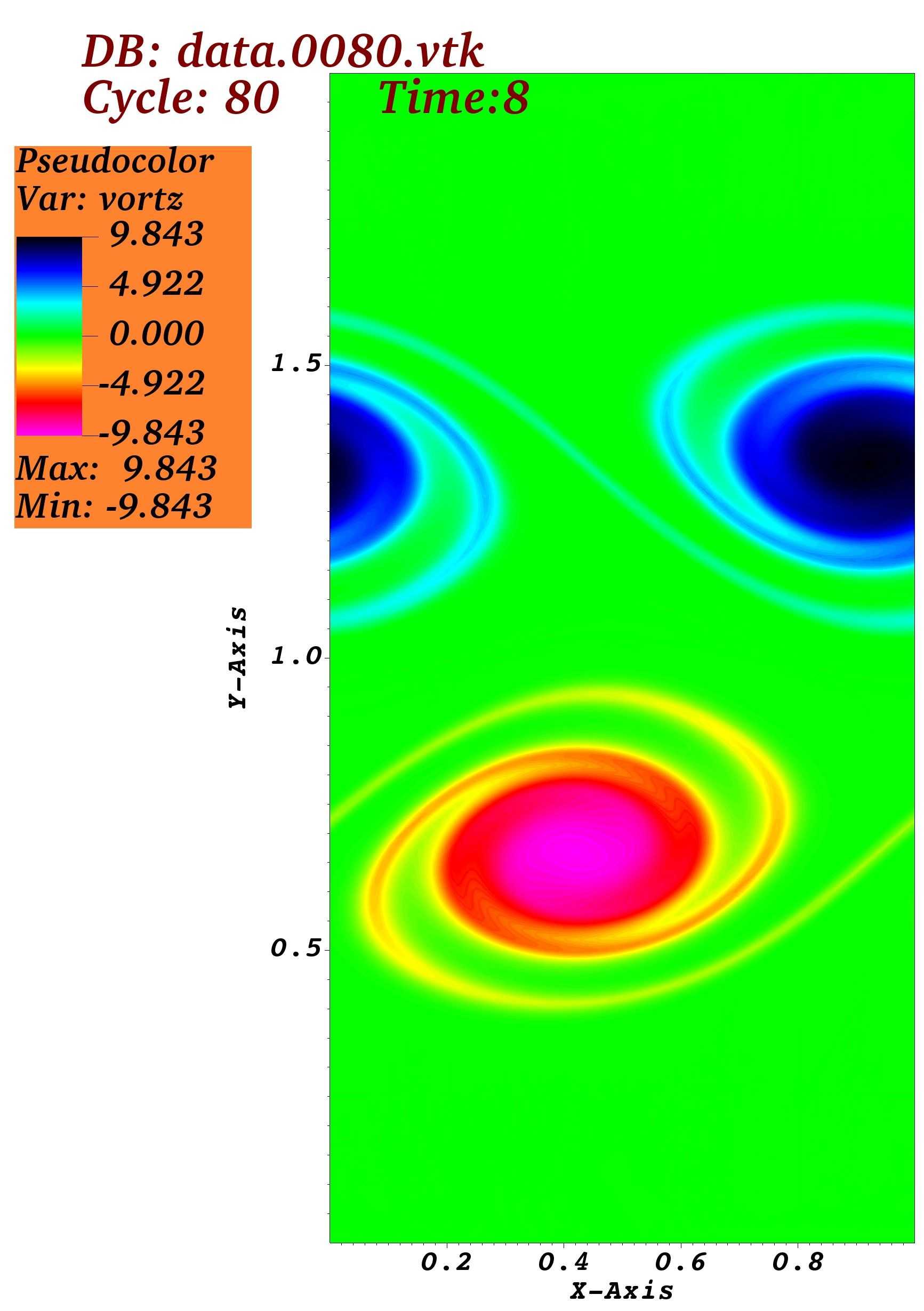}
		\caption{}
	\end{subfigure}
	\caption{Time evolution of vorticity for two oppositely directed KH unstable jets (broken jet) from GMHD3D code [upper panel (a--d)] and PLUTO4.4 code [lower panel (e--h)]. Both the solver has captured identical dynamics. \textcolor{black}{Simulation Details: Grid resolution $N = 512^2$, Time stepping $dt = 10^{-4}$.}}
		\label{KH GMHD3D PLUTO}
\end{figure*}

We have also shown the evolution of the vorticity profile of KH instability to conduct a more in-depth comparison between the GMHD3D and PLUTO4.4 codes. It is evident from Fig. \ref{KH GMHD3D PLUTO} that the outcomes of both solvers are identical.

\subsection{Test 2 [Hydrodynamics]: Dynamics of 3-dimensional Taylor-Green (TG) vortex} \label{3D TG}

\begin{figure*}
	\centering
	\begin{turn}{90} 
		\normalsize{\textbf{\textcolor{blue}{GMHD3D}}}
	\end{turn}
	\begin{subfigure}{0.48\textwidth}
		\centering
		\includegraphics[scale=0.0750]{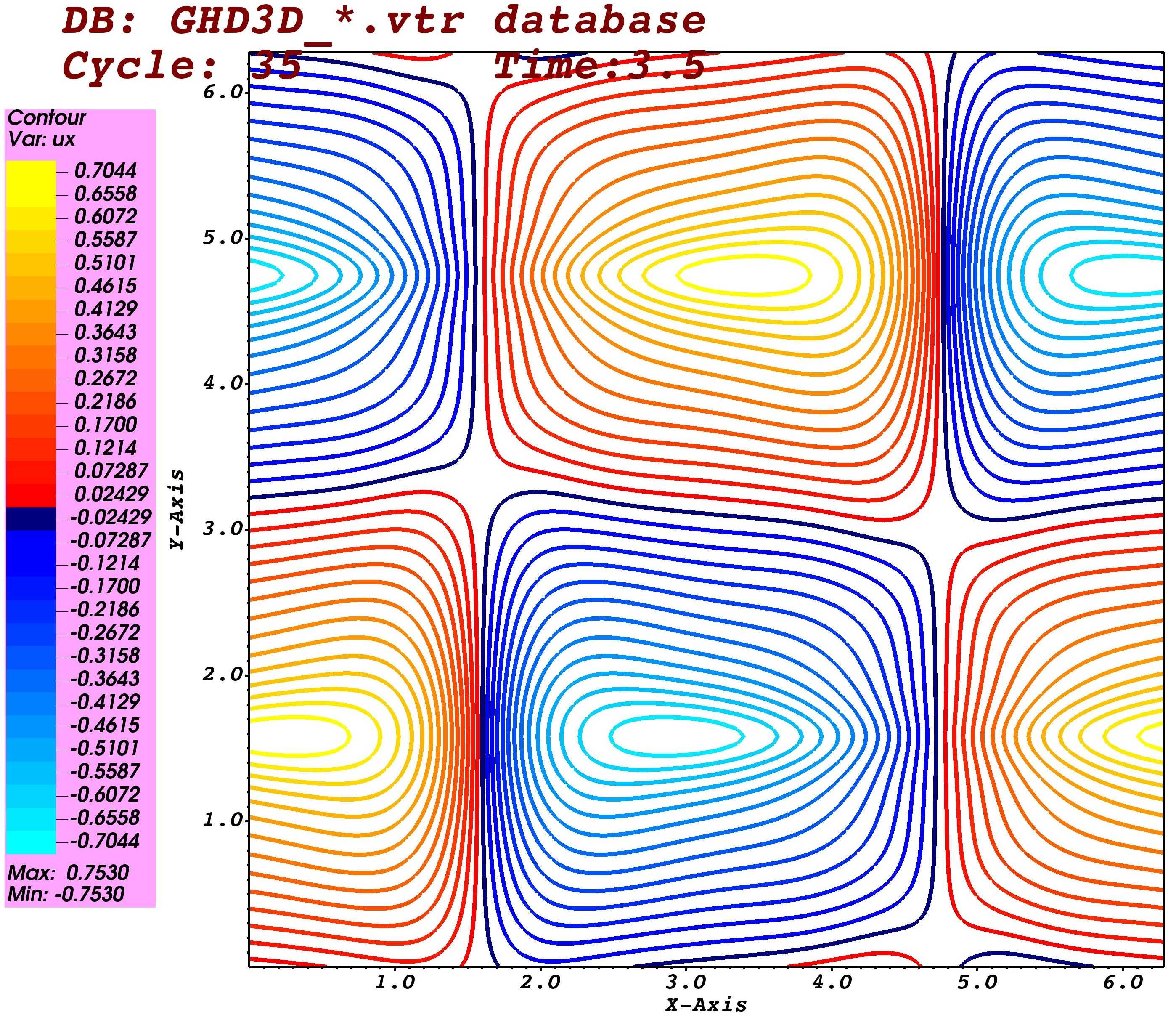}
		\caption{}
	\end{subfigure}
	\begin{subfigure}{0.48\textwidth}
		\centering
		\includegraphics[scale=0.0750]{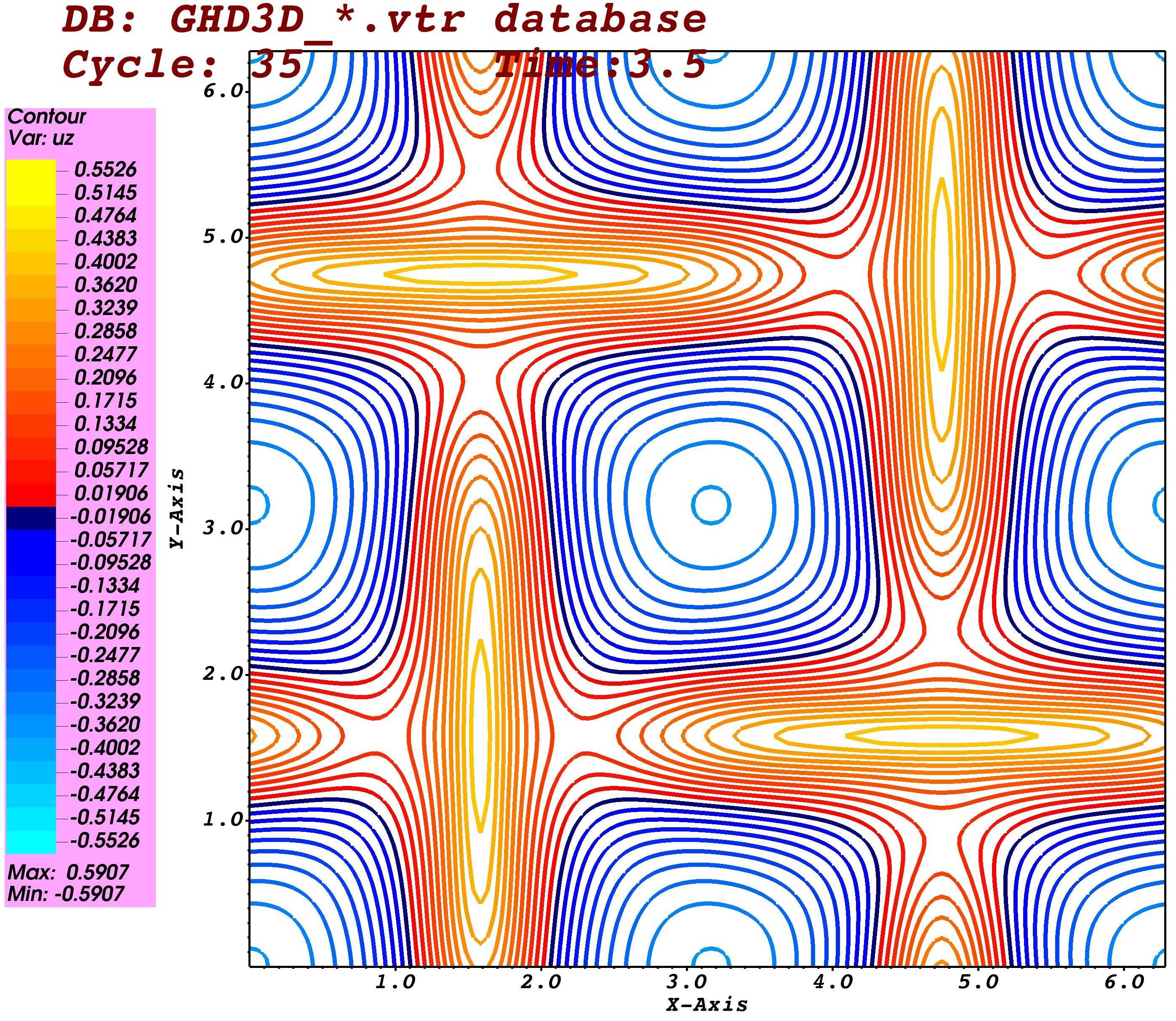}
		\caption{}
	\end{subfigure}
	\begin{turn}{90} 
		\normalsize{\textbf{\textcolor{blue}{PLUTO4.4}}}
	\end{turn}
	\begin{subfigure}{0.48\textwidth}
		\centering
		\includegraphics[scale=0.0750]{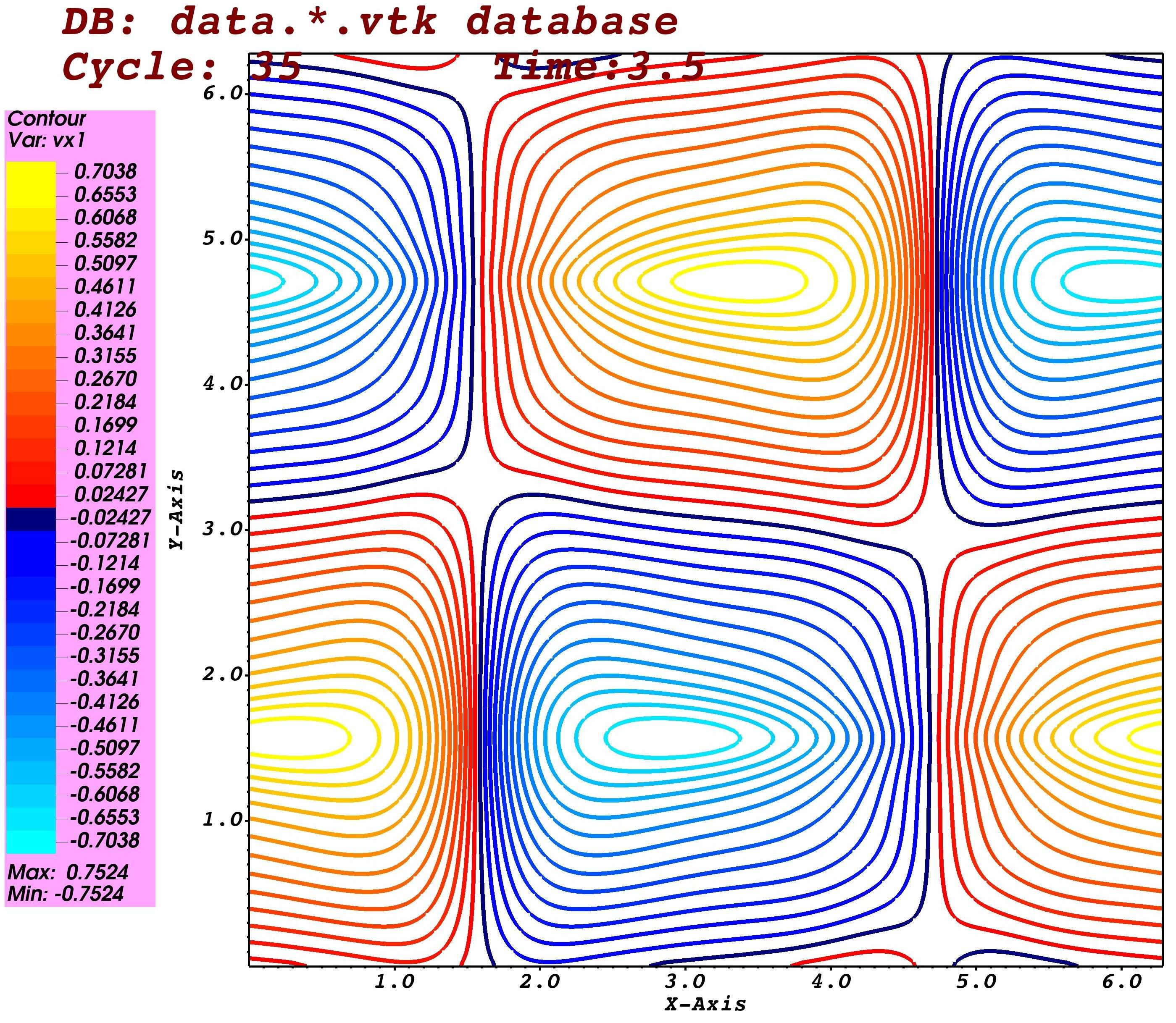}
		\caption{}
	\end{subfigure}
	\begin{subfigure}{0.48\textwidth}
		\centering
		\includegraphics[scale=0.0750]{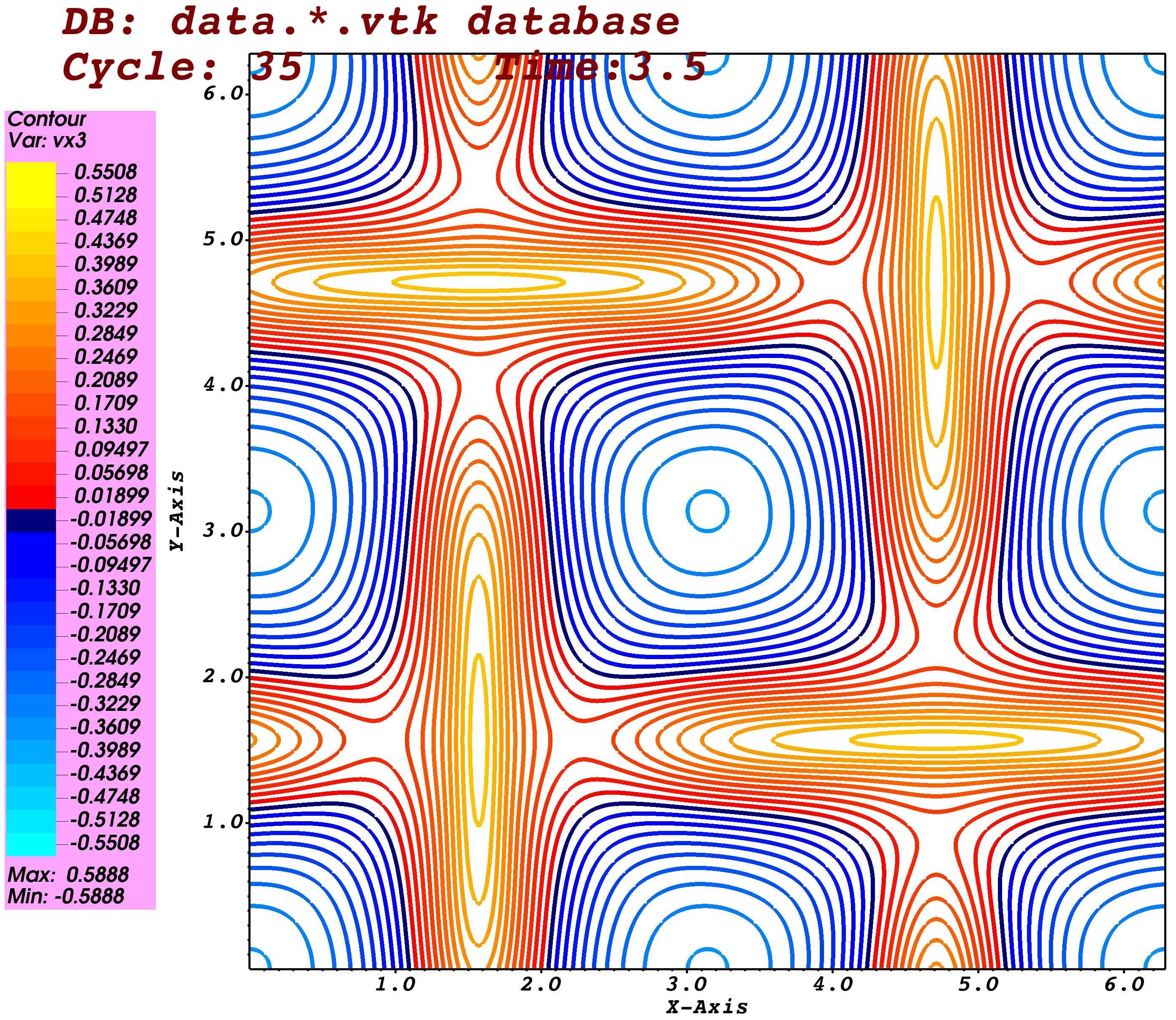}
		\caption{}
	\end{subfigure}
	\caption{Contour plot of $x$-component of velocity ($u_x$) and $z$-component of velocity ($u_z$) at $z = \frac{\pi}{4}$ plane from GMHD3D code (a \& b) and PLUTO4.4 code (c \& d). Our numerical observation from both the codes match well with the observation of Orszag \cite{orszag:1983} and Sharma et al \cite{Sharma_Sengupta:2019}. Simulation Details: Reynolds number $R_e = 100$, Grid resolution $N = 128^3$, Time stepping $dt = 10^{-4}$.}
		\label{ux and uz from GMHD3D and PLUTO}
\end{figure*}

In this subsection, we investigate the conventional 3-dimensional Taylor-Green (TG) vortex problem in the incompressible. For the purpose of testing the accuracy of numerical solvers and algorithms, Taylor-Green vortex flow is frequently employed as a usual benchmark problem. At time $t = 0$, the components of velocity are as follows, 
	\begin{eqnarray}
	\begin{aligned}
	u_x &=   \cos x \sin y \cos z\\
	u_y &=  - \sin x \cos y \cos z \\
	u_y &= 0
	\end{aligned}
	\end{eqnarray}
Even though the initial $z$-component of the velocity field is zero, the flow evaluated over time is three-dimensional. At time $t = 3.5$ and in the $z = \frac{\pi}{4}$ plane, we have calculated the $x$-component ($u_x$) and the $z$-component ($u_z$) of velocity using the GMHD3D code and the PLUTO4.4 code, respectively [See Fig. \ref{ux and uz from GMHD3D and PLUTO}]. As can be seen in Fig. \ref{ux and uz from GMHD3D and PLUTO}, our findings from both codes are identical. When compared to Orszag \cite{orszag:1983} and Sharma et al. \cite{Sharma_Sengupta:2019}, our numerical observation using the GMHD3D code and the PLUTO4.4 code shows good agreement.

\begin{figure*}
 	\centering
 	\begin{subfigure}{0.49\textwidth}
 		\centering
 		\includegraphics[scale=0.55]{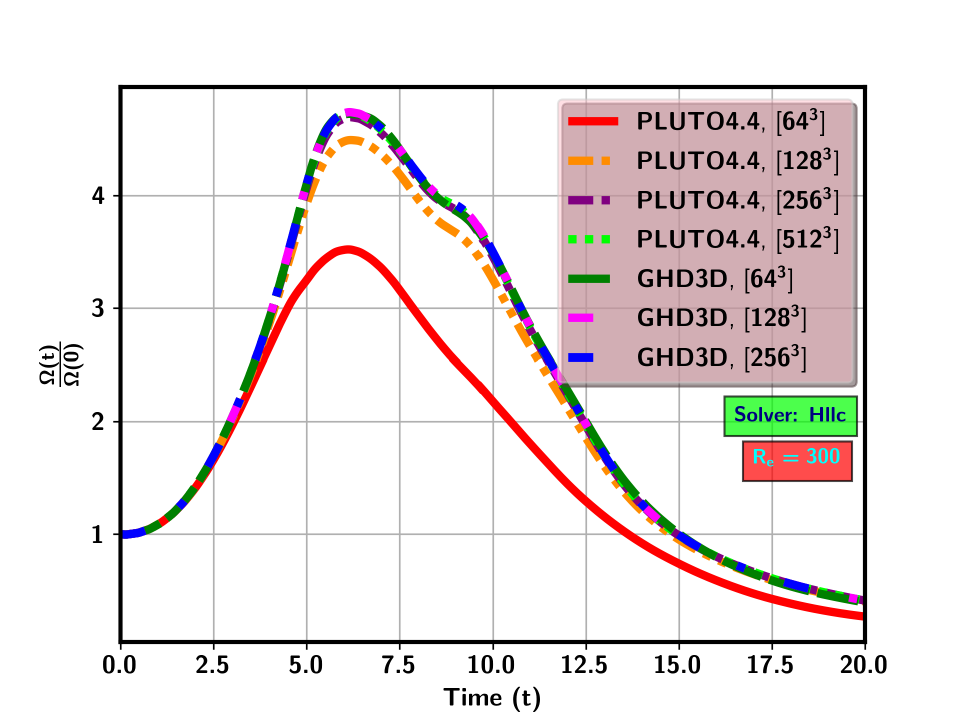}
 		\caption{}
 		\label{Normalised Enstropy for Re = 300}
 	\end{subfigure}
 	\begin{subfigure}{0.49\textwidth}
 		\centering
 		\includegraphics[scale=0.55]{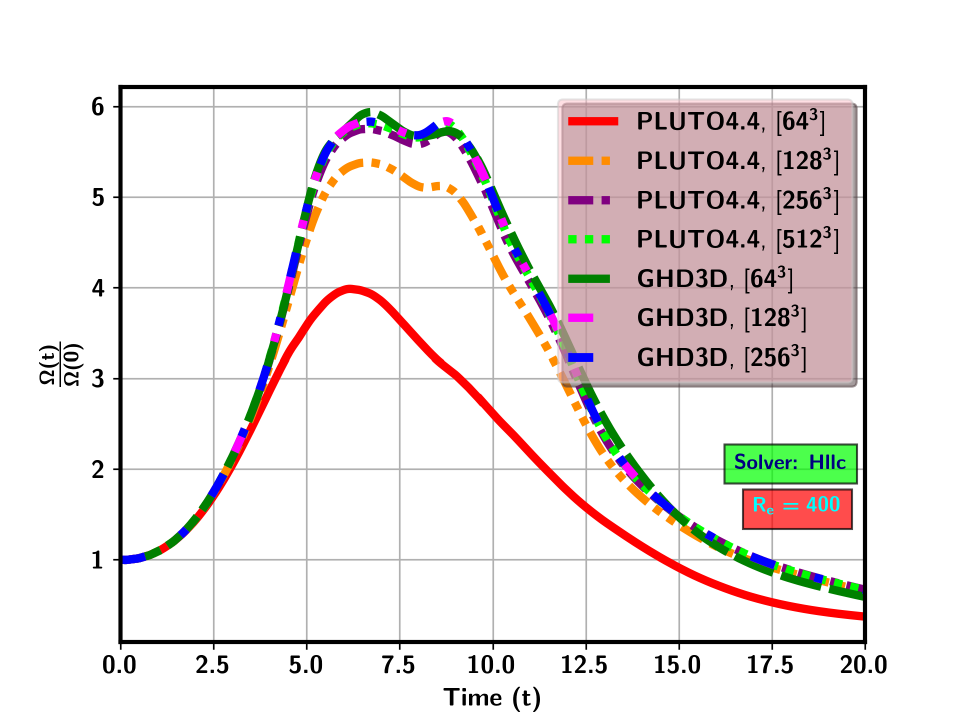}
 		\caption{}
 	\label{Normalised Enstropy for Re = 400}
 	\end{subfigure}
 	\caption{Growth of normalized mean square vorticity $\left(\frac{\Omega(t)}{\Omega(0)}\right)$ calculated as function of time for Reynolds number $R_e = 300$ \& $400$. We observe a secondary peak for $R_e = 400$, at around $t = 9$ from both the solver. \textcolor{black}{Simulation Details: Time stepping $dt = 10^{-4}$}.}
 	\label{Normalised Enstropy}
 \end{figure*}

We compute the growth of normalized mean square vorticity, denoted by $\left(\frac{\Omega(t)}{\Omega(0)}\right)$ [where $\Omega(t) = \frac{1}{N}\int \int \int \omega^2 (x,y,z,t) dx dy dz$; $N$ is total number of grid points], using the GMHD3D code and the PLUTO4.4 code for two distinct values of Reynolds numbers [See Fig. \ref{Normalised Enstropy for Re = 300} \& \ref{Normalised Enstropy for Re = 400}]. The results are in agreement with the earlier observation by Sharma et al. \cite{Sharma_Sengupta:2019}. From Fig. \ref{Normalised Enstropy for Re = 300} \& \ref{Normalised Enstropy for Re = 400}, we observe that the GMHD3D code effectively reproduces the growth of normalized men square vorticity at a grid resolution of $64^3$, whereas the PLUTO4.4 code requires at least $256^3$. The \textcolor{black}{accuracy} of the pseudo-spectral solver over a grid-based solution is evident from this finding.

\begin{figure*}
	\centering
	\begin{turn}{90} 
		\normalsize{\textbf{\textcolor{blue}{GMHD3D}}}
	\end{turn}
	\begin{subfigure}{0.19\textwidth}
		\centering
		\includegraphics[scale=0.03550]{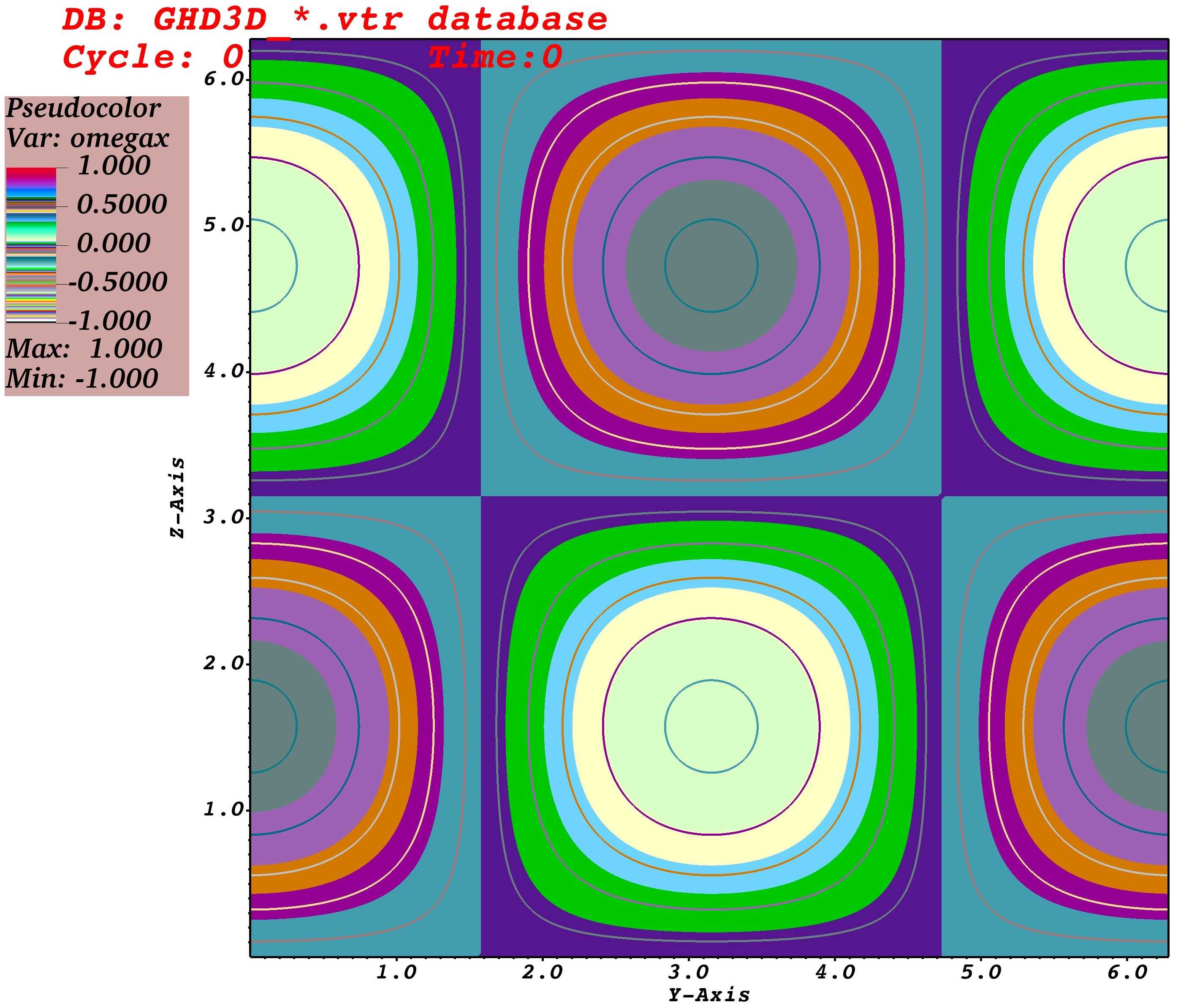}
		\caption{Time = 0.0}
	\end{subfigure}
	\begin{subfigure}{0.19\textwidth}
		\centering
		\includegraphics[scale=0.03550]{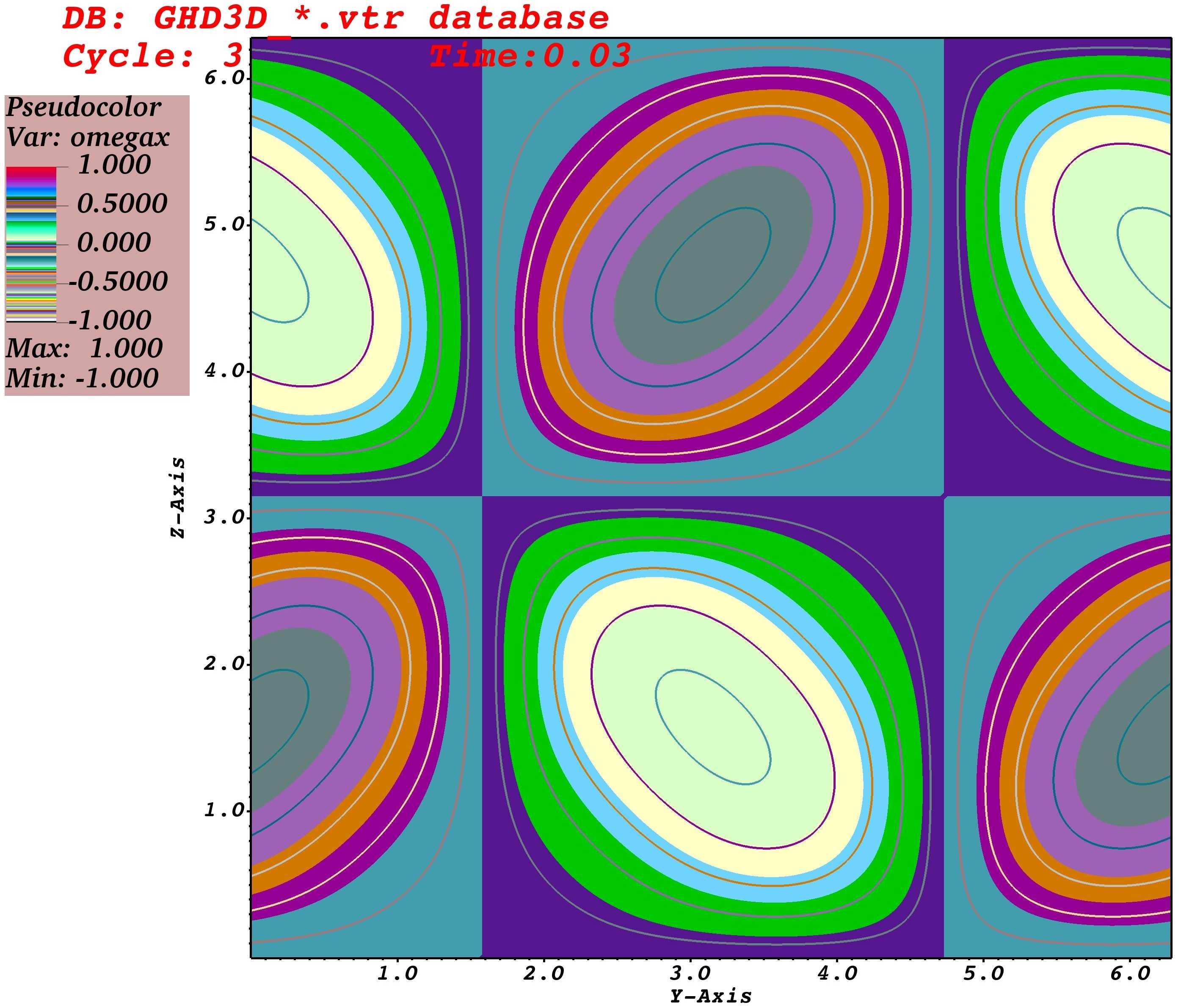}
		\caption{Time = 0.03}
	\end{subfigure}
	\begin{subfigure}{0.19\textwidth}
		\centering
		\includegraphics[scale=0.03550]{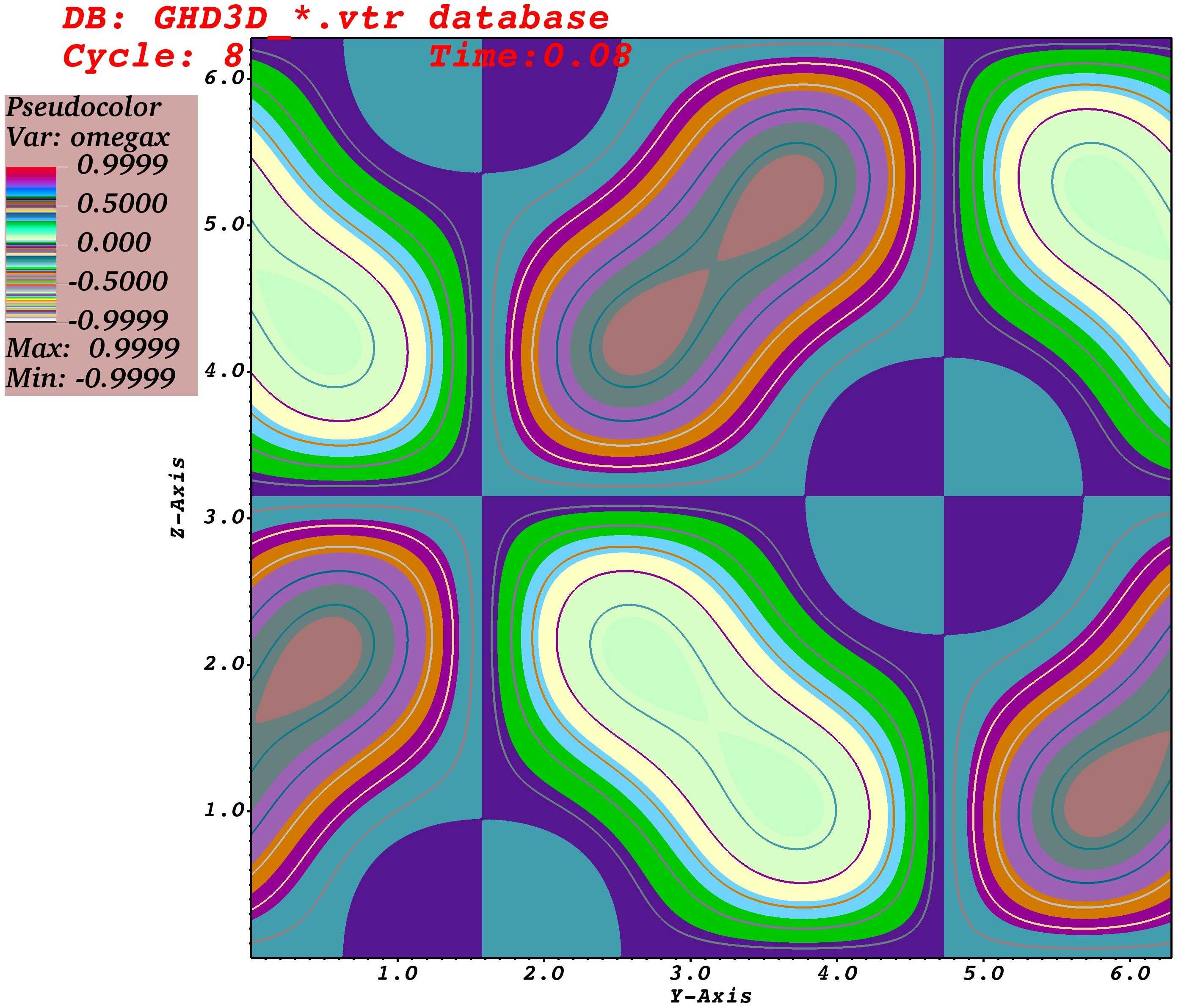}
		\caption{Time = 0.08}
	\end{subfigure}
	\begin{subfigure}{0.19\textwidth}
		\centering
		\includegraphics[scale=0.03550]{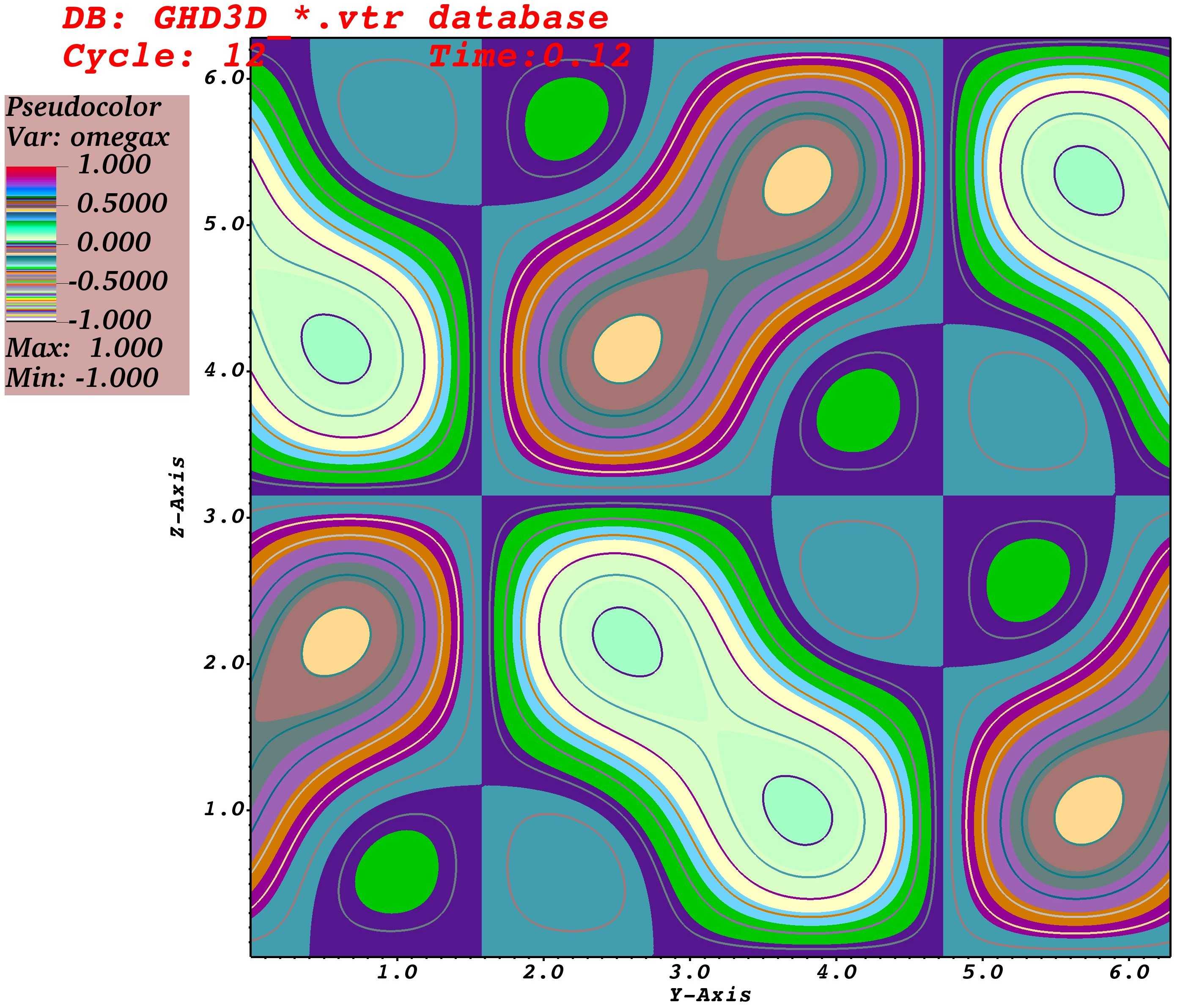}
		\caption{Time = 0.12}
	\end{subfigure}
	\begin{subfigure}{0.19\textwidth}
		\centering
		\includegraphics[scale=0.03550]{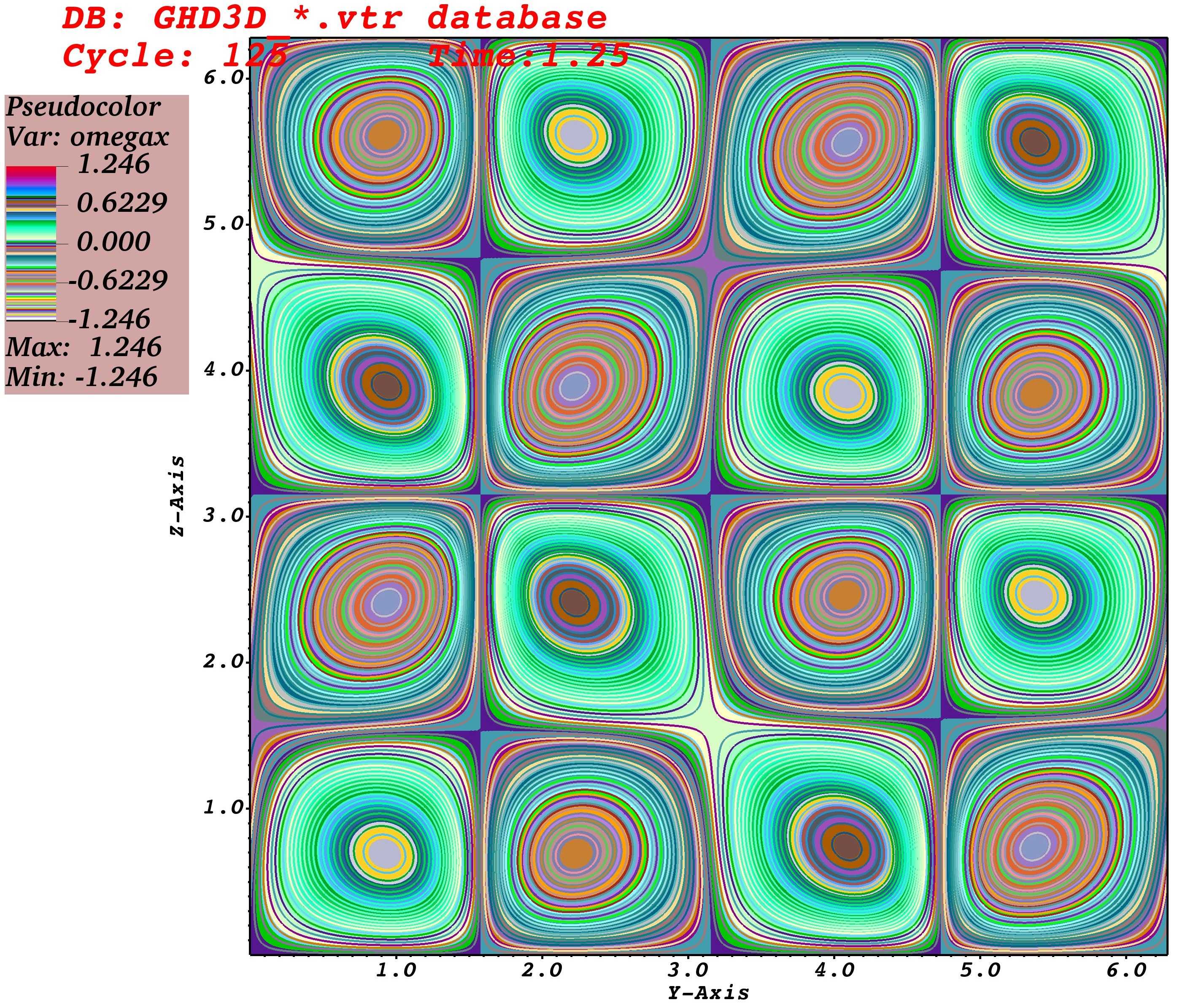}
		\caption{Time = 1.25}
	\end{subfigure}
\begin{turn}{90} 
	\normalsize{\textbf{\textcolor{blue}{PLUTO4.4}}}
\end{turn}
\begin{subfigure}{0.19\textwidth}
	\centering
	\includegraphics[scale=0.03550]{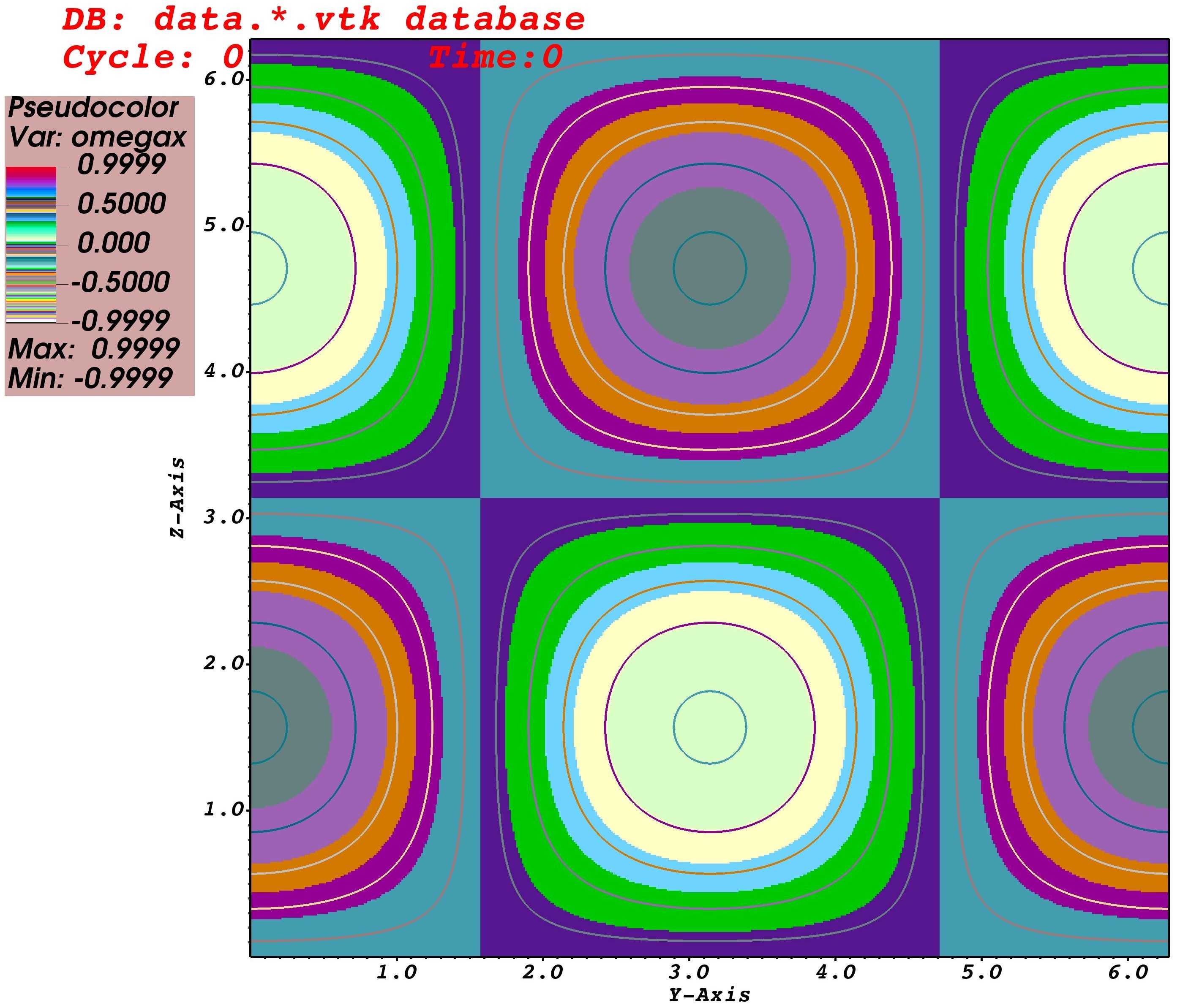}
	\caption{Time = 0.0}
\end{subfigure}
\begin{subfigure}{0.19\textwidth}
	\centering
	\includegraphics[scale=0.03550]{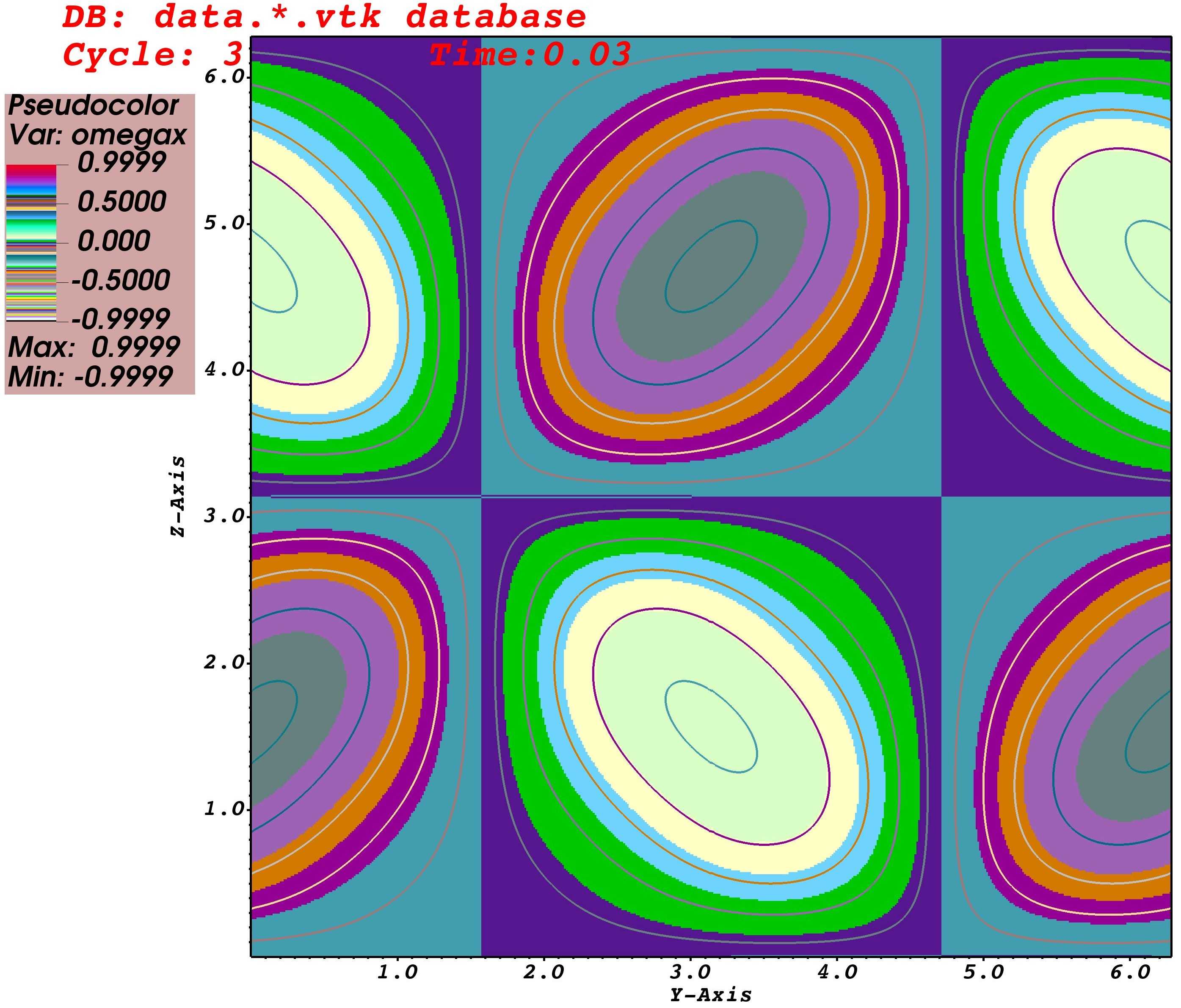}
	\caption{Time = 0.03}
\end{subfigure}
\begin{subfigure}{0.19\textwidth}
	\centering
	\includegraphics[scale=0.03550]{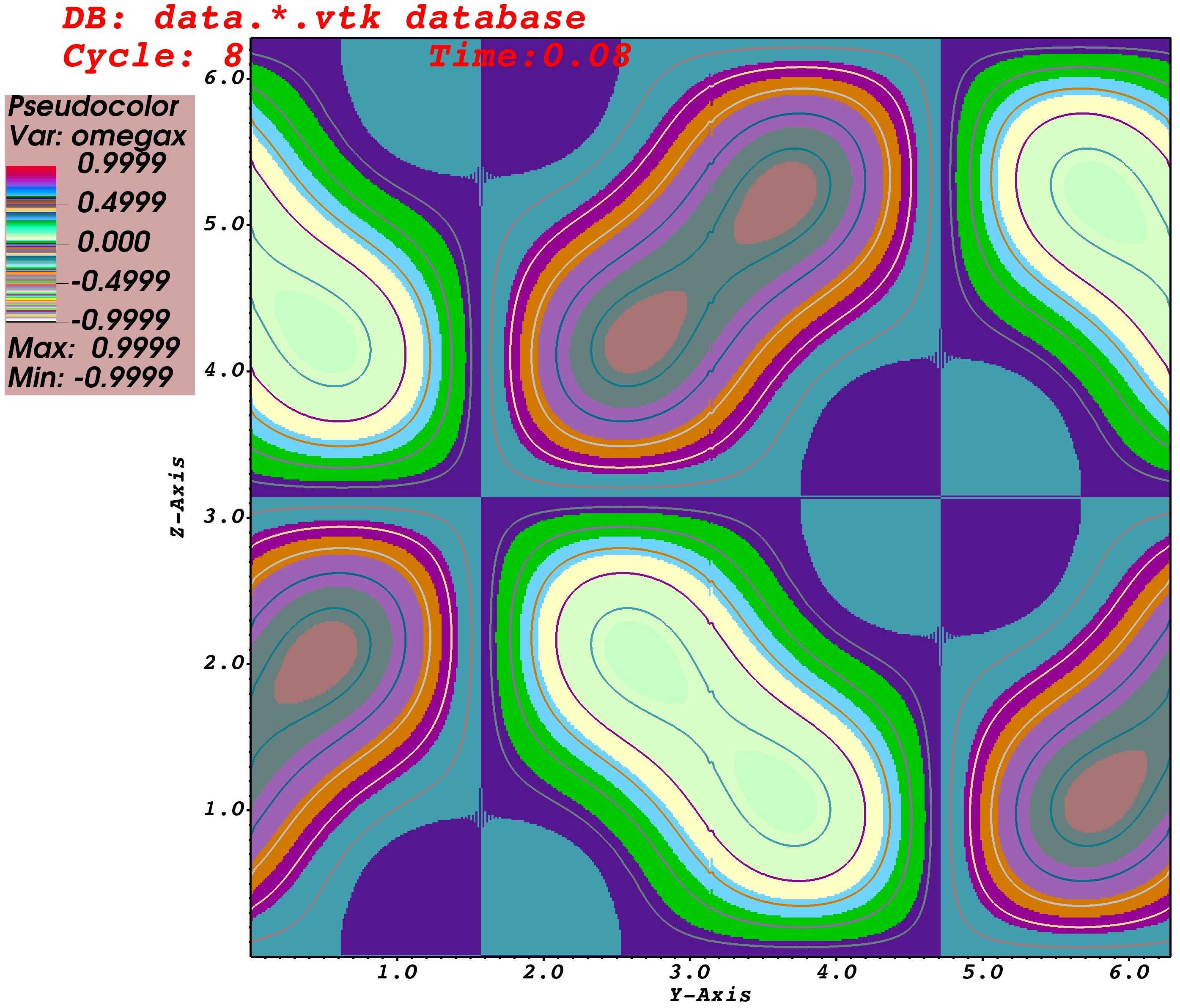}
	\caption{Time = 0.08}
\end{subfigure}
\begin{subfigure}{0.19\textwidth}
	\centering
	\includegraphics[scale=0.03550]{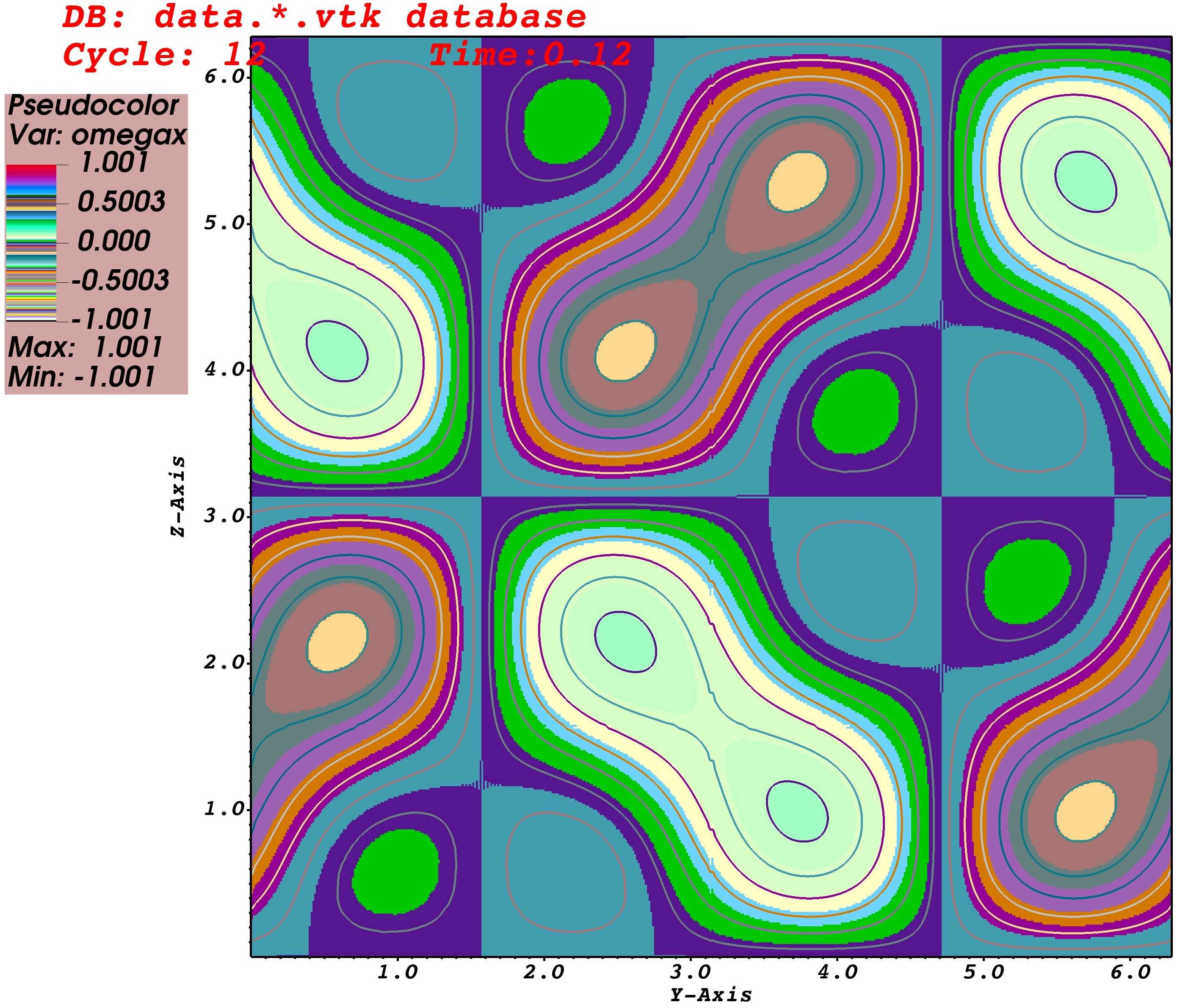}
	\caption{Time = 0.12}
\end{subfigure}
\begin{subfigure}{0.19\textwidth}
	\centering
	\includegraphics[scale=0.03550]{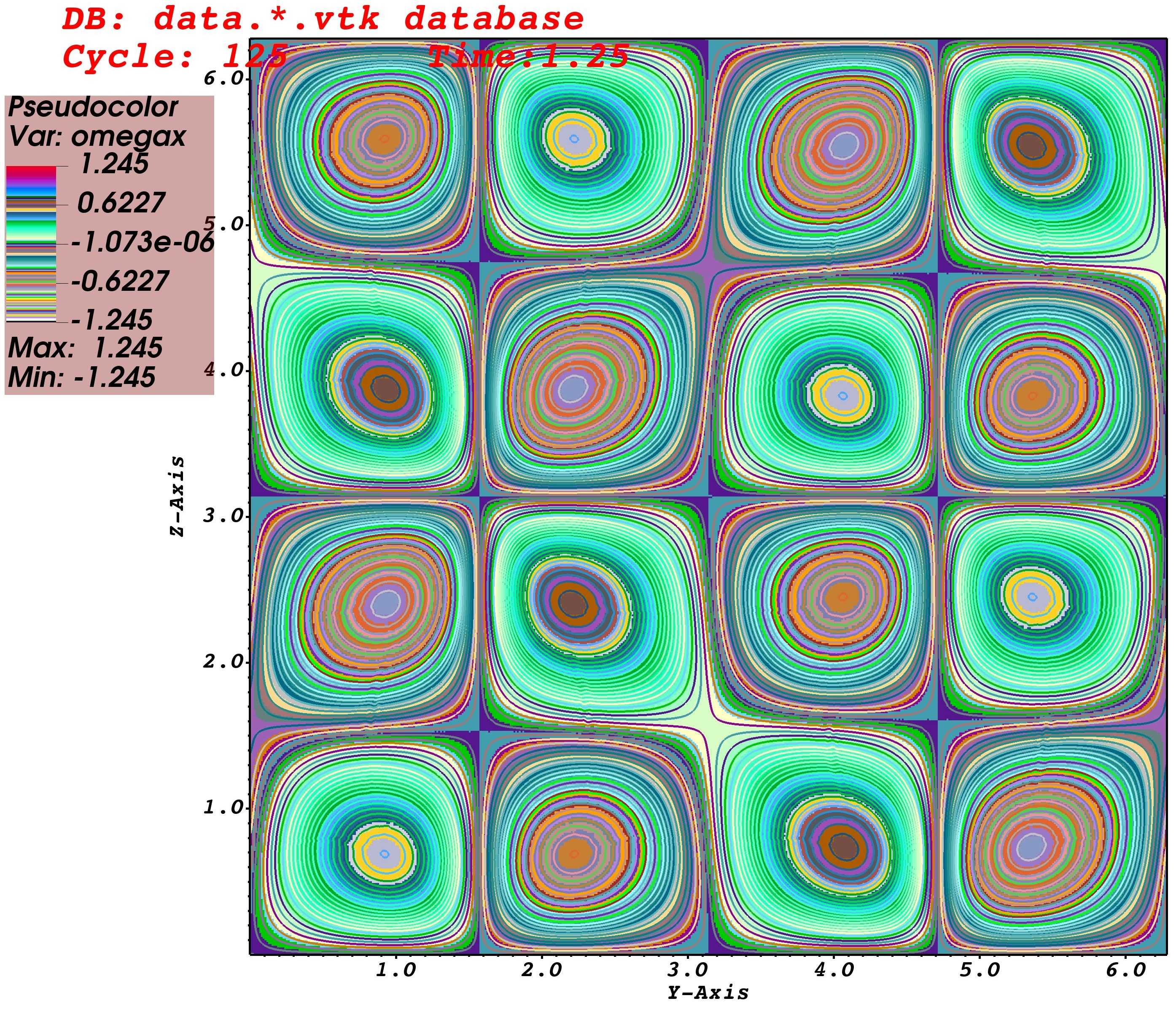}
	\caption{Time = 1.25}
\end{subfigure}
	\caption{Contours of the $x$-component of vorticity ($\omega_x$) at the $x = 0.03\pi$ plane for Reynolds number $R_e = 2000$ from GMHD3D code [upper panel (a--e)] and PLUTO4.4 code [lower panel (f--j)]. Each Taylor-Green vortex cell forms into four new smaller cells due to period doubling bifurcation. Simulation Details: Reynolds number $R_e = 2000$, Grid resolution $N = 512^3$, Time stepping $dt = 10^{-4}$.}
	\label{Period doubling bifurcation}
\end{figure*}

For Reynolds number $R_e = 2000$, we plot $\omega_x$ contours in the $x = 0.03\pi$ plane. Based on Fig. \ref{Period doubling bifurcation}, we see that the vortices start stretching at an early time ($t = 0.03$), and that this vortex stretching causes a period doubling bifurcation at a later time ($t \ge 1.25$). Each Taylor-Green vortex cell bifurcates into four daughter cells due to period doubling bifurcation. We report identical observation from both the numerical solver (GMHD3D \& PLUTO4.4) at grid resolution $512^3$ [See Fig. \ref{Period doubling bifurcation}].

\begin{figure*}
	\centering
	\begin{turn}{90} 
		\normalsize{\textbf{\textcolor{blue}{GMHD3D}}}
	\end{turn}
	\begin{subfigure}{0.19\textwidth}
		\centering
		\includegraphics[scale=0.03570]{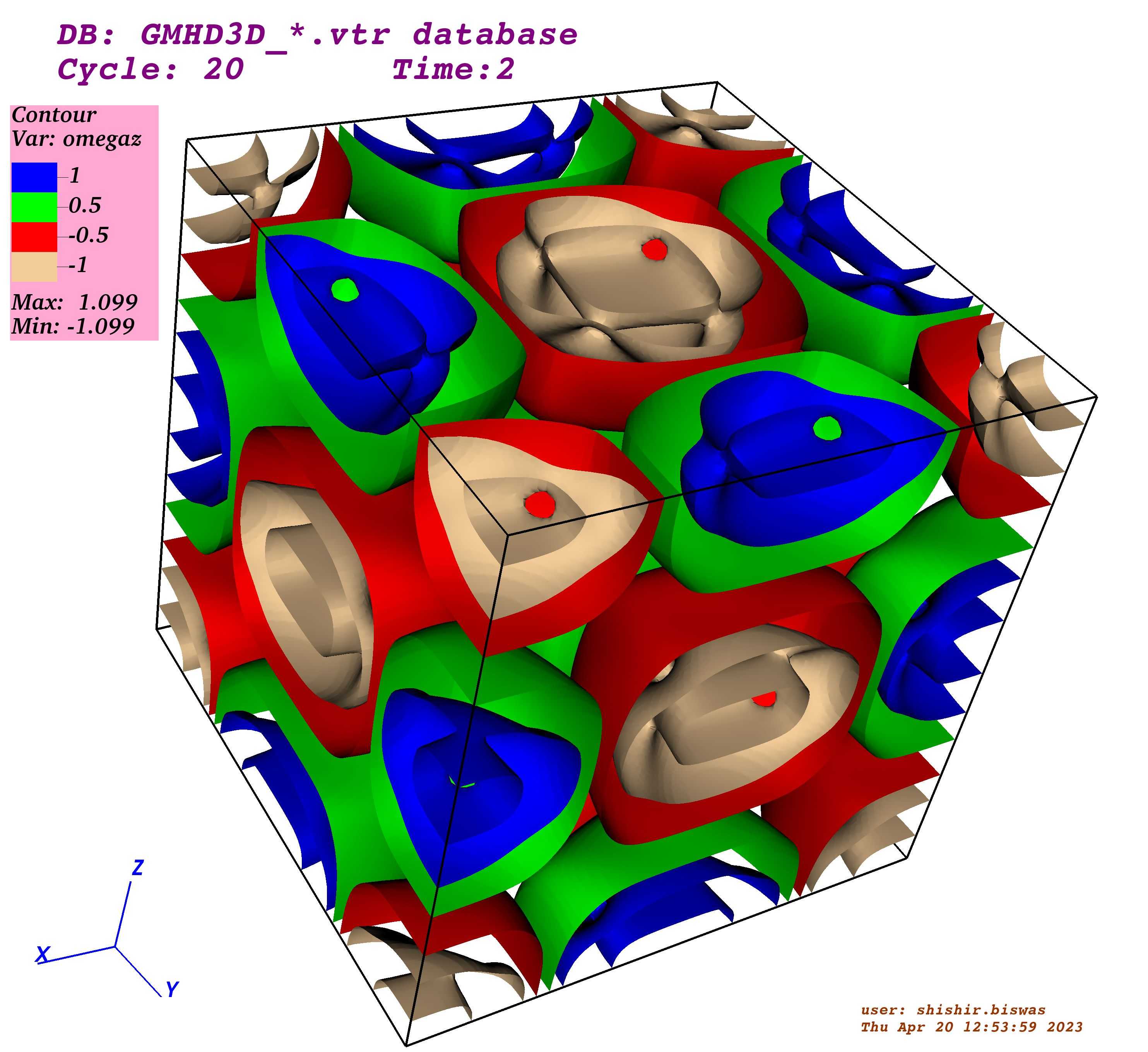}
		\caption{Time = 2.0}
			\label{GMHD3D 0p2}
	\end{subfigure}
	\begin{subfigure}{0.19\textwidth}
		\centering
		\includegraphics[scale=0.03570]{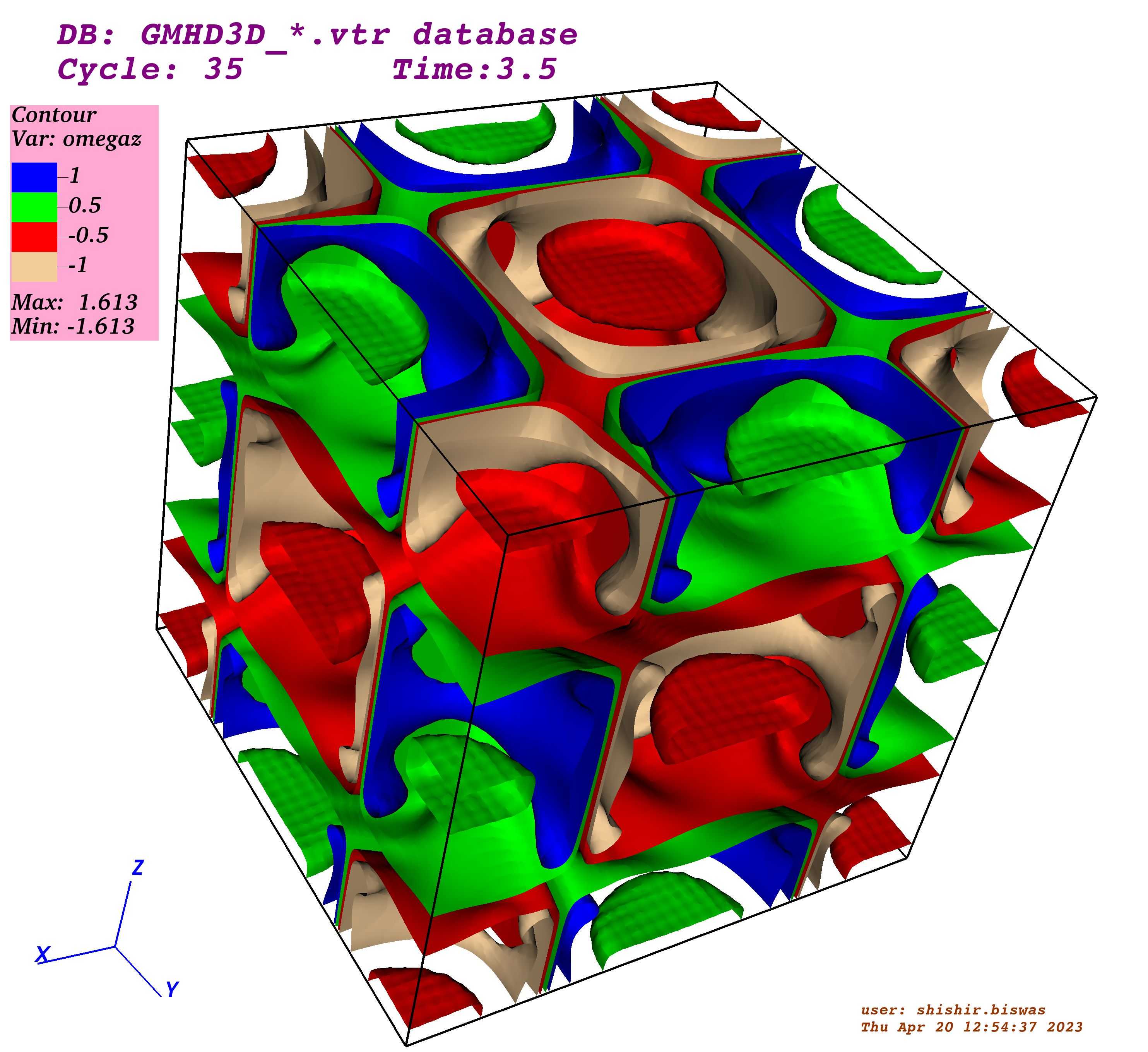}
		\caption{Time = 3.5}
		  \label{GMHD3D 3p5}
	\end{subfigure}
	\begin{subfigure}{0.19\textwidth}
		\centering
		\includegraphics[scale=0.03570]{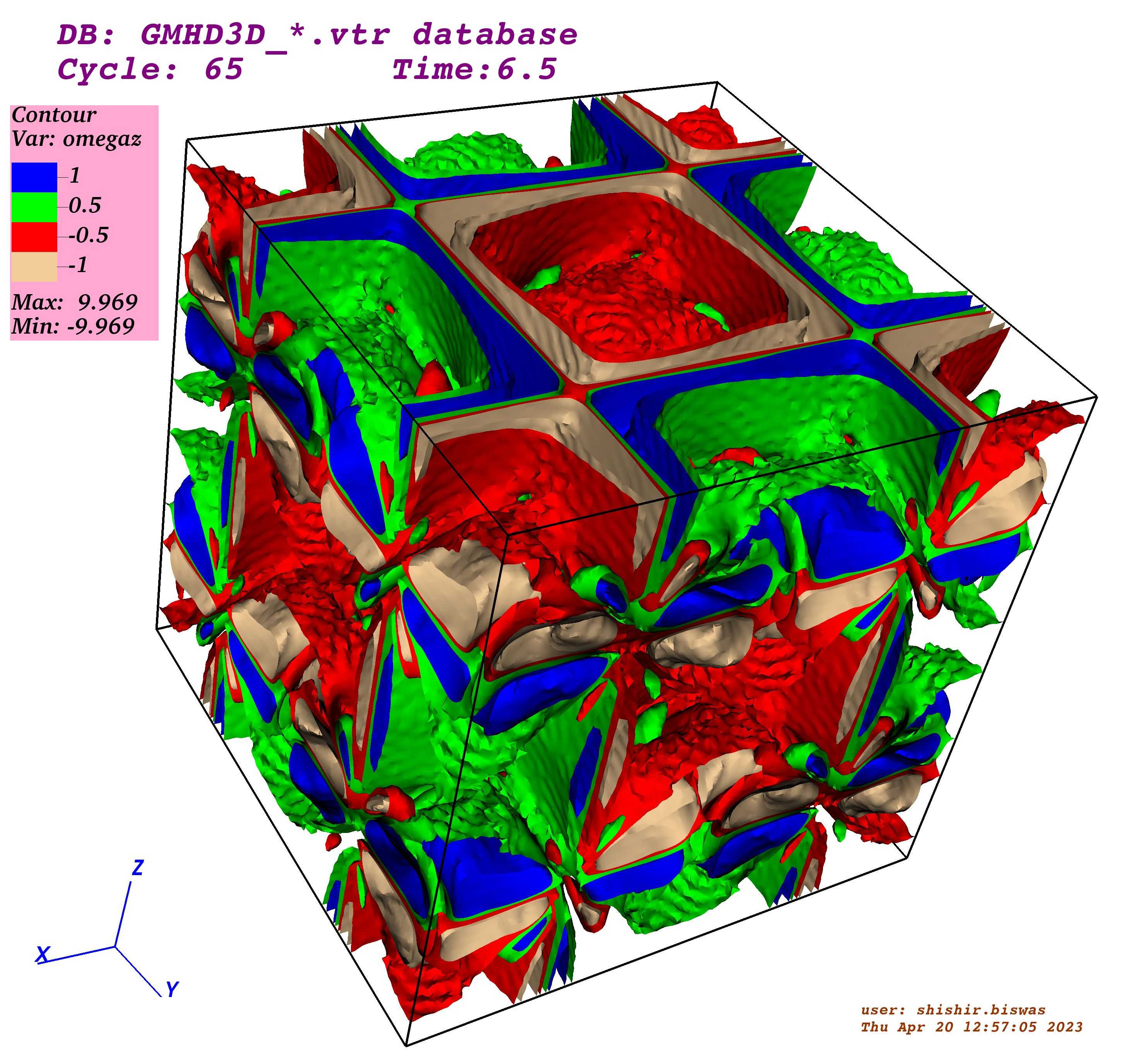}
		\caption{Time = 6.5}
			\label{GMHD3D 6p5}
	\end{subfigure}
	\begin{subfigure}{0.19\textwidth}
		\centering
		\includegraphics[scale=0.03570]{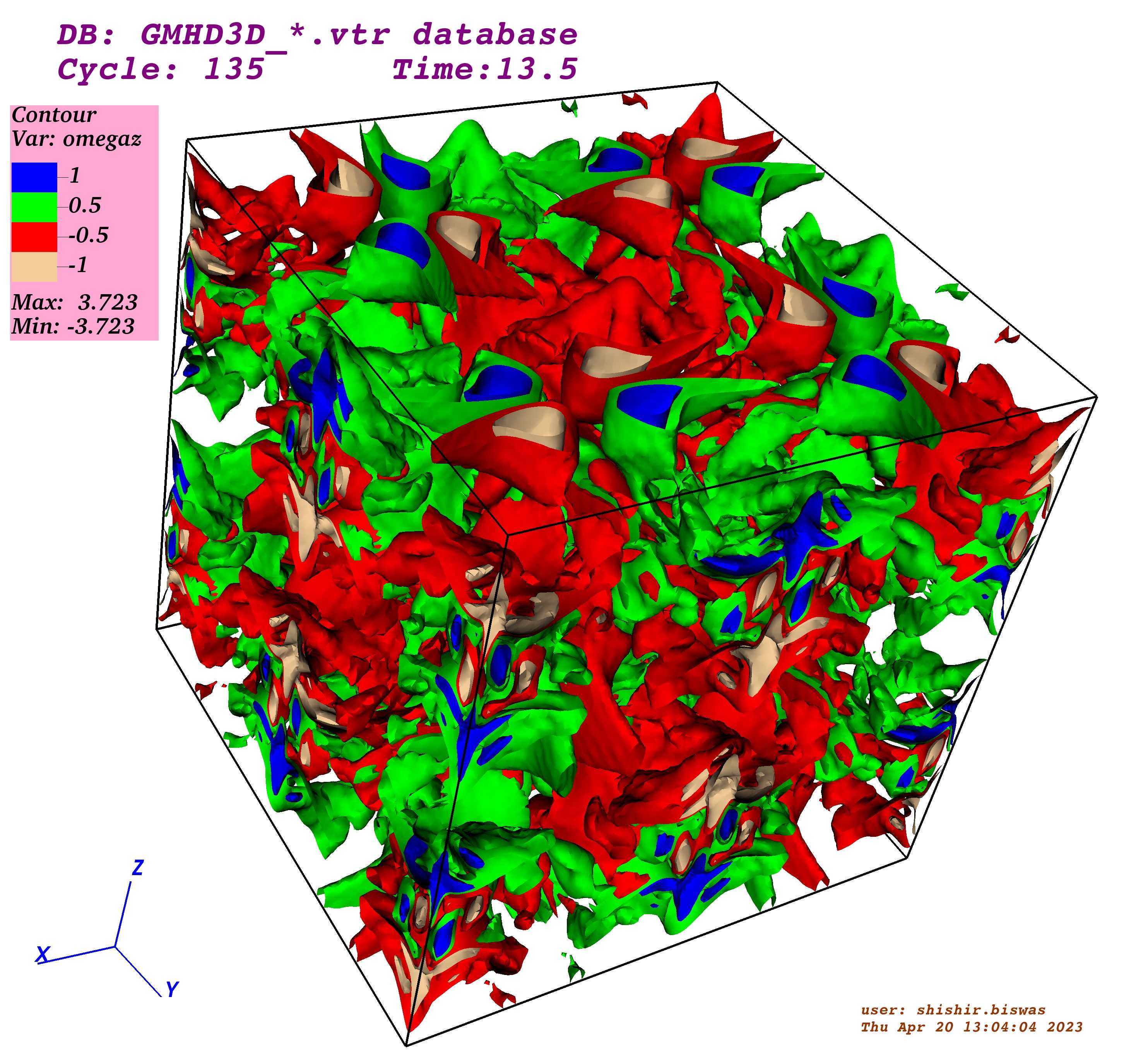}
		\caption{Time = 13.5}
			\label{GMHD3D 13p5}
	\end{subfigure}
	\begin{subfigure}{0.19\textwidth}
		\centering
		\includegraphics[scale=0.03570]{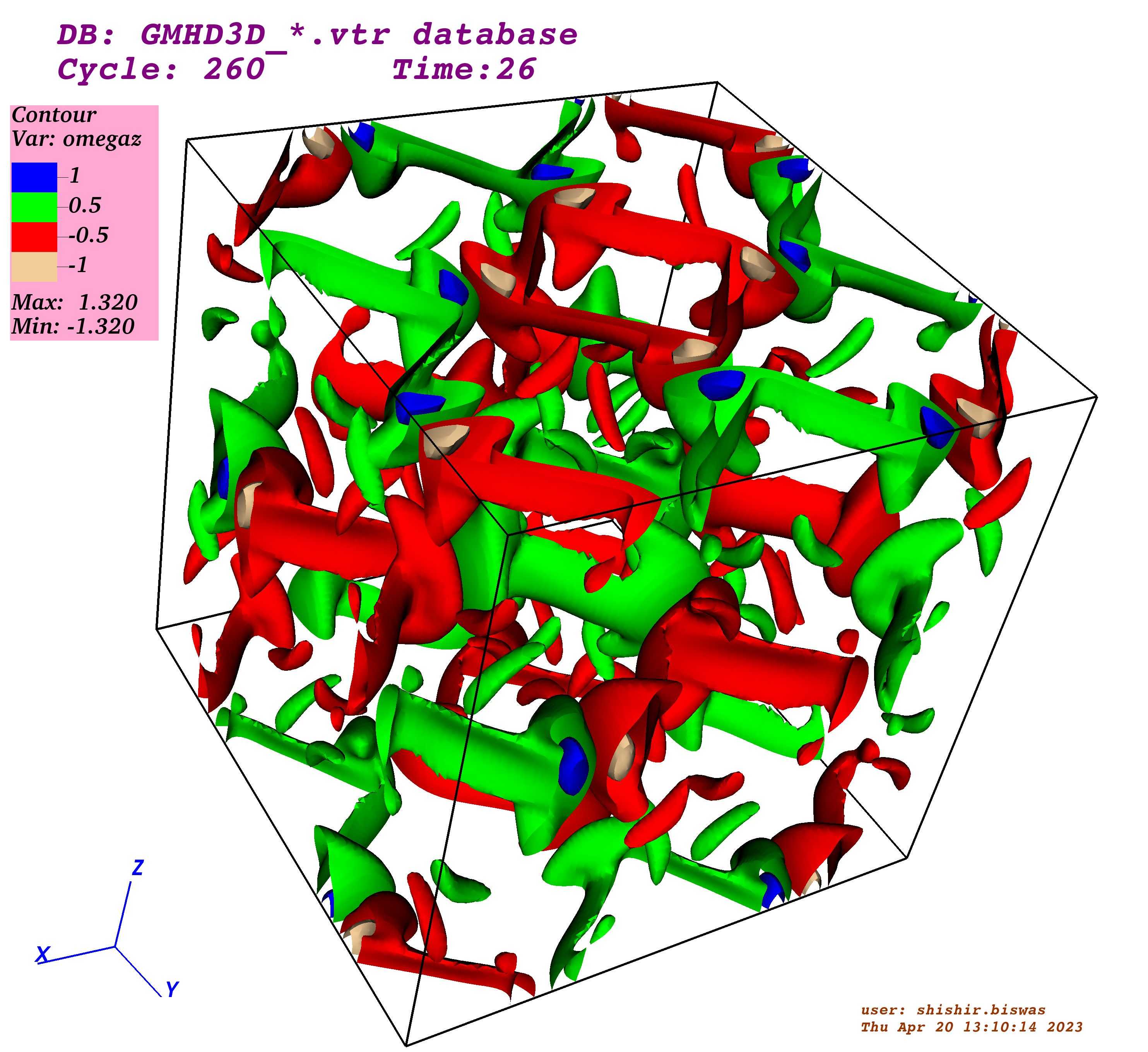}
		\caption{Time = 26.0}
			\label{GMHD3D 26p0}
	\end{subfigure}
	\begin{turn}{90} 
		\normalsize{\textbf{\textcolor{blue}{PLUTO4.4}}}
	\end{turn}
	\begin{subfigure}{0.19\textwidth}
		\centering
		\includegraphics[scale=0.03570]{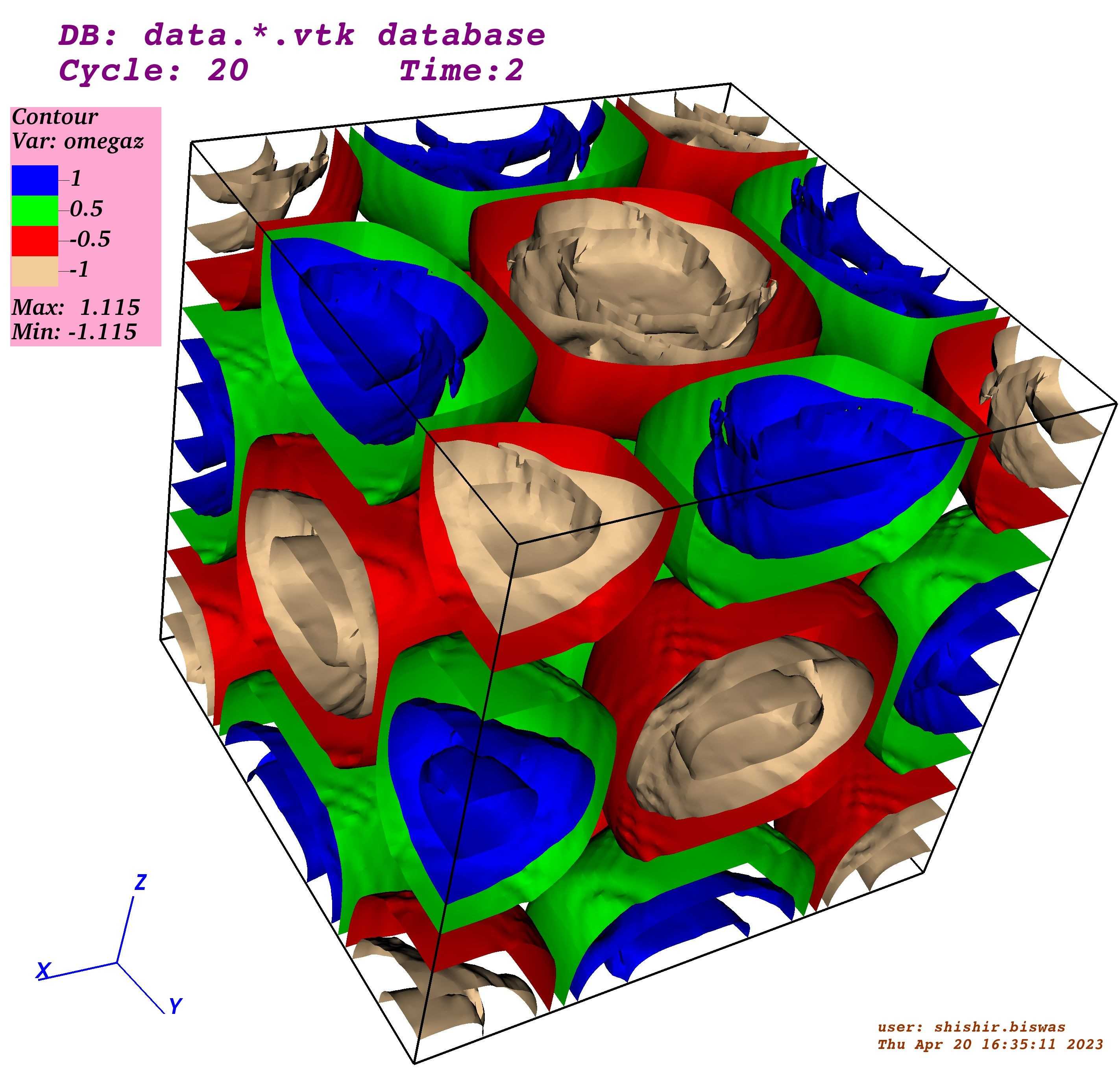}
		\caption{Time = 2.0}
			\label{PLUTO 0p2}
	\end{subfigure}
	\begin{subfigure}{0.19\textwidth}
		\centering
		\includegraphics[scale=0.03570]{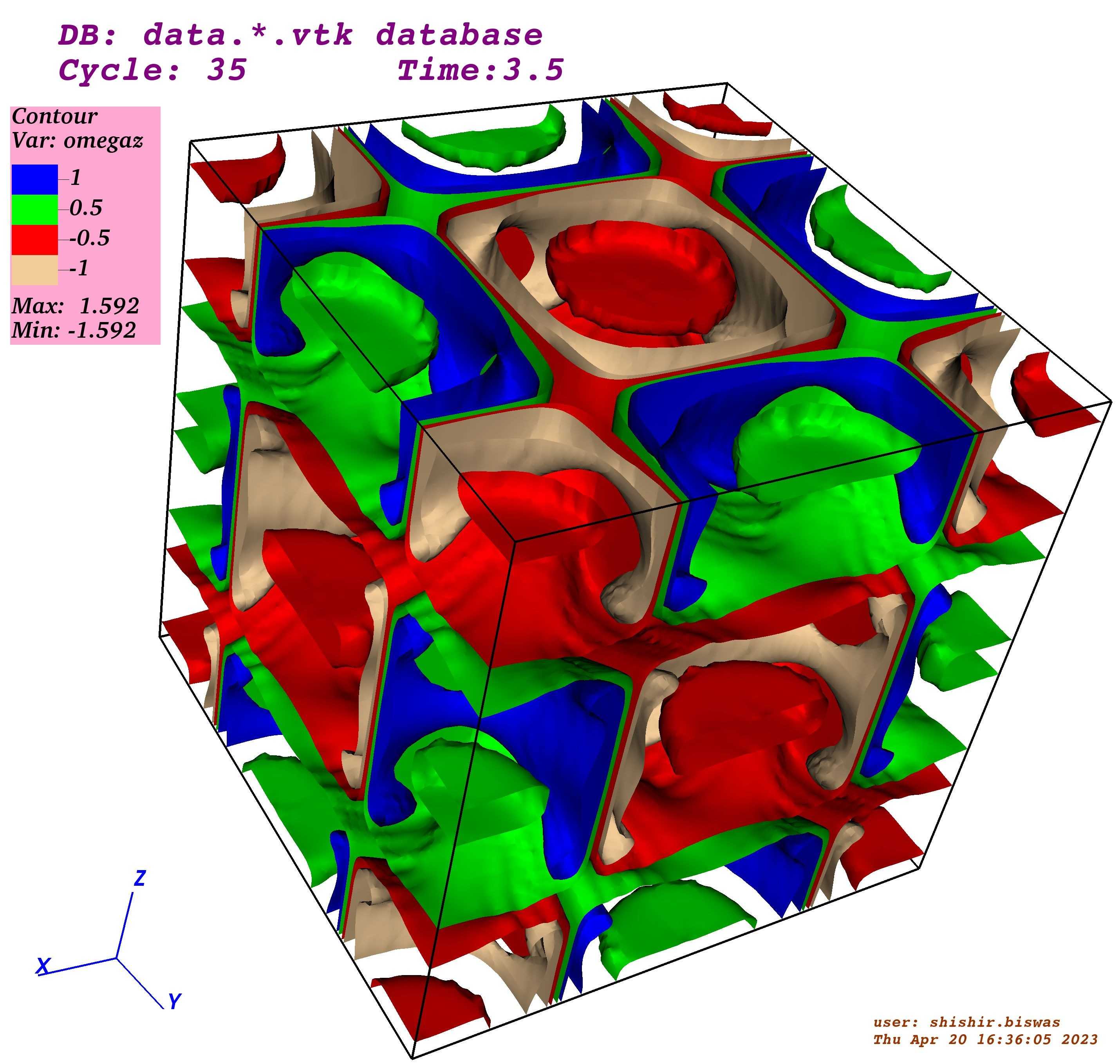}
		\caption{Time = 3.5}
			\label{PLUTO 3p5}
	\end{subfigure}
	\begin{subfigure}{0.19\textwidth}
		\centering
		\includegraphics[scale=0.03570]{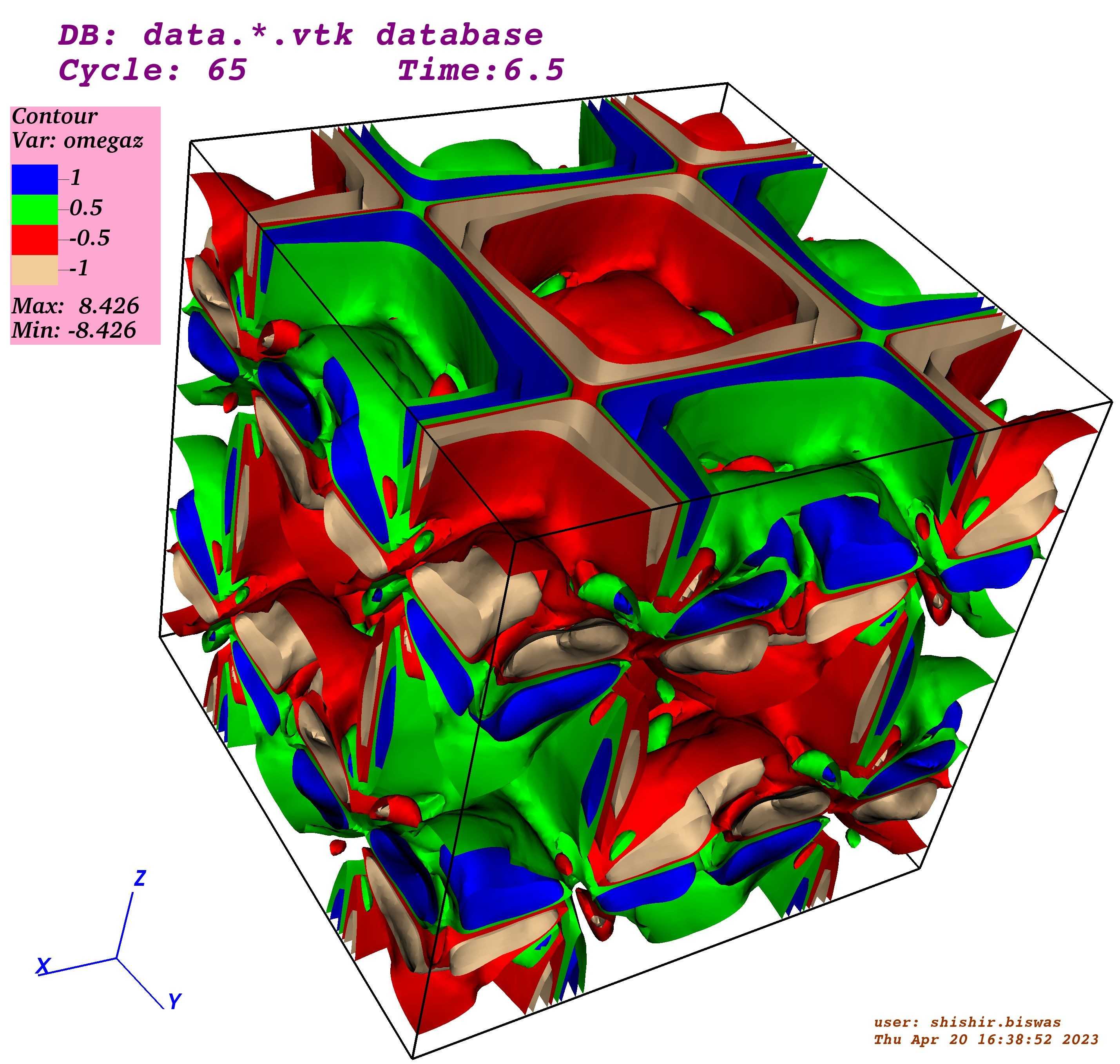}
		\caption{Time = 6.5}
			\label{PLUTO 6p5}
	\end{subfigure}
	\begin{subfigure}{0.19\textwidth}
		\centering
		\includegraphics[scale=0.03570]{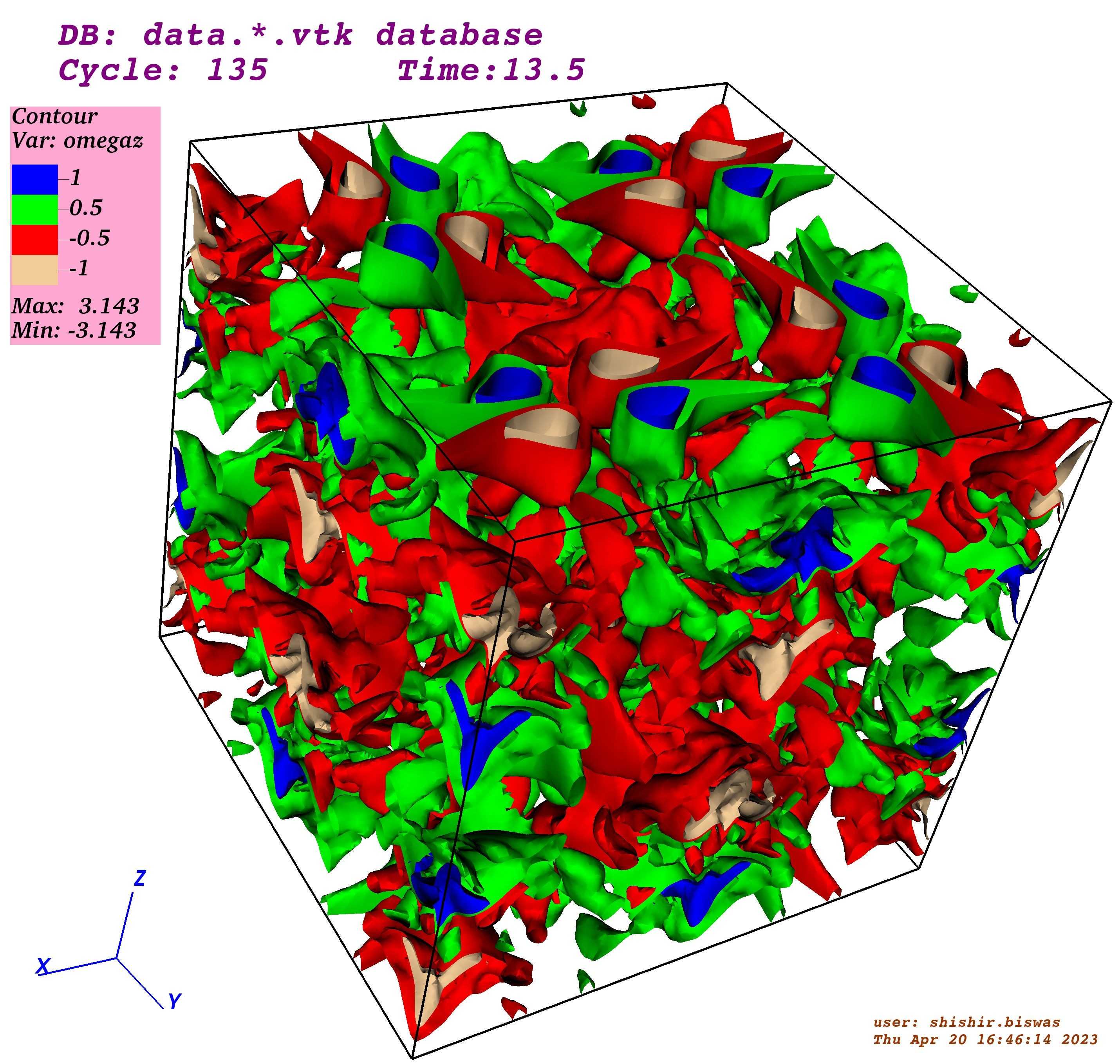}
		\caption{Time = 13.5}
			\label{PLUTO 13p5}
	\end{subfigure}
	\begin{subfigure}{0.19\textwidth}
		\centering
		\includegraphics[scale=0.03570]{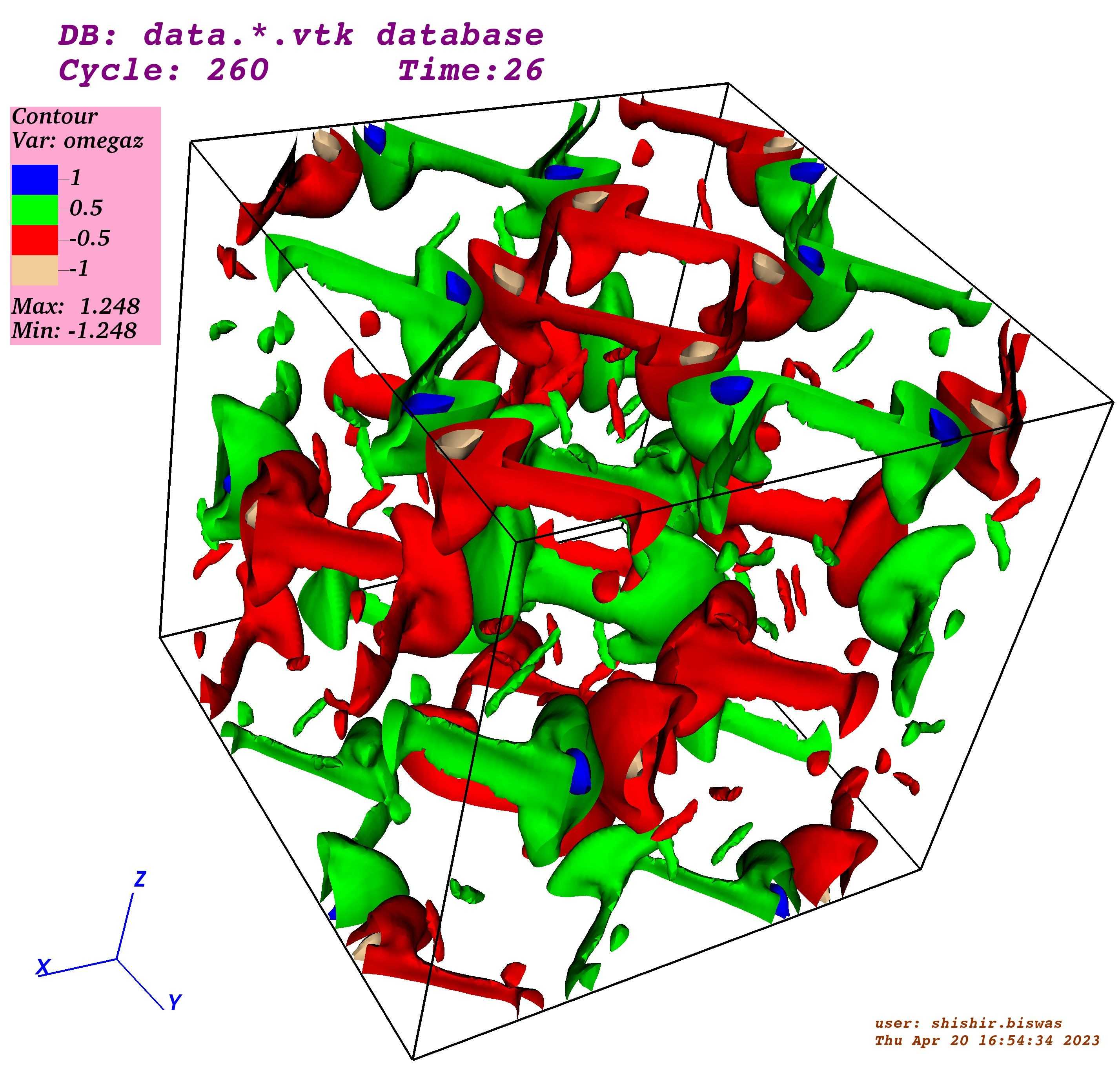}
		\caption{Time = 26.0}
			\label{PLUTO 26p0}
	\end{subfigure}
	\caption{3-dimensional Iso-surfaces of the $z$-component of vorticity ($\omega_z$) from GMHD3D code [upper panel (a--e)] and PLUTO4.4 code [lower panel (f--j)]. The flow evolution dynamics consists of several process like vortex stretching (a \& f), vortex roll up (b \& g), vortex break down (c \& h), turbulence (d \& i) and finally turbulent decay (e \& j). Simulation Details: Reynolds number $R_e = 400$, Grid resolution $N = 256^3$, Time stepping $dt = 10^{-4}$.}
		\label{TG flow evolution}
\end{figure*}

We have created a 3D iso-contour visualization of $\omega_z$ using data from both GMHD3D and PLUTO4.4 to obtain a more comprehensive look at the flow evolution. Fig. \ref{TG flow evolution} shows that the Taylor-Green vortex is laminar and anisotropic at early stages and that energy is transported to smaller scales as a result of vortex stretching. The impact of the vortex stretching term is clearly seen in Fig.  \ref{GMHD3D 0p2} \& \ref{PLUTO 0p2}. As it is seen in Fig. \ref{GMHD3D 3p5} \& \ref{PLUTO 3p5}, the vortices roll up at an intermediate time, followed by a coherent breakdown [See Fig. \ref{GMHD3D 6p5} \& \ref{PLUTO 6p5}] and the creation of small-scale structures. Later on, as shown in Fig. \ref{GMHD3D 13p5} \& \ref{PLUTO 13p5}, the flow becomes completely turbulent, eventually ending in a turbulent decay [See Fig. \ref{GMHD3D 26p0} \& \ref{PLUTO 26p0}]. Using the GMHD3D code and the PLUTO4.4 code, we are able to capture the whole dynamics of a Taylor-Green vortex, including its stretching, rolling up, dividing, and reconnecting (vortex break down), turbulence, turbulent decays, etc. We find that our numerical observation is consistent with other, earlier studies \cite{Sharma_Sengupta:2019}.

In the following, some standard magnetohydrodynamic problems are considered to test the MHD module. We present the same in the coming subsection. 

 \subsection{Test 3 [Magnetohydrodynamics]: Coherent  nonlinear oscillations using 2D Orszag-Tang (OT) Flow} \label{2D OT}

\begin{figure*}
 	\centering
 	\begin{subfigure}{0.49\textwidth}
 		\centering
 		\includegraphics[scale=0.55]{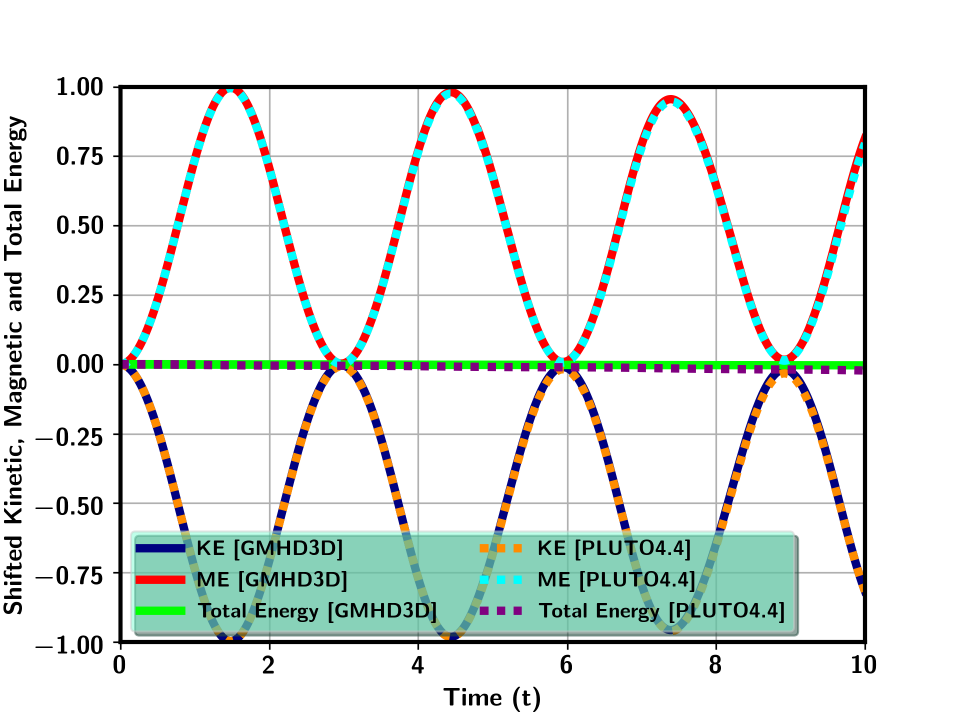}
 		\caption{}
 		\label{2D OT KE and EMF no EMF}
 	\end{subfigure}
 	\begin{subfigure}{0.49\textwidth}
 		\centering
 		\includegraphics[scale=0.55]{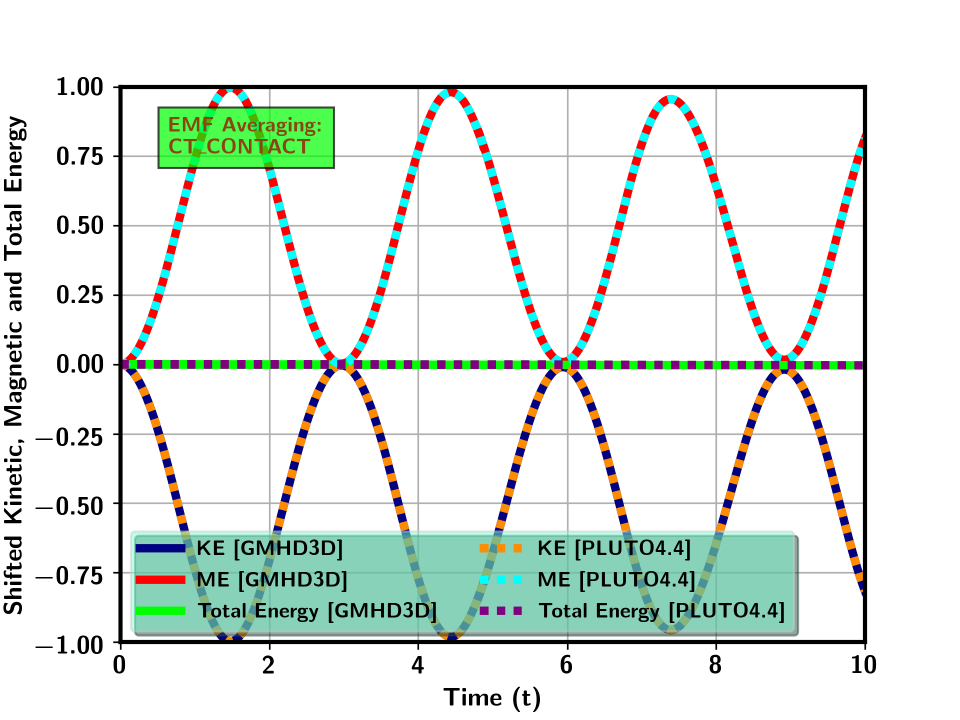}
 		\caption{}
 		\label{2D OT KE and EMF with EMF}
 	\end{subfigure}
 	\caption{The shifted kinetic and magnetic energies for 2D Orszag-Tang Flow from GMHD3D and PLUTO4.4  code (a) with out CT\_CONTACT scheme (b) with CT\_CONTACT scheme at grid resolution $128^2$. PLUTO4.4 data is perfectly matched with GMHD3D data by employing appropriate electric field reconstruction algorithms. \textcolor{black}{Simulation Details: Time stepping $dt = 10^{-4}$}.}
 \end{figure*}

 Orszag and Tang were the first to examine the Orszag-Tang flow \cite{Orszag_Tang:1979}. Since then, it has been tested and compared a lot in numerical MHD simulation models. 2D Orszag-Tang [OT] Flow is known as the divergence free flow. For 2D Orszag-Tang [OT] flow, the velocity profile takes the form,
 	\begin{equation}\label{2D OT Flow}
 	\begin{aligned}
 	u_x &= - u_0 [ A \sin(k_0y)]\\
 	u_y &=   u_0 [ A \sin(k_0x)] 
 	\end{aligned}
 	\end{equation}
with $A = 1.0$ and $k_0 = 1.0$. The initial magnetic field is assumed to be homogeneous and ambient, with a value determined by the Alfven Mach number ($M_A$) and the initial fluid velocity ($u_0$). Sonic mach ($M_s$) for our model is equal to $0.01$. In this simulation, we consider a grid resolution of $128^2$ for both codes, and we find that kinetic energy is converted to magnetic energy and vise-versa at regular intervals, in the form of coherent non-linear oscillations [See Fig. \ref{2D OT KE and EMF no EMF}].  Now, a close review of Fig. \ref{2D OT KE and EMF no EMF} reveals that, substantial differences between the PLUTO4.4 and GMHD3D data do exist. Even though both sets of code have identical parameters, the data does not match very well. Using the appropriate electric field reconstruction techniques [CT\_EMF\_AVG] available in the PLUTO4.4 code, we are able to improve the results obtained from the code. We utilize the CT\_CONTACT scheme, which is the least dissipative EMF AVG scheme available in PLUTO4.4 \cite{CT:2005}. Using this method, it is seen that the kinetic and magnetic energy oscillations from the PLUTO4.4 code are accurately recreated, and that it matches the GMHD3D data perfectly [See Fig. \ref{2D OT KE and EMF with EMF}]. The period of oscillation is seen to be $T = 2.971$ from both the code.

\begin{figure*}
	\centering
	\begin{turn}{90} 
		\normalsize{\textbf{\textcolor{blue}{GMHD3D}}}
	\end{turn}
	\begin{subfigure}{0.48\textwidth}
		\centering
		\includegraphics[scale=0.07500]{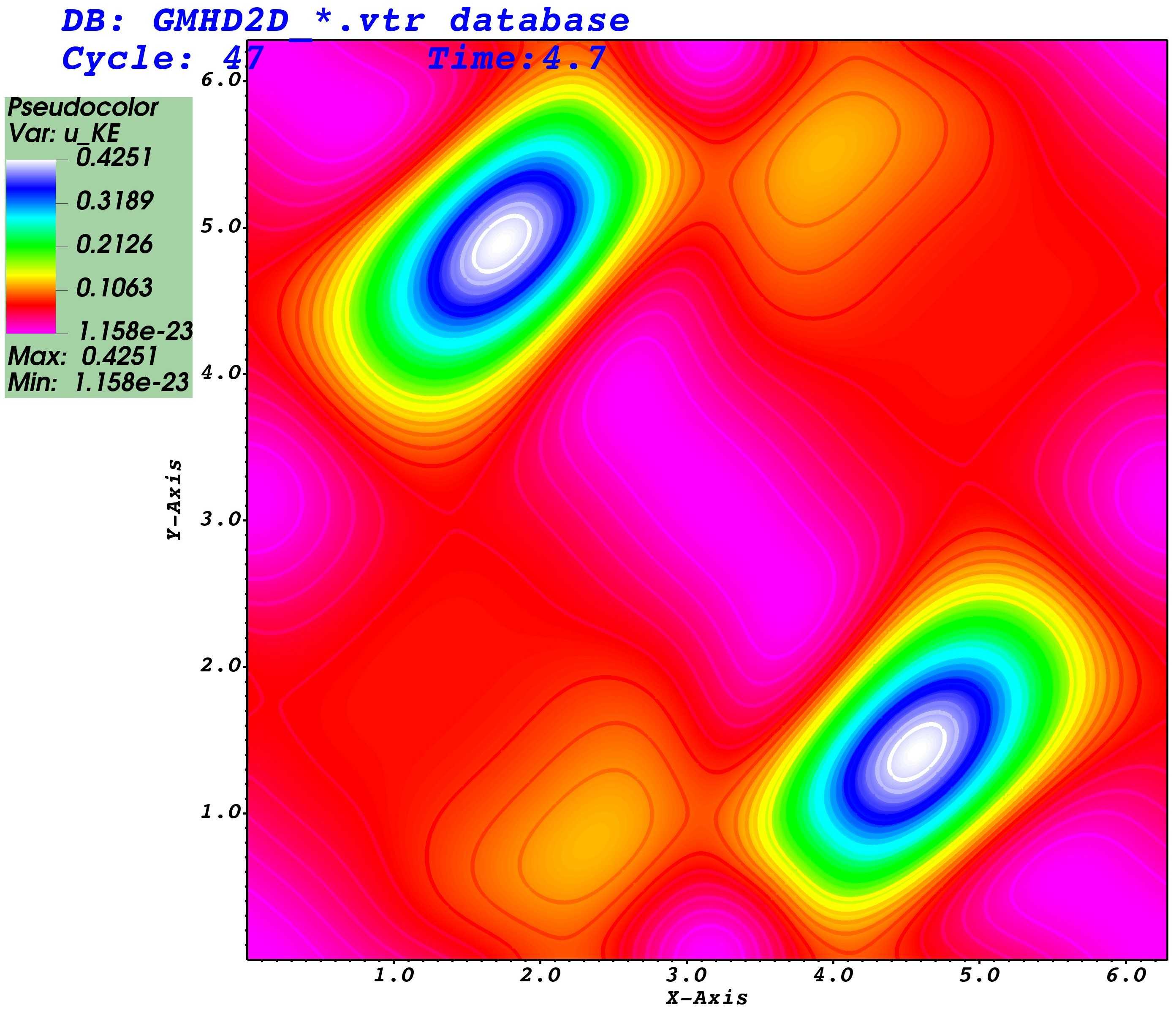}
		\caption{}
		\label{2D OT KE GMHD2D}
	\end{subfigure}
	\begin{subfigure}{0.48\textwidth}
		\centering
		\includegraphics[scale=0.07500]{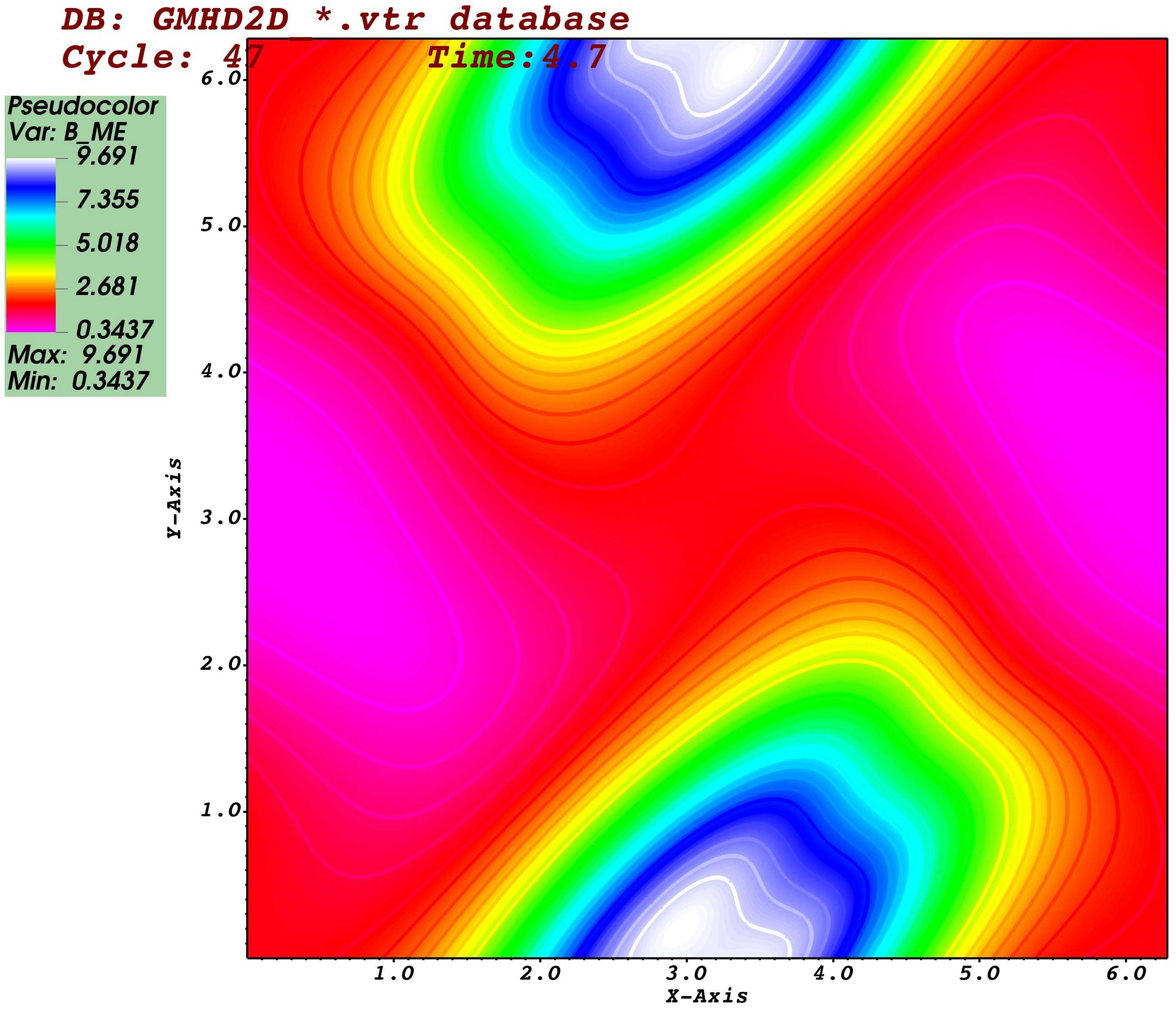}
		\caption{}
		\label{2D OT ME GMHD2D}
	\end{subfigure}
	\begin{turn}{90} 
		\normalsize{\textbf{\textcolor{blue}{PLUTO4.4}}}
	\end{turn}
    \begin{subfigure}{0.48\textwidth}
    	\centering
    	\includegraphics[scale=0.07500]{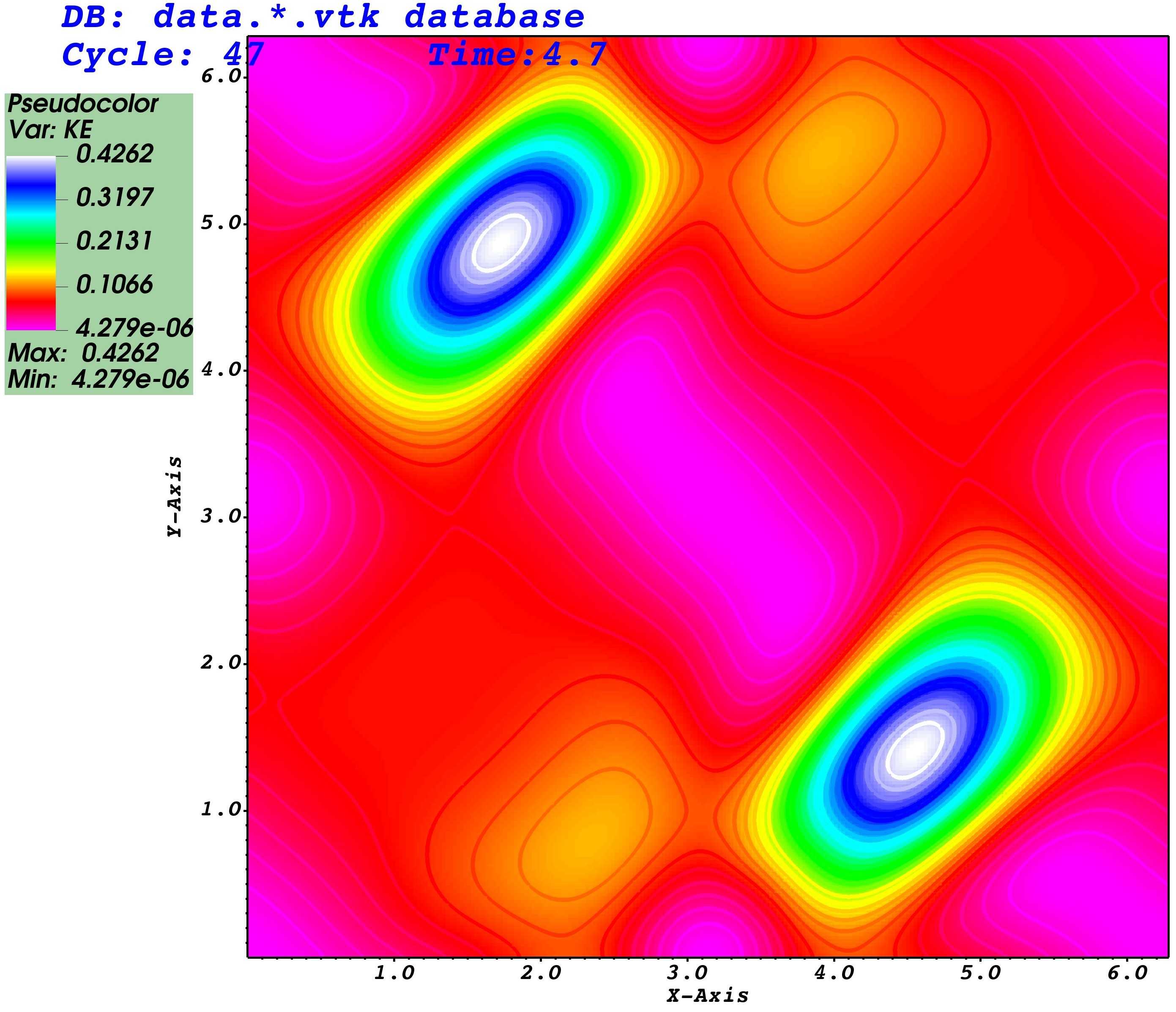}
    	\caption{}
    	\label{2D OT KE PLUTO}
    \end{subfigure}
    \begin{subfigure}{0.48\textwidth}
    	\centering
    	\includegraphics[scale=0.07500]{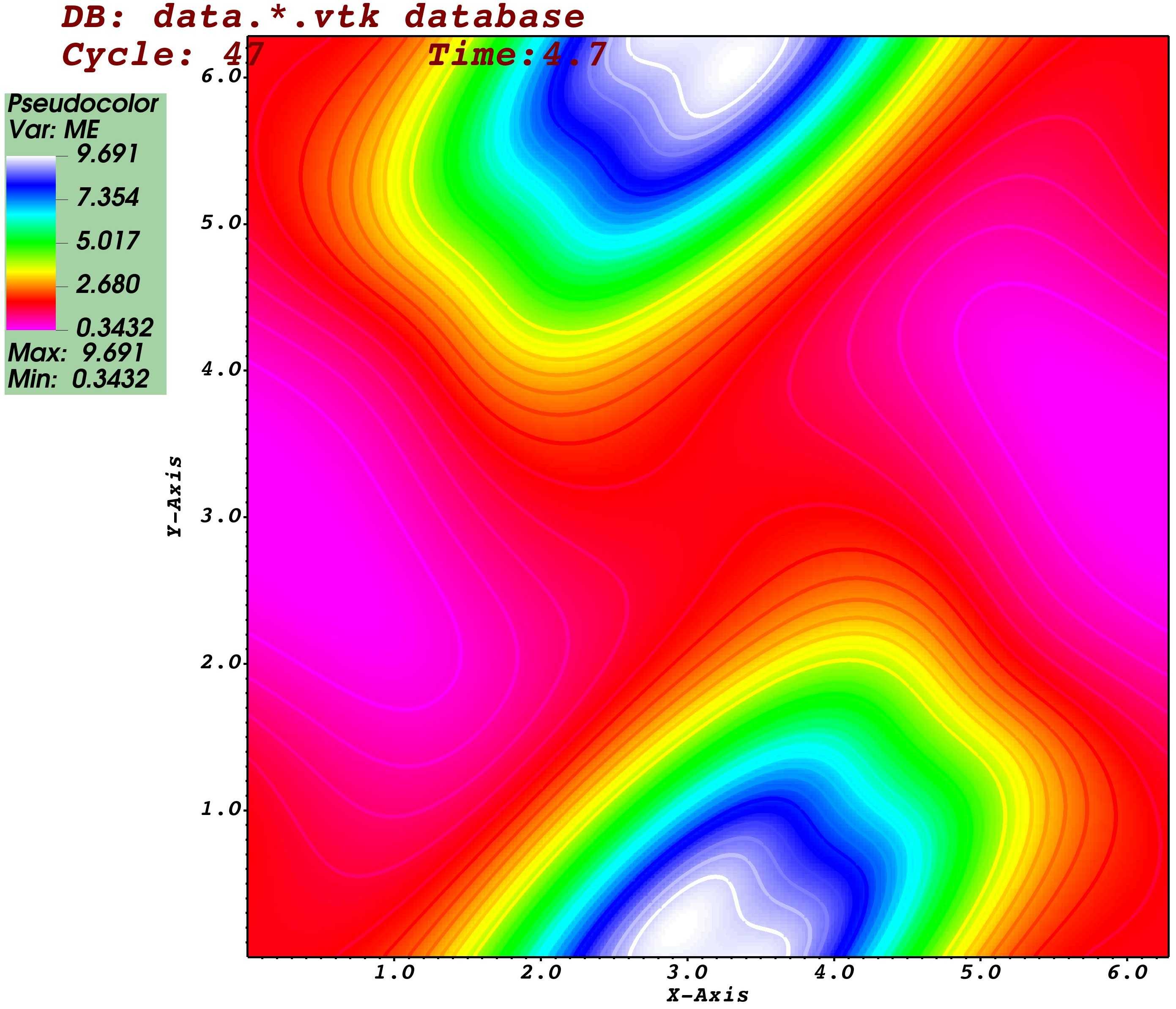}
    	\caption{}
    	\label{2D OT ME PLUTO}
    \end{subfigure}
	\caption{2D Orszag-Tang Flow kinetic energy contour and magnetic energy contour from GMHD3D code (a \& b) and PLUTO4.4 code (c \& d). Simulation Details: Reynolds number $R_e = R_m = 10^{5}$, Grid resolution $N = 512^2$, Time stepping $dt = 10^{-4}$, initial fluid velocity $u_0 = 1.0$, Alfven Mach number $M_A = 1.0$.}
	\label{2D OT Energy Contour}
\end{figure*}

We also visualize the kinetic energy contour [See Fig. \ref{2D OT KE GMHD2D} \& \ref{2D OT KE PLUTO}] and the magnetic energy contour [See Fig. \ref{2D OT ME GMHD2D} \& \ref{2D OT ME PLUTO}] using GMHD3D and PLUTO4.4 data, and we find that they are similar.

%
 
  \subsection{Test 4 [Magnetohydrodynamics]: Coherent  nonlinear oscillations using 2D Cats Eye (CE) Flow} \label{2D CE flow}
  
\begin{figure*}
		\centering
		\begin{subfigure}{0.49\textwidth}
			\centering
			\includegraphics[scale=0.55]{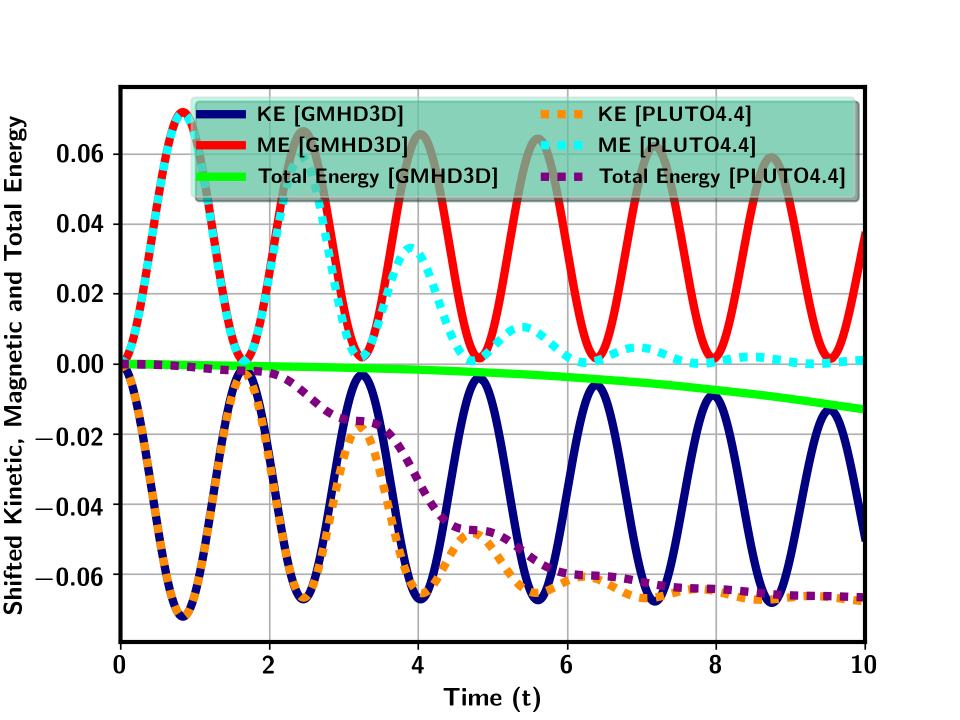}
			\caption{}
			\label{2D CE}
		\end{subfigure}
		\begin{subfigure}{0.49\textwidth}
			\centering
			\includegraphics[scale=0.55]{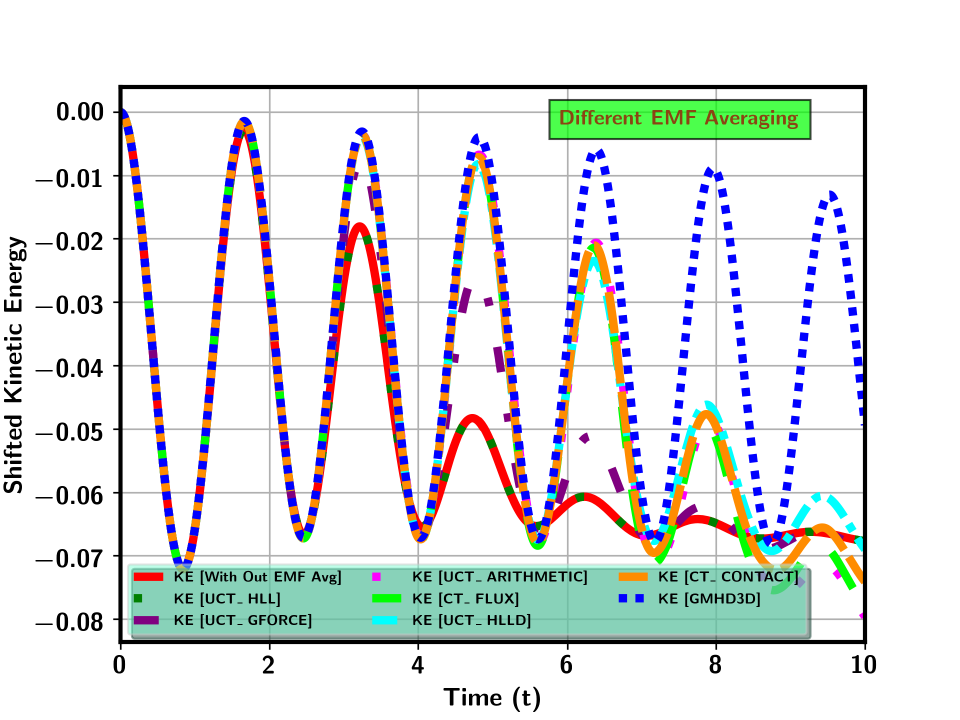}
			\caption{}
			\label{2D CE All EMF}
		\end{subfigure}
		\caption{(a) The shifted kinetic and magnetic energies for 2D Cats Eye Flow from GMHD3D and PLUTO4.4 at grid resolution $128^2$. (b) The shifted kinetic energy for 2D Cats Eye Flow from GMHD2D and PLUTO4.4 (with all CT\_EMF\_AVG schemes)at grid resolution $128^2$. Among all the CT\_EMF\_AVG schemes UCT\_HLLD (shown by the cyan line) and CT\_CONTACT (represented by the orange line) shows the least amount of dissipation. \textcolor{black}{Simulation Details: Time stepping $dt = 10^{-4}$}}
	\end{figure*}

  The velocity profile for 2D Cats Eye [CE] flow is given by,
  	\begin{equation}\label{2D CE Flow}
  		\begin{aligned}
  			u_x &=  u_0 [ \sin(k_0x) \cos(k_0y) - A \cos(k_0x) \sin(k_0y)]\\
  			u_y &=  - u_0 [ \cos(k_0x) \sin(k_0y) + A \sin(k_0x) \cos(k_0y)]
  		\end{aligned}
  	\end{equation}
 with $A$ equal to $0.5$ and $k_0$ equal to $1$. Here, we consider that the Alfven Mach number $M_A = 1.0$ and the initial speed of the fluid, $u_0 = 1.0$, from which the initial magnetic field strength is determined. For this simulation, we use a grid resolution of $128^2$ and set the sonic Mach number ($M_s$) to $0.01$. The conversion of energy from the kinetic to the magnetic mode is shown clearly in Fig. \ref{2D CE}. It is also evident that, the oscillations are significantly dampened for the PLUTO4.4 solver [See Fig. \ref{2D CE}].



  We employ various electric field averaging approach (CT\_EMF\_AVG) in PLUTO4.4, similar to the previously stated instance, in order to improve the precision of the results. If we adopt the UCT\_HLL  \cite{UCT_HLL:2003, UCT_HLL:2004} technique, we find that the kinetic and magnetic energy oscillations are similarly dampened. As seen in Fig. \ref{2D CE All EMF}, if we do not employ any of the CT\_EMF\_AVG algorithms, the results appear to be the same with UCT\_HLL scheme. While investigating the other schemes like, UCT\_GFORCE \cite{GFORCE:2021}, ARITHMETIC \cite{ARITHMETIC:1999}, CT\_FLUX, UCT\_HLLD \cite{GFORCE:2021}, and CT\_CONTACT \cite{CT:2005}, we notice that UCT\_HLLD (shown by the cyan line) and CT\_CONTACT (represented by the orange line) exhibit the least amount of dissipation [See Fig. \ref{2D CE All EMF}]. The results from GMHD3D and PLUTO4.4 still differ significantly from one another. This discrepancy may be owing to the fact that PLUTO4.4 has a higher numerical viscosity.


\begin{figure*}
 	\centering
 	\begin{subfigure}{0.49\textwidth}
 		\centering
 		\includegraphics[scale=0.55]{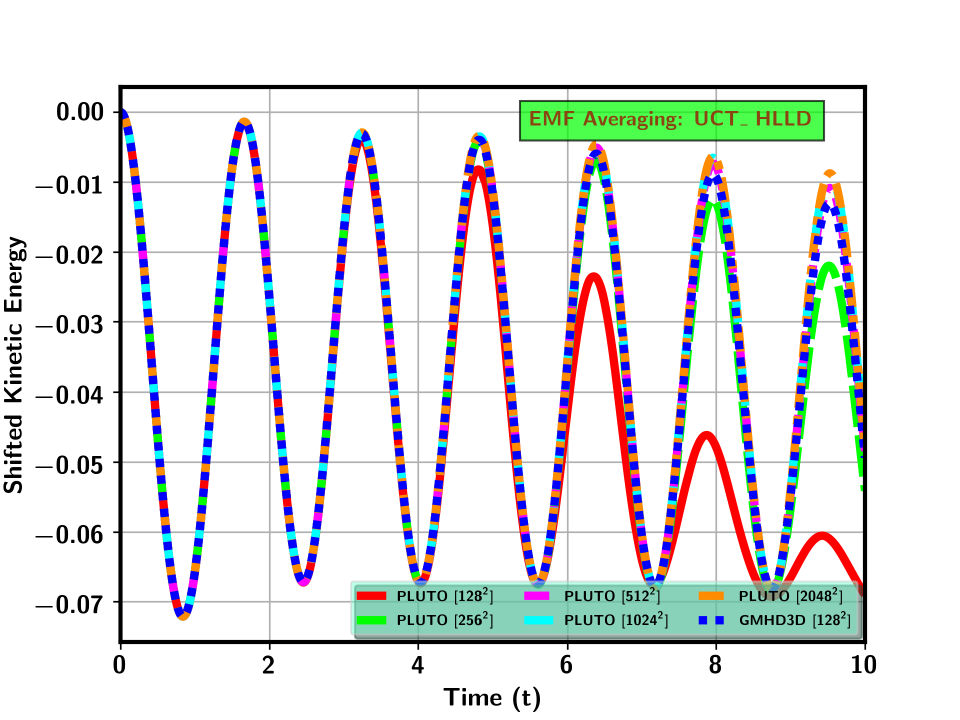}
 		\caption{}
 		\label{2D CE UCT HLLD EMF High}
 	\end{subfigure}
 	\begin{subfigure}{0.49\textwidth}
 		\centering
 		\includegraphics[scale=0.55]{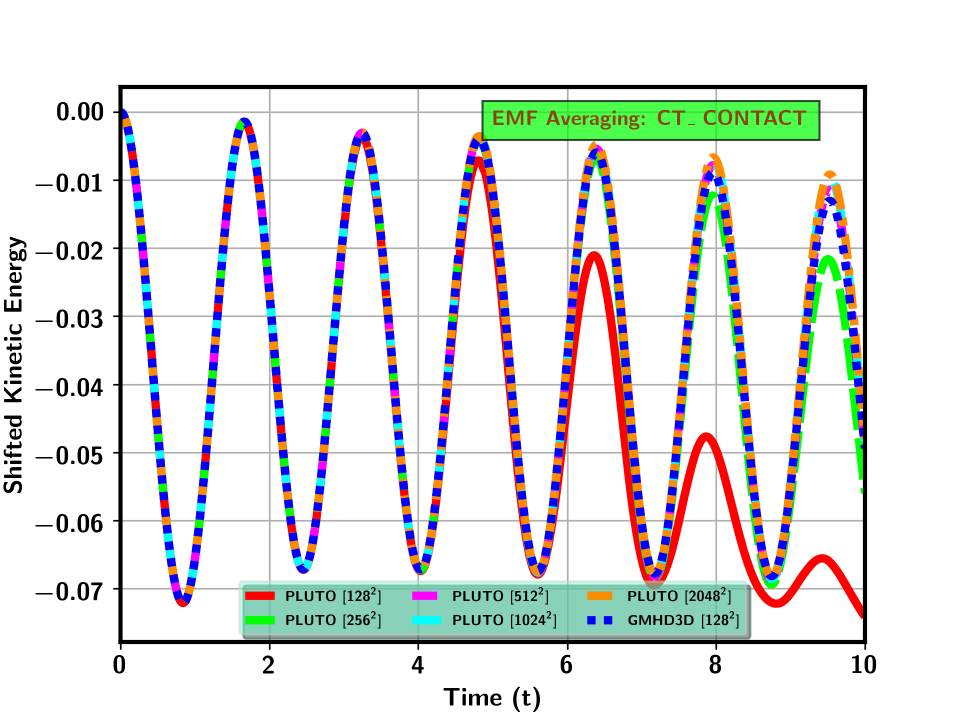}
 		\caption{}
 		 	\label{2D CE CT CONTACT EMF High}
 	\end{subfigure}
 	\caption{The shifted kinetic energy for 2D Cats Eye Flow from GMHD3D at grid resolution: $128^2$ and PLUTO4.4 at grid resolution: $128^2$, $256^2$, $512^2$, $1024^2$ and $2048^2$ using (a) UCT\_HLLD scheme (b) CT\_CONTACT scheme. PLUTO4.4 data at grid resolution $512^2$ (magenta line) agrees with the GMHD3D data at grid resolution $128^2$ (blue dotted line). \textcolor{black}{Simulation Details: Time stepping $dt = 10^{-4}$}}
 \end{figure*}

  To eliminate the impact of numerical viscosity and double-check the resolution effect, we have increased the grid resolution in PLUTO4.4 from $128^2$ to $2048^2$. Since UCT\_HLLD [See Fig. \ref{2D CE UCT HLLD EMF High}] and CT\_CONTACT [See Fig. \ref{2D CE CT CONTACT EMF High}] are the least dissipative with respect to others [See Fig. \ref{2D CE All EMF}] at grid resolution $128^2$, we focus our attention only on these two schemes for our higher resolution analysis. We can see that the PLUTO4.4 data at grid resolution $512^2$ (represented by magenta line) agrees with the GMHD3D data at grid resolution $128^2$ (shown by blue dotted line) by comparing Fig. \ref{2D CE UCT HLLD EMF High} \& Fig. \ref{2D CE CT CONTACT EMF High}.

\begin{figure*}
 	\centering
 	\begin{turn}{90} 
 		\normalsize{\textbf{\textcolor{blue}{GMHD3D}}}
 	\end{turn}
 	\begin{subfigure}{0.48\textwidth}
 		\centering
 		\includegraphics[scale=0.07500]{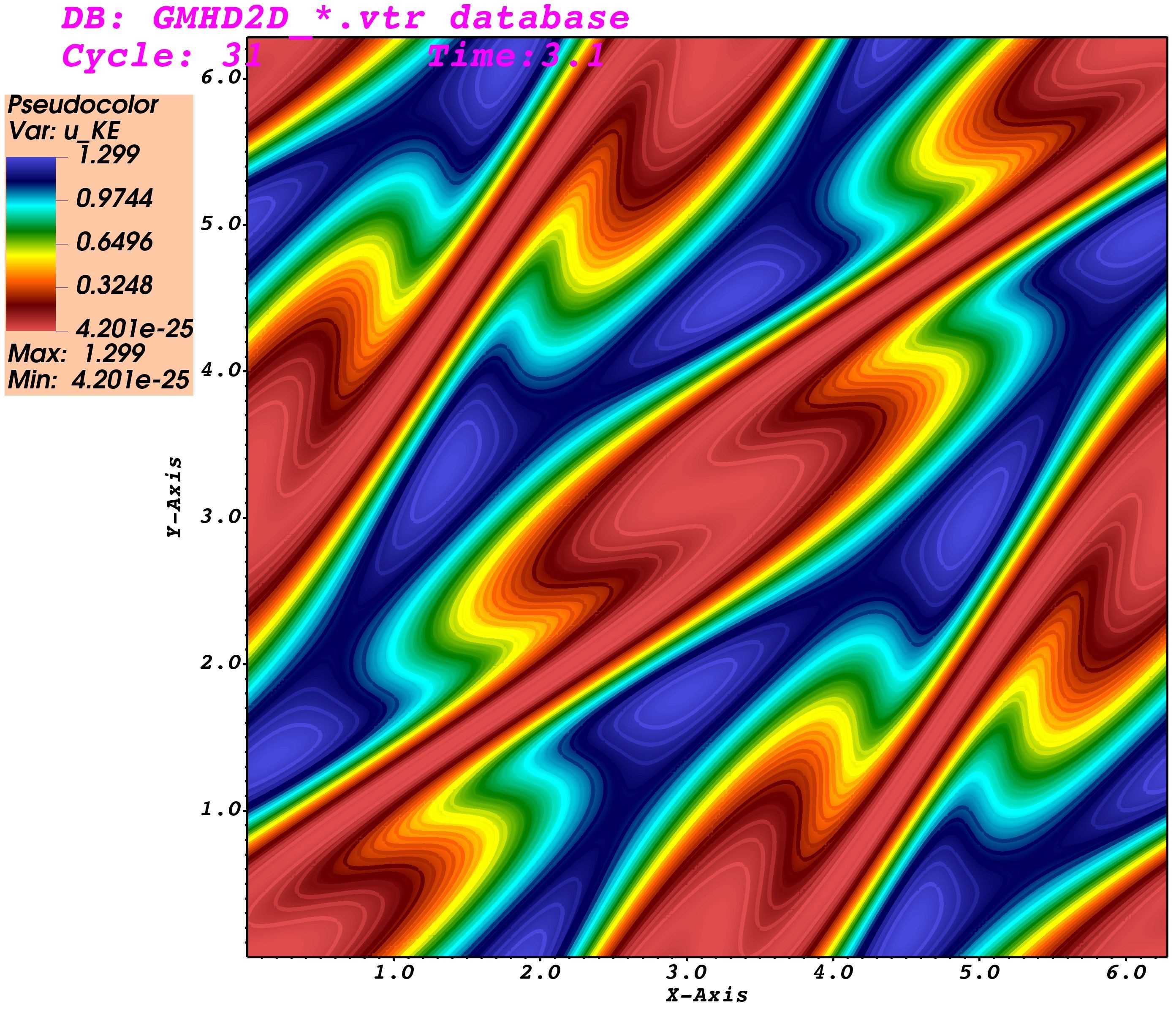}
 		\caption{}
 		\label{2D CE KE GMHD2D}
 	\end{subfigure}
 	\begin{subfigure}{0.48\textwidth}
 		\centering
 		\includegraphics[scale=0.07500]{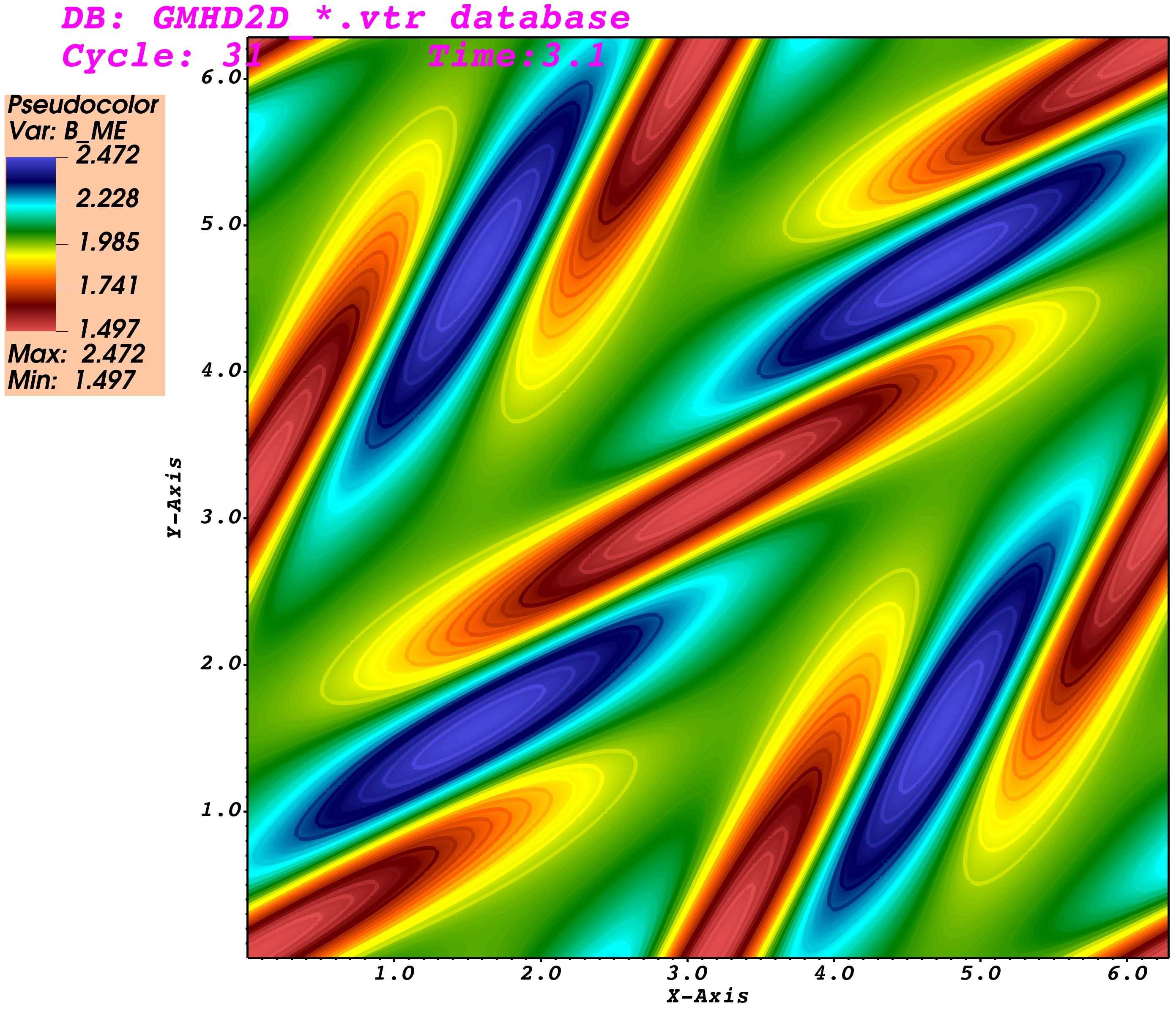}
 		\caption{}
 		\label{2D CE ME GMHD2D}
 	\end{subfigure}
 	\begin{turn}{90} 
 		\normalsize{\textbf{\textcolor{blue}{PLUTO4.4}}}
 	\end{turn}
 	\begin{subfigure}{0.48\textwidth}
 		\centering
 		\includegraphics[scale=0.07500]{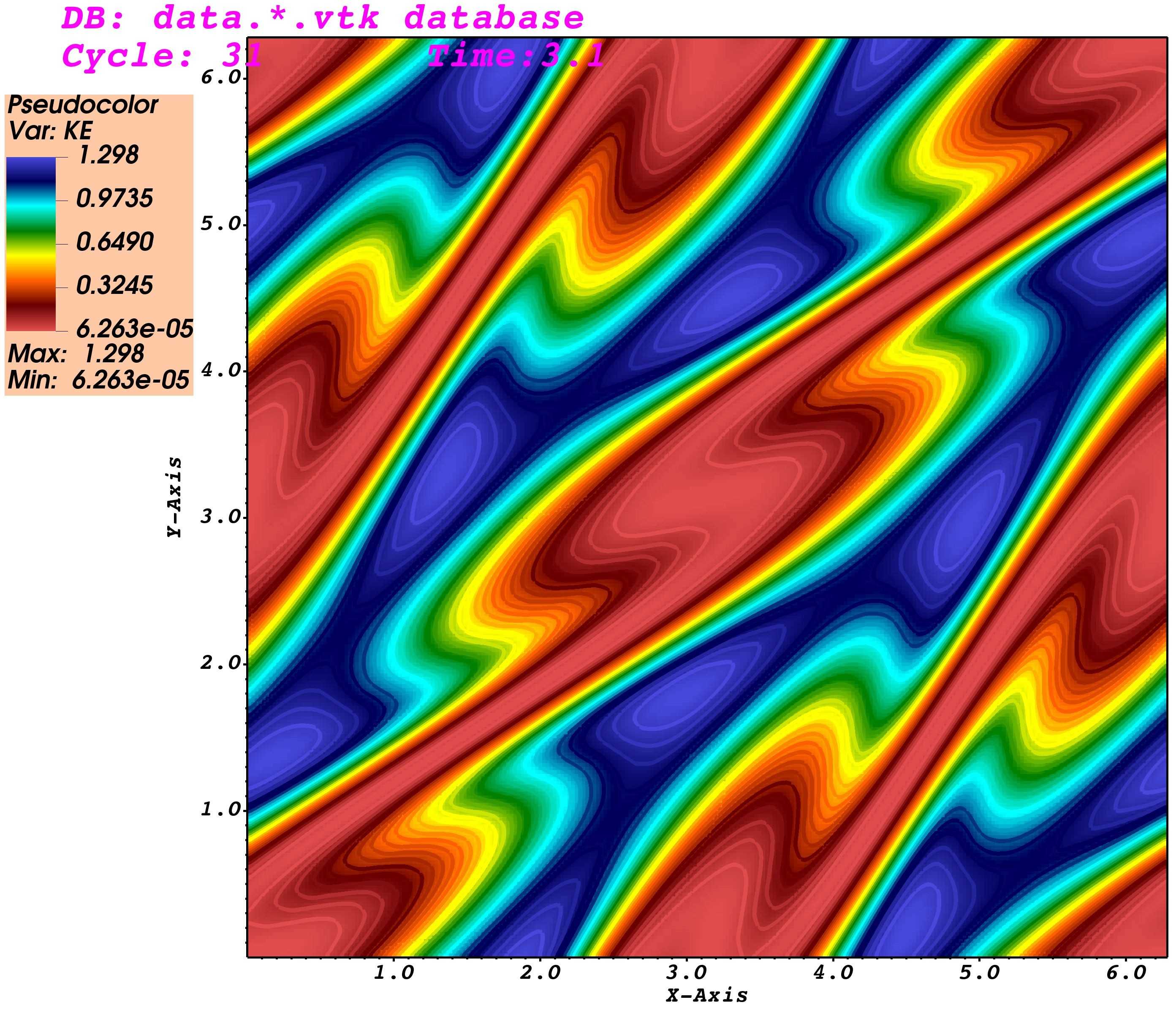}
 		\caption{}
 		\label{2D CE KE PLUTO}
 	\end{subfigure}
 	\begin{subfigure}{0.48\textwidth}
 		\centering
 		\includegraphics[scale=0.07500]{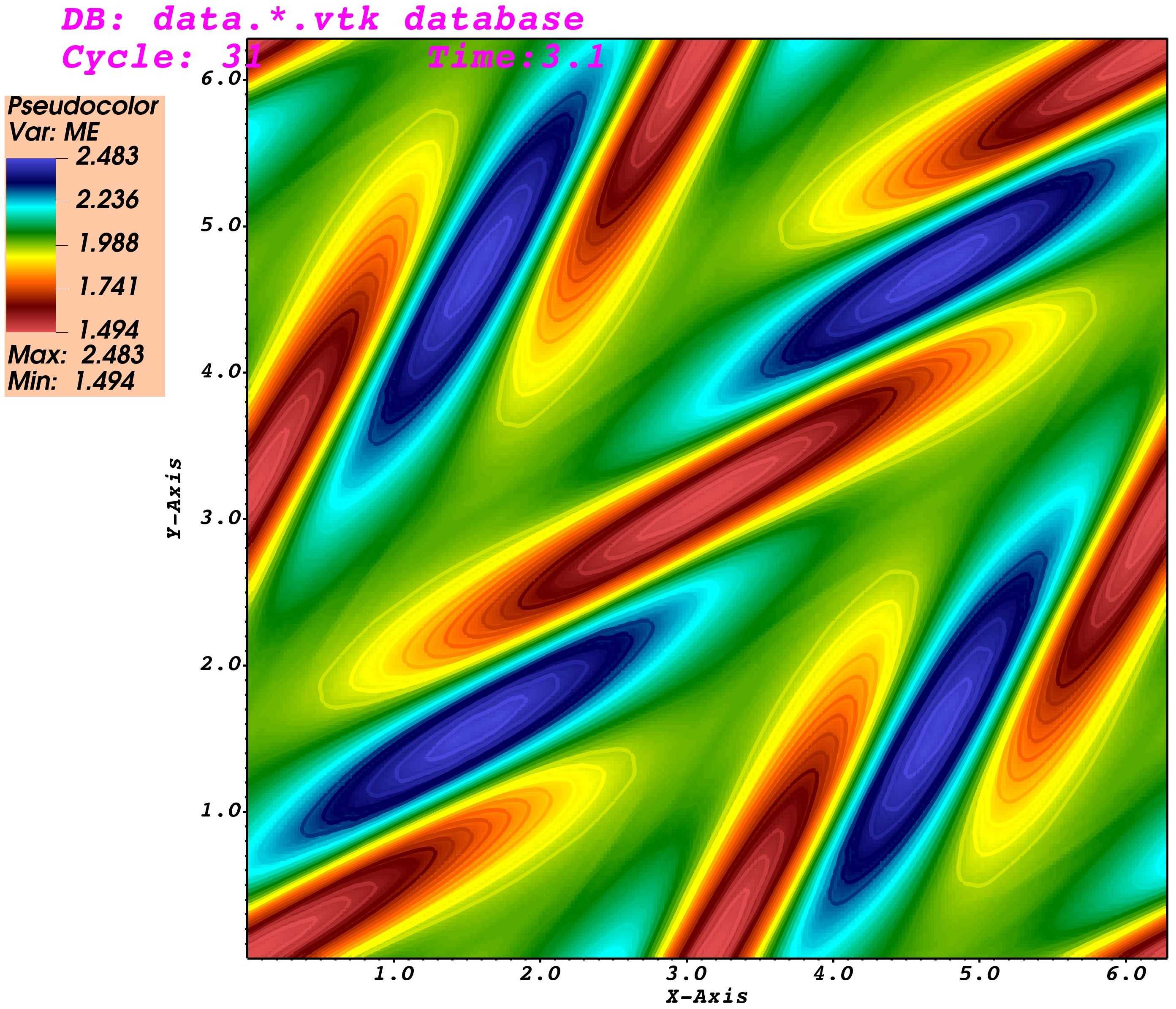}
 		\caption{}
 		\label{2D CE ME PLUTO}
 	\end{subfigure}
 	\caption{2D Cats Eye Flow kinetic energy contour and magnetic energy contour from GMHD3D code (a \& b) and PLUTO4.4 code (c \& d). Simulation Details: Reynolds number $R_e = R_m = 10^{5}$, Grid resolution $N = 512^2$, Time stepping $dt = 10^{-4}$, initial fluid velocity $u_0 = 1.0$, Alfven Mach number $M_A = 1.0$.}
 \end{figure*}

Moreover, we compare the kinetic [See Fig. \ref{2D CE KE GMHD2D} \& \ref{2D CE KE PLUTO}] and magnetic [See Fig. \ref{2D CE ME GMHD2D} \& \ref{2D CE ME PLUTO}] energy contours from GMHD3D and PLUTO4.4 data. It has been established that the contours for both codes are identical.

%


 \subsection{Test 5 [Magnetohydrodynamics]: Coherent  nonlinear oscillations using 3-dimensional astrophysical Flows}\label{3D flows}
In this subsection, we discuss the dynamics of some well-known three-dimensional astrophysical flows \textcolor{black}{for example: Taylor-Green flow (See Appendix \ref{Appen A} for details), Archontis flow (See Appendix \ref{Appen A} for details), Cats Eye flow (See Appendix \ref{Appen A} for details) \& Arnold–Beltrami–Childress Flow}. 

\subsubsection{3D Arnold–Beltrami–Childress [ABC] Flow}

\begin{figure*}
	\centering
	\begin{subfigure}{0.32\textwidth}
		\centering
		\includegraphics[scale=0.4]{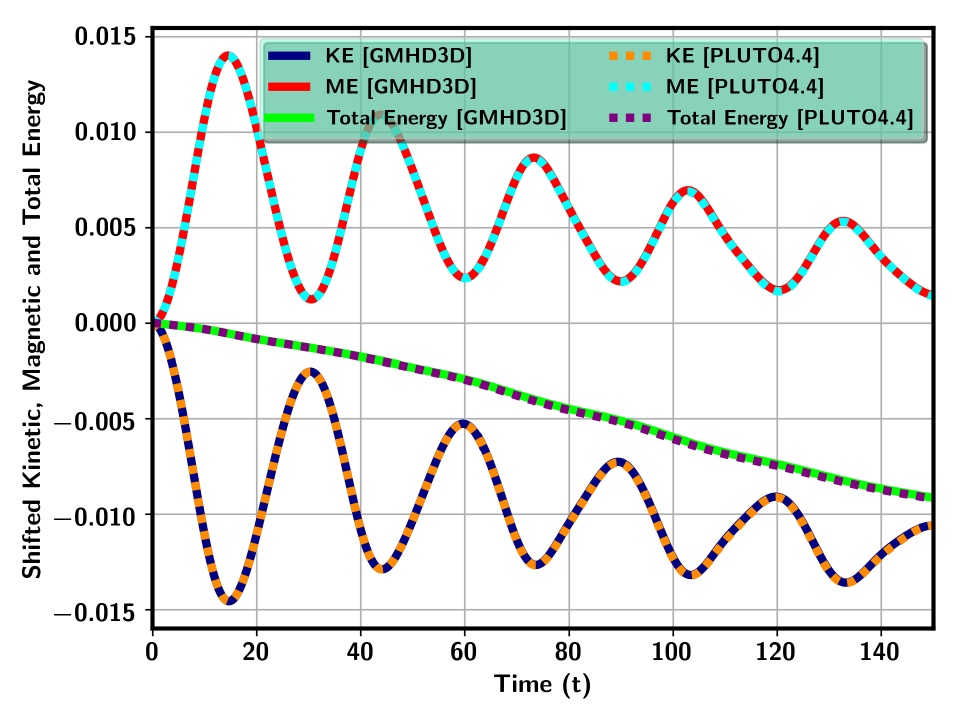}
		\caption{}
		\label{3D ABC flow energy}
	\end{subfigure}
	\begin{subfigure}{0.32\textwidth}
		\centering
		\includegraphics[scale=0.047]{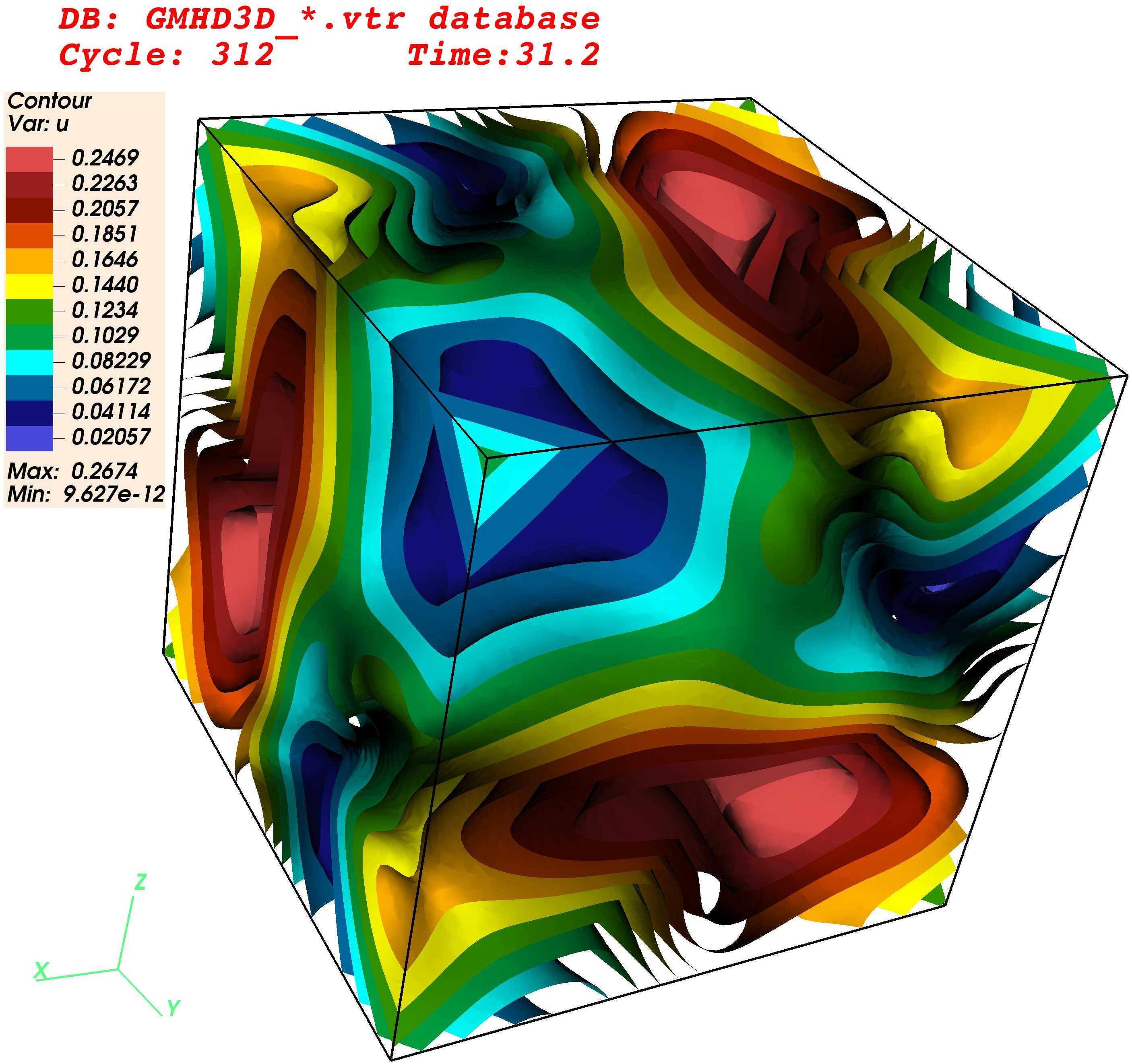}
		\caption{}
		\label{3D ABC ISOV GMHD3D}
	\end{subfigure}
	\begin{subfigure}{0.32\textwidth}
		\centering
		\includegraphics[scale=0.047]{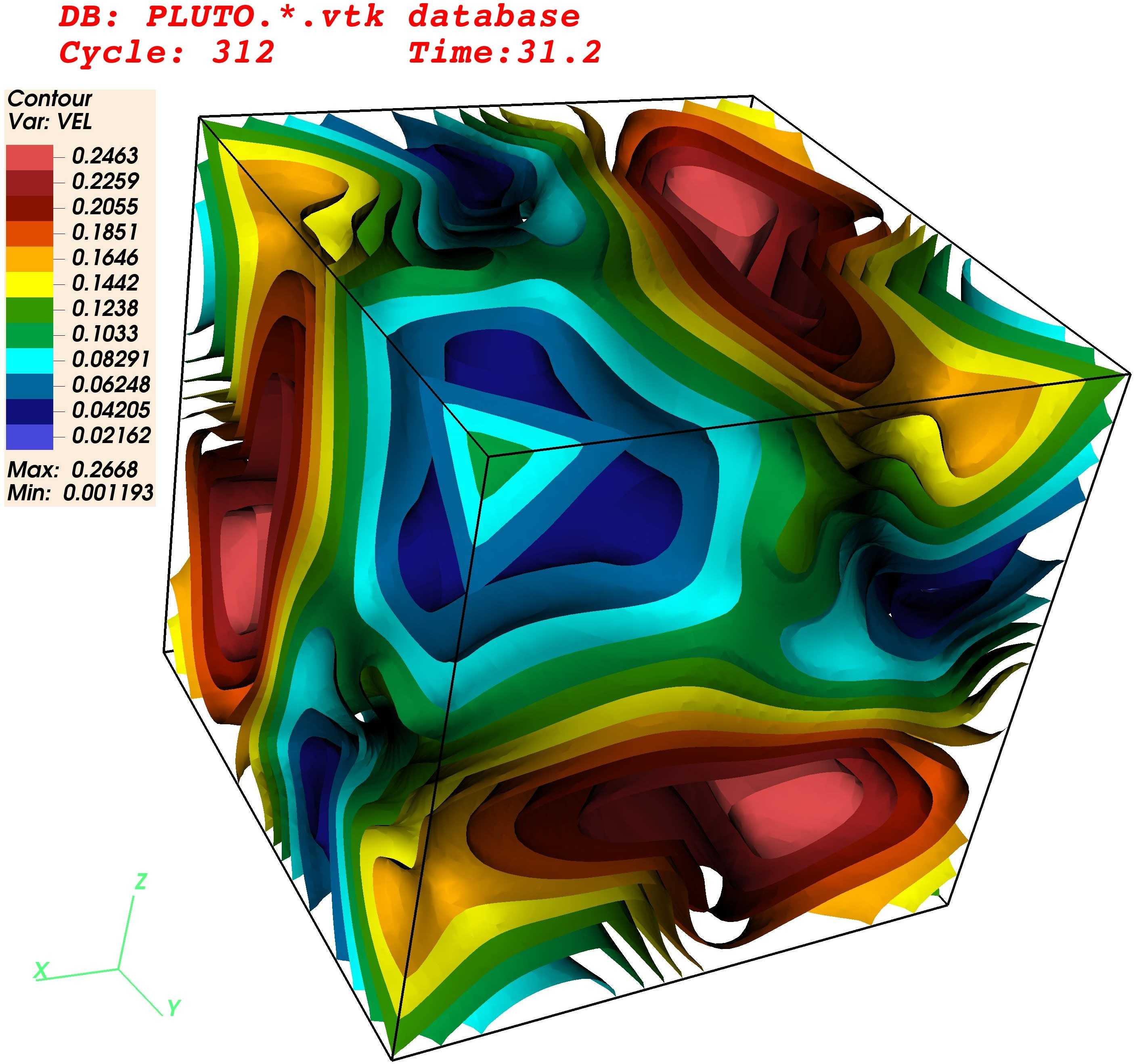}
		\caption{}
		\label{3D ABC ISOV PLUTO}
	\end{subfigure}
	\caption{(a) The shifted kinetic and magnetic energies for 3D Arnold-Beltrami-Childress [ABC] flow from GMHD3D and PLUTO4.4 code. The visualization of velocity iso-surface (Iso-V) for 3D Arnold-Beltrami-Childress [ABC] flow at any arbitrary time from (b) GMHD3D code  and (c) PLUTO4.4 code. Simulation Details: Reynolds number $R_e = R_m = 1000$, Grid resolution $N = 128^3$, Time stepping $dt = 10^{-4}$, initial fluid velocity $u_0 = 1.0$, Alfven Mach number $M_A = 1.0$.}
	\label{3D ABC Iso V}
\end{figure*}

Lastly, we look at the most well-known flow in astrophysics, which is called the 3D Arnold-Beltrami-Childress flow, or 3D ABC flow in short. The flow is divergence-free, and it is widely acknowledged in the astrophysical research area for its complicated nature and numerous symmetries. The velocity profile for 3D ABC flow is given by,
		\begin{equation}\label{ABC}
			\begin{aligned}
				u_x &= u_0 [ A \sin(k_0z) + C \cos(k_0y) ]\\
				u_y &= u_0 [ B \sin(k_0x) + A \cos(k_0z) ]\\
				u_z &= u_0 [ C \sin(k_0y) + B \cos(k_0x) ]
			\end{aligned}
		\end{equation}
with $A = B = C = 1.0$ and $k_0 = 1.0$. The remaining parameters are identical to those used in previous numerical experiments. The consistent and periodic exchange of energy between kinetic and magnetic modes is shown in Fig. \ref{3D ABC flow energy} in the form of coherent non-linear oscillation. The oscillation periods are measured to be $T = 30.171$ for both codes.

We also visualize the velocity iso-surface (Iso-V) using data from both codes and confirm that the two iso-surfaces (Iso-V) are identical [See Fig. \ref{3D ABC ISOV GMHD3D} \& \ref{3D ABC ISOV PLUTO}].
Using 3D ABC flow from both codes, we further study some parameter scanning.

\begin{figure*}
	\centering
	\begin{subfigure}{0.32\textwidth}
		\centering
		\includegraphics[scale=0.37]{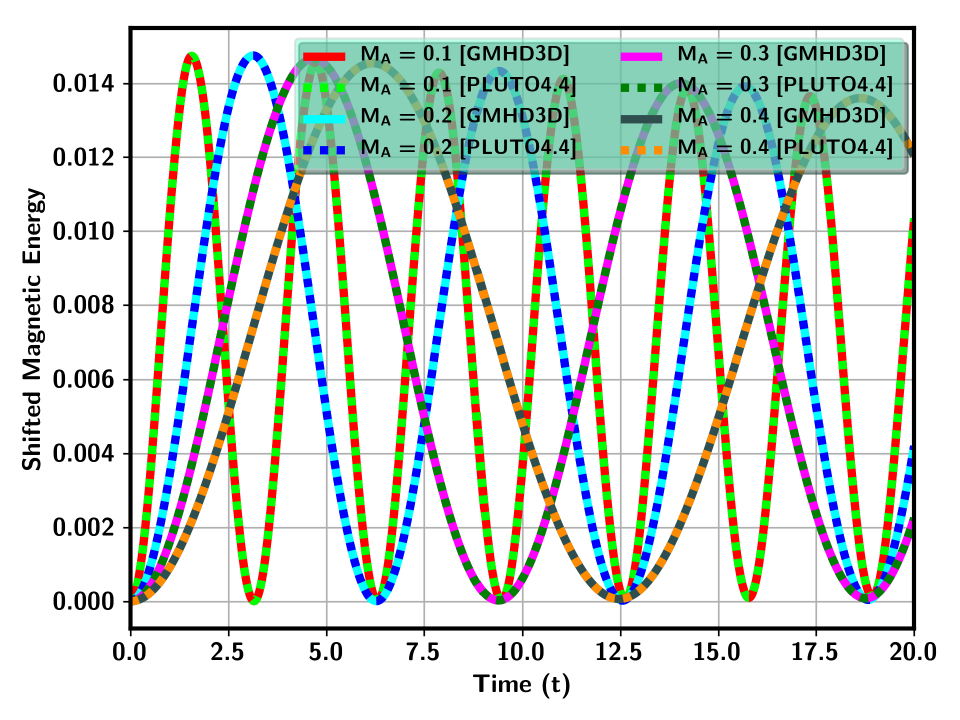}
		\caption{}
		\label{3D ABC Ma 0p1 to 0p4}
	\end{subfigure}
	\begin{subfigure}{0.32\textwidth}
		\centering
		\includegraphics[scale=0.37]{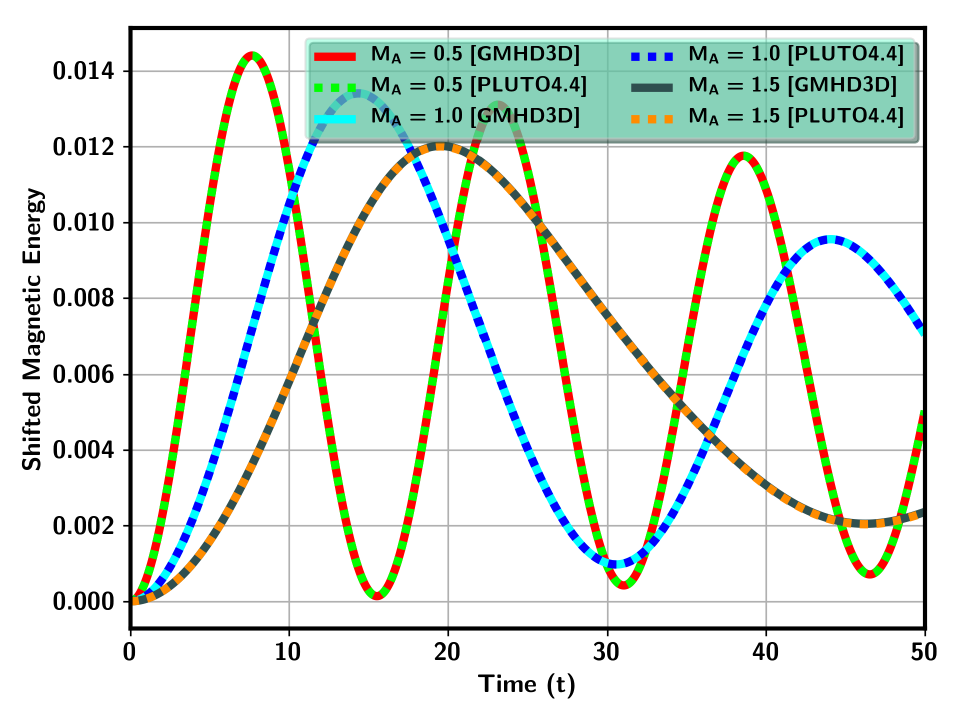}
		\caption{}
		\label{3D ABC Ma 0p5 to 1p5}
	\end{subfigure}
	\begin{subfigure}{0.32\textwidth}
		\centering
		\includegraphics[scale=0.37]{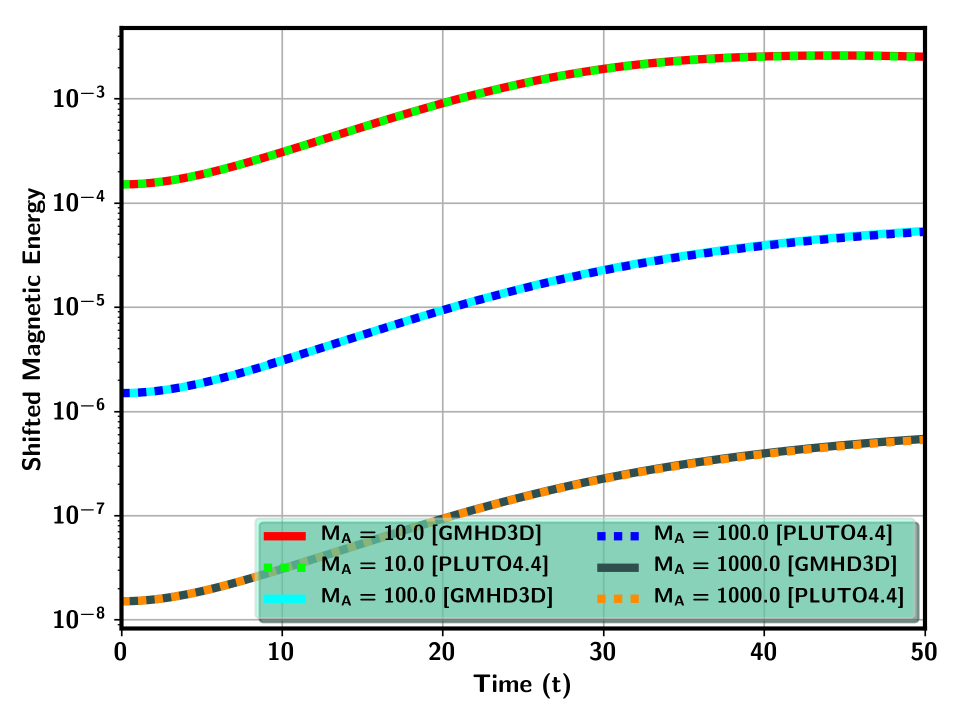}
		\caption{}
		\label{3D ABC Ma 10p0 to 1000p0}
	\end{subfigure}
	\caption{Shifted magnetic energy for 3D Arnold–Beltrami–Childress [ABC] flow with Alfven mach number (a) $M_a$ = 0.1, 0.2, 0.3, 0.4, (b) $M_A$ = 0.5, 1.0, 1.5 (c) growth of  magnetic energies with Alfven Mach number ($M_A$) = 10.0, 100.0, 1000.0 from GMHD3D code and PLUTO4.4 code at grid resolution $64^3$. It appears that the results from both codes are identical. \textcolor{black}{Simulation Details: Time stepping $dt = 10^{-4}$}.}
	\label{3D ABC different Ma}
\end{figure*}

We begin by investigating the impact of Alfven speed on coherent non-linear oscillations. As shown in Figures \ref{3D ABC Ma 0p1 to 0p4} \& \ref{3D ABC Ma 0p5 to 1p5}, the period of oscillation of energy (kinetic and magnetic) linearly increases with increasing of Alfven Mach number ($M_A$) from both codes when the initial wave number ($k_0$) remains constant at $1.0$ and the Alfven Mach number ($M_A$) is varied over the range $0.1, 0.2, 0.3, 0.4, 0.5, 1.0, 1.5$. According to recently published works, these results are quite predictable and consistent \cite{RM_Conh:2019}.

Moreover, we see that the oscillation completely disappears as the Alfven Mach number ($M_A$) is increased to very high limits such as $M_A = 10, 100, 1000$. In addition, a notable saturation of magnetic energy is observed, followed by a growth [See Fig. \ref{3D ABC Ma 10p0 to 1000p0}]. The concept of ``dynamo action'' is used to describe this process of magnetic energy growth.


In the following, we present findings from both codes for a wide range of the initial wave number ($k_0$).

\begin{figure*}
	\centering
	\begin{subfigure}{0.49\textwidth}
		\centering
		\includegraphics[scale=0.50]{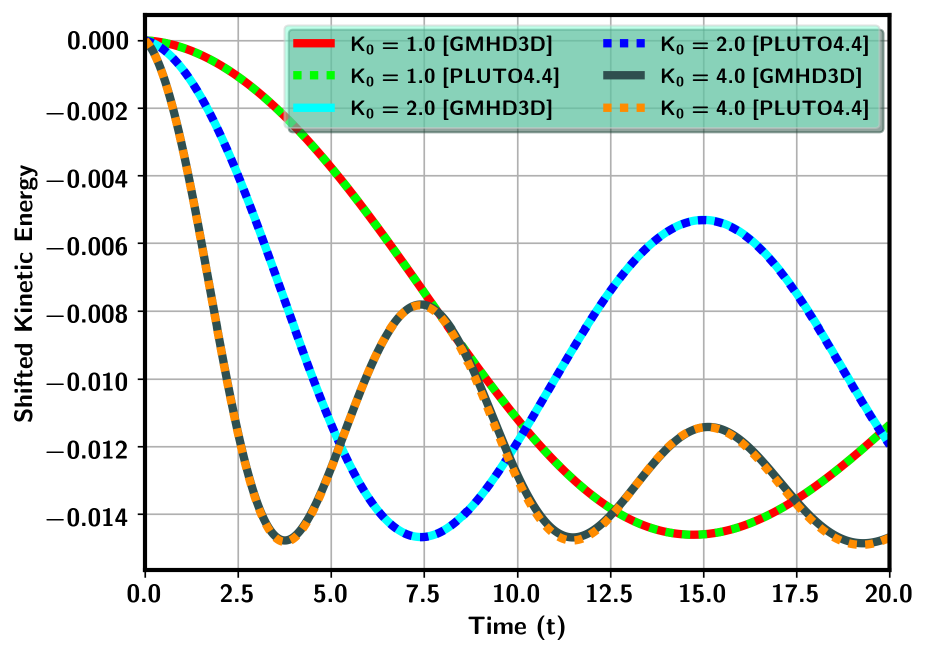}
		\caption{}
		\label{3D ABC k0 1 to 4}
	\end{subfigure}
	\begin{subfigure}{0.49\textwidth}
		\centering
		\includegraphics[scale=0.50]{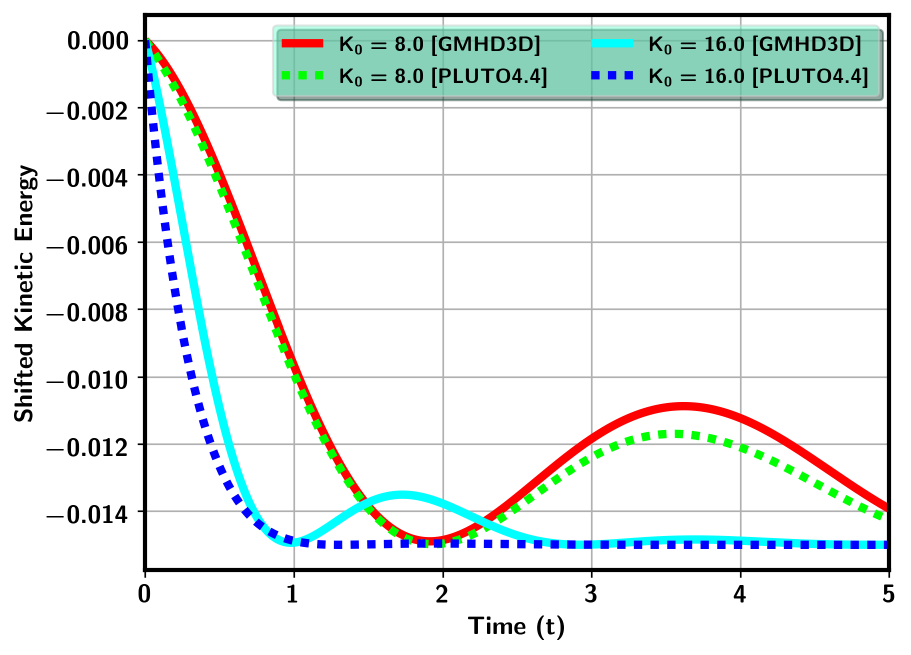}
		\caption{}
		\label{3D ABC k0 8 to 16}
	\end{subfigure}
	\caption{Shifted kinetic energy for 3D Arnold–Beltrami–Childress [ABC] Flow with initial wave number (a) $k_0 = 1.0, 2.0, 4.0$ and (b) $k_0 =8.0, 16.0$ from GMHD3D code and PLUTO4.4 code at grid resolution $64^3$. For lower wave numbers, the oscillations are exactly reproduced from both codes but for higher wave numbers the oscillations are heavily damped for PLUTO4.4 code. \textcolor{black}{Simulation Details: Time stepping $dt = 10^{-4}$}.}
	\label{3D ABC different k0}
\end{figure*}

We fix the Alfven Mach number ($M_A$) at $1.0$ and vary $k_0$ value in the range $k_0 = 1, 2, 4, 8, 16$ to analyze the impact of the initial mode number on coherent nonlinear oscillation. It can be seen in Fig. \ref{3D ABC k0 1 to 4} \& \ref{3D ABC k0 8 to 16} that when the initial wave number ($k_0$) increases, the frequency of oscillation increases, i.e. the time period of oscillation reduces. For lower wave numbers, such as $k_0 = 1.0, 2.0, 4.0$, the energy oscillation is reproduced identically by both codes [See Fig. \ref{3D ABC k0 1 to 4}]; however, for higher wave numbers, such as $k_0 = 8.0, 16.0$, the oscillation of energy is heavily damped for the PLUTO4.4 code [See Fig. \ref{3D ABC k0 8 to 16}].

In order to improve PLUTO4.4 results, we first adopt the same approach as we conducted for the 2D Cats Eye [CE] flow, i.e., we investigate all of the available CT\_EMF\_AVG schemes.

\begin{figure*}
	\centering
	\begin{subfigure}{0.49\textwidth}
		\centering
		\includegraphics[scale=0.50]{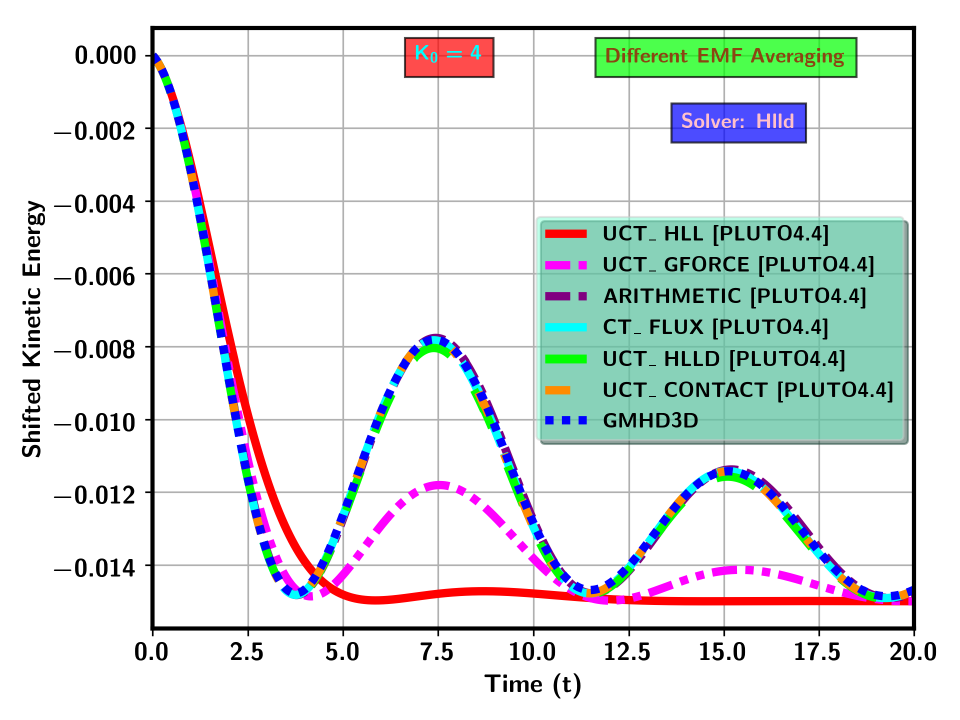}
		\caption{}
		\label{3D ABC k0_4 all emf}
	\end{subfigure}
	\begin{subfigure}{0.49\textwidth}
		\centering
		\includegraphics[scale=0.50]{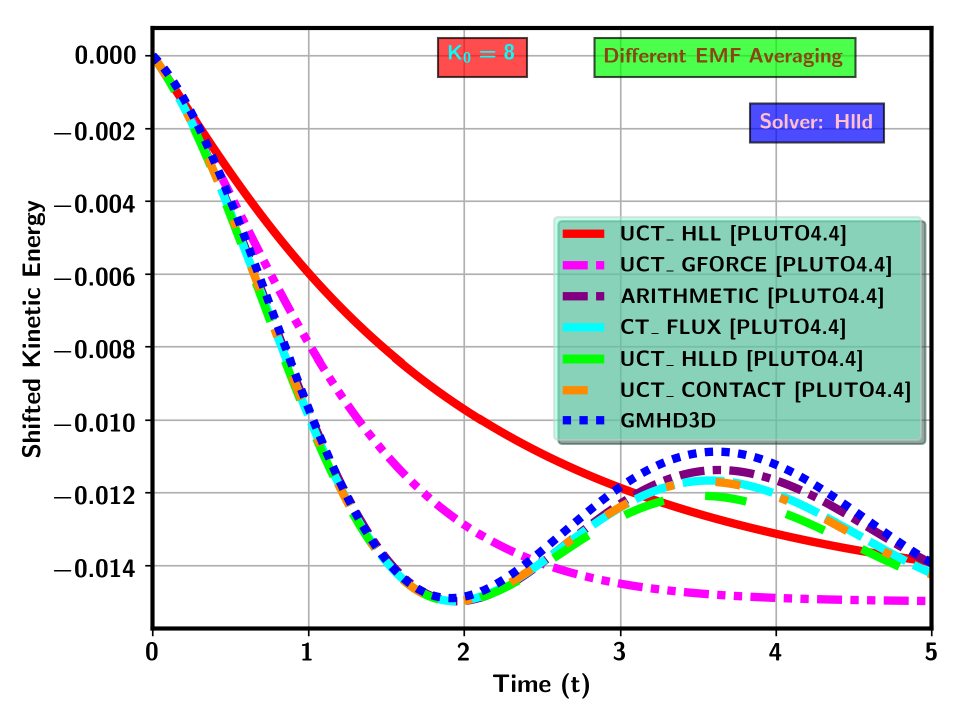}
		\caption{}
		\label{3D ABC k0_8 all emf}
	\end{subfigure}
	\caption{Shifted kinetic energy using 3D Arnold–Beltrami–Childress [ABC] Flow with initial wave number (a) $k_0 = 8.0$ and (b) $k_0 = 16.0$ from GMHD3D code and PLUTO4.4 code (with all available  CT\_EMF\_AVG schemes) at grid resolution $64^3$. Among all the schemes UCT\_HLL and UCT\_GFORCE have height diffusivity. \textcolor{black}{Simulation Details: Time stepping $dt = 10^{-4}$}.}
\end{figure*}

From Fig. \ref{3D ABC k0 1 to 4}, we can see that when $k_0 = 1, 2, 4$, the data for both codes are exactly the same. For the sake of completeness, we investigate all electric field averaging techniques for the $k_0 = 4.0$ scenario and find that, with the exception of the UCT\_HLL \cite{UCT_HLL:2003,UCT_HLL:2004} and UCT\_GFORCE \cite{GFORCE:2021} schemes, the findings are identical for all schemes [See Fig. \ref{3D ABC k0_4 all emf}]. In comparison to other schemes, these two are discovered to have a height diffusive effect.

Figure \ref{3D ABC k0 8 to 16} shows that the oscillations are significantly dampened for larger wave numbers ($k_0 = 8, 16$). We now look into the most efficient emf averaging techniques for $k_0 = 8.0$. From Fig. \ref{3D ABC k0_8 all emf}, we can see that, of all the possible schemes, only the ARITHMETIC \cite{ARITHMETIC:1999} and CT\_CONTACT \cite{CT:2005} ones are the best ones for this present case. It is important to notice that the PLUTO4.4 data does not perfectly correspond with the GMHD3D data at the $64^3$ grid resolution, even if we are utilizing ARITHMETIC \cite{ARITHMETIC:1999} and CT\_CONTACT \cite{CT:2005} schemes.

\begin{figure*}
 	\centering
 	\begin{subfigure}{0.49\textwidth}
 		\centering
 		\includegraphics[scale=0.50]{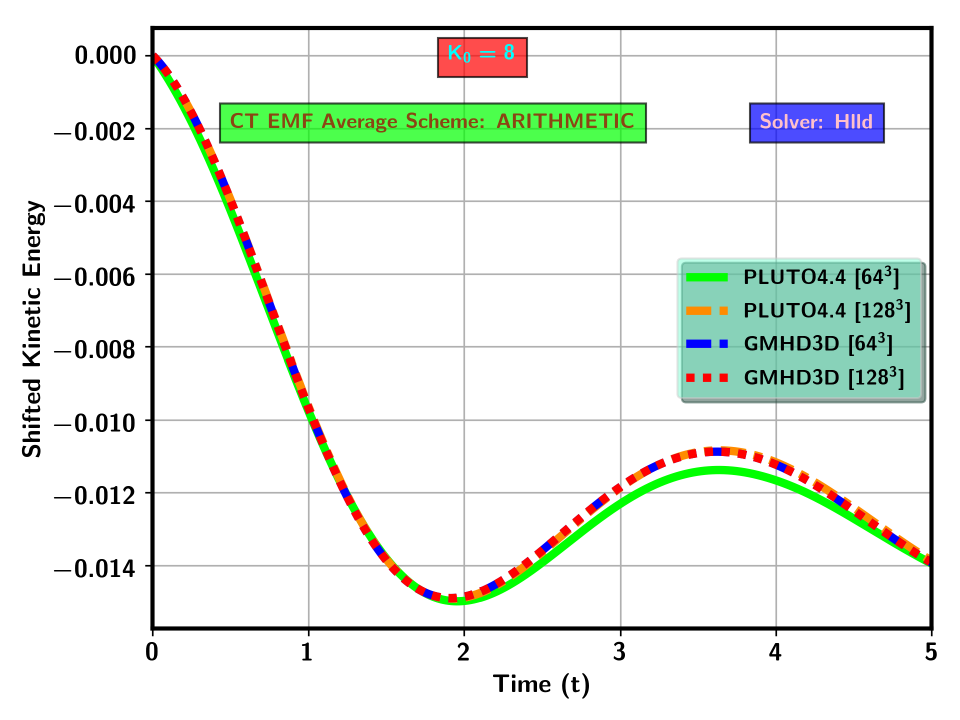}
 		\caption{}
 	\end{subfigure}
 	\begin{subfigure}{0.49\textwidth}
 		\centering
 		\includegraphics[scale=0.50]{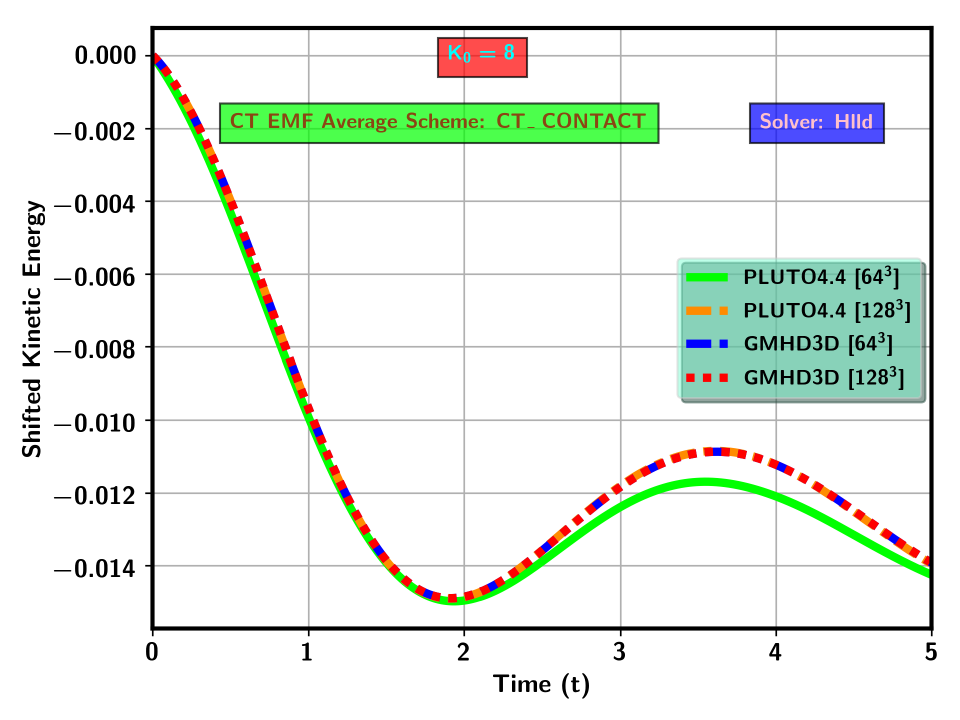}
 		\caption{}
 	\end{subfigure}
 	\caption{Shifted kinetic energy using 3D Arnold–Beltrami–Childress [ABC] Flow for $k_0 = 8.0$ from GMHD3D code and PLUTO4.4 code with (a) ARITHMETIC scheme (b) CT\_CONTACT scheme. PLUTO4.4 code needs atleast $128^3$ grid resolution to reproduce the oscillation that we get from GMHD3D code at $64^3$ grid resolution. \textcolor{black}{Simulation Details: Time stepping $dt = 10^{-4}$}.}
 	\label{3D ABC different k0_8 higher resolution}
 \end{figure*}

 To further improve our results, we provide higher resolution runs for PLUTO4.4 with ARITHMETIC and CT\_CONTACT averaging schemes. It is easy to observe from Fig. \ref{3D ABC different k0_8 higher resolution} that the PLUTO4.4 data with a greater resolution, i.e. $128^3$, matches the GMHD3D data with a lower resolution, $64^3$, exactly [See Fig. \ref{3D ABC different k0_8 higher resolution}]. The results are quite encouraging, as it demonstrate the superior accuracy of the spectral solver compared to that of the grid-based solver in a triply periodic domain.
 
\begin{figure*}
	\centering
	\begin{subfigure}{0.32\textwidth}
		\centering
		\includegraphics[scale=0.37]{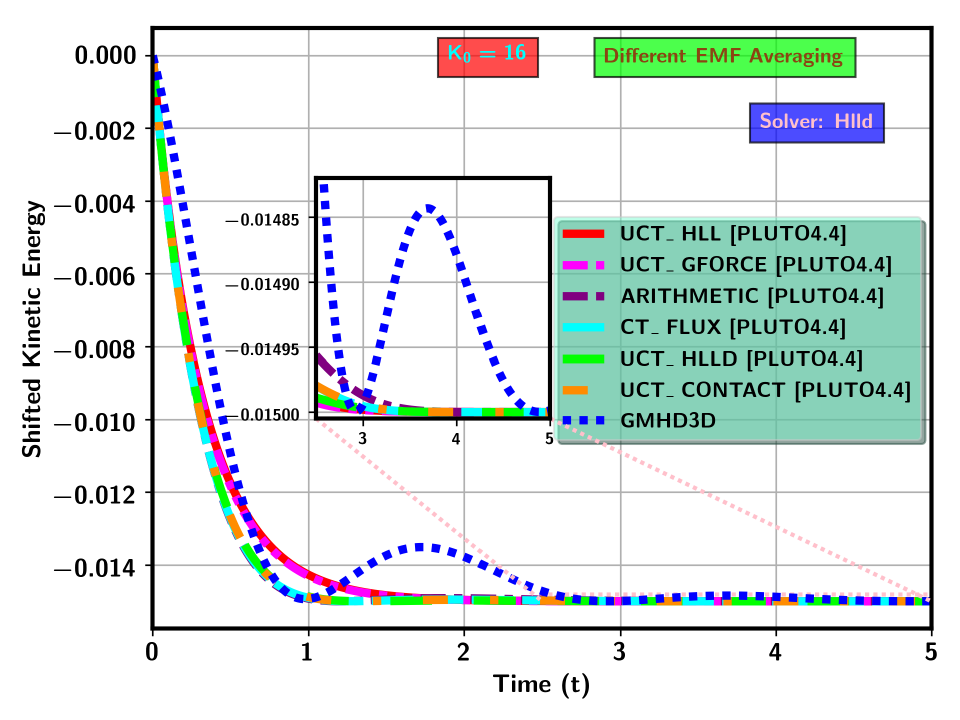}
		\caption{}
			\label{3D ABC k0_16 all emf}
	\end{subfigure}
	\begin{subfigure}{0.32\textwidth}
		\centering
		\includegraphics[scale=0.37]{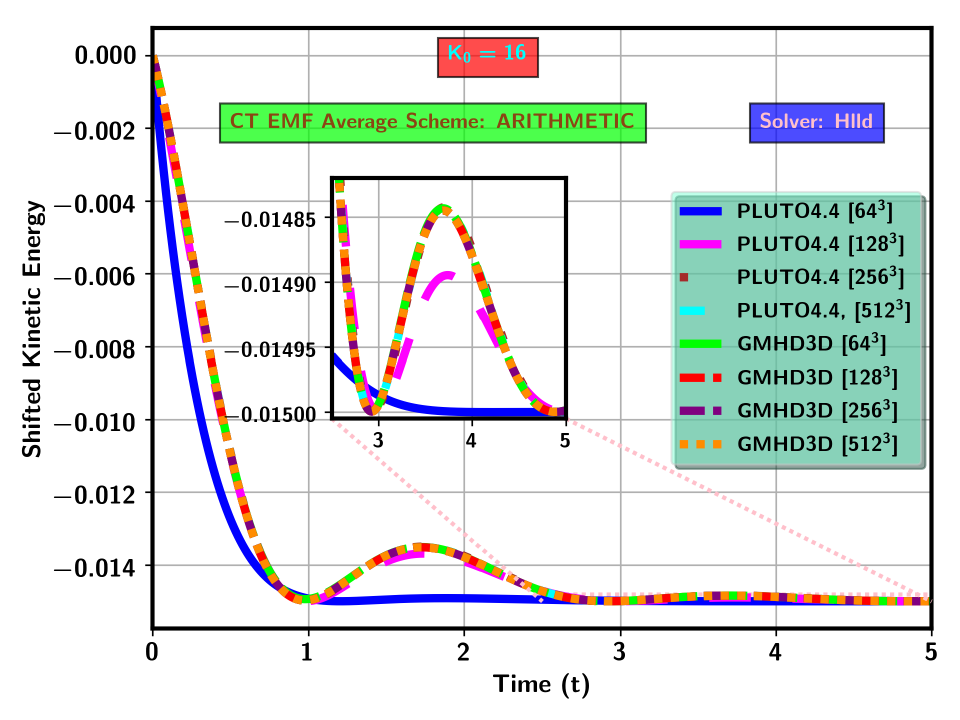}
		\caption{}
			\label{3D ABC arithmetic k0_16 higher resolution}
	\end{subfigure}
	\begin{subfigure}{0.32\textwidth}
		\centering
		\includegraphics[scale=0.37]{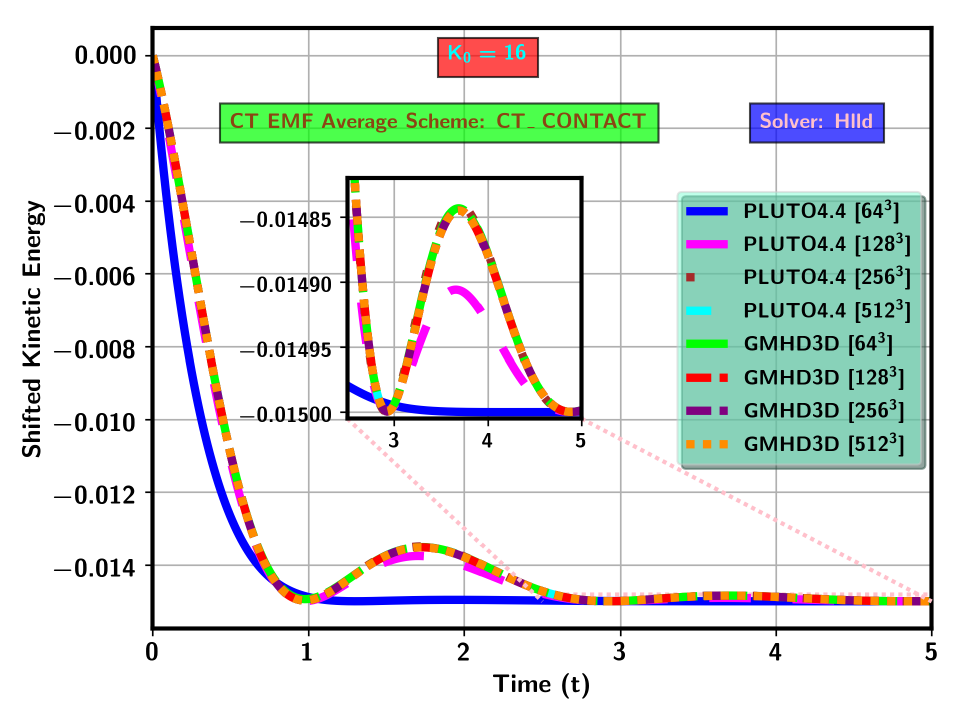}
		\caption{}
			\label{3D ABC ctcontact k0_16 higher resolution}
	\end{subfigure}
	\caption{(a) Shifted kinetic energy using 3D Arnold–Beltrami–Childress [ABC] Flow for $k_0 = 16.0$ from GMHD3D code and PLUTO4.4 code (with all available  CT\_EMF\_AVG schemes). High resolution study of 3D Arnold–Beltrami–Childress Flow for $k_0 = 16.0$ from GMHD3D code [at grid resolution $64^3$, $128^3$, $256^3$ \& $512^3$] and PLUTO4.4 code [at grid resolution $64^3$, $128^3$, $256^3$ \& $512^3$] with (b) ARITHMETIC scheme (c) CT\_CONTACT scheme. PLUTO4.4 code requires atleast $256^3$ grid resolution to reproduce the oscillation that we get from GMHD3D code at $64^3$ grid resolution. \textcolor{black}{Simulation Details: Time stepping $dt = 10^{-4}$}.}
\end{figure*}

Based on our findings in the $k_0 = 8$ case, we investigate all of the electric field averages techniques in the $k_0 = 16$ case as well [See Fig. \ref{3D ABC k0_16 all emf}]. We can see that none of the available algorithms work sufficiently well to make the oscillation we are looking for, which we obtained from GMHD3D at $64^3$. All of the emf averaging approaches demonstrate that the oscillations are significantly damped. So, the best way to resolve this right now is to improve the grid resolution. As in the previously stated scenario, we increase the grid resolution while keeping the ARITHMETIC \cite{ARITHMETIC:1999} and CT\_CONTACT \cite{CT:2005} averaging schemes in place.  In conjunction with the ARITHMETIC averages scheme, it is evident from Fig. \ref{3D ABC arithmetic k0_16 higher resolution} that the high resolution simulation at grid resolution $128^3$ is unable to reproduce the oscillation in its entirety.  It is likely to be reproduced with a $256^3$ grid resolution. The same issue is seen when using the CT\_CONTACT scheme [See Fig. \ref{3D ABC ctcontact k0_16 higher resolution}].

These findings are also quite promising. Our analysis shows that PLUTO4.4 requires $256^3$ grid resolution, but GMHD3D only need $64^3$ grid resolution, in order to resolve the highest initial wave number ($k_0 = 16.0$). This finding once again demonstrates the superior accuracy of a spectral solver over a grid-based solver.


\textcolor{black}{So far, we have discussed about several of the well-known flows without a driving mechanism. The impact of an external driver on these flows are also investigated both in two and three dimensions (See Appendix \ref{Appen B} for details)}.

\subsection{Test 6 [Magnetohydrodynamics]: Recurrence Dynamics in 3D MHD plasma}\label{Recurrence}

Coherent nonlinear oscillations of kinetic and magnetic energy in the form of Alfven waves have been demonstrated in 3D single fluid magnetohydrodynamic plasmas, as detailed above. A periodic reconstruction of the initial fluid flow and magnetic variables mediated by coherent non-linear oscillations is predicted when the energy alternates between kinetic and magnetic forms. This phenomenon is called recurrence. It is recently discovered that astrophysical plasmas exhibit two distinct types of flow \cite{RM_Recurrence:2019}. Unlike regular 3D Arnold-Beltrami-Childress [ABC] flow, the initial velocity and magnetic field surface of these flows cannot be reconstructed, hence they are characterized as non-recurring \cite{RM_Recurrence:2019}. Another type of flow is the 3D Taylor-Green [TG] flow, which shows full recurrence by reconstructing the structure of the isosurfaces of kinetic and magnetic energy \cite{RM_Recurrence:2019}. These flows are called Recurring flows.


Here, we use the recently developed code GMHD3D and the open-source code PLUTO4.4 to investigate the recurrent phenomenon.

\subsubsection{Non-Recurring 3D ABC flow}
%

\begin{figure*}
	\centering
	\begin{turn}{90} 
		\LARGE{\textbf{\textcolor{blue}{\hspace{-4.0cm} $\longleftarrow$ GMHD3D $\longrightarrow$}}}
	\end{turn}
	\begin{subfigure}{0.23\textwidth}
		\centering
		\includegraphics[scale=0.044]{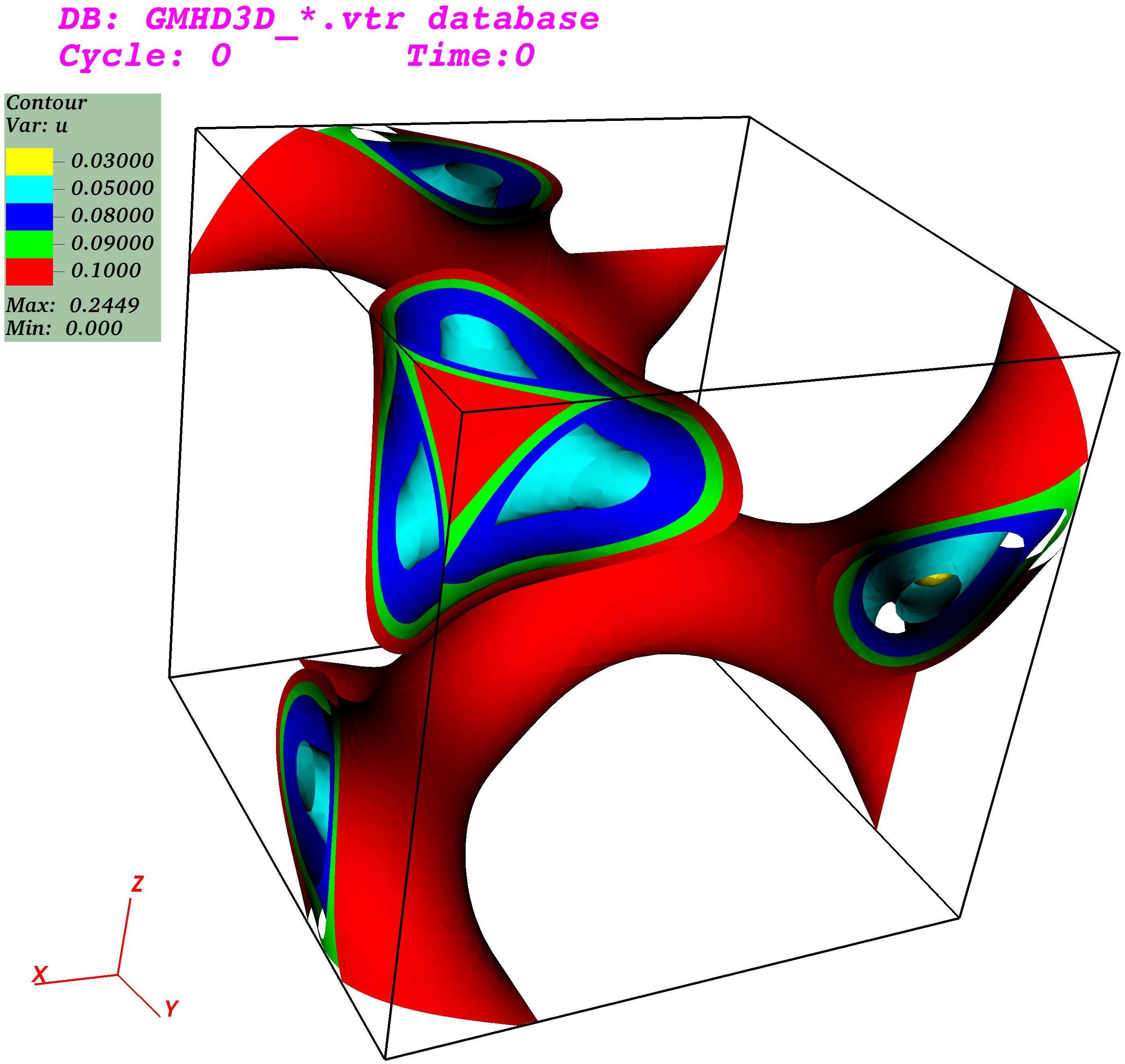}
		\caption{Time = 0.0}
	\end{subfigure}
	\begin{subfigure}{0.23\textwidth}
		\centering
		\includegraphics[scale=0.044]{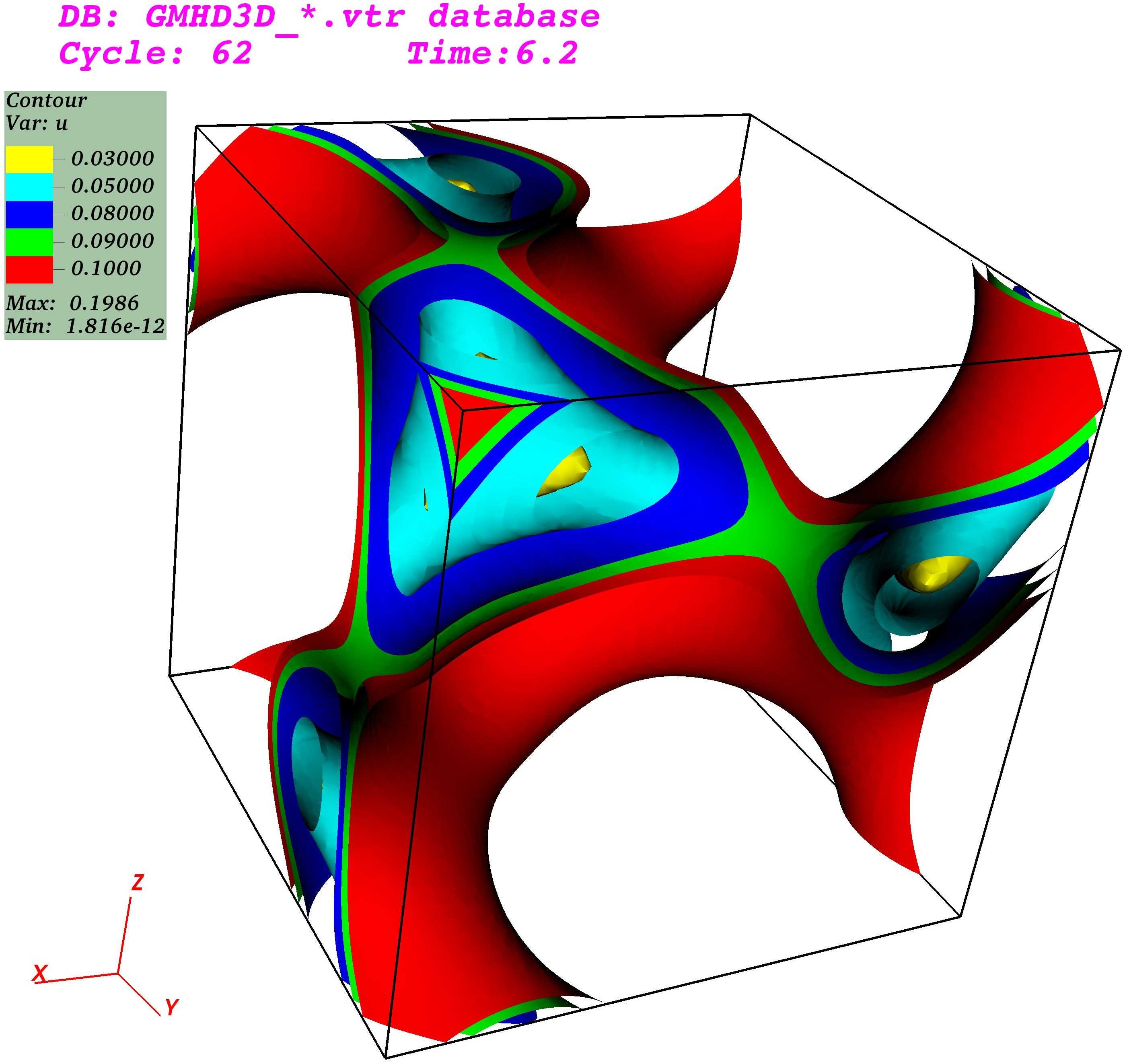}
		\caption{Time = 6.2}
	\end{subfigure}
	\begin{subfigure}{0.23\textwidth}
		\centering
		\includegraphics[scale=0.044]{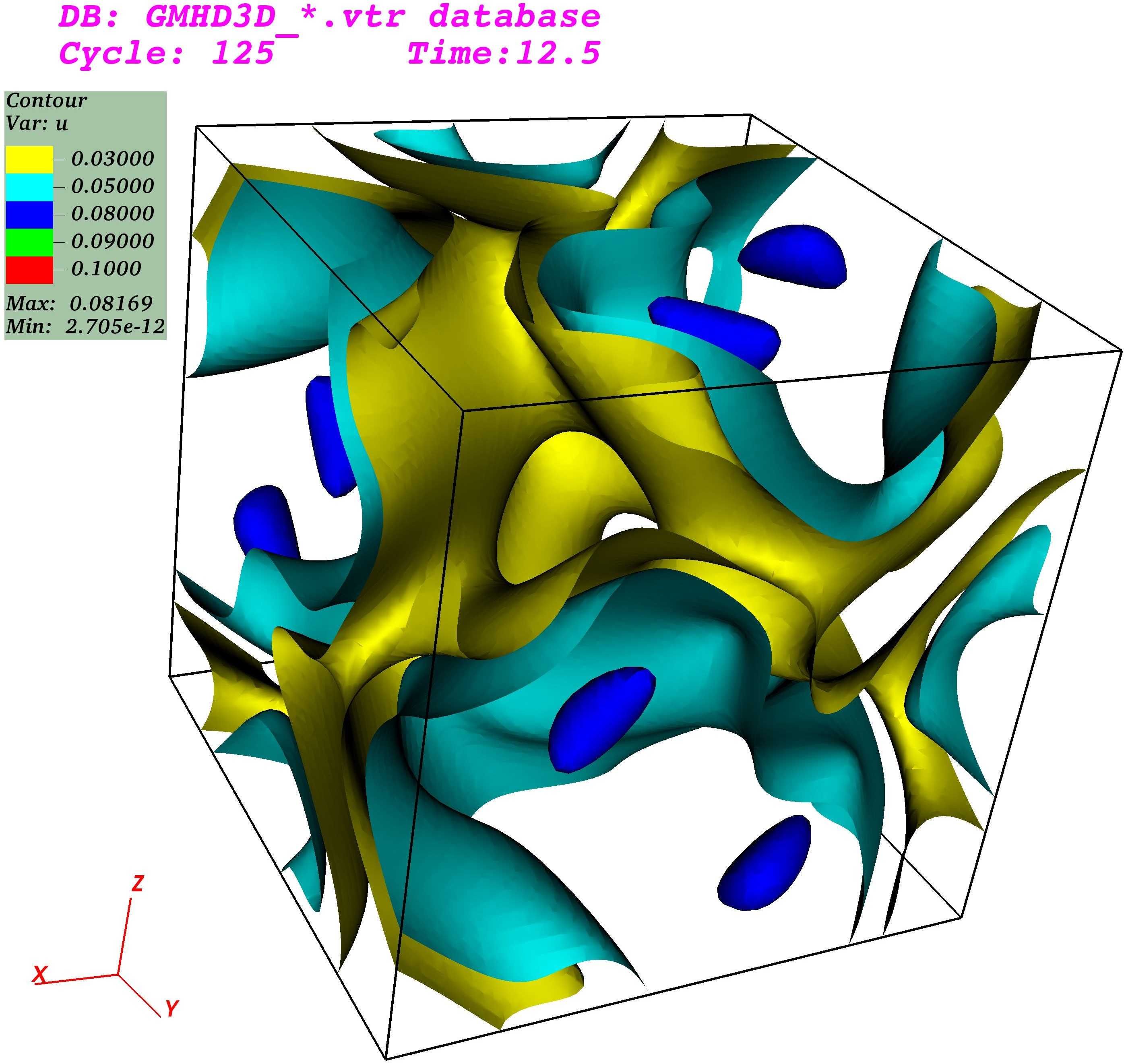}
		\caption{Time = 12.5}
	\end{subfigure}
	\begin{subfigure}{0.23\textwidth}
		\centering
		\includegraphics[scale=0.044]{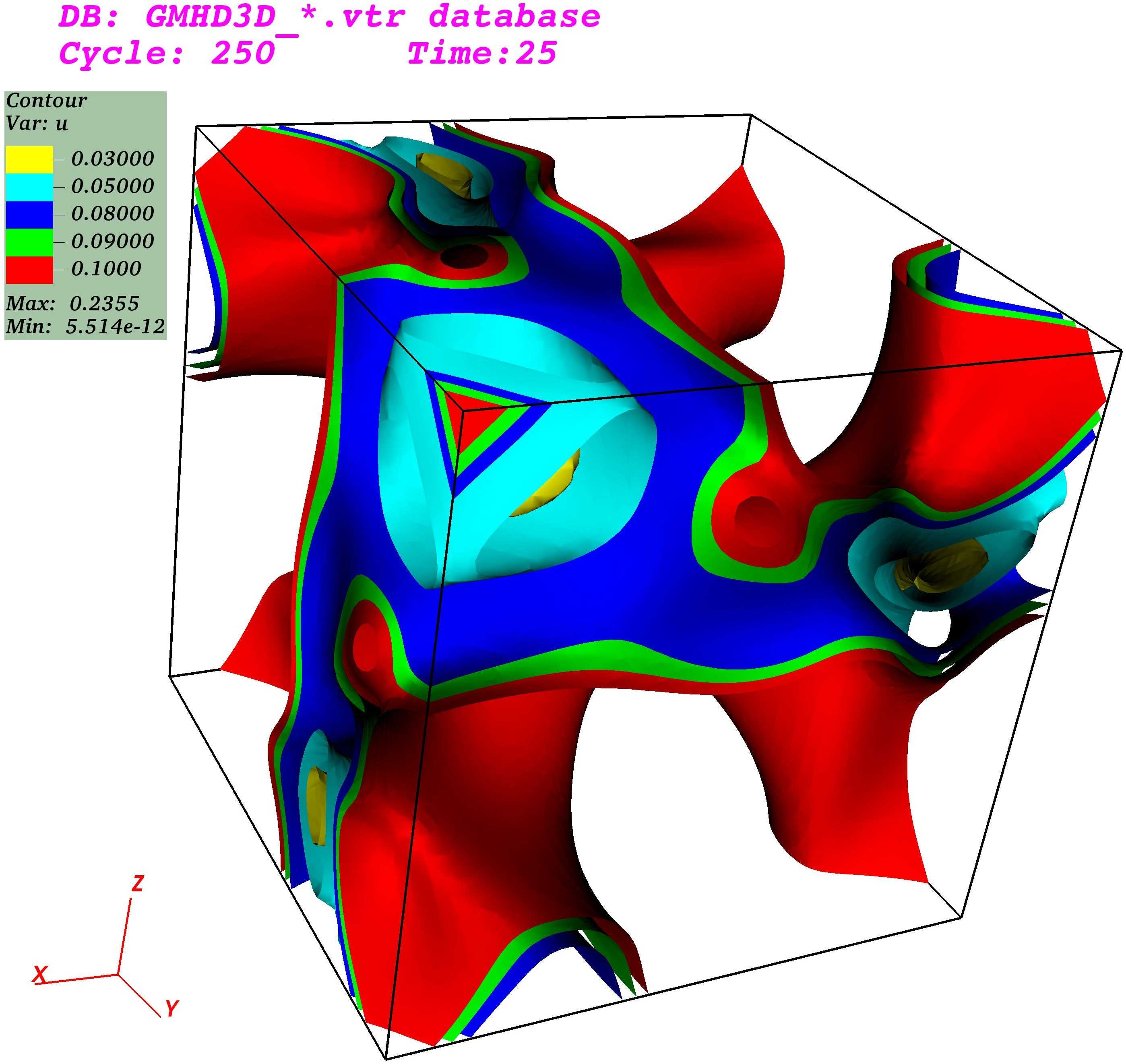}
		\caption{Time = 25.0}
	\end{subfigure}
	\begin{subfigure}{0.23\textwidth}
		\centering
		\includegraphics[scale=0.044]{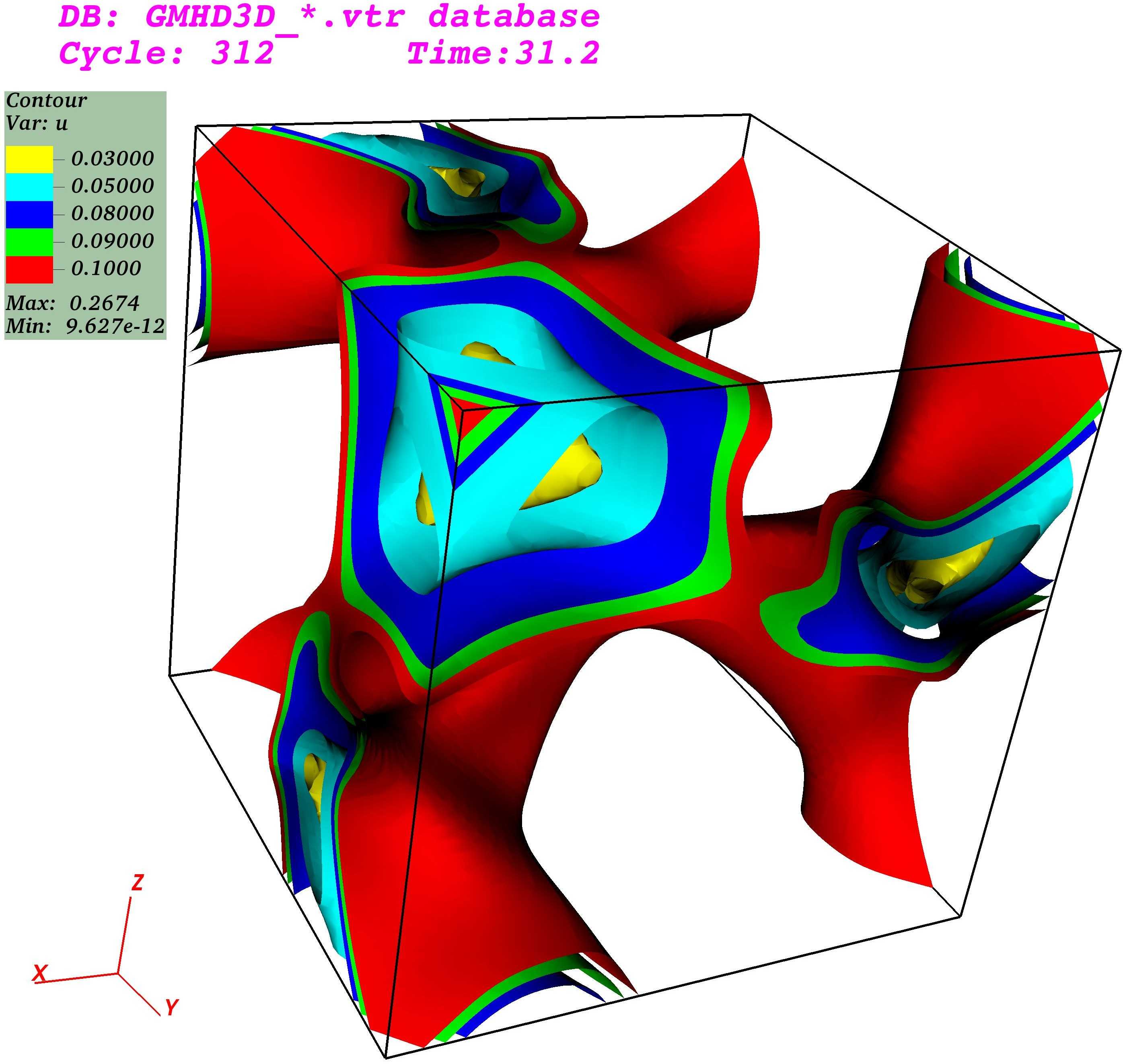}
		\caption{Time = 31.2}
	\end{subfigure}
	\begin{subfigure}{0.23\textwidth}
		\centering
		\includegraphics[scale=0.044]{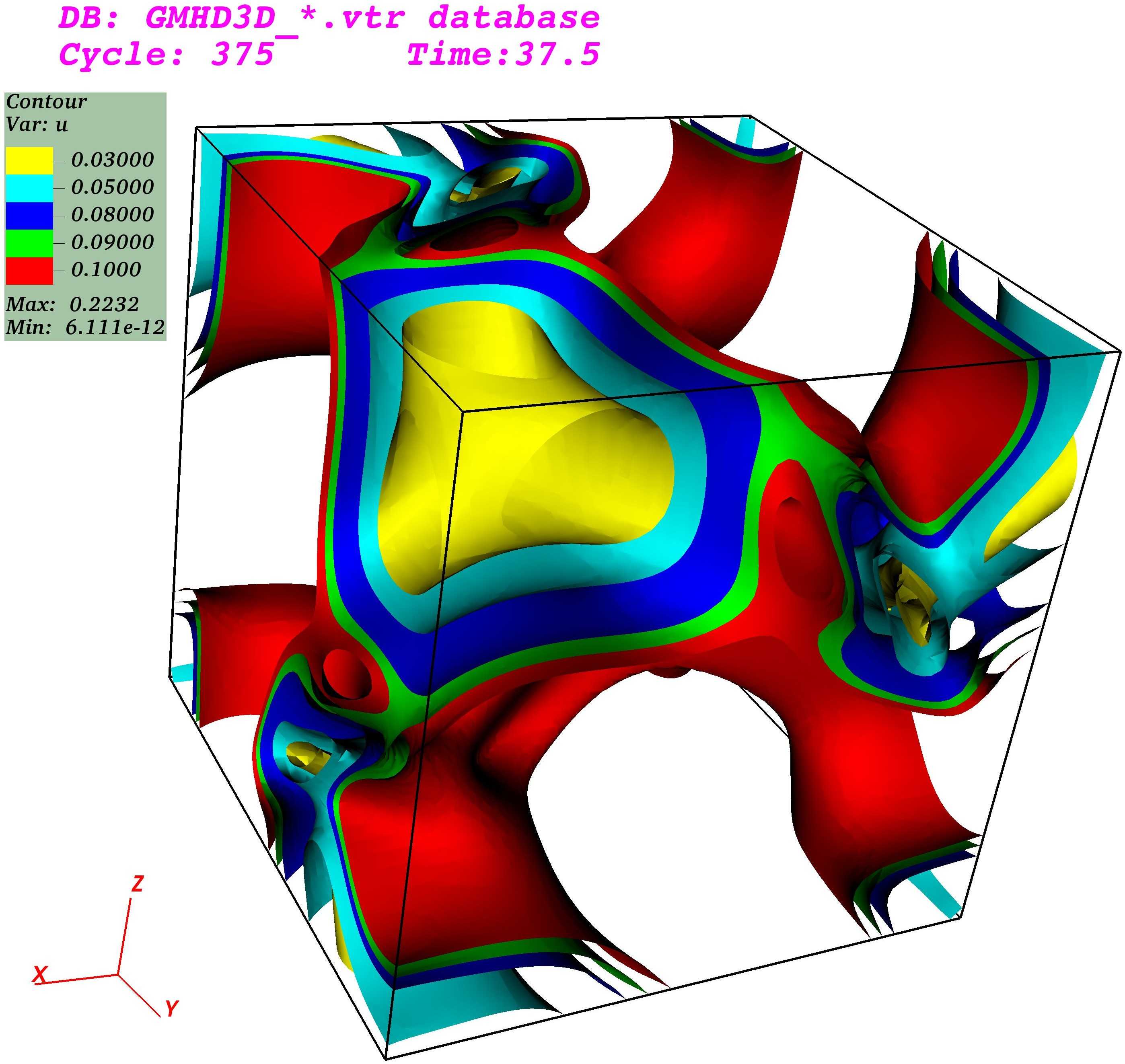}
		\caption{Time = 37.5}
	\end{subfigure}
	\begin{subfigure}{0.22\textwidth}
		\centering
		\includegraphics[scale=0.044]{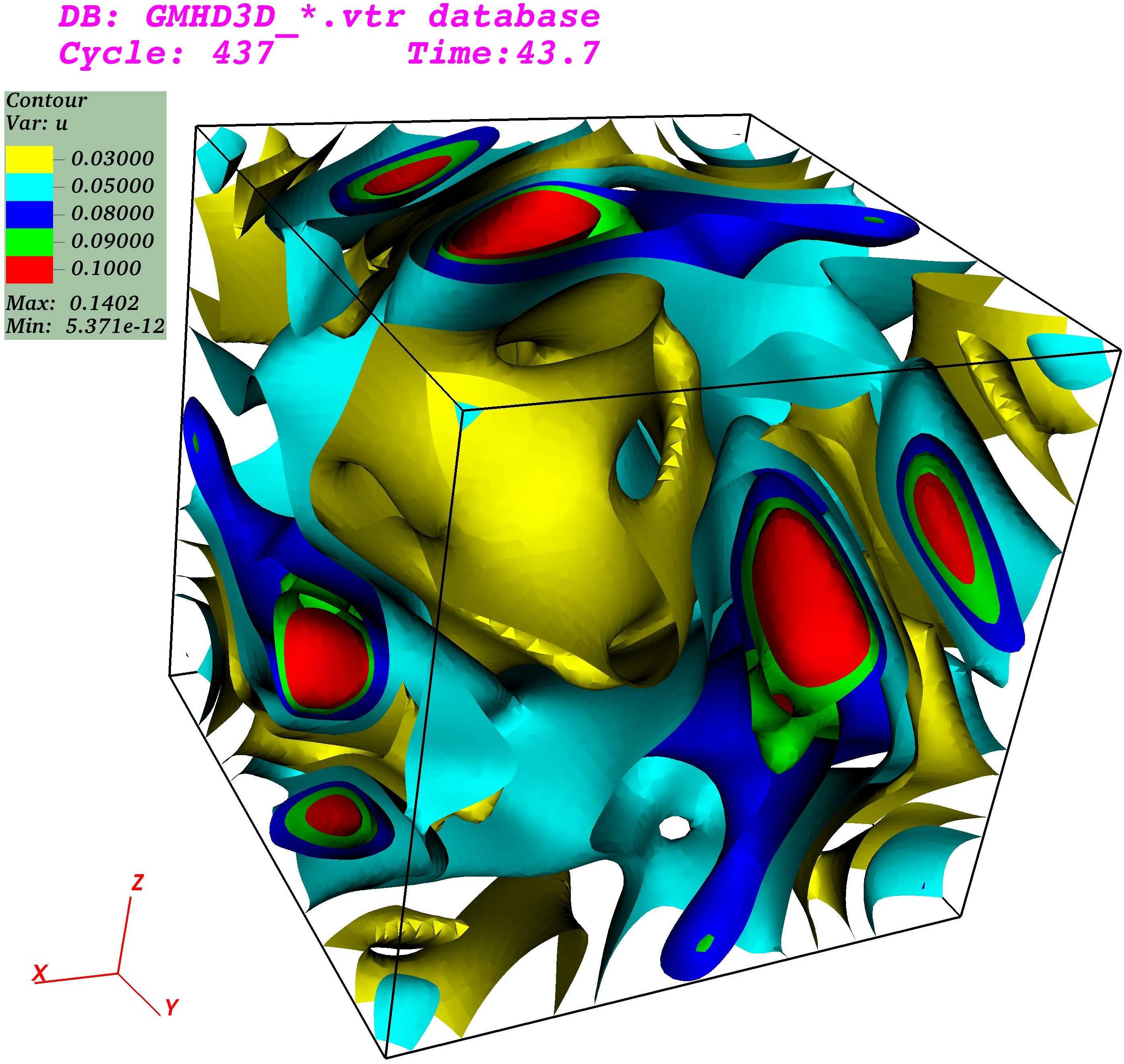}
		\caption{Time = 43.7}
	\end{subfigure}
	\begin{subfigure}{0.22\textwidth}
		\centering
		\includegraphics[scale=0.044]{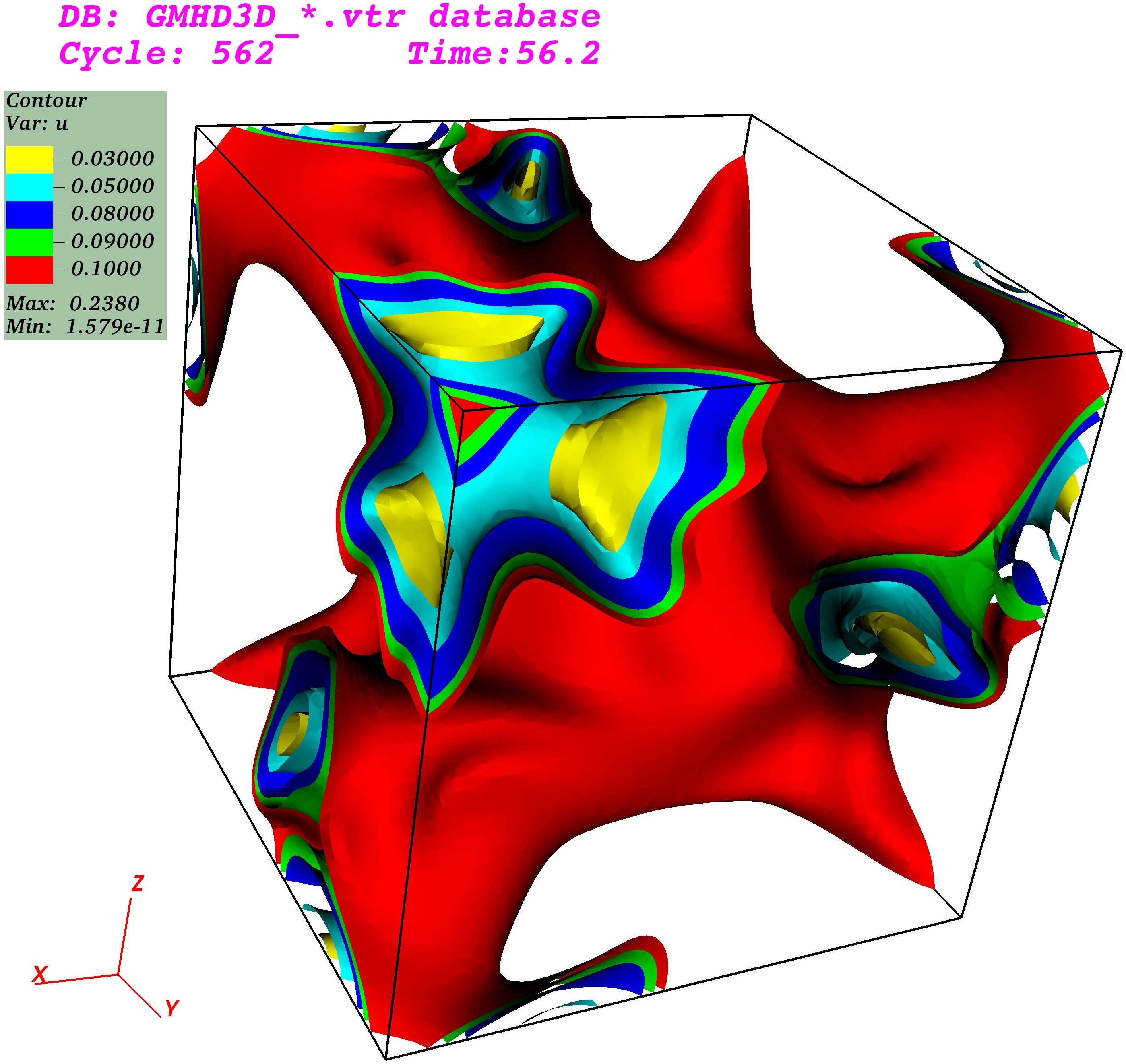}
		\caption{Time = 56.2}
	\end{subfigure}\\
	\centering
	\begin{turn}{90} 
		\LARGE{\textbf{\textcolor{blue}{\hspace{-4.0cm} $\longleftarrow$ PLUTO4.4 $\longrightarrow$}}}
	\end{turn}
	\begin{subfigure}{0.23\textwidth}
		\centering
		\includegraphics[scale=0.044]{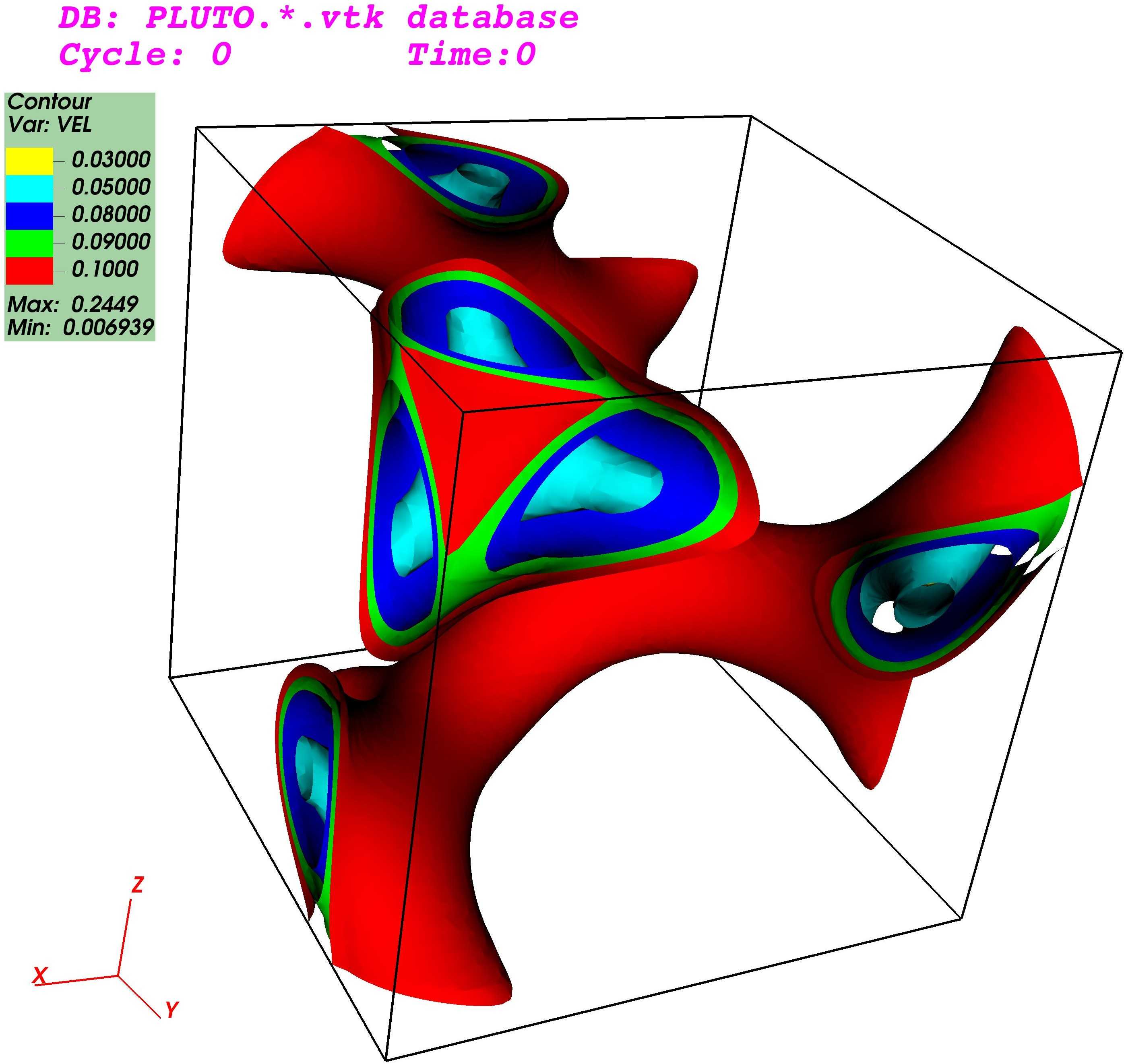}
		\caption{Time = 0.0}
	\end{subfigure}
	\begin{subfigure}{0.23\textwidth}
		\centering
		\includegraphics[scale=0.044]{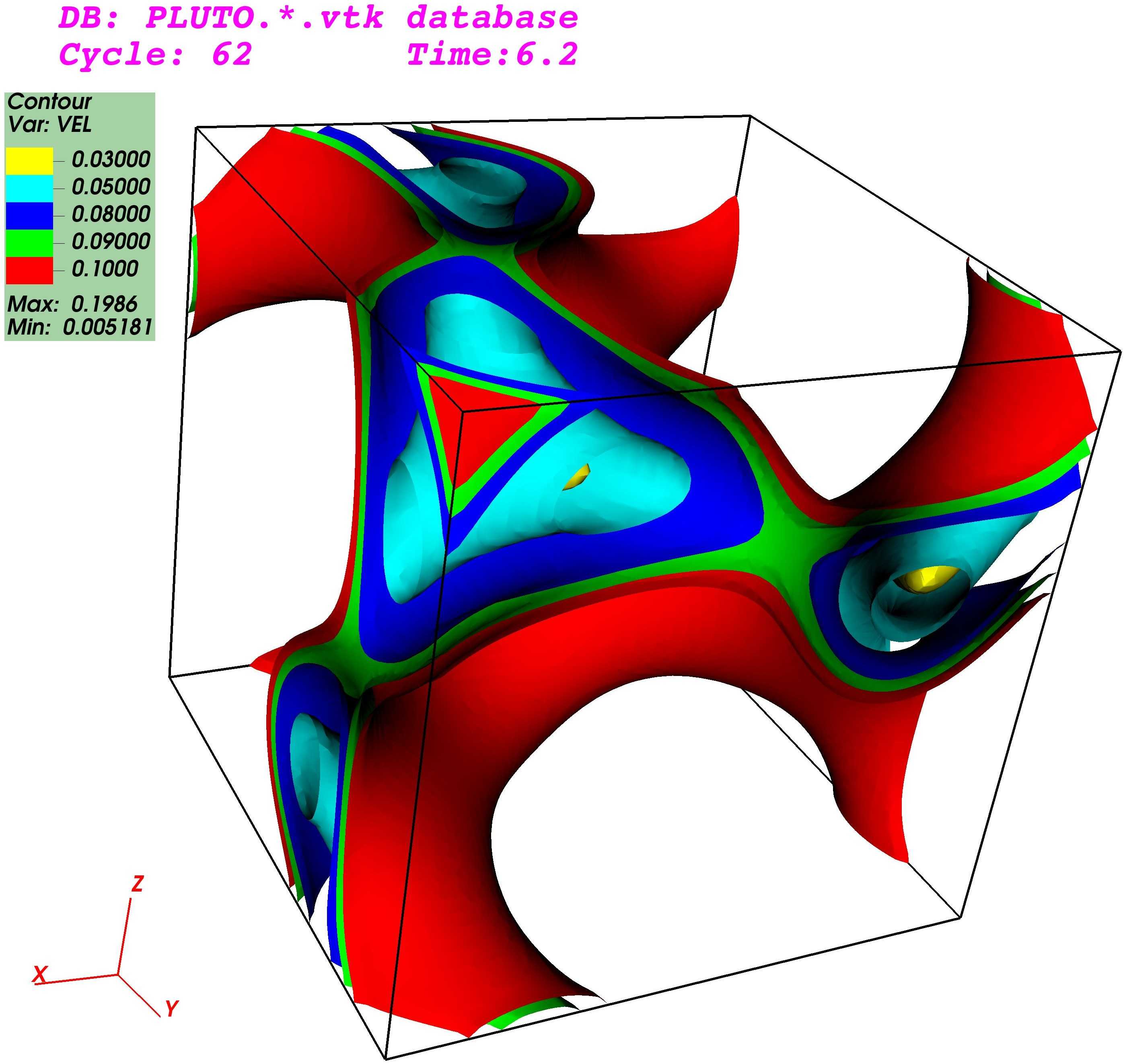}
		\caption{Time = 6.2}
	\end{subfigure}
	\begin{subfigure}{0.23\textwidth}
		\centering
		\includegraphics[scale=0.044]{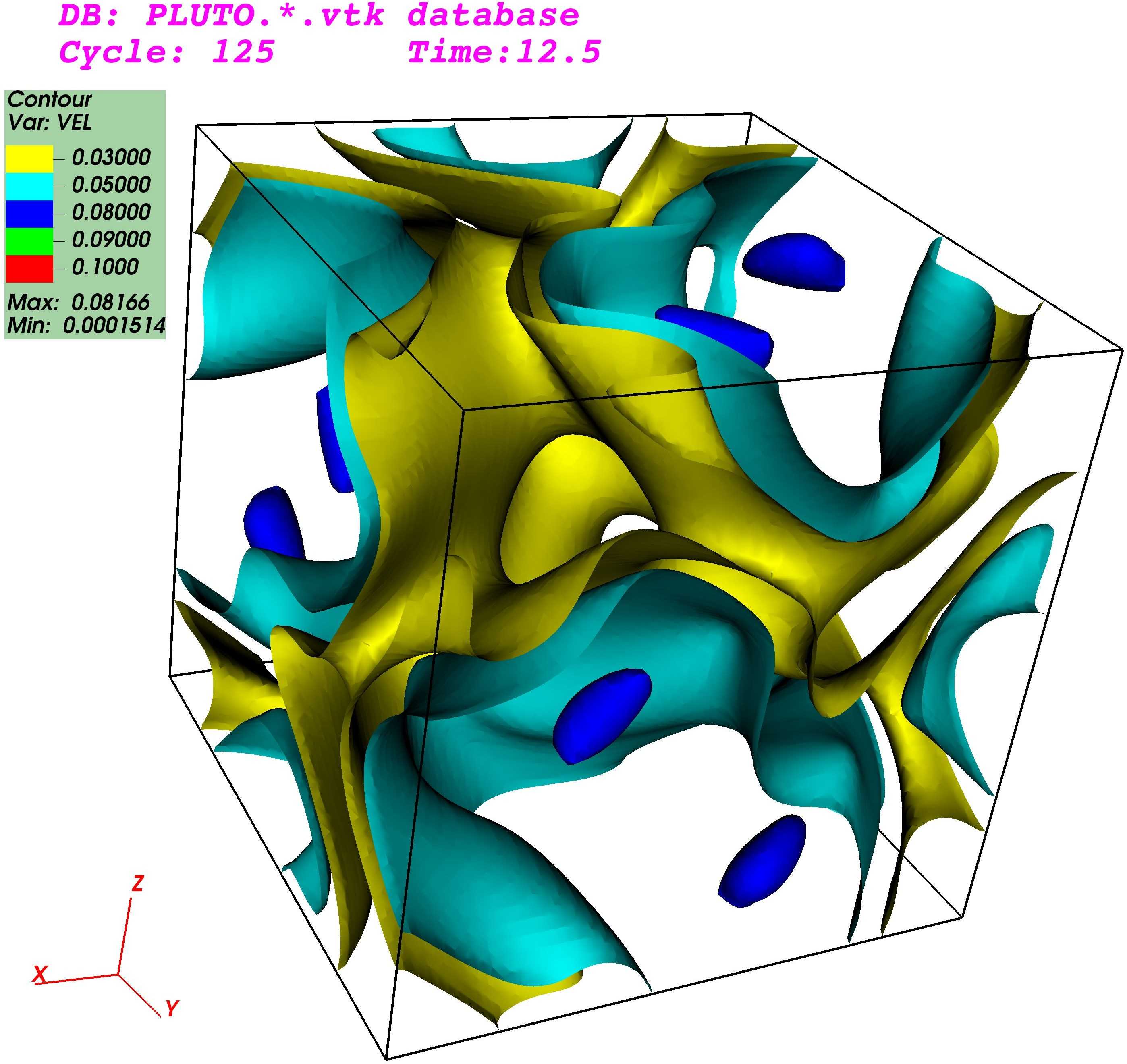}
		\caption{Time = 12.5}
	\end{subfigure}
	\begin{subfigure}{0.23\textwidth}
		\centering
		\includegraphics[scale=0.044]{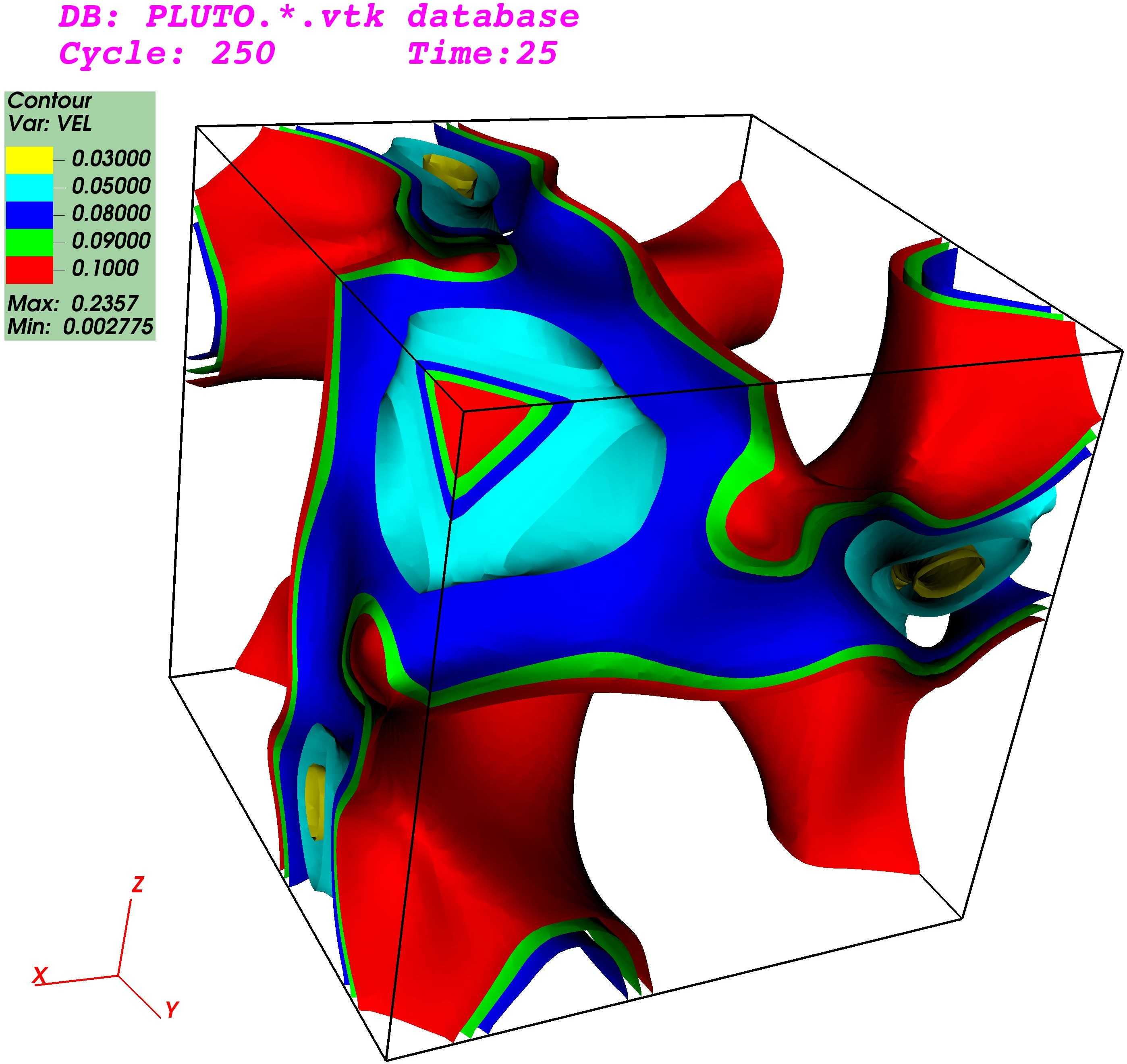}
		\caption{Time = 25.0}
	\end{subfigure}
	\begin{subfigure}{0.23\textwidth}
		\centering
		\includegraphics[scale=0.044]{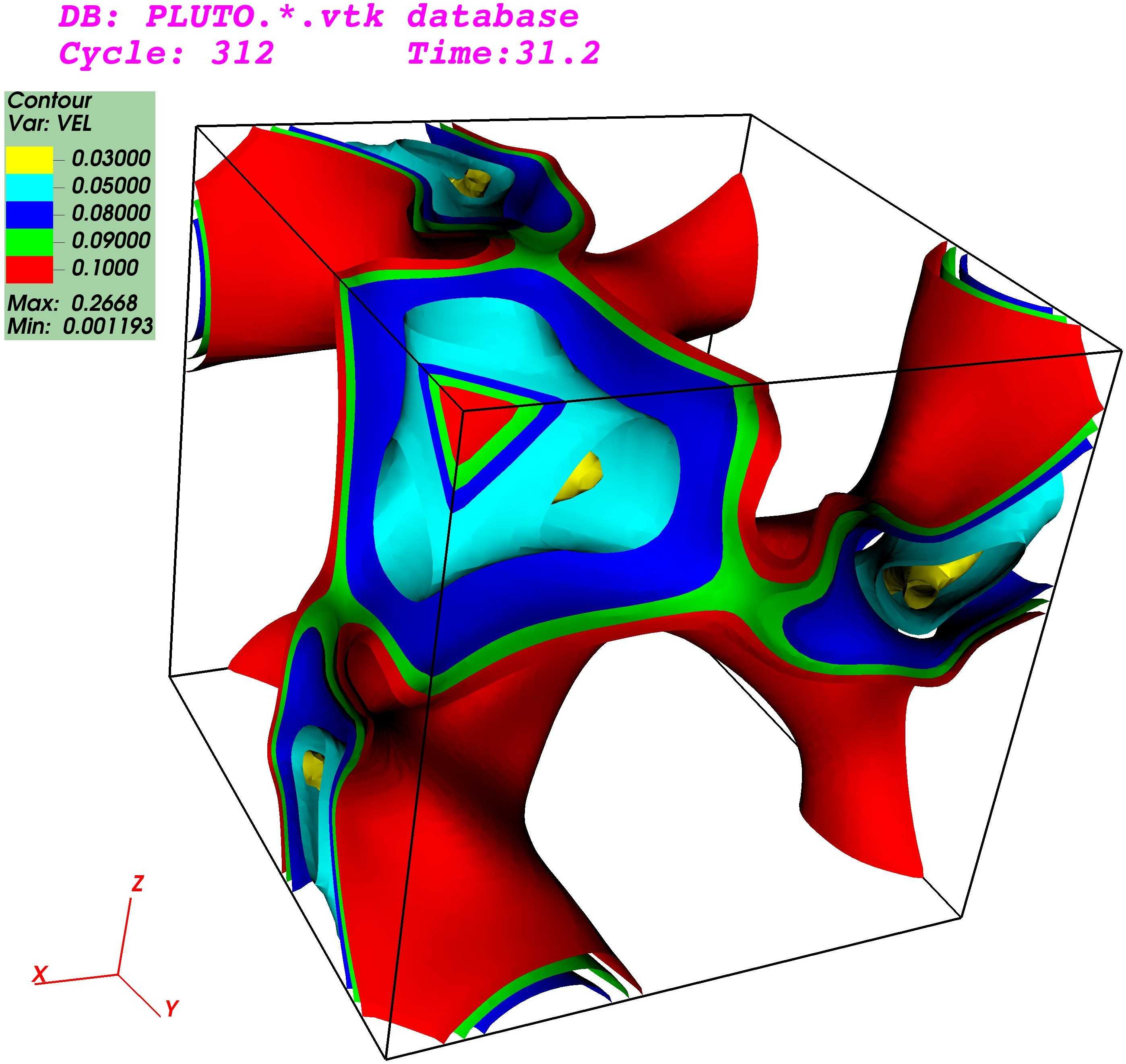}
		\caption{Time = 31.2}
	\end{subfigure}
	\begin{subfigure}{0.23\textwidth}
		\centering
		\includegraphics[scale=0.044]{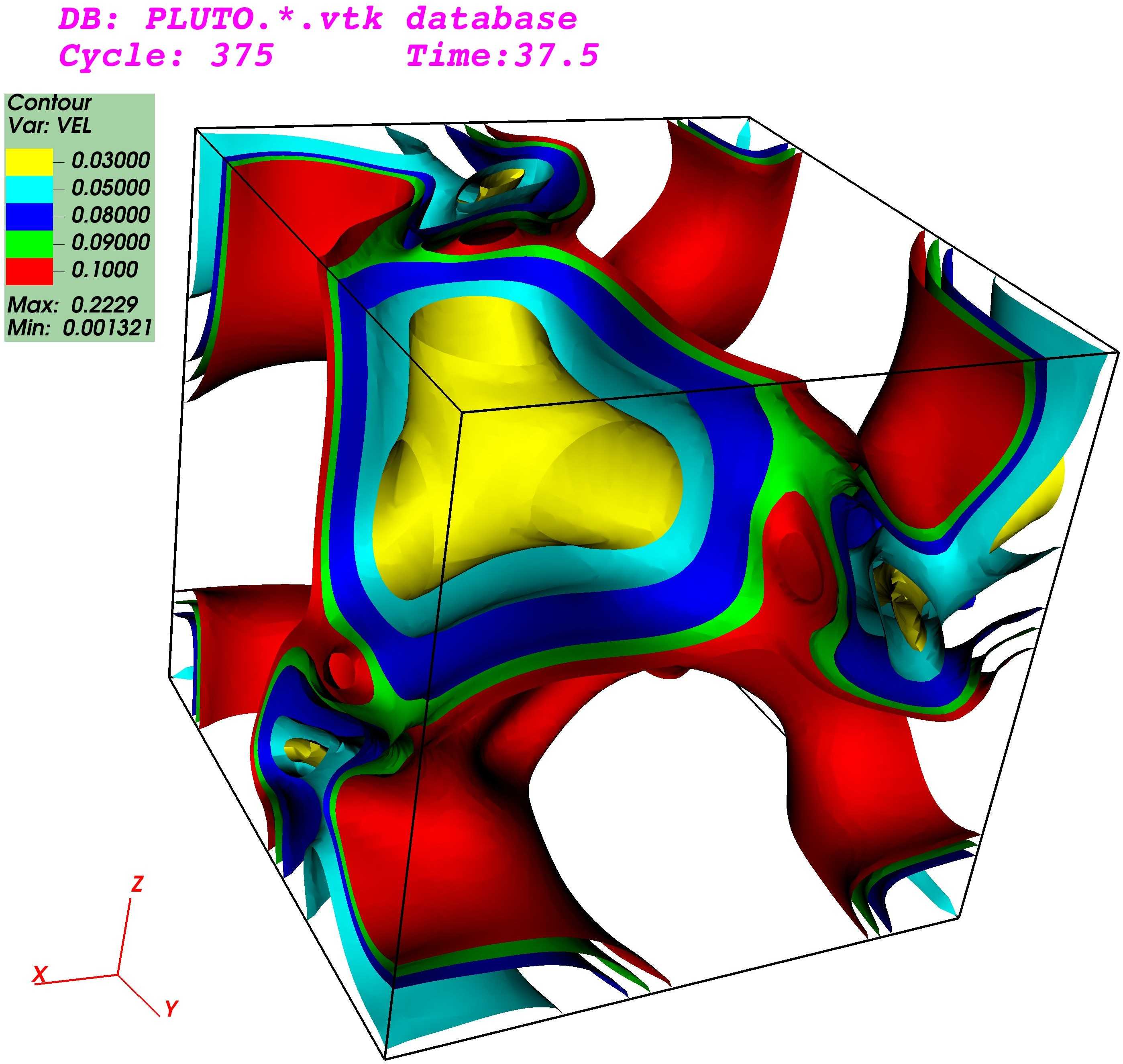}
		\caption{Time = 37.5}
	\end{subfigure}
	\begin{subfigure}{0.22\textwidth}
		\centering
		\includegraphics[scale=0.044]{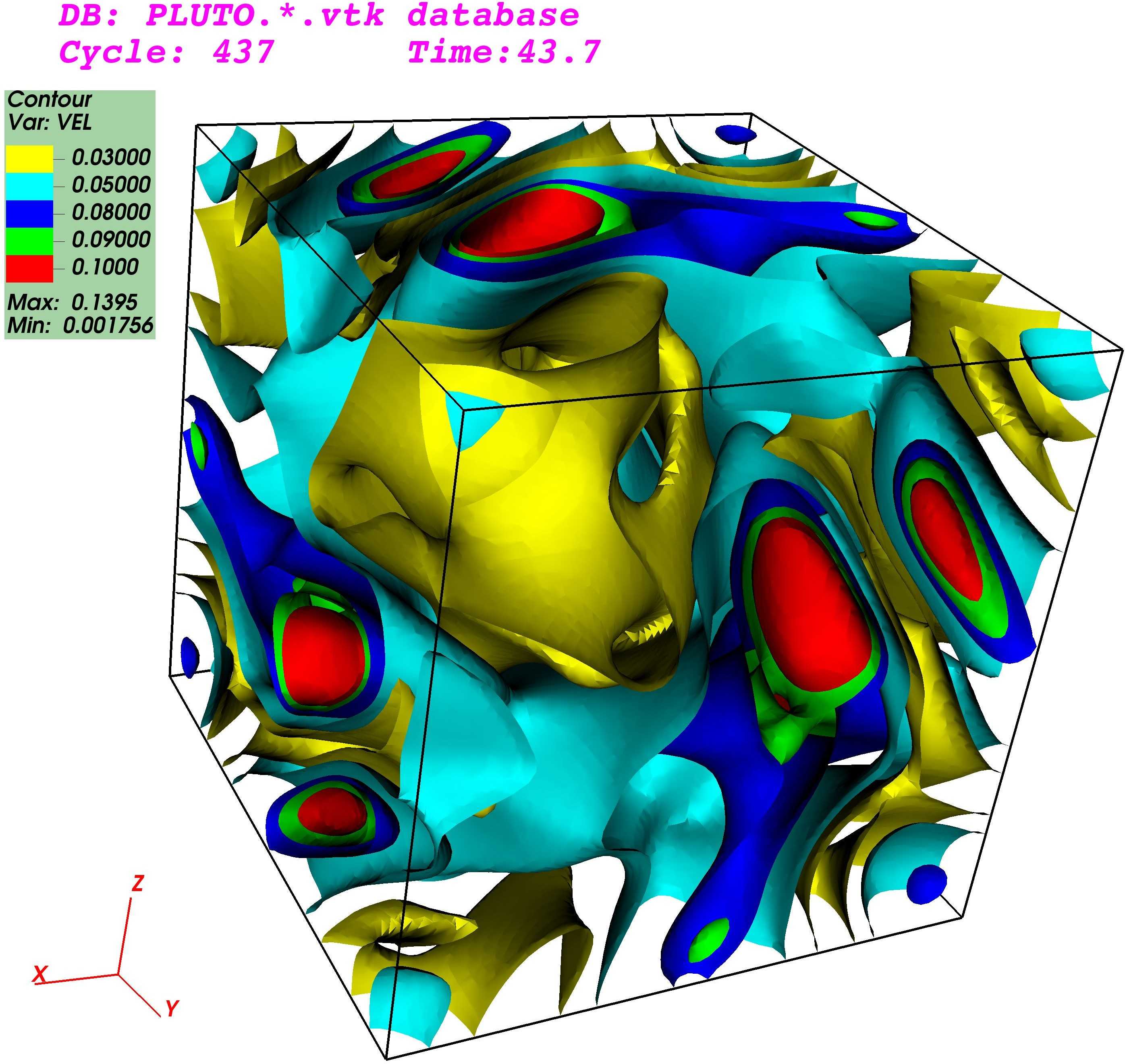}
		\caption{Time = 43.7}
	\end{subfigure}
	\begin{subfigure}{0.22\textwidth}
		\centering
		\includegraphics[scale=0.044]{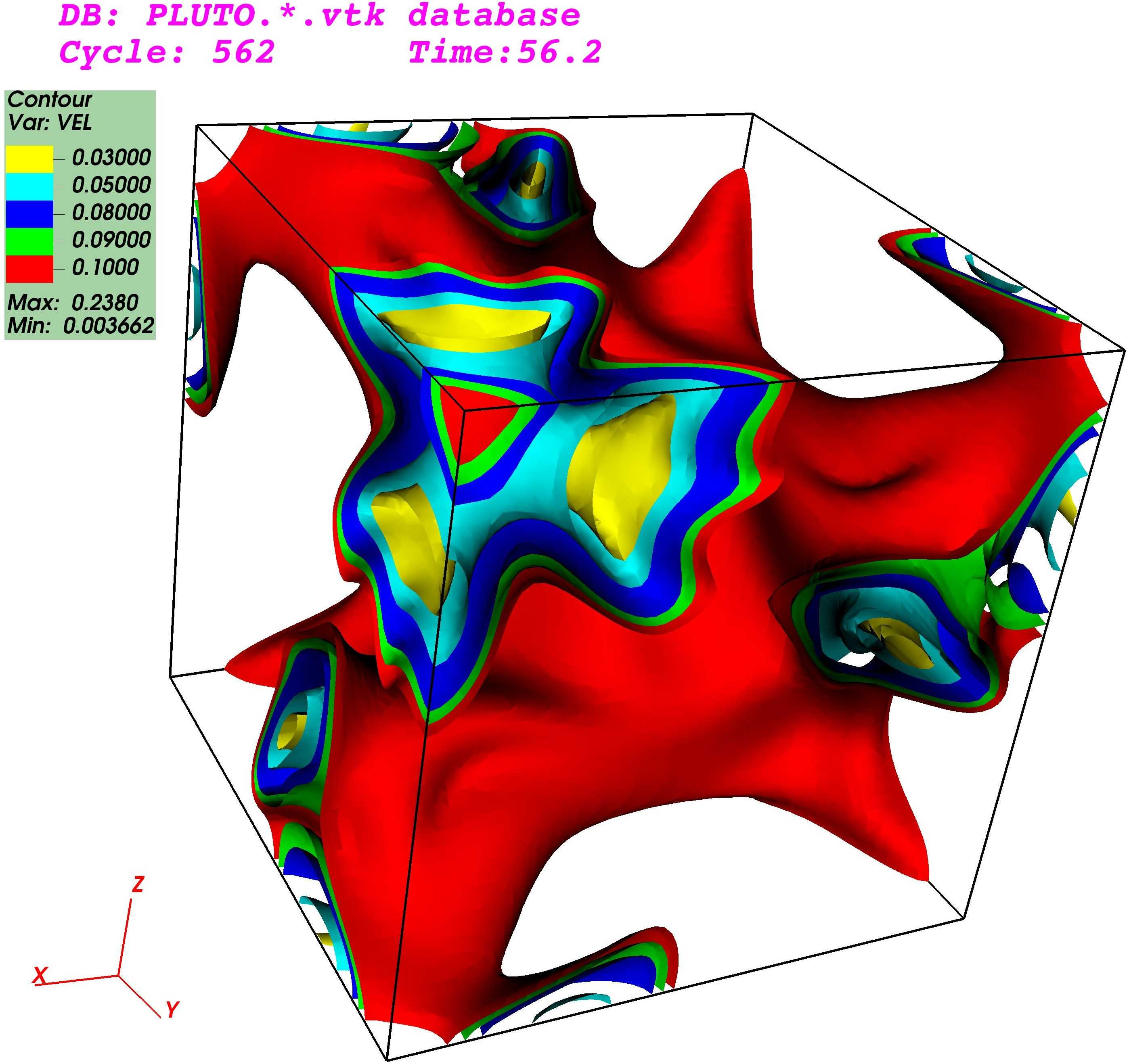}
		\caption{Time = 56.2}
	\end{subfigure}
	\caption{Non-Recurring 3D  Arnold–Beltrami–Childress [ABC] Flow Velocity iso-surface from GMHD3D code [Top two rows (a--h)] and PLUTO4.4 code [Bottom two rows (i--p)]. Values of iso-surface: \textbf{ 0.1 (Red), 0.09 (Green), 0.08 (Blue), 0.05 (Cyan) and 0.03 (Yellow)}. Simulation Details: Reynolds number $R_e = R_m = 1000$, Grid resolution $N = 128^3$, Time stepping $dt = 10^{-4}$, initial fluid velocity $u_0 = 1.0$, Alfven Mach number $M_A = 1.0$.}
	\label{Non Recurr 3D ABC Iso V}
\end{figure*}

\begin{figure*}
	\centering
	\begin{turn}{90} 
		\LARGE{\textbf{\textcolor{blue}{\hspace{-4.0cm} $\longleftarrow$ GMHD3D $\longrightarrow$}}}
	\end{turn}
	\begin{subfigure}{0.23\textwidth}
		\centering
		\includegraphics[scale=0.044]{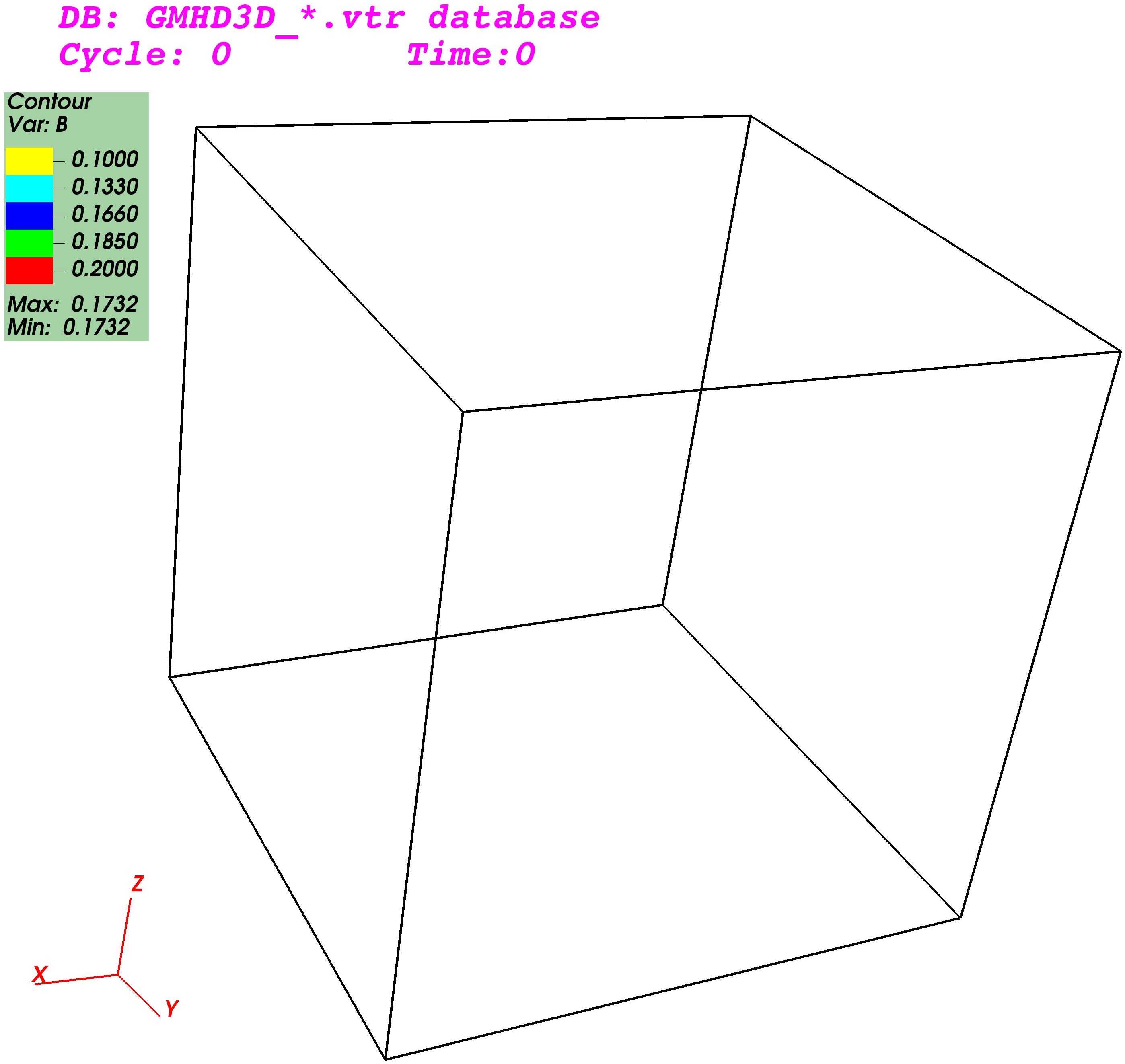}
		\caption{Time = 0.0}
	\end{subfigure}
	\begin{subfigure}{0.23\textwidth}
		\centering
		\includegraphics[scale=0.044]{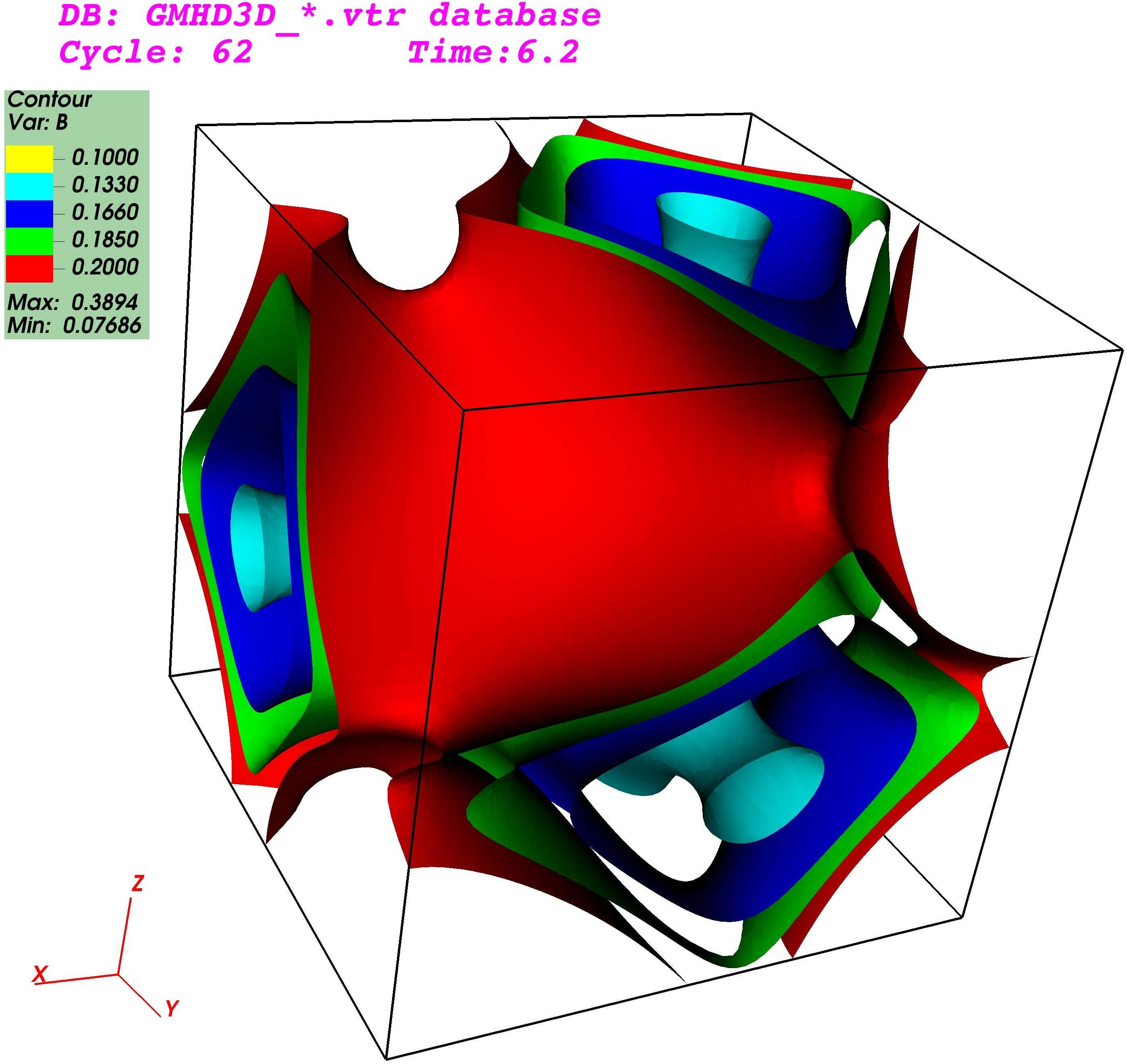}
		\caption{Time = 6.2}
	\end{subfigure}
	\begin{subfigure}{0.23\textwidth}
		\centering
		\includegraphics[scale=0.044]{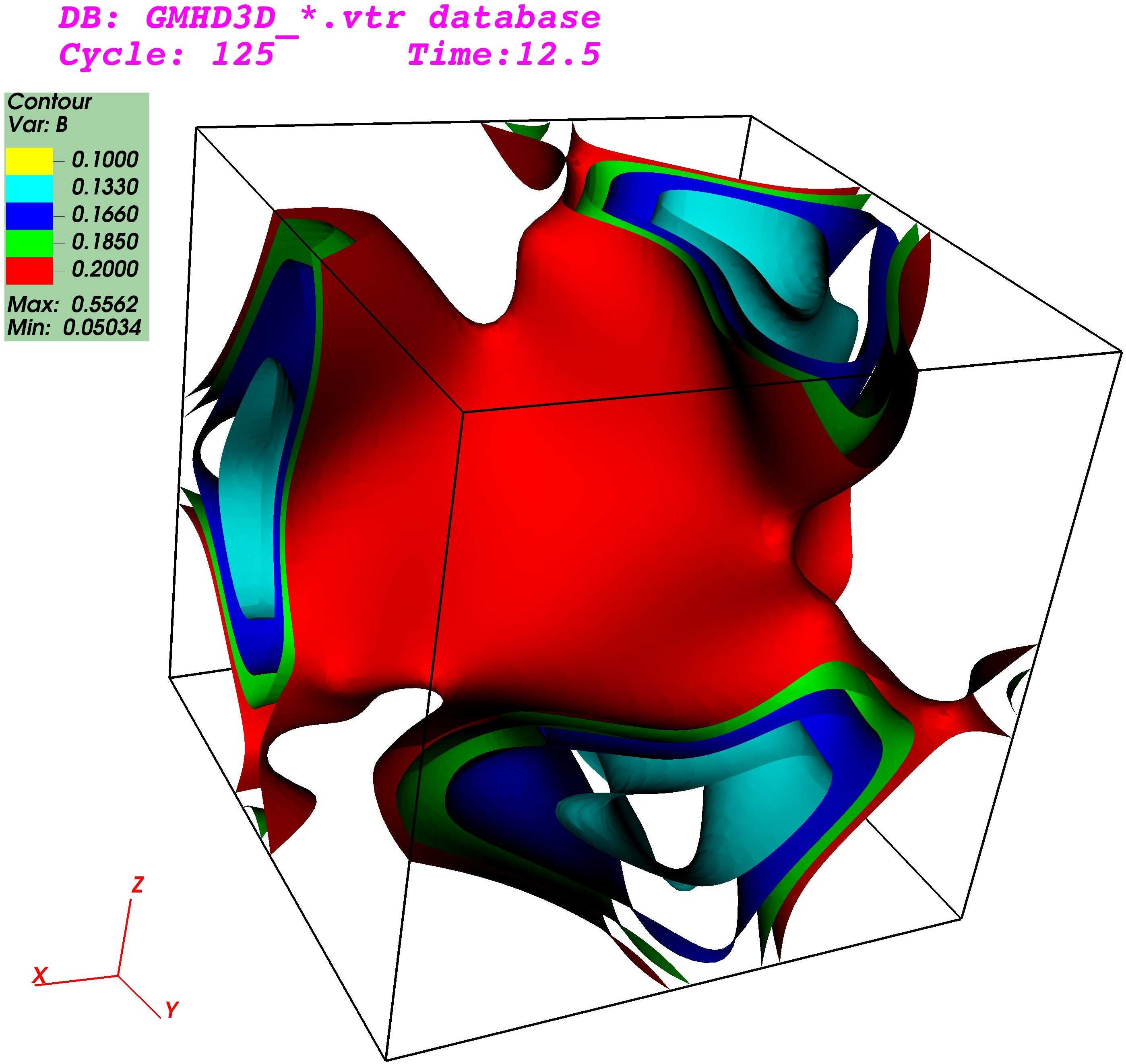}
		\caption{Time = 12.5}
	\end{subfigure}
	\begin{subfigure}{0.23\textwidth}
		\centering
		\includegraphics[scale=0.044]{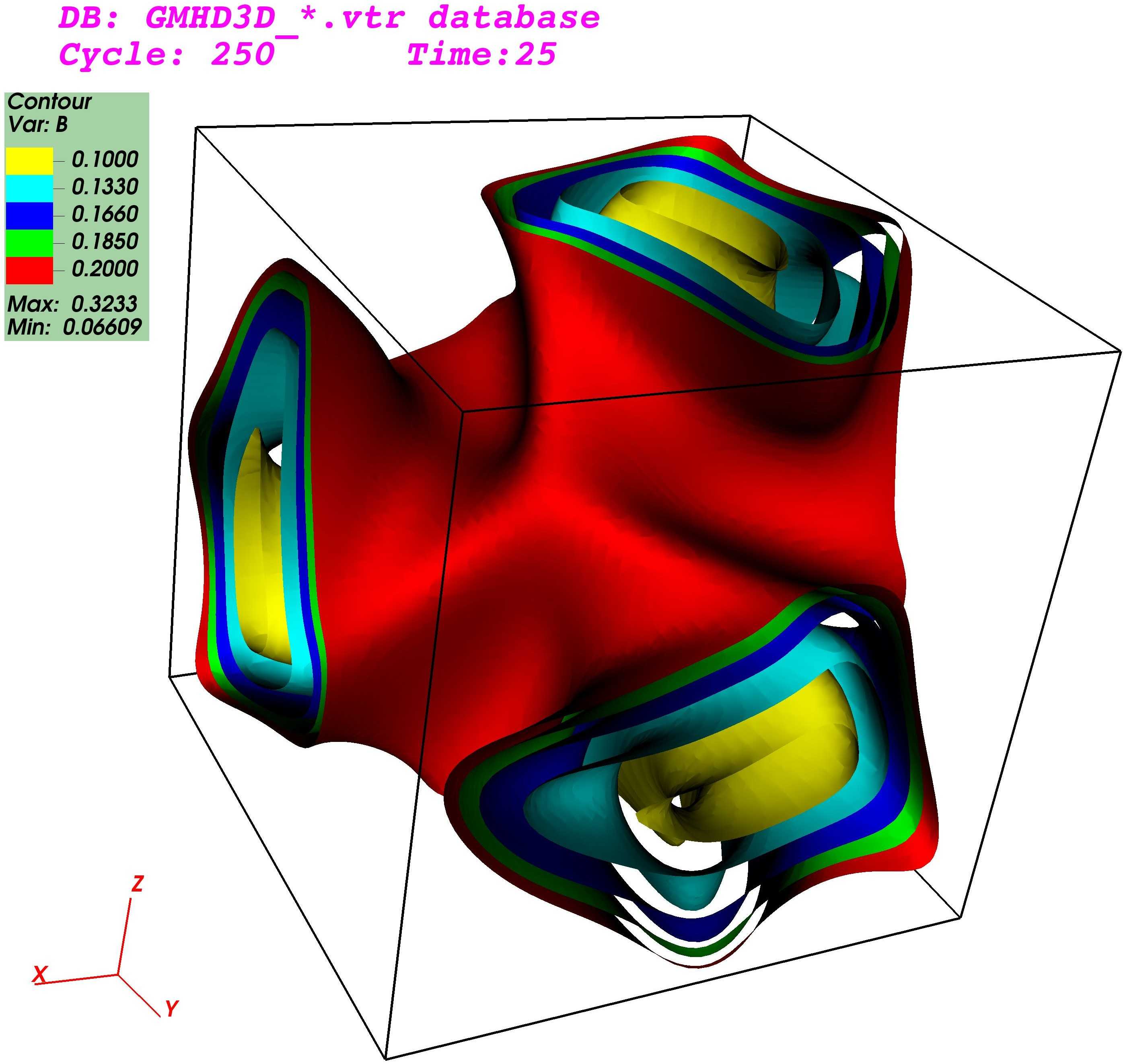}
		\caption{Time = 25.0}
	\end{subfigure}
	\begin{subfigure}{0.23\textwidth}
		\centering
		\includegraphics[scale=0.044]{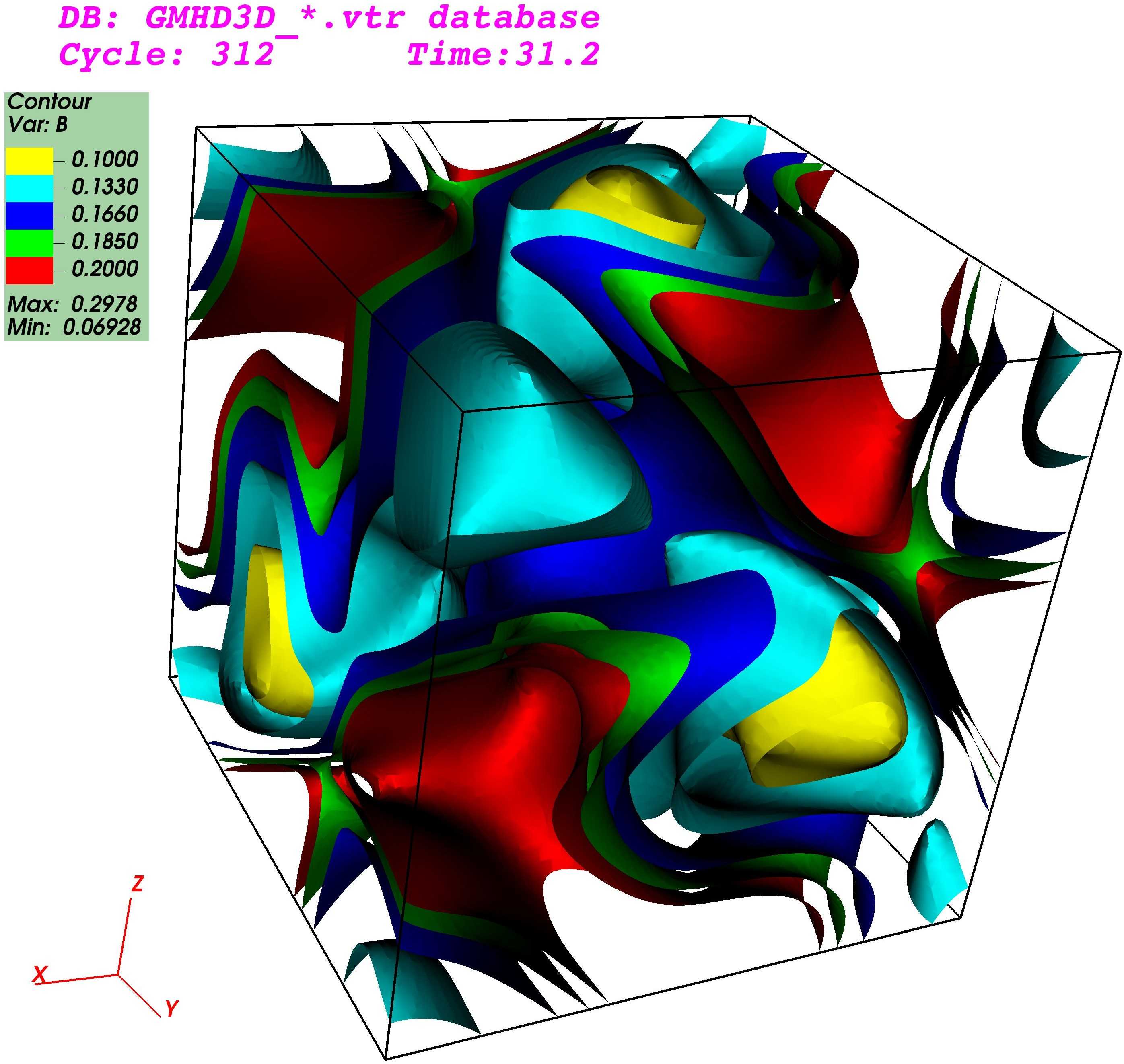}
		\caption{Time = 31.2}
	\end{subfigure}
	\begin{subfigure}{0.23\textwidth}
		\centering
		\includegraphics[scale=0.044]{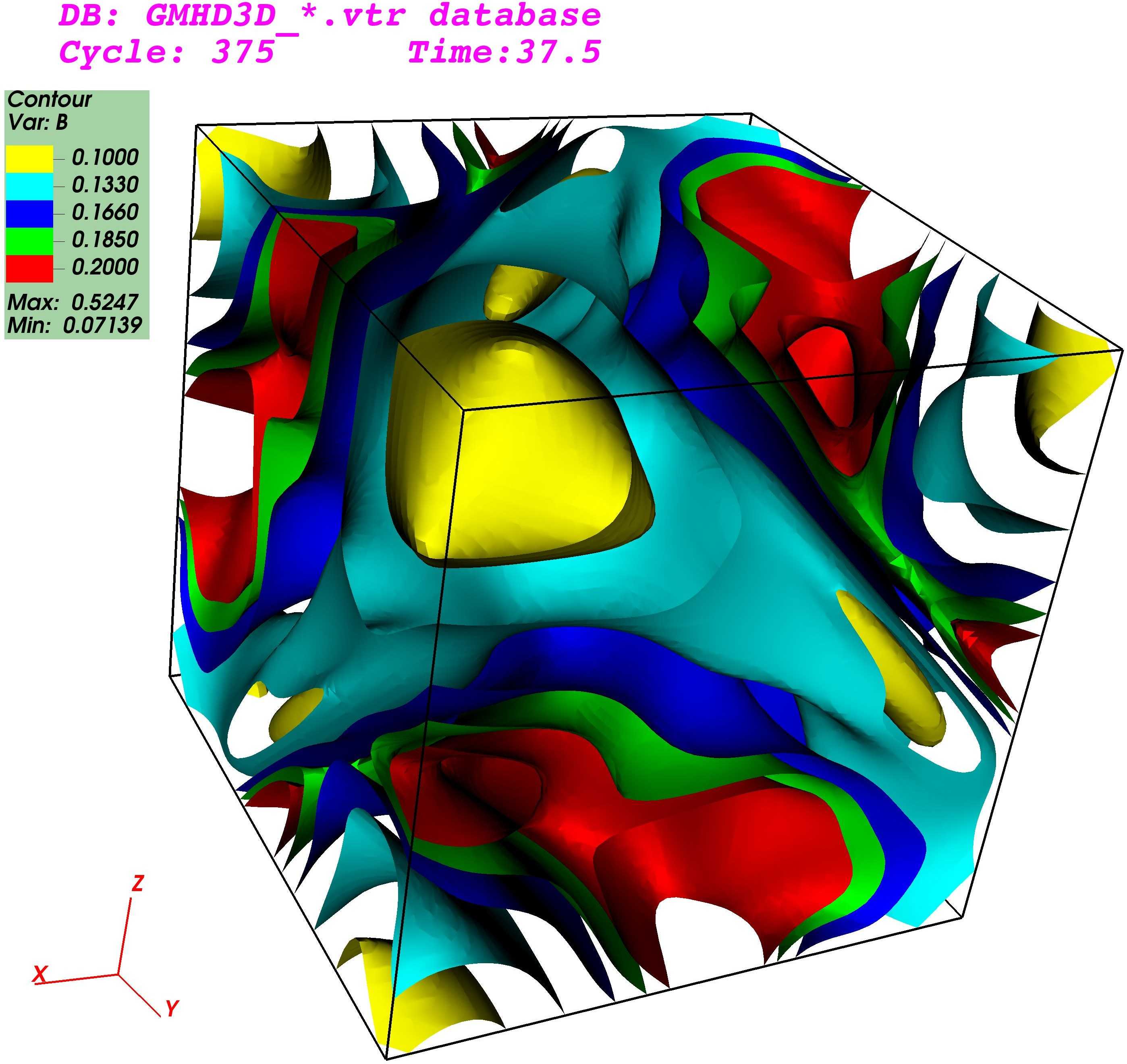}
		\caption{Time = 37.5}
	\end{subfigure}
	\begin{subfigure}{0.22\textwidth}
		\centering
		\includegraphics[scale=0.044]{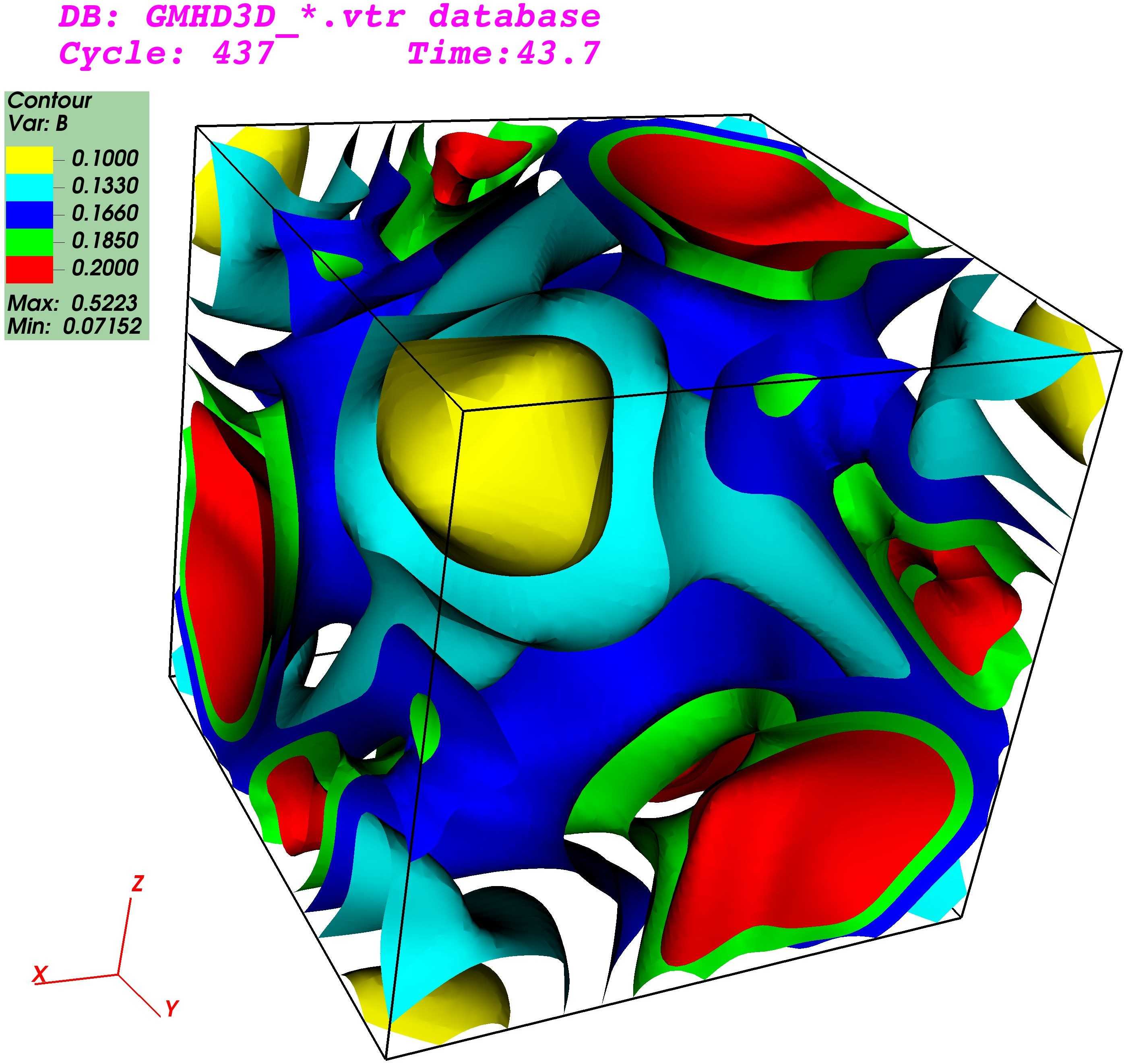}
		\caption{Time = 43.7}
	\end{subfigure}
	\begin{subfigure}{0.22\textwidth}
		\centering
		\includegraphics[scale=0.044]{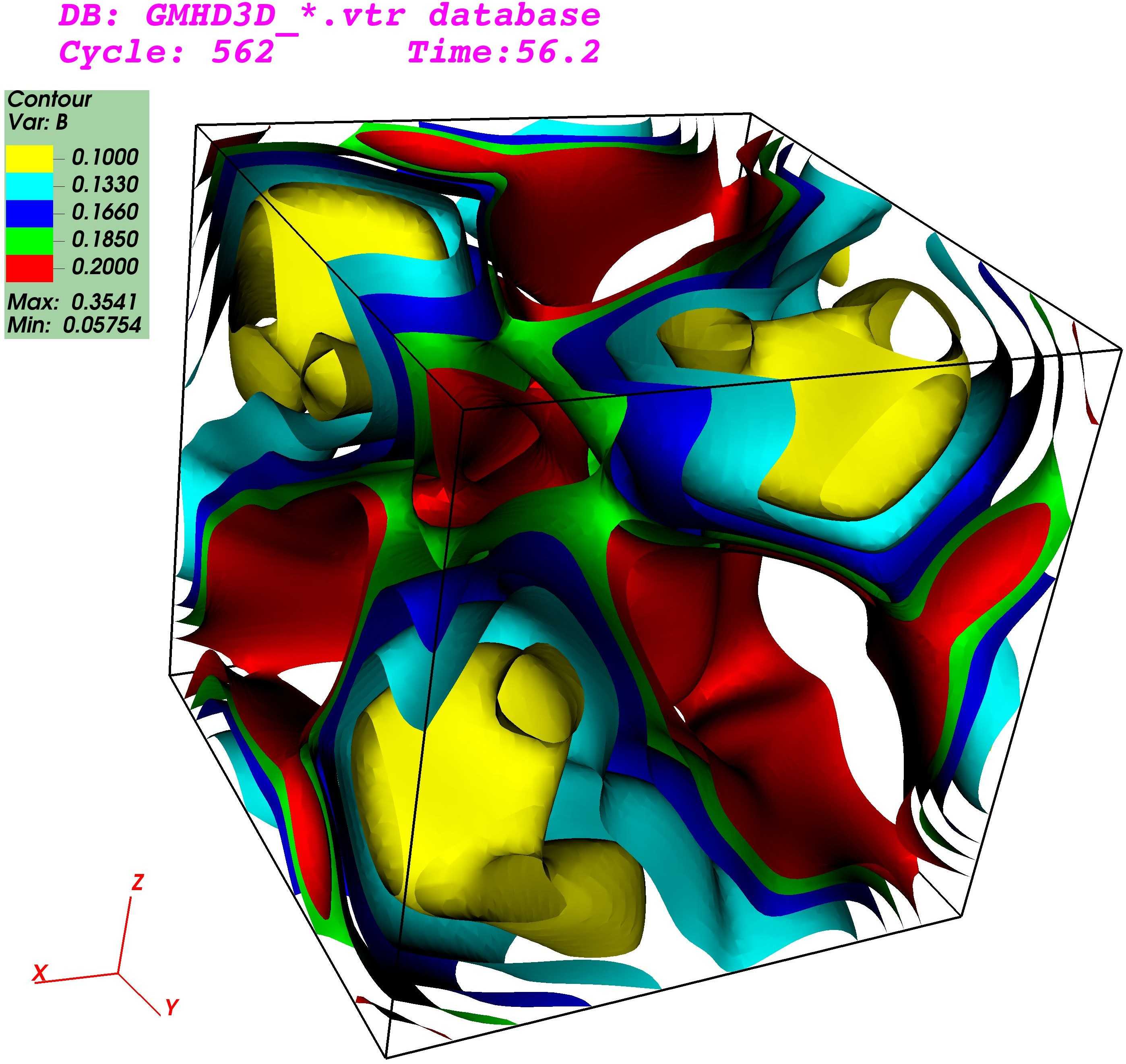}
		\caption{Time = 56.2}
	\end{subfigure}\\
	\centering
	\begin{turn}{90} 
		\LARGE{\textbf{\textcolor{blue}{\hspace{-4.0cm} $\longleftarrow$ PLUTO4.4 $\longrightarrow$}}}
	\end{turn}
	\begin{subfigure}{0.23\textwidth}
		\centering
		\includegraphics[scale=0.044]{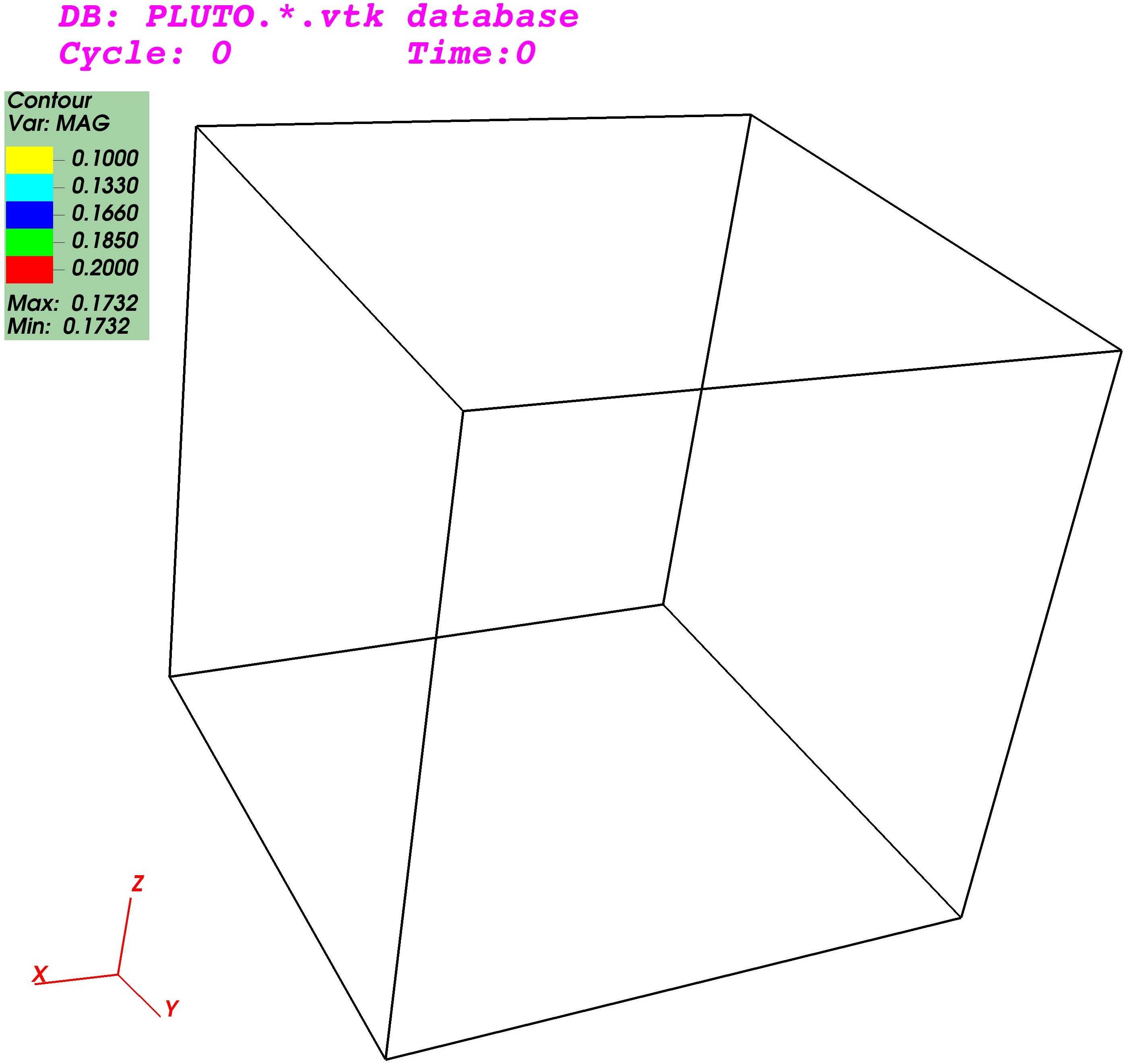}
		\caption{Time = 0.0}
	\end{subfigure}
	\begin{subfigure}{0.23\textwidth}
		\centering
		\includegraphics[scale=0.044]{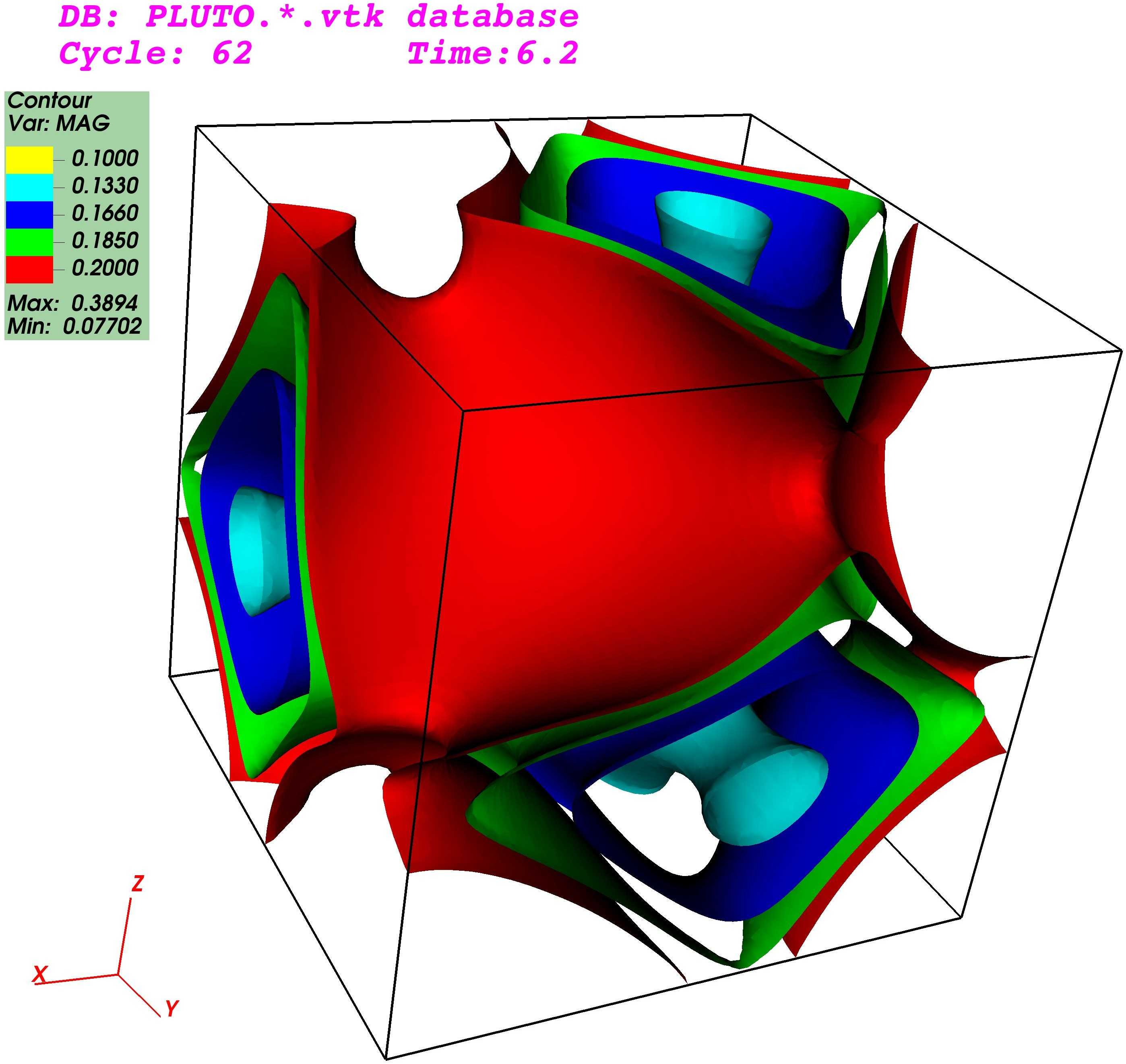}
		\caption{Time = 6.2}
	\end{subfigure}
	\begin{subfigure}{0.23\textwidth}
		\centering
		\includegraphics[scale=0.044]{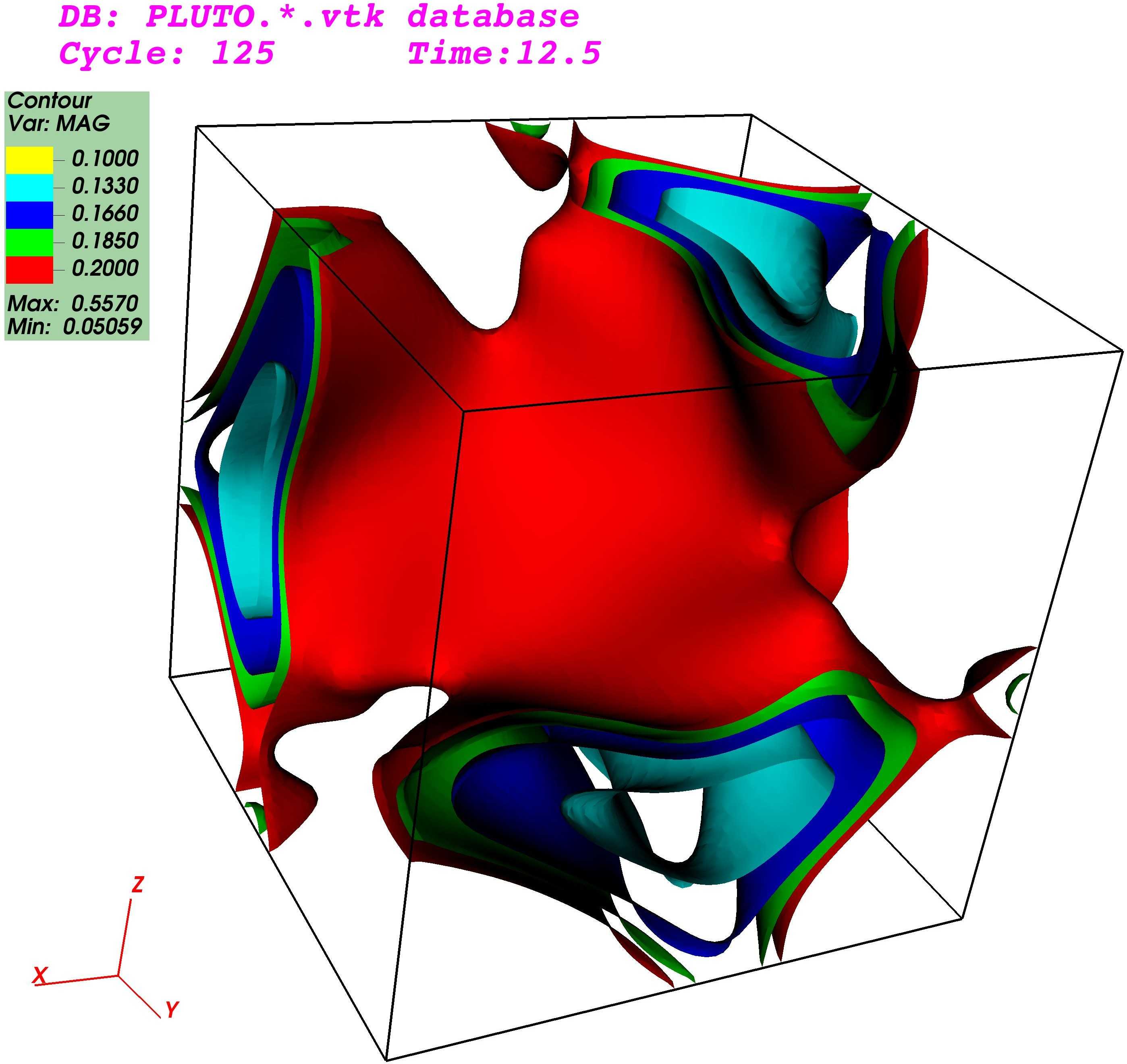}
		\caption{Time = 12.5}
	\end{subfigure}
	\begin{subfigure}{0.23\textwidth}
		\centering
		\includegraphics[scale=0.044]{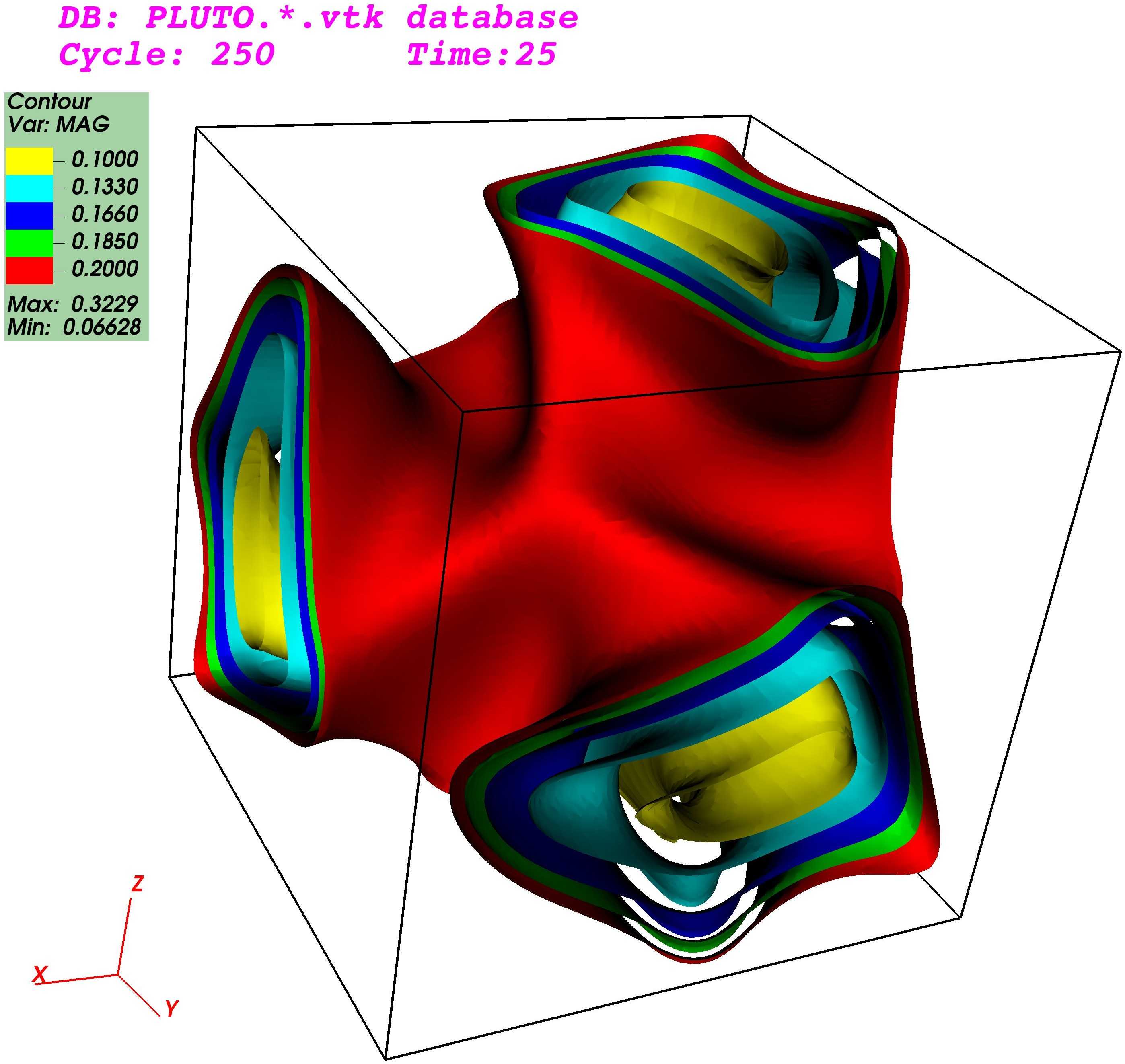}
		\caption{Time = 25.0}
	\end{subfigure}
	\begin{subfigure}{0.23\textwidth}
		\centering
		\includegraphics[scale=0.044]{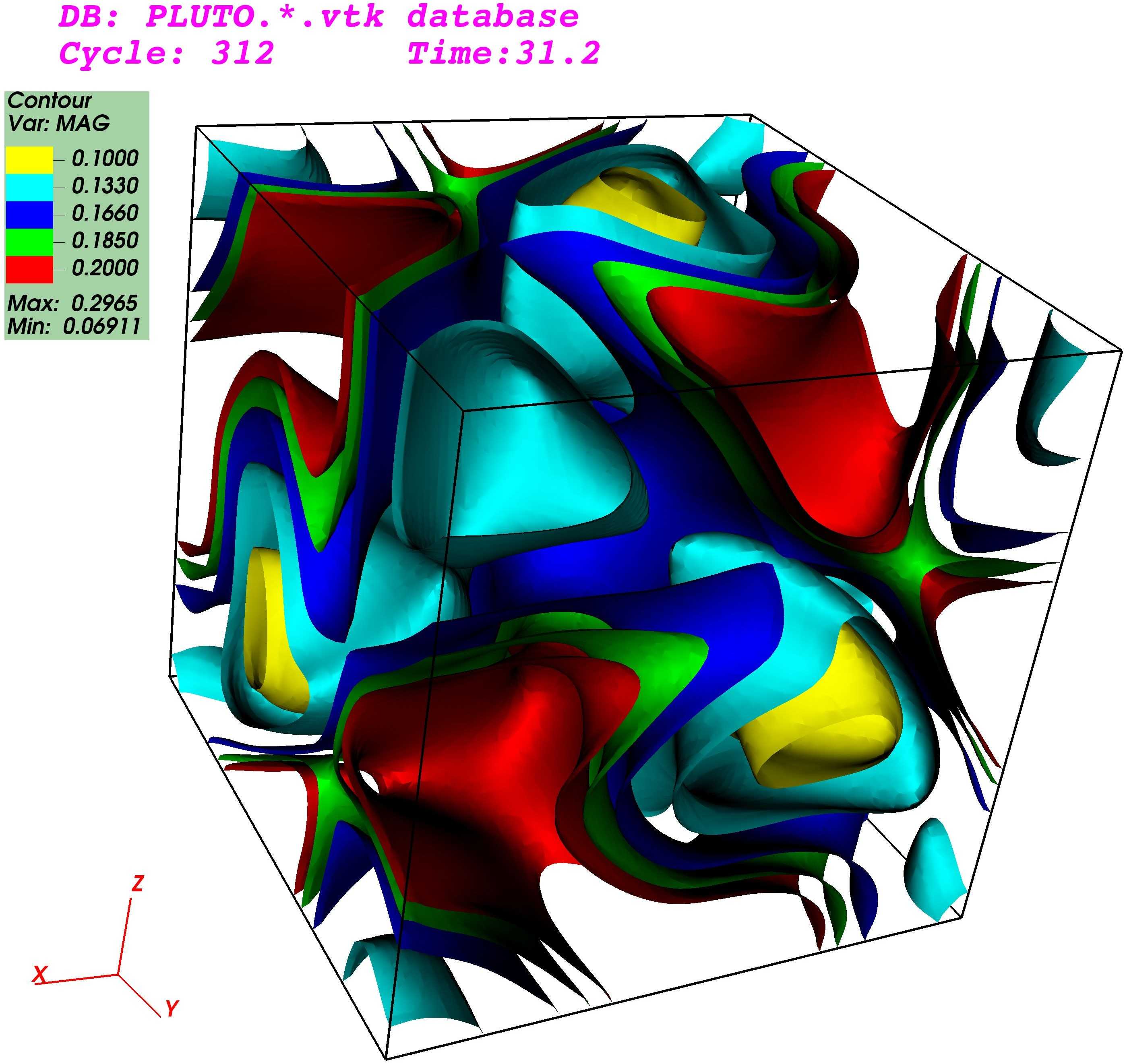}
		\caption{Time = 31.2}
	\end{subfigure}
	\begin{subfigure}{0.23\textwidth}
		\centering
		\includegraphics[scale=0.044]{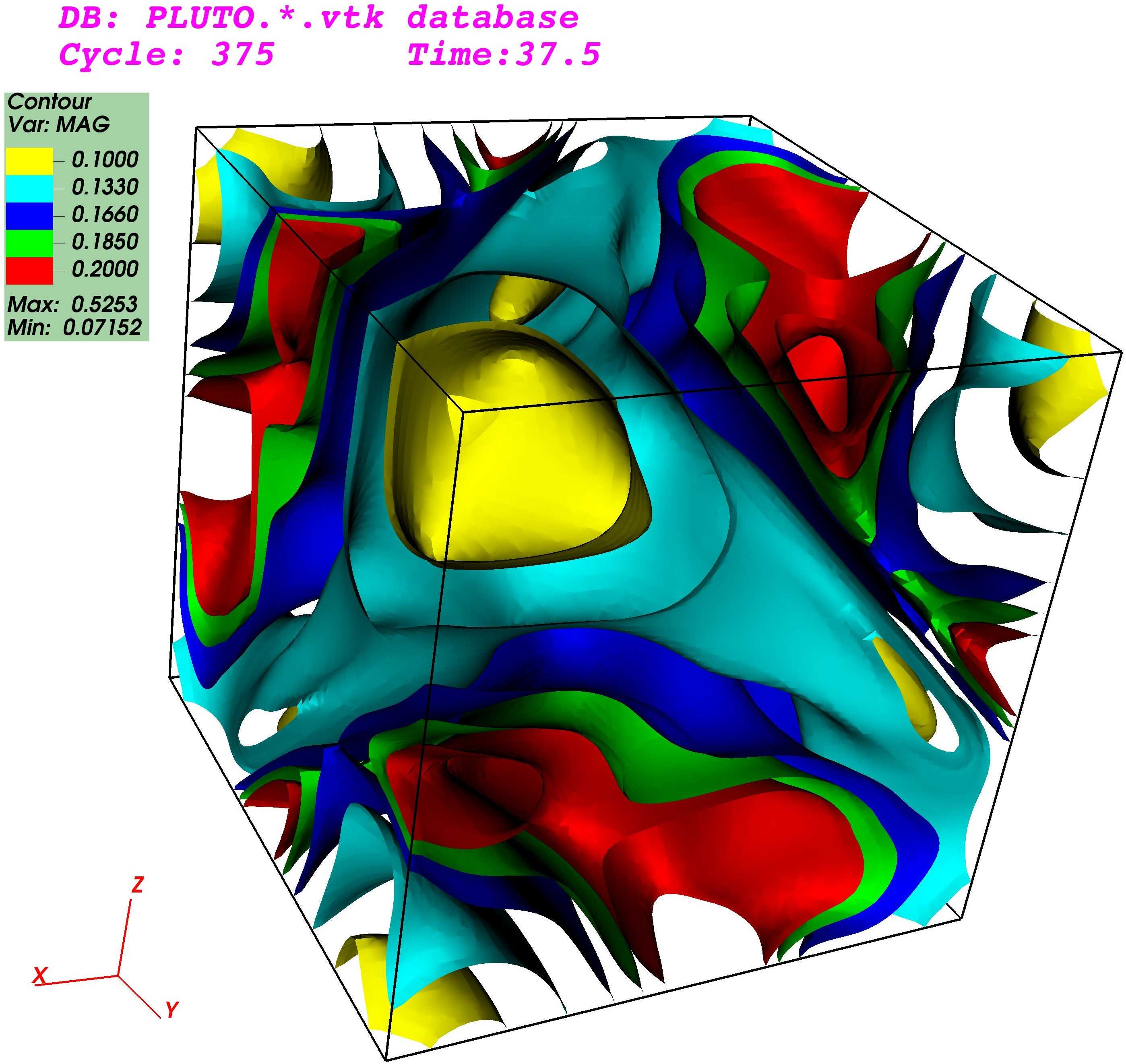}
		\caption{Time = 37.5}
	\end{subfigure}
	\begin{subfigure}{0.22\textwidth}
		\centering
		\includegraphics[scale=0.044]{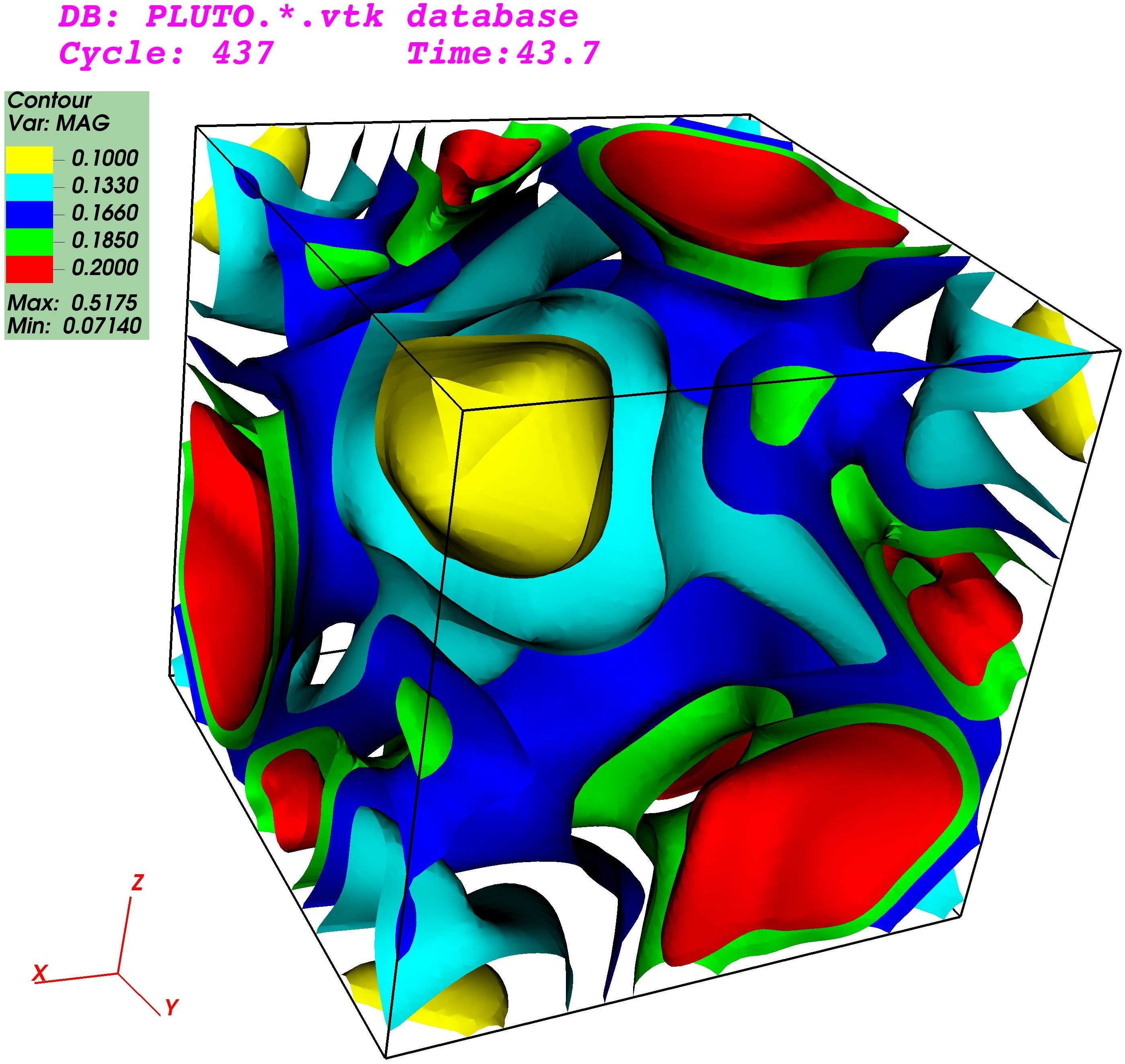}
		\caption{Time = 43.7}
	\end{subfigure}
	\begin{subfigure}{0.22\textwidth}
		\centering
		\includegraphics[scale=0.044]{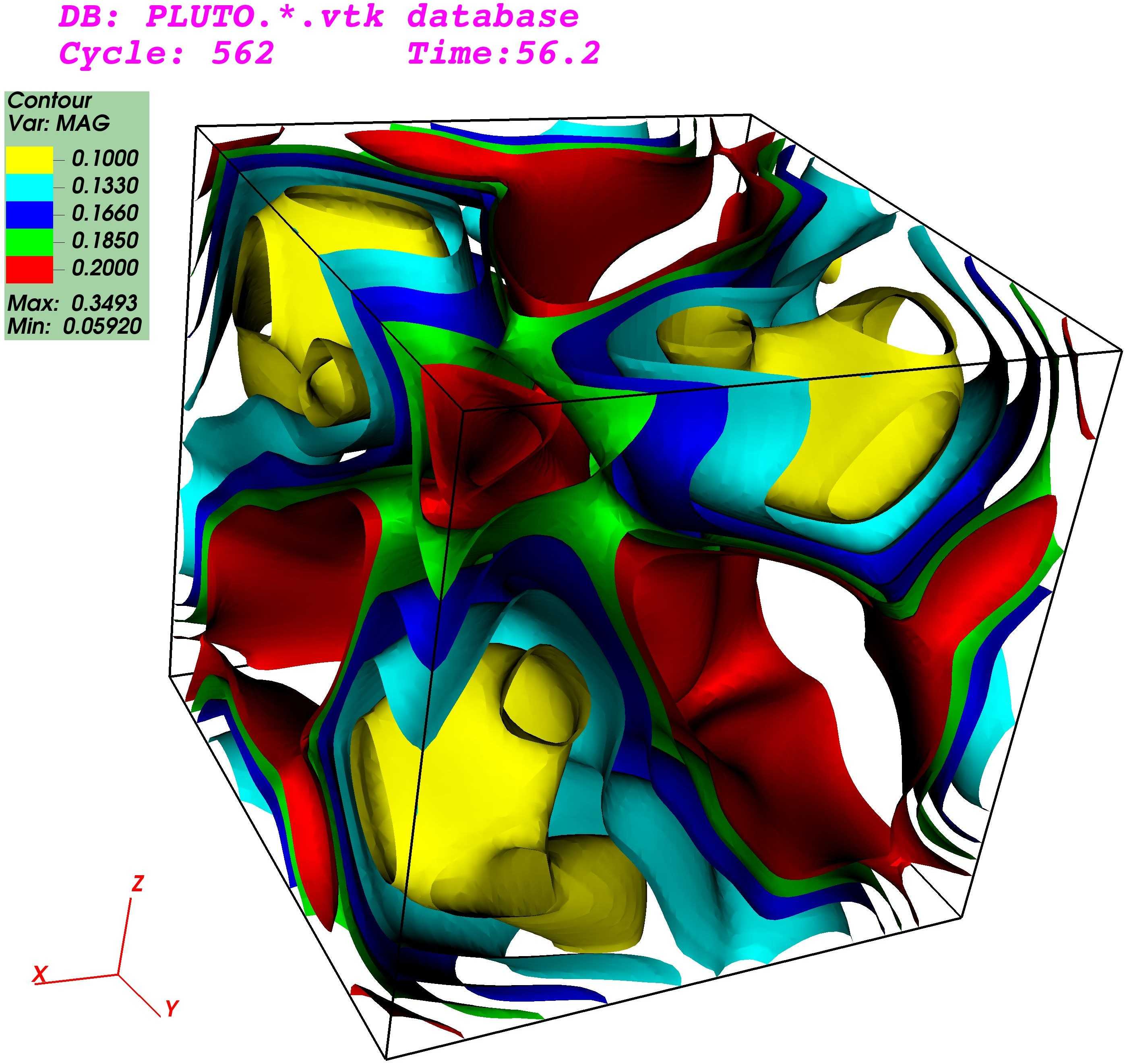}
		\caption{Time = 56.2}
	\end{subfigure}
	\caption{Non-Recurring 3D  ABC Flow Magnetic Field iso-surface from from GMHD3D code [Top two rows (a--h)] and PLUTO4.4 code [Bottom two rows (i--p)]. Values of iso-surface: \textbf{0.02 (Red), 0.185 (Green), 0.166 (Blue), 0.133 (Cyan) and 0.1 (Yellow)}. Simulation Details: Reynolds number $R_e = R_m = 1000$, Grid resolution $N = 128^3$, Time stepping $dt = 10^{-4}$, initial fluid velocity $u_0 = 1.0$, Alfven Mach number $M_A = 1.0$.}
	\label{Non Recurr 3D ABC Iso B}
\end{figure*}

For the recurrence study, we first focus on the divergence-free 3D Arnold-Beltrami-Childress [ABC] flow. In this case, the profile of the flow is given by 
		\begin{equation}\label{Recurrence ABC_like}
			\begin{aligned}
				u_x &= u_0 [ A \sin(k_0z) + C \cos(k_0y) ]\\
				u_y &= u_0 [ B \sin(k_0x) + A \cos(k_0z) ]\\
				u_z &= u_0 [ C \sin(k_0y) + B \cos(k_0x) ]
			\end{aligned}
		\end{equation}
 where $A=B=C=1$ and $k_0=1$. For the 3D ABC flow, which we have previously investigated, oscillations of the kinetic and magnetic energy in the form of coherent non-linear oscillation have been observed. We now plot the velocity and magnetic field iso-surfaces from the GMHD3D code and the PLUTO4.4 code, respectively, as illustrated in Figs. \ref{Non Recurr 3D ABC Iso V} \& \ref{Non Recurr 3D ABC Iso B}. As depicted in fig. \ref{Non Recurr 3D ABC Iso V}, the values of the velocity isosurfaces are 0.1 (Red), 0.09 (Green), 0.08 (Blue), 0.05 (Cyan), and 0.03 (Yellow). 

Also, the values for the magnetic field iso-surface are as follows: 0.02 (Red), 0.185 (Green), 0.166 (Blue), 0.133 (Cyan), and 0.10 (Yellow) [See Fig. \ref{Non Recurr 3D ABC Iso B}]. From Fig. \ref{Non Recurr 3D ABC Iso V} \& \ref{Non Recurr 3D ABC Iso B} it can be seen that neither the velocity isosurface nor the magnetic field isosurface are reconstructed back. This is a clear sign of non-recurrence for both the velocity and the magnetic field \cite{RM_Recurrence:2019}. The same thing has been observed using both GMHD3D code and PLUTO4.4 code.

\subsubsection{Recurring 3D Taylor-Green flow}
%

\begin{figure*}
	\centering
	\begin{turn}{90} 
		\LARGE{\textbf{\textcolor{blue}{\hspace{-4.0cm} $\longleftarrow$ GMHD3D $\longrightarrow$}}}
	\end{turn}
	\begin{subfigure}{0.23\textwidth}
		\centering
		\includegraphics[scale=0.044]{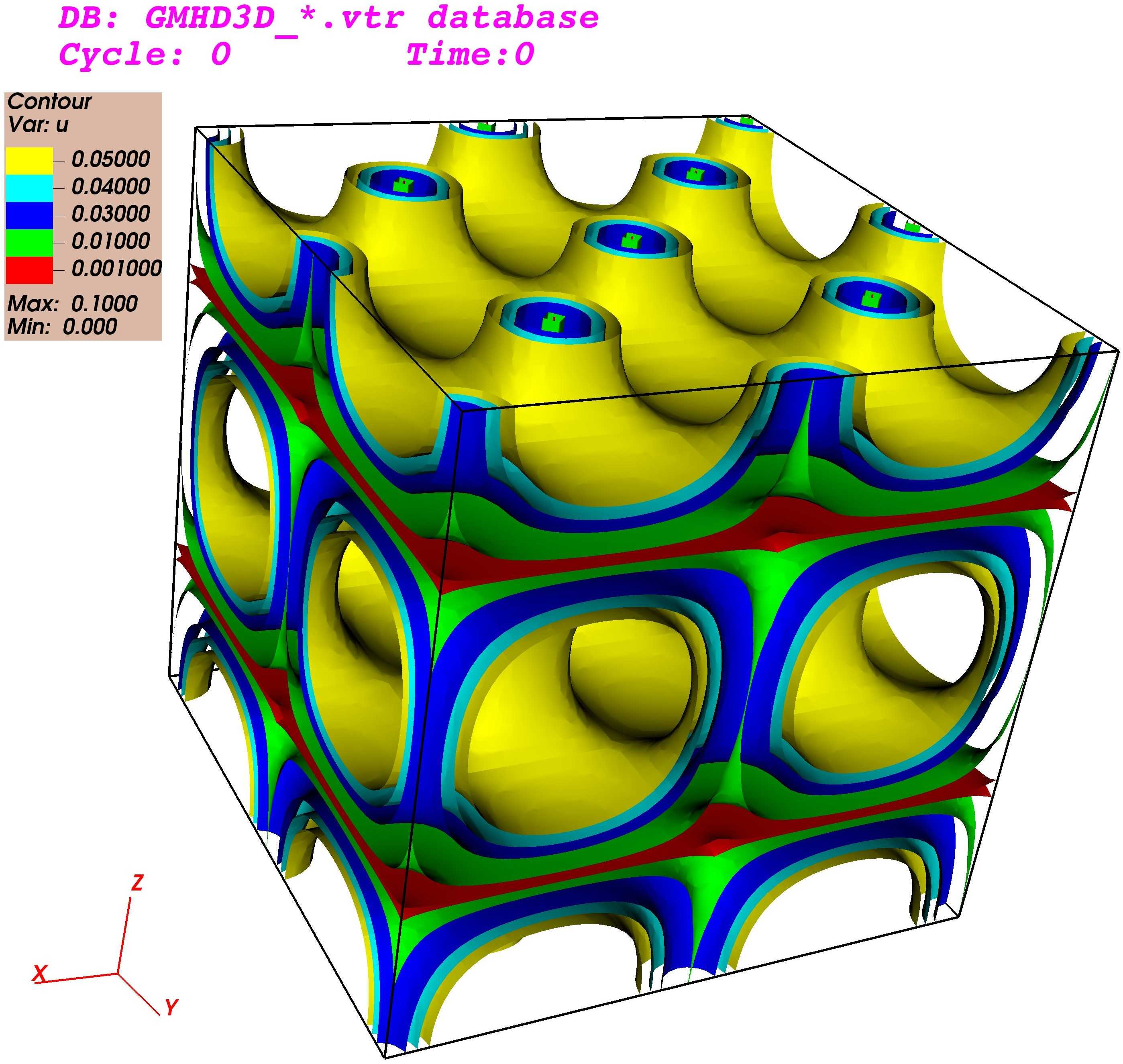}
		\caption{Time = 0.0}
	\end{subfigure}
	\begin{subfigure}{0.23\textwidth}
		\centering
		\includegraphics[scale=0.044]{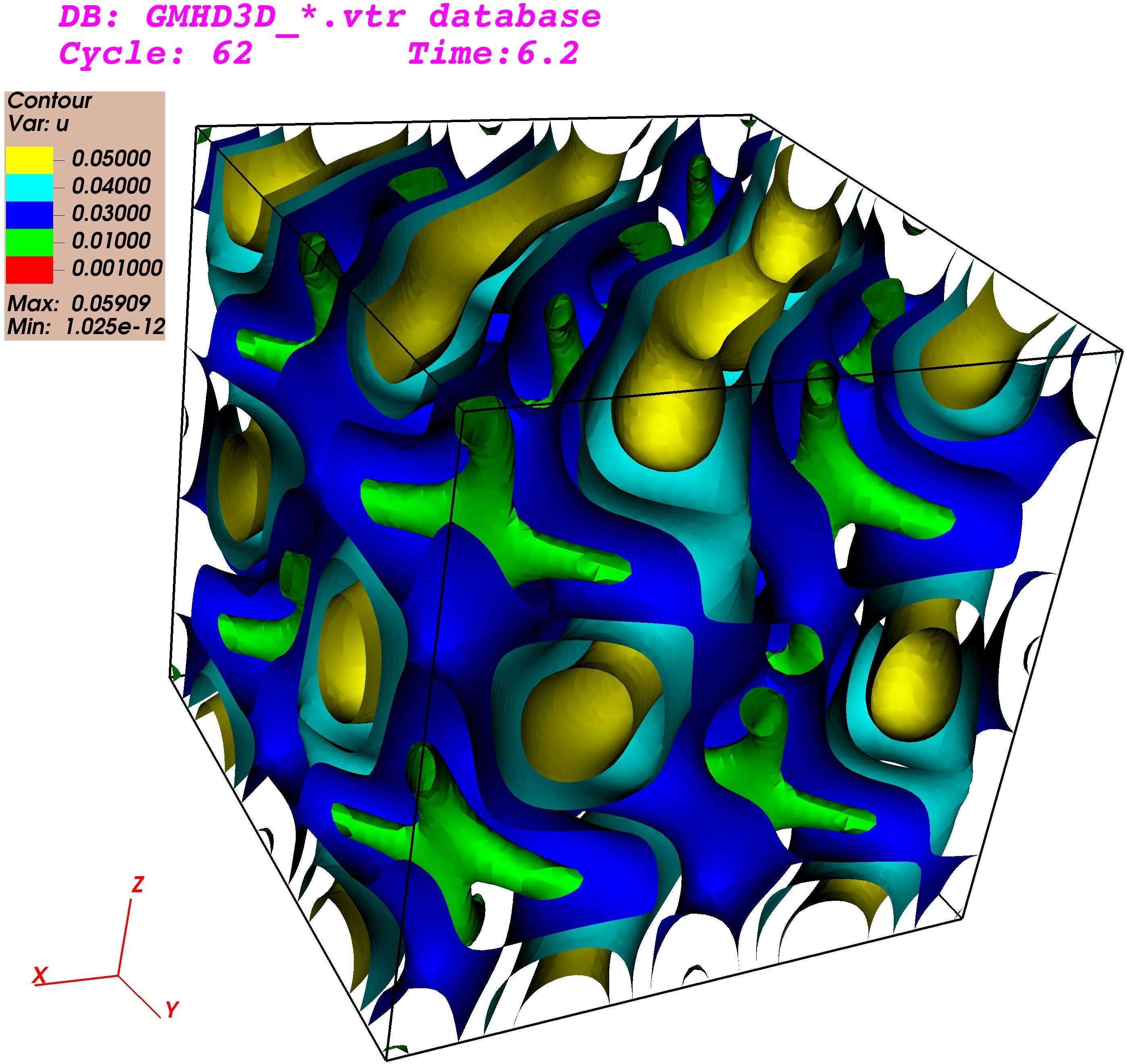}
		\caption{Time = 6.2}
	\end{subfigure}
	\begin{subfigure}{0.23\textwidth}
		\centering
		\includegraphics[scale=0.044]{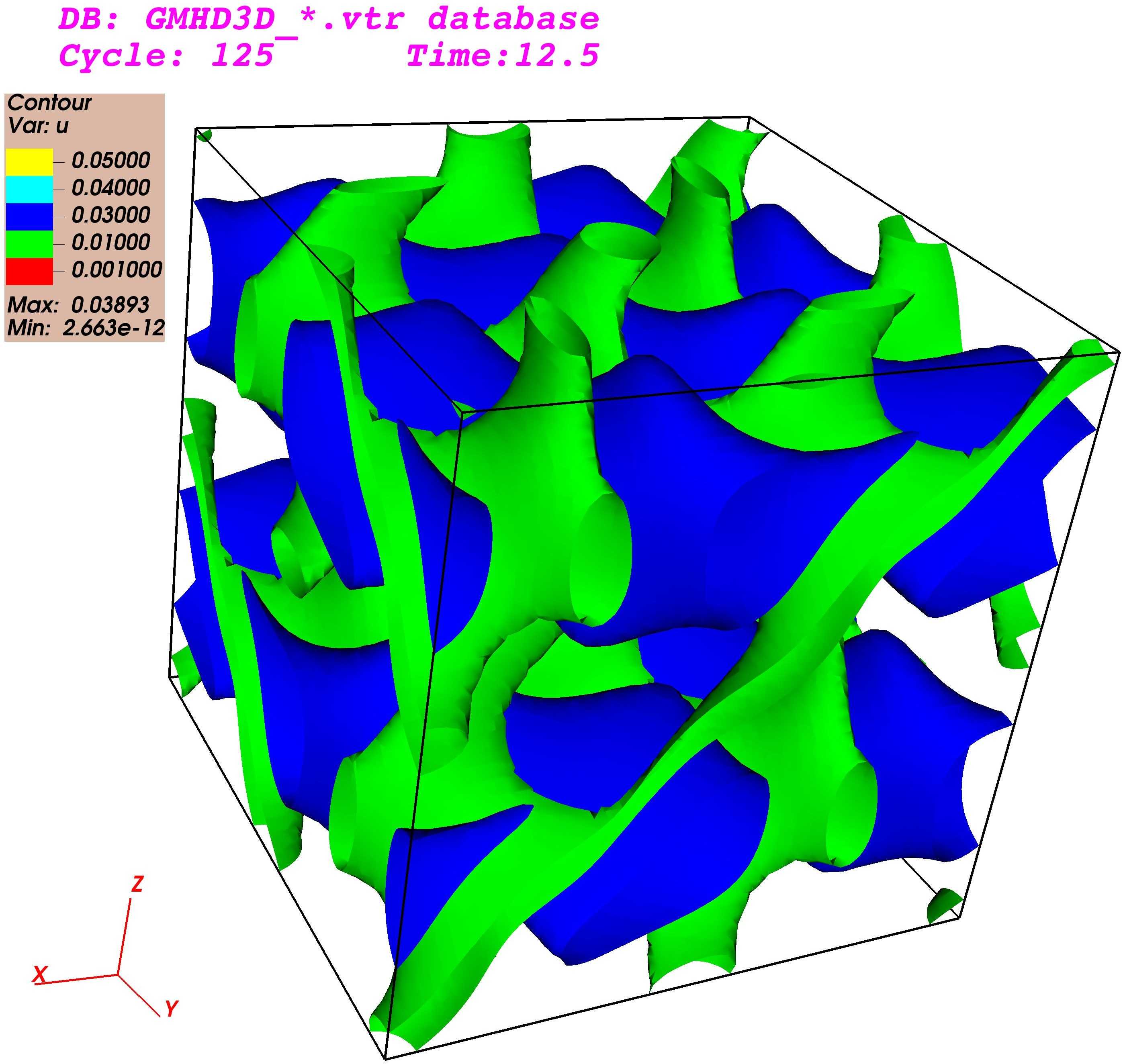}
		\caption{Time = 12.5}
	\end{subfigure}
	\begin{subfigure}{0.23\textwidth}
		\centering
		\includegraphics[scale=0.044]{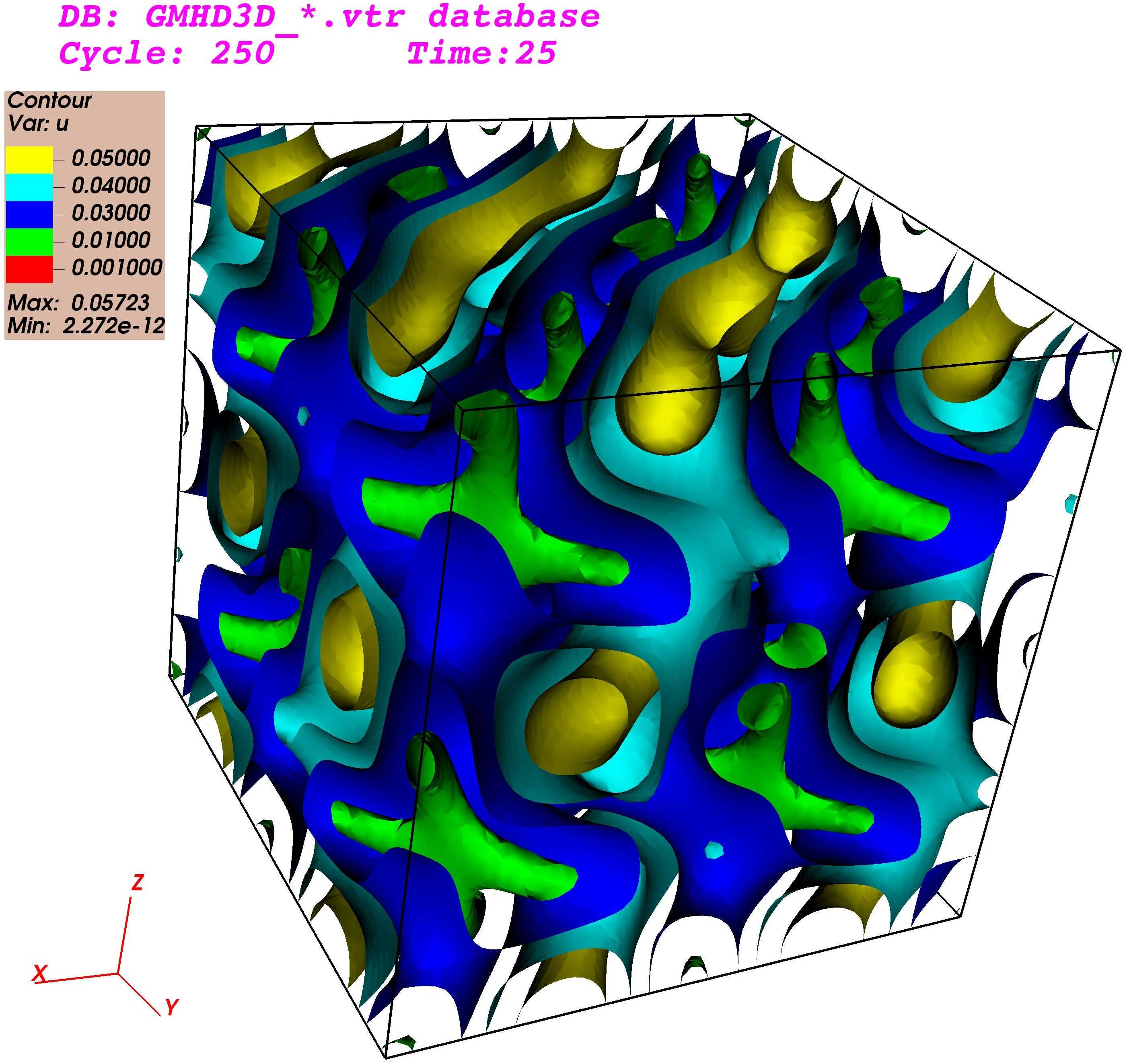}
		\caption{Time = 25.0}
	\end{subfigure}
	\begin{subfigure}{0.23\textwidth}
		\centering
		\includegraphics[scale=0.044]{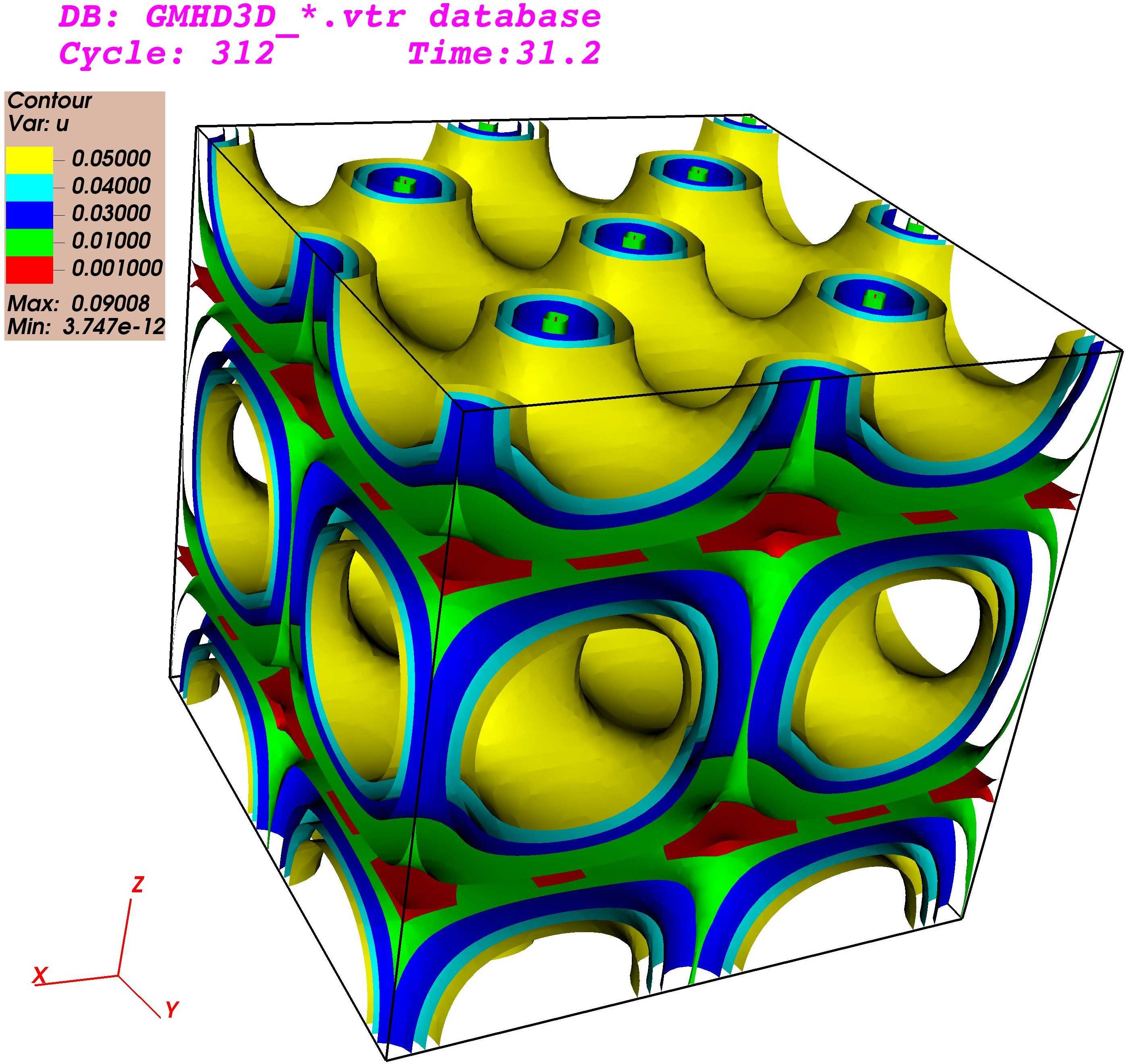}
		\caption{Time = 31.2}
	\end{subfigure}
	\begin{subfigure}{0.23\textwidth}
		\centering
		\includegraphics[scale=0.044]{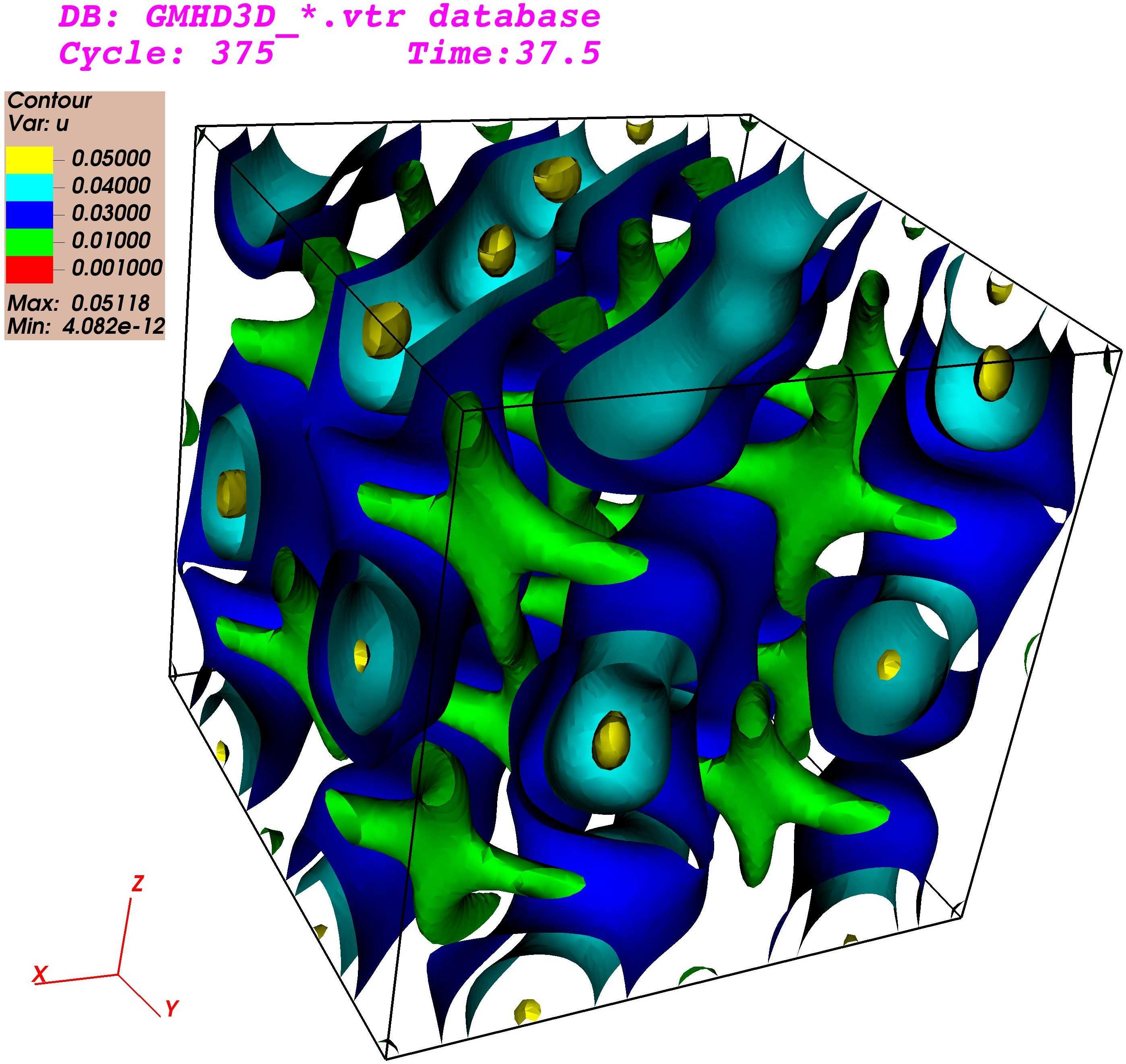}
		\caption{Time = 37.5}
	\end{subfigure}
	\begin{subfigure}{0.22\textwidth}
		\centering
		\includegraphics[scale=0.044]{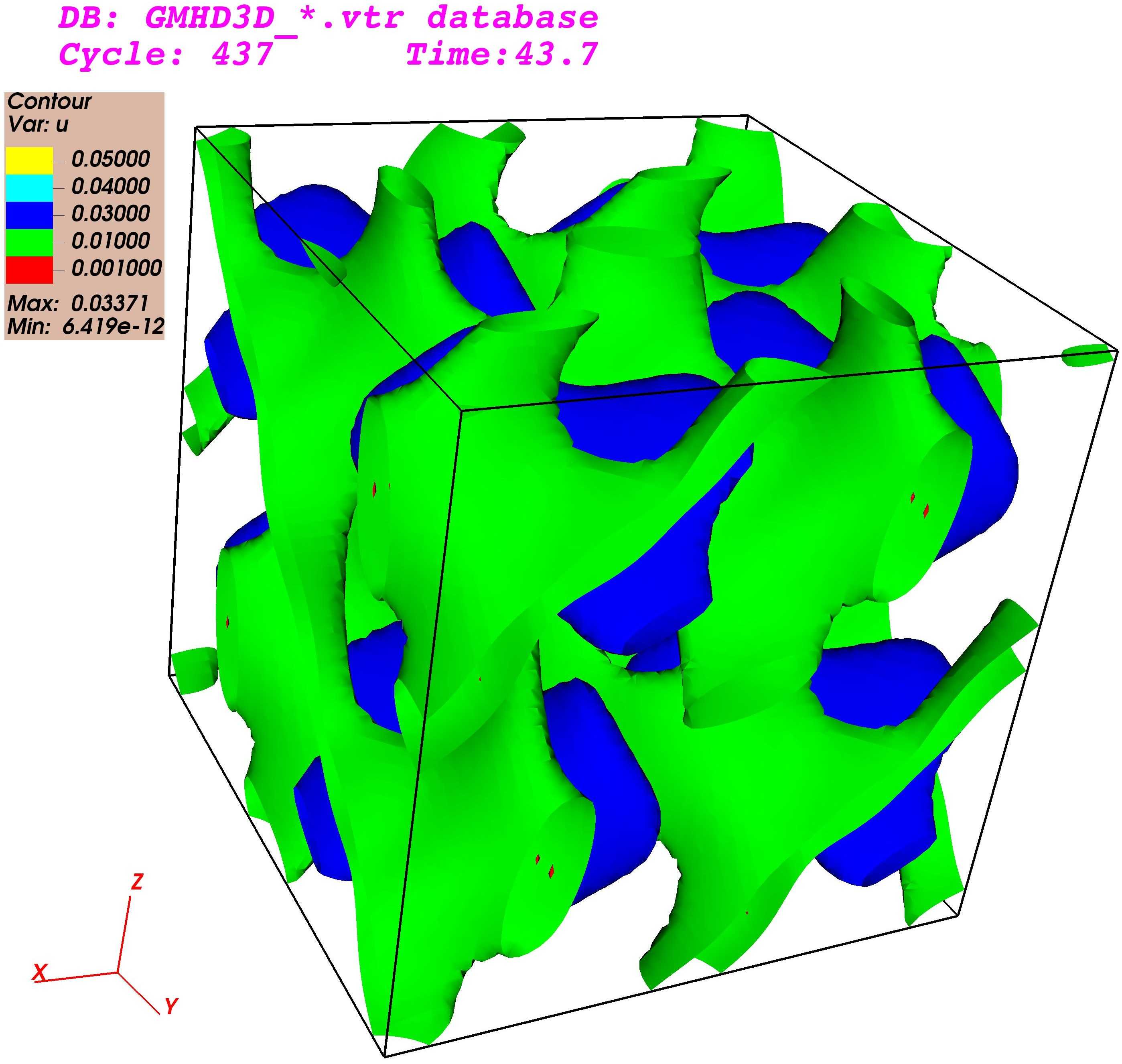}
		\caption{Time = 43.7}
	\end{subfigure}
	\begin{subfigure}{0.22\textwidth}
		\centering
		\includegraphics[scale=0.044]{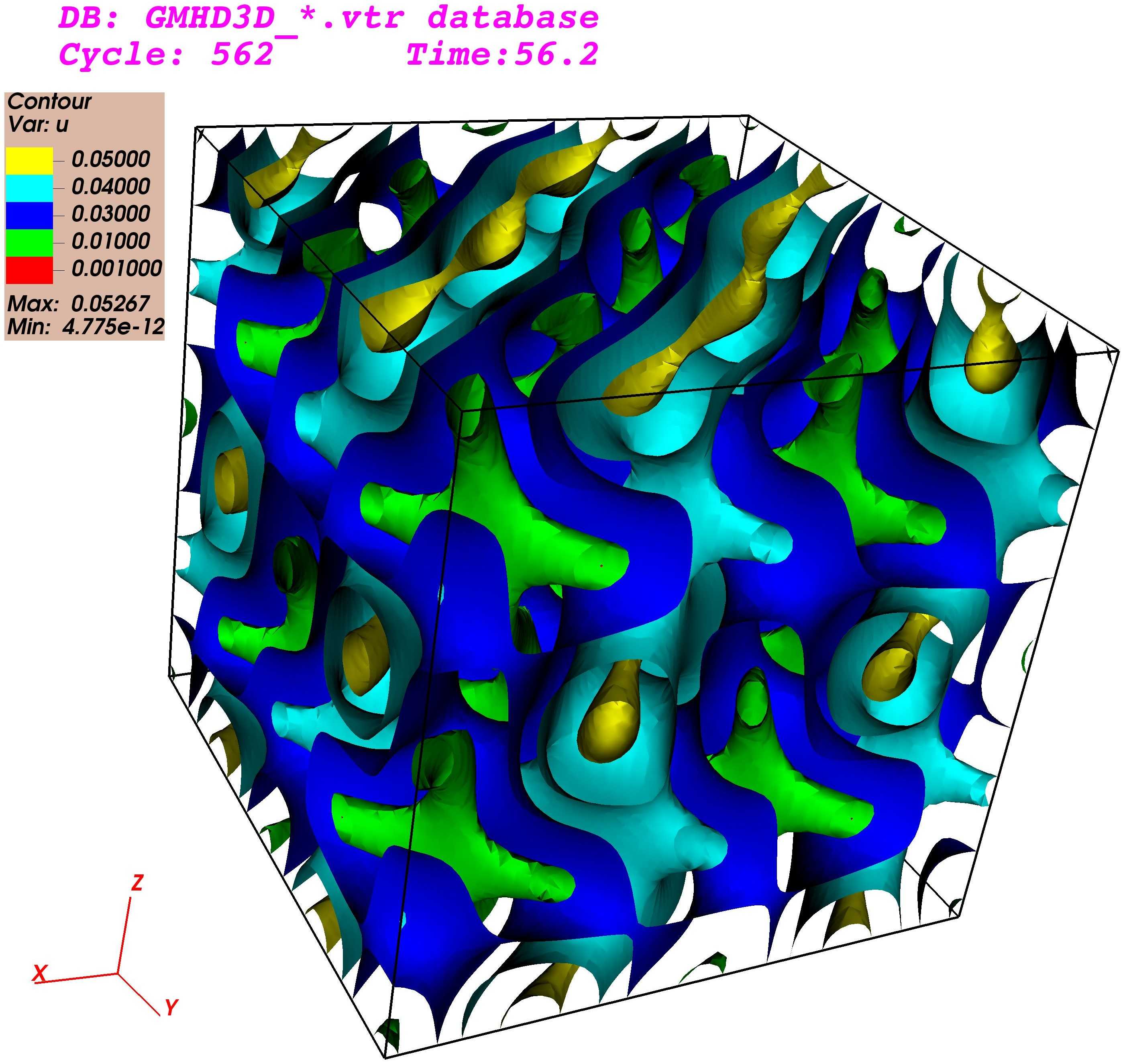}
		\caption{Time = 56.2}
	\end{subfigure}\\
	\centering
	\begin{turn}{90} 
		\LARGE{\textbf{\textcolor{blue}{\hspace{-4.0cm} $\longleftarrow$ PLUTO4.4 $\longrightarrow$}}}
	\end{turn}
	\begin{subfigure}{0.23\textwidth}
		\centering
		\includegraphics[scale=0.044]{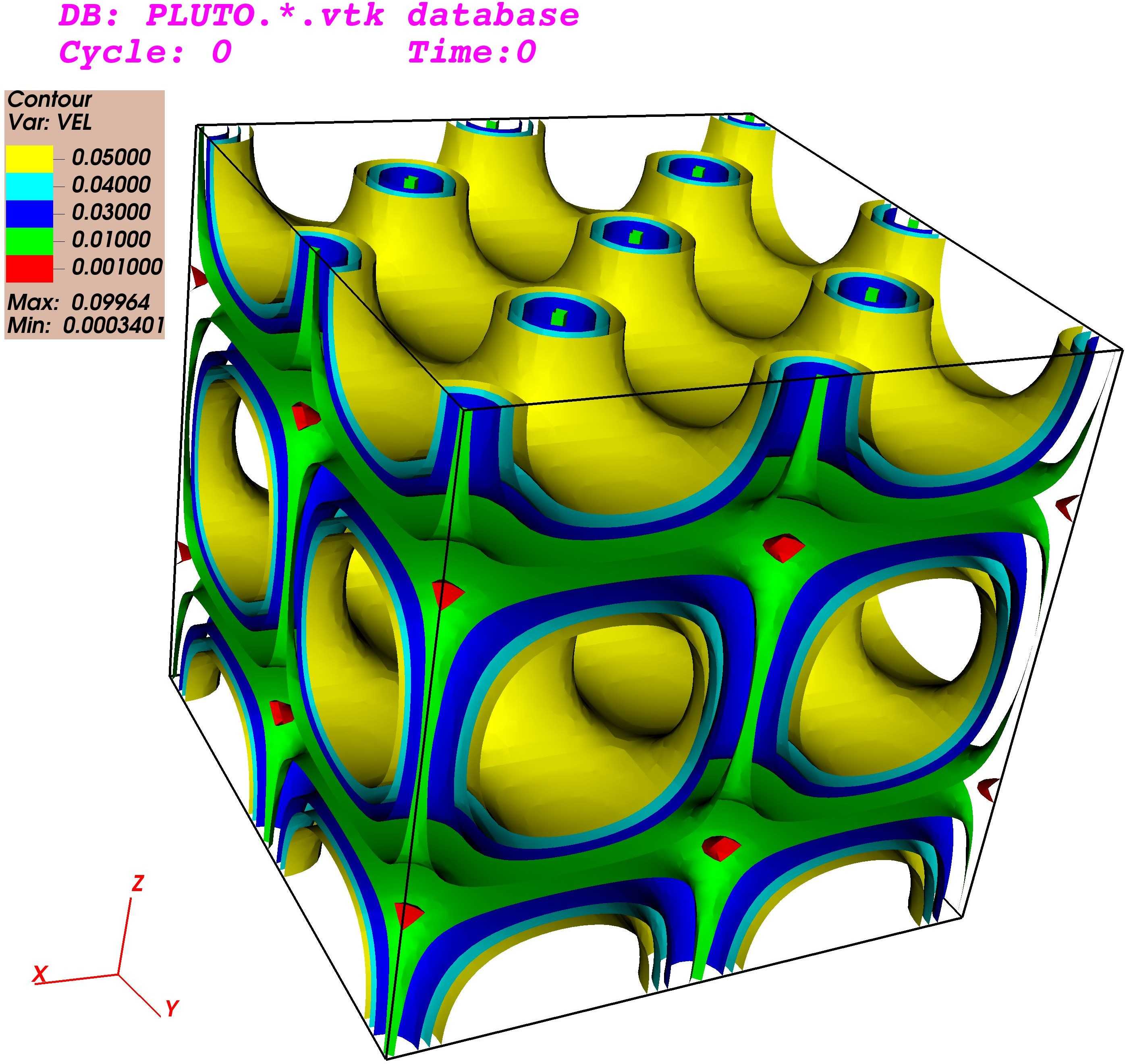}
		\caption{Time = 0.0}
	\end{subfigure}
	\begin{subfigure}{0.23\textwidth}
		\centering
		\includegraphics[scale=0.044]{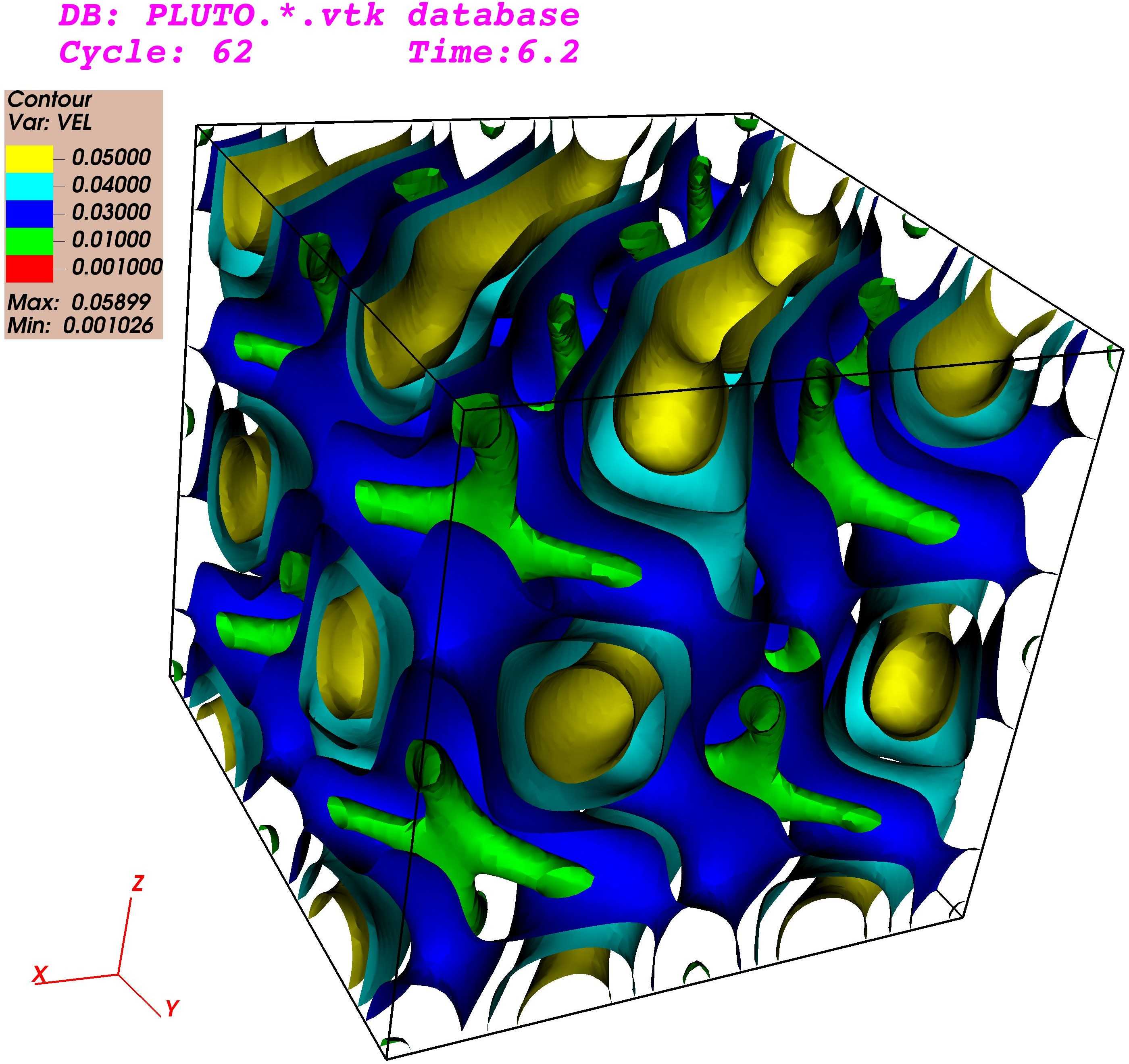}
		\caption{Time = 6.2}
	\end{subfigure}
	\begin{subfigure}{0.23\textwidth}
		\centering
		\includegraphics[scale=0.044]{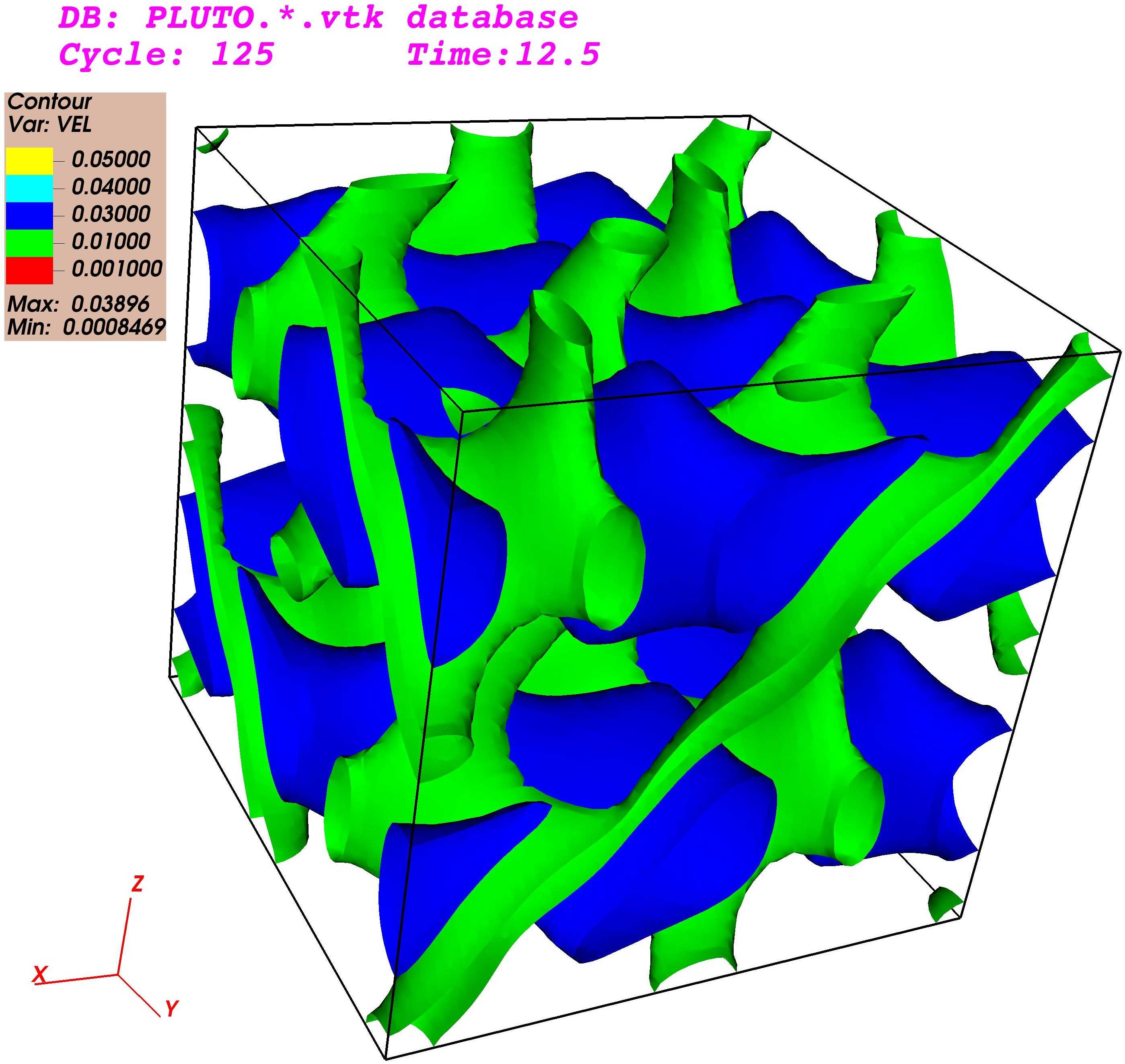}
		\caption{Time = 12.5}
	\end{subfigure}
	\begin{subfigure}{0.23\textwidth}
		\centering
		\includegraphics[scale=0.044]{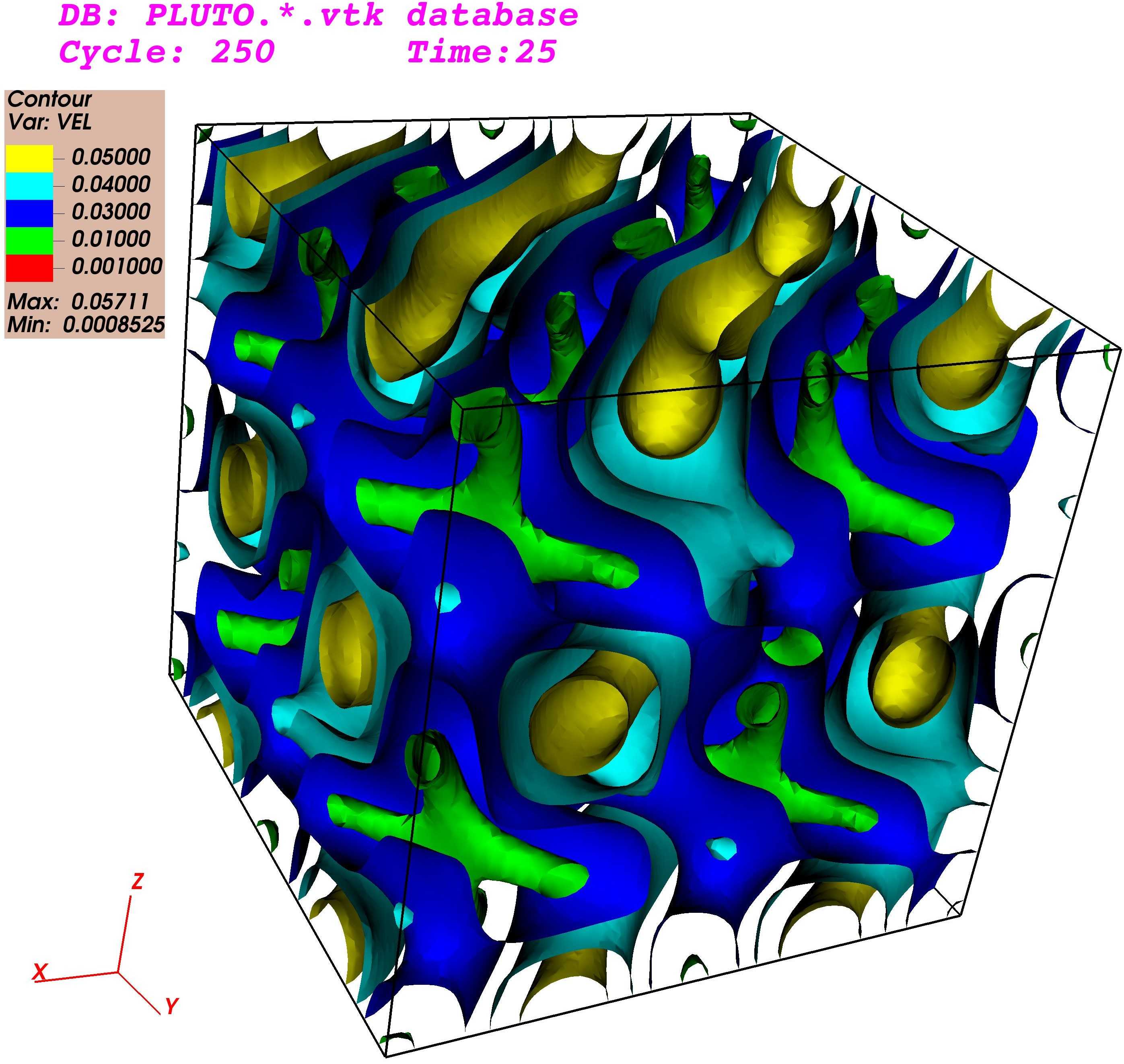}
		\caption{Time = 25.0}
	\end{subfigure}
	\begin{subfigure}{0.23\textwidth}
		\centering
		\includegraphics[scale=0.044]{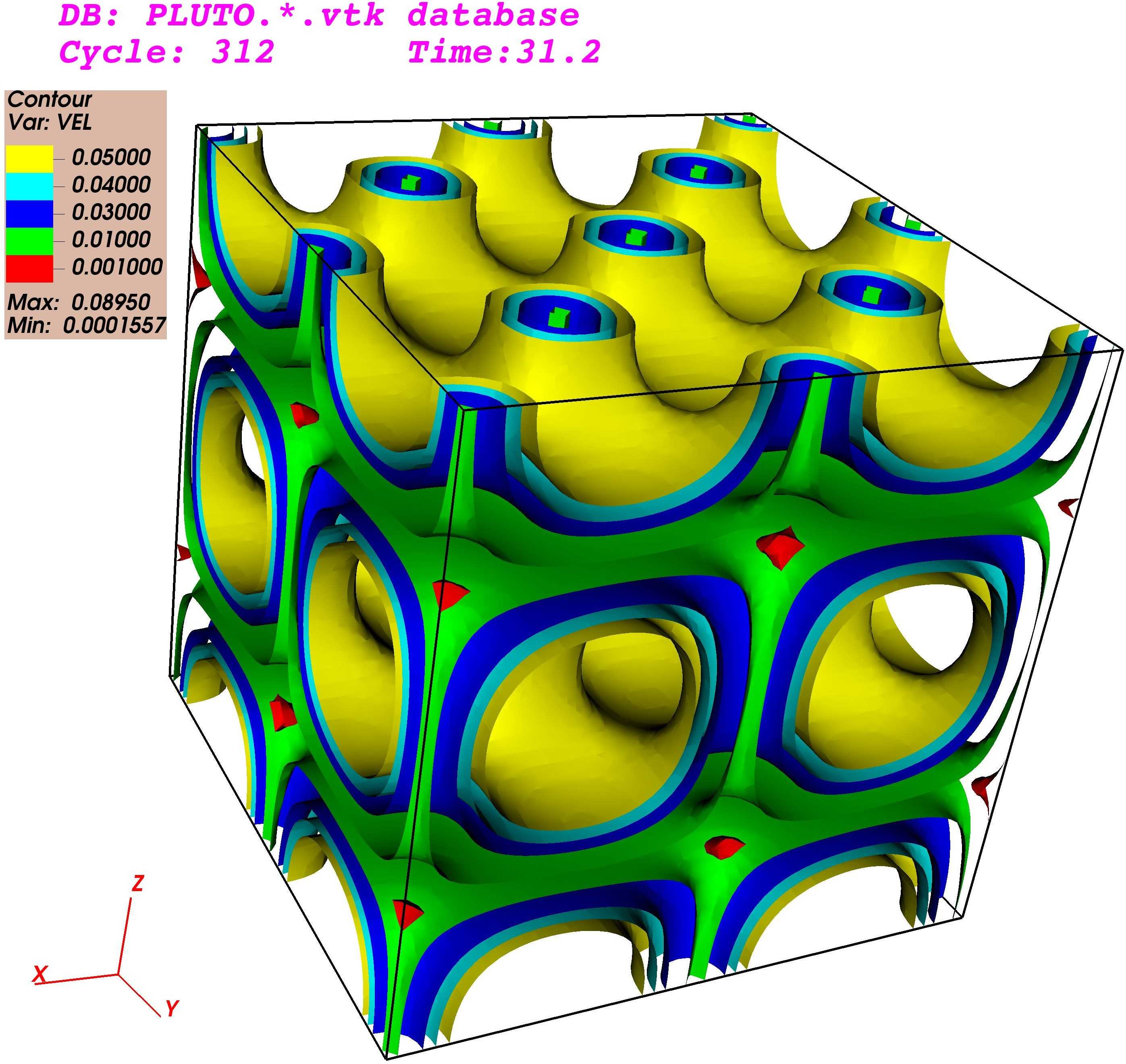}
		\caption{Time = 31.2}
	\end{subfigure}
	\begin{subfigure}{0.23\textwidth}
		\centering
		\includegraphics[scale=0.044]{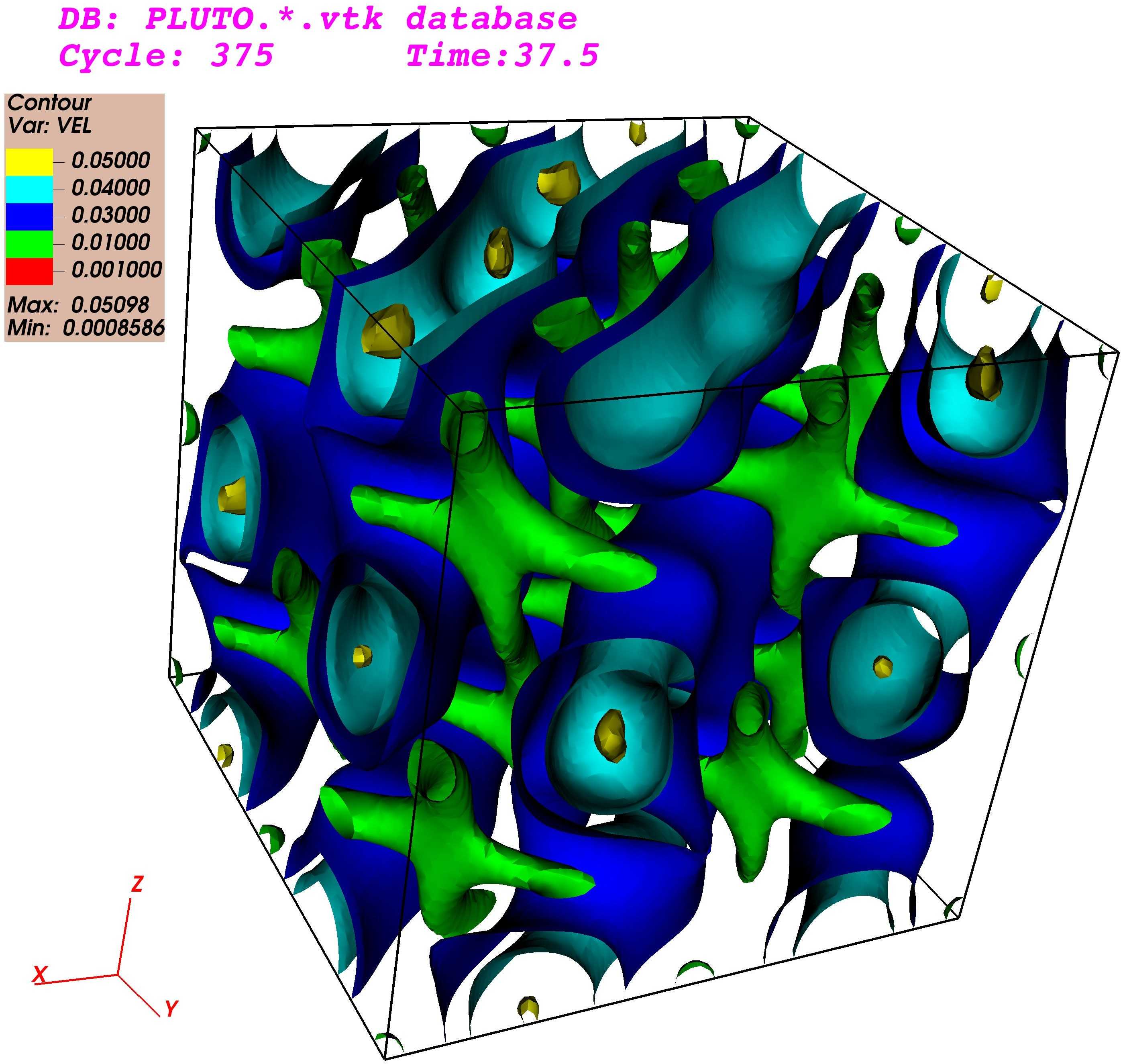}
		\caption{Time = 37.5}
	\end{subfigure}
	\begin{subfigure}{0.22\textwidth}
		\centering
		\includegraphics[scale=0.044]{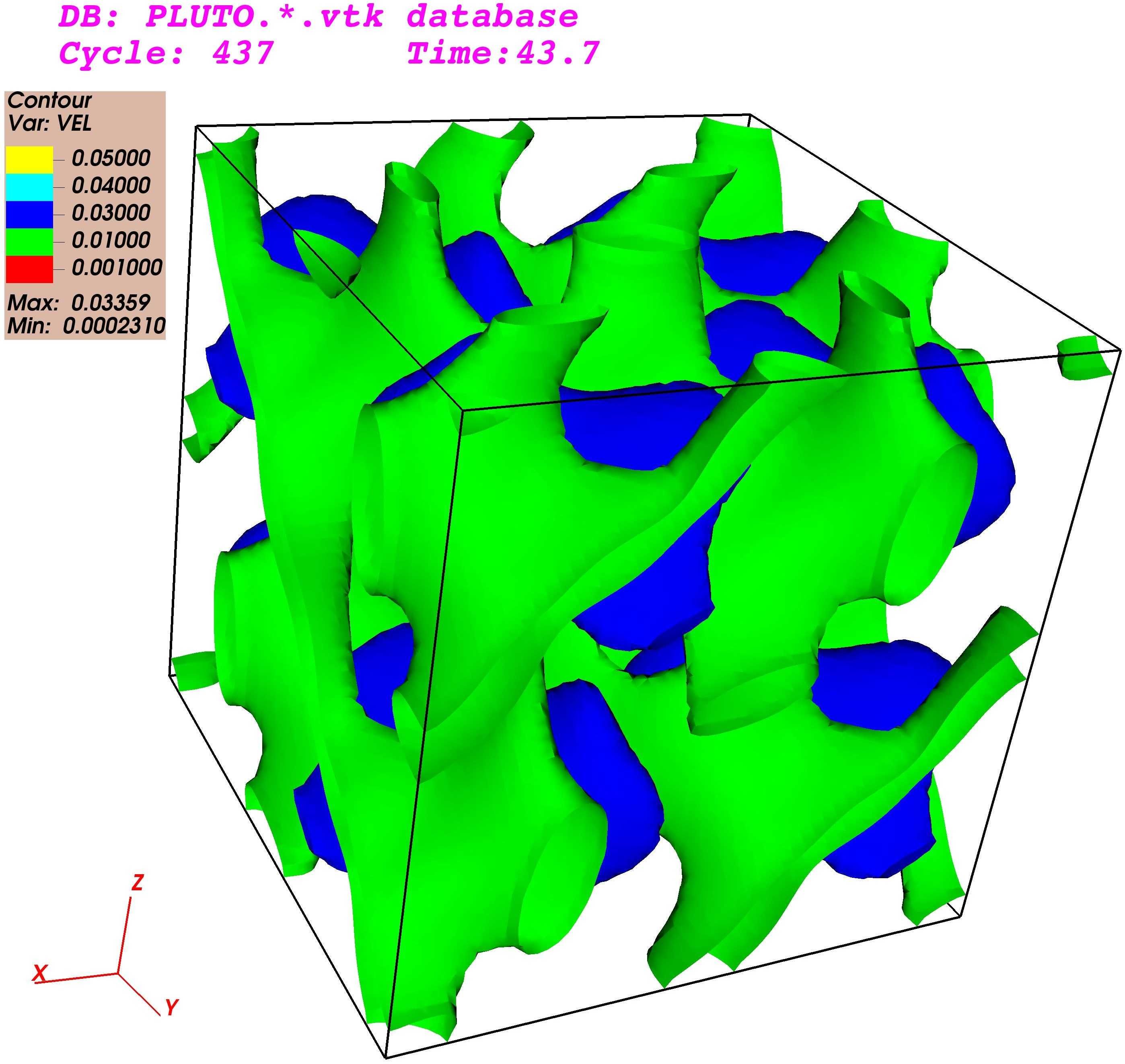}
		\caption{Time = 43.7}
	\end{subfigure}
	\begin{subfigure}{0.22\textwidth}
		\centering
		\includegraphics[scale=0.044]{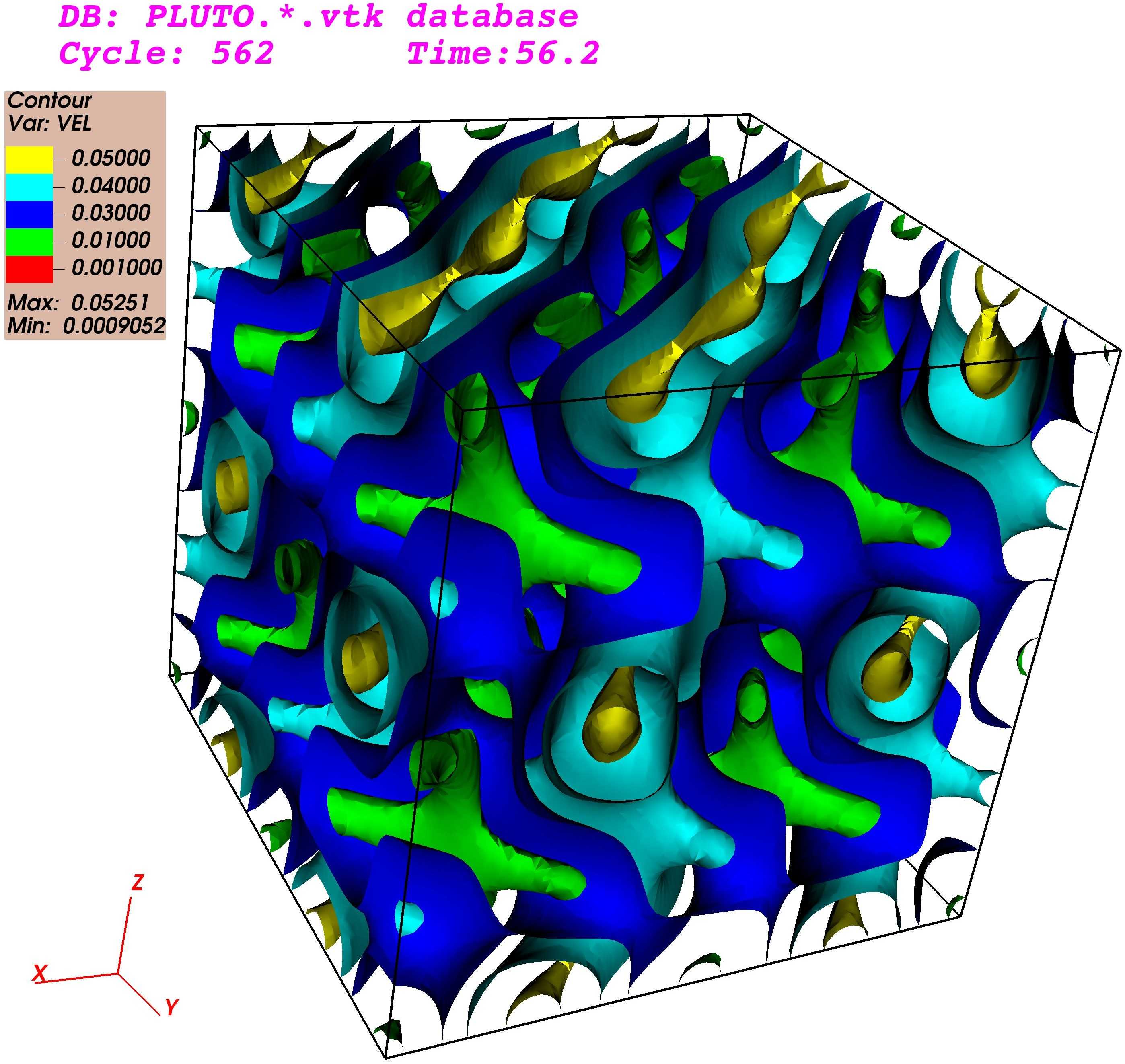}
		\caption{Time = 56.2}
	\end{subfigure}
	\caption{Recurring 3D  Taylor-Green Flow Velocity iso-surface from GMHD3D code [Top two rows (a--h)] and PLUTO4.4 code [Bottom two rows (i--p)]. Values of iso-surface: \textbf{0.001 (red), 0.01 (Green), 0.03 (Blue),  0.04 (Cyan), 0.05 (Yellow)}. Simulation Details: Reynolds number $R_e = R_m = 1000$, Grid resolution $N = 128^3$, Time stepping $dt = 10^{-4}$, initial fluid velocity $u_0 = 1.0$, Alfven Mach number $M_A = 1.0$.}
	\label{3D TG Iso V}
\end{figure*}

\begin{figure*}
	\centering
	\begin{turn}{90} 
		\LARGE{\textbf{\textcolor{blue}{\hspace{-4.0cm} $\longleftarrow$ GMHD3D $\longrightarrow$}}}
	\end{turn}
	\begin{subfigure}{0.23\textwidth}
		\centering
		\includegraphics[scale=0.044]{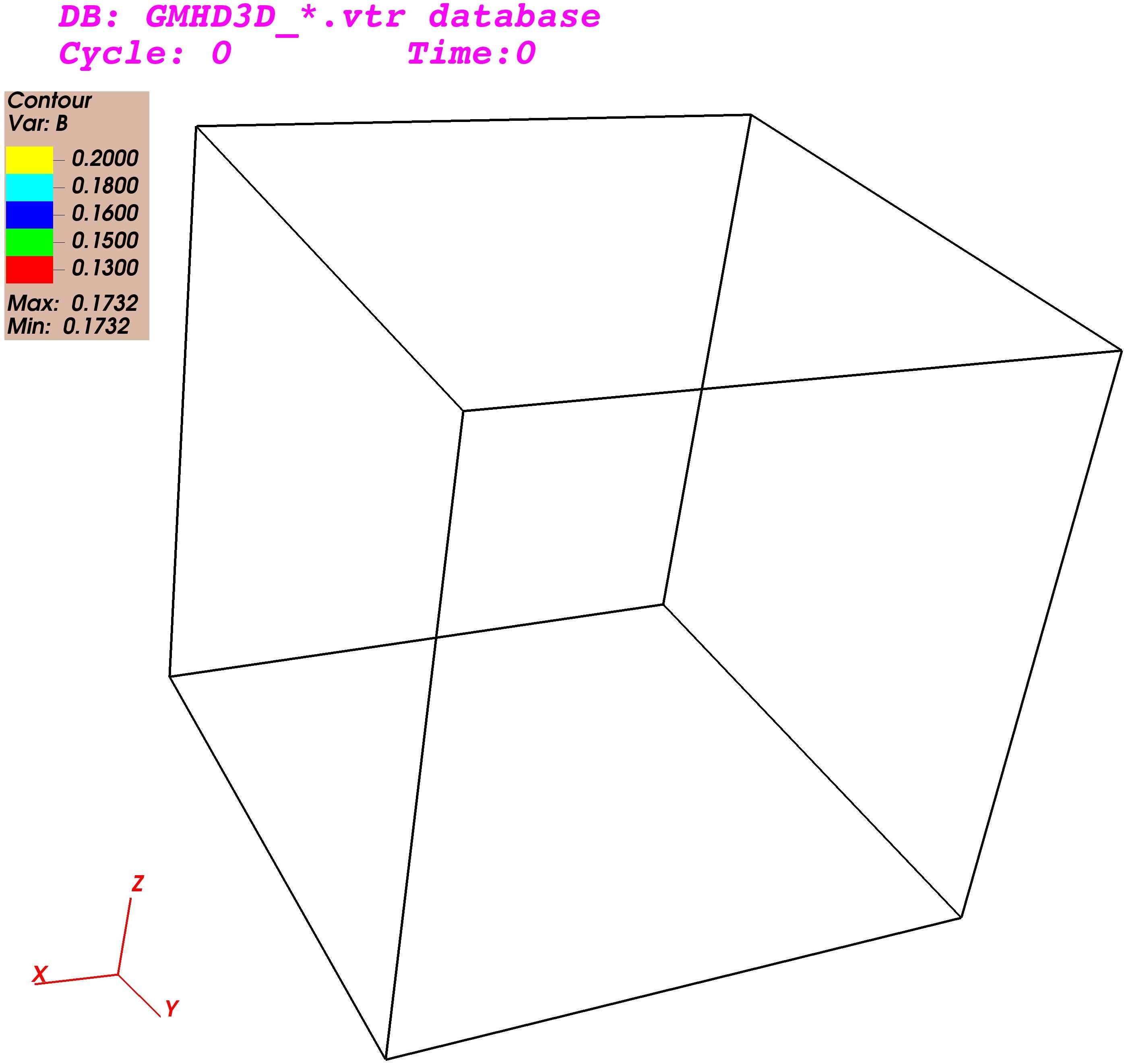}
		\caption{Time = 0.0}
	\end{subfigure}
	\begin{subfigure}{0.23\textwidth}
		\centering
		\includegraphics[scale=0.044]{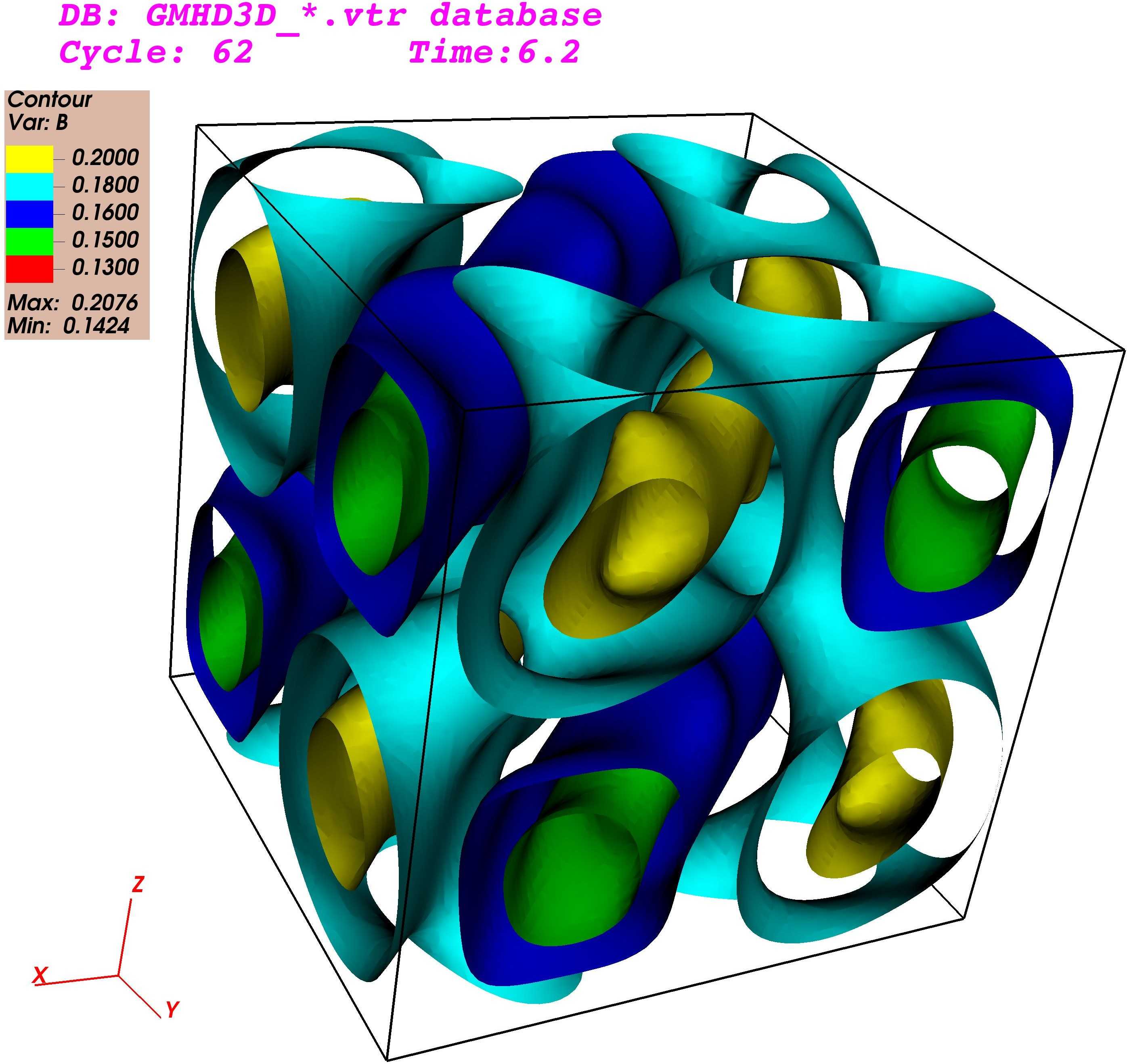}
		\caption{Time = 6.2}
	\end{subfigure}
	\begin{subfigure}{0.23\textwidth}
		\centering
		\includegraphics[scale=0.044]{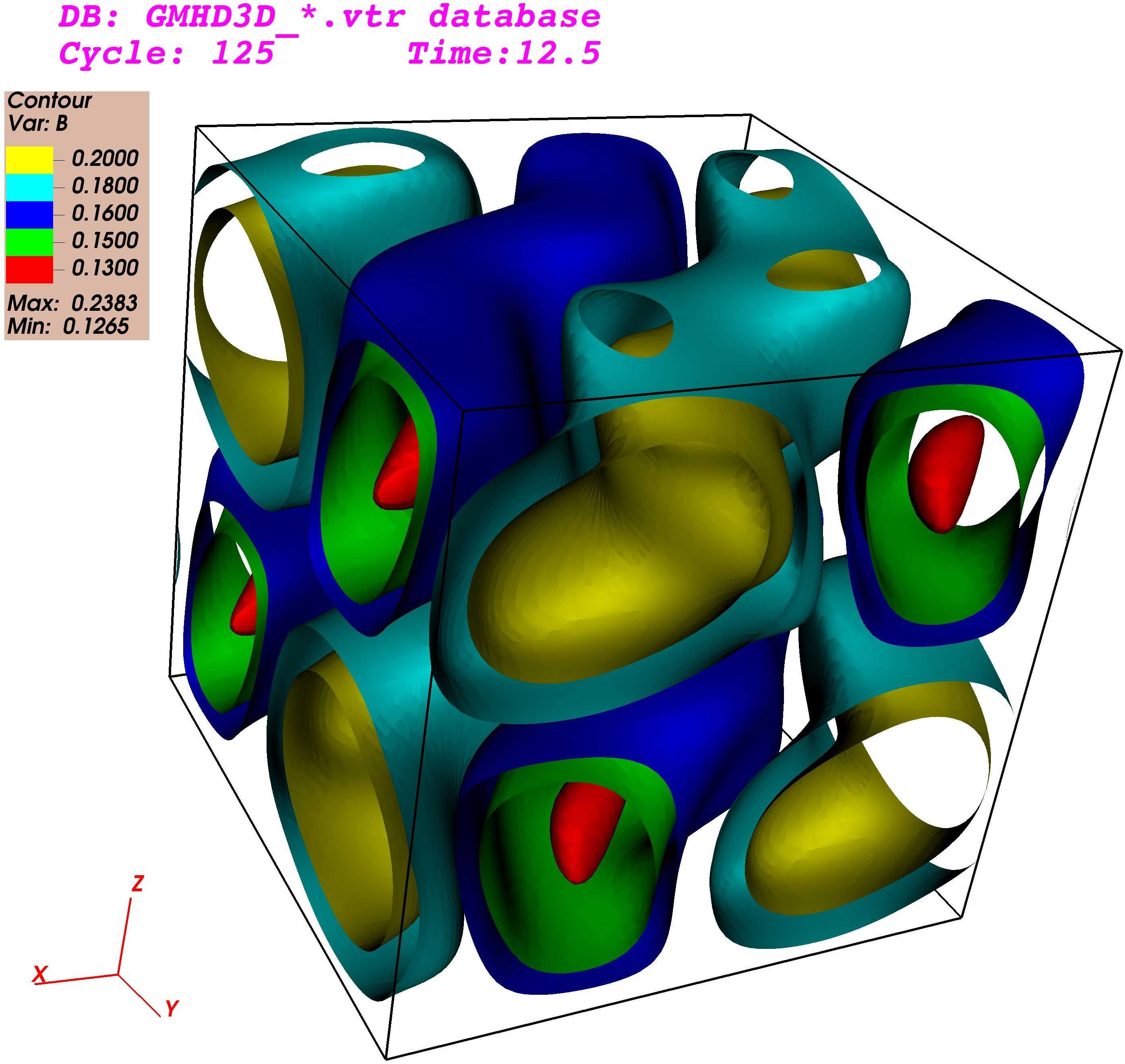}
		\caption{Time = 12.5}
	\end{subfigure}
	\begin{subfigure}{0.23\textwidth}
		\centering
		\includegraphics[scale=0.044]{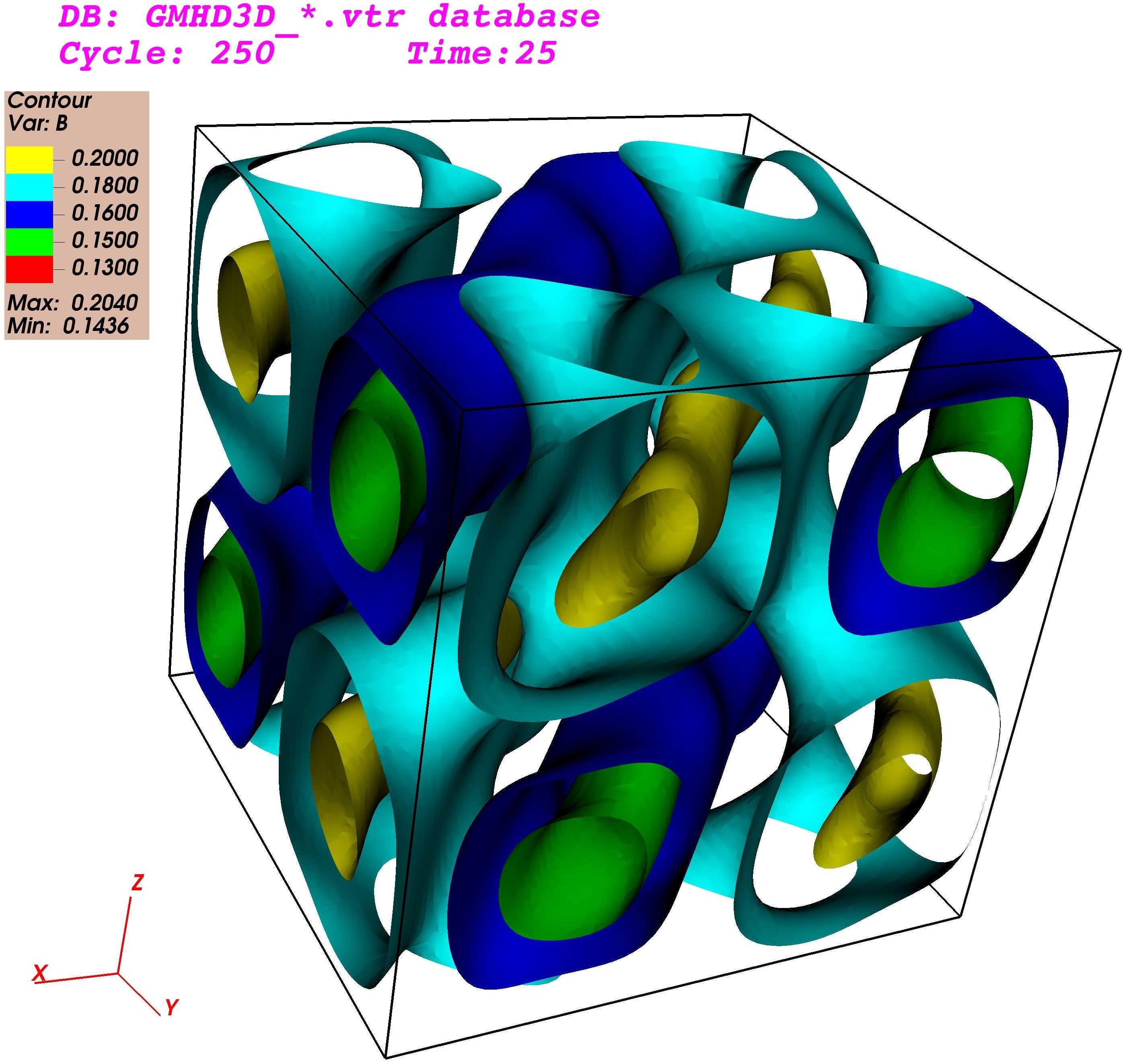}
		\caption{Time = 25.0}
	\end{subfigure}
	\begin{subfigure}{0.23\textwidth}
		\centering
		\includegraphics[scale=0.044]{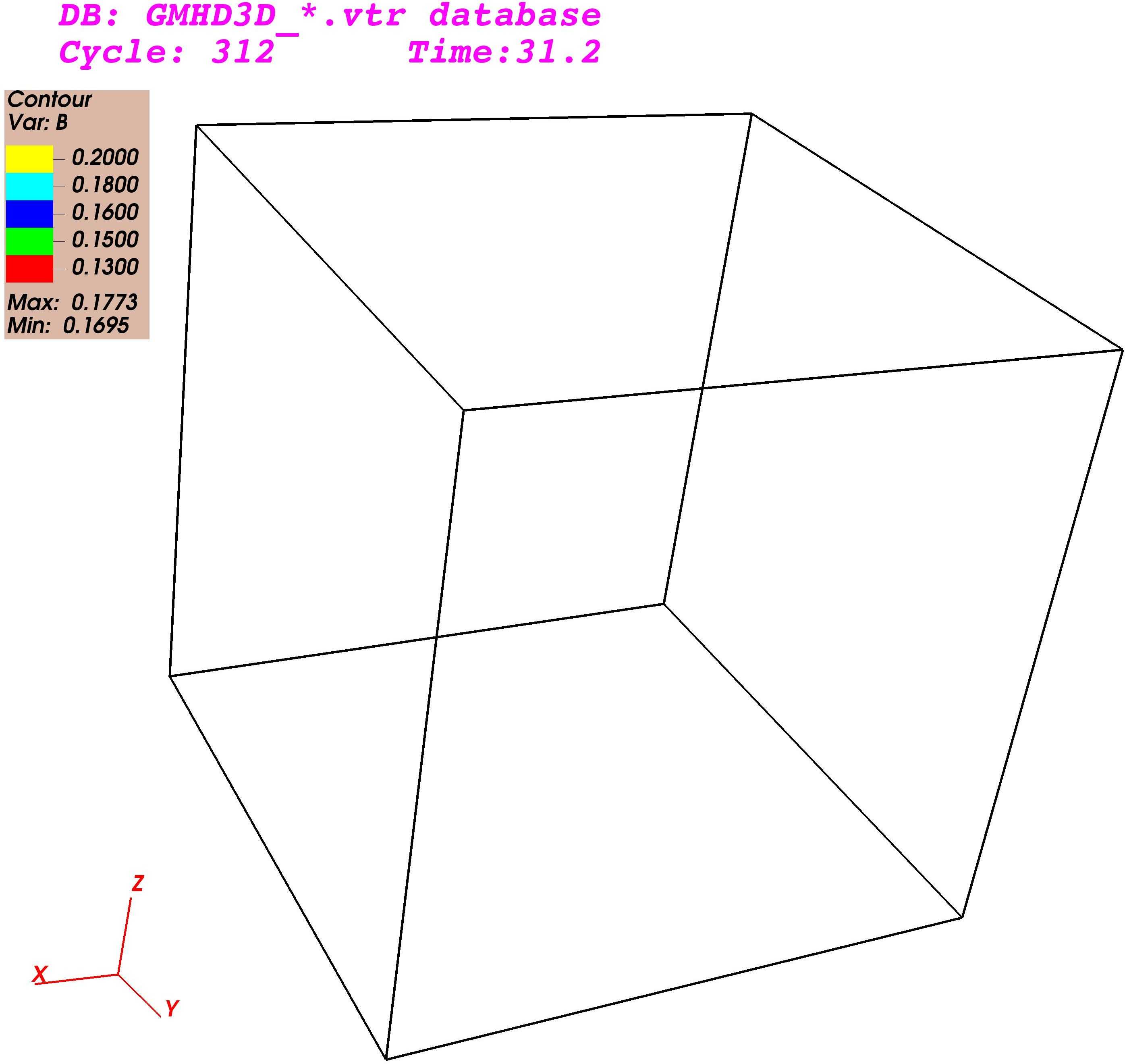}
		\caption{Time = 31.2}
	\end{subfigure}
	\begin{subfigure}{0.23\textwidth}
		\centering
		\includegraphics[scale=0.044]{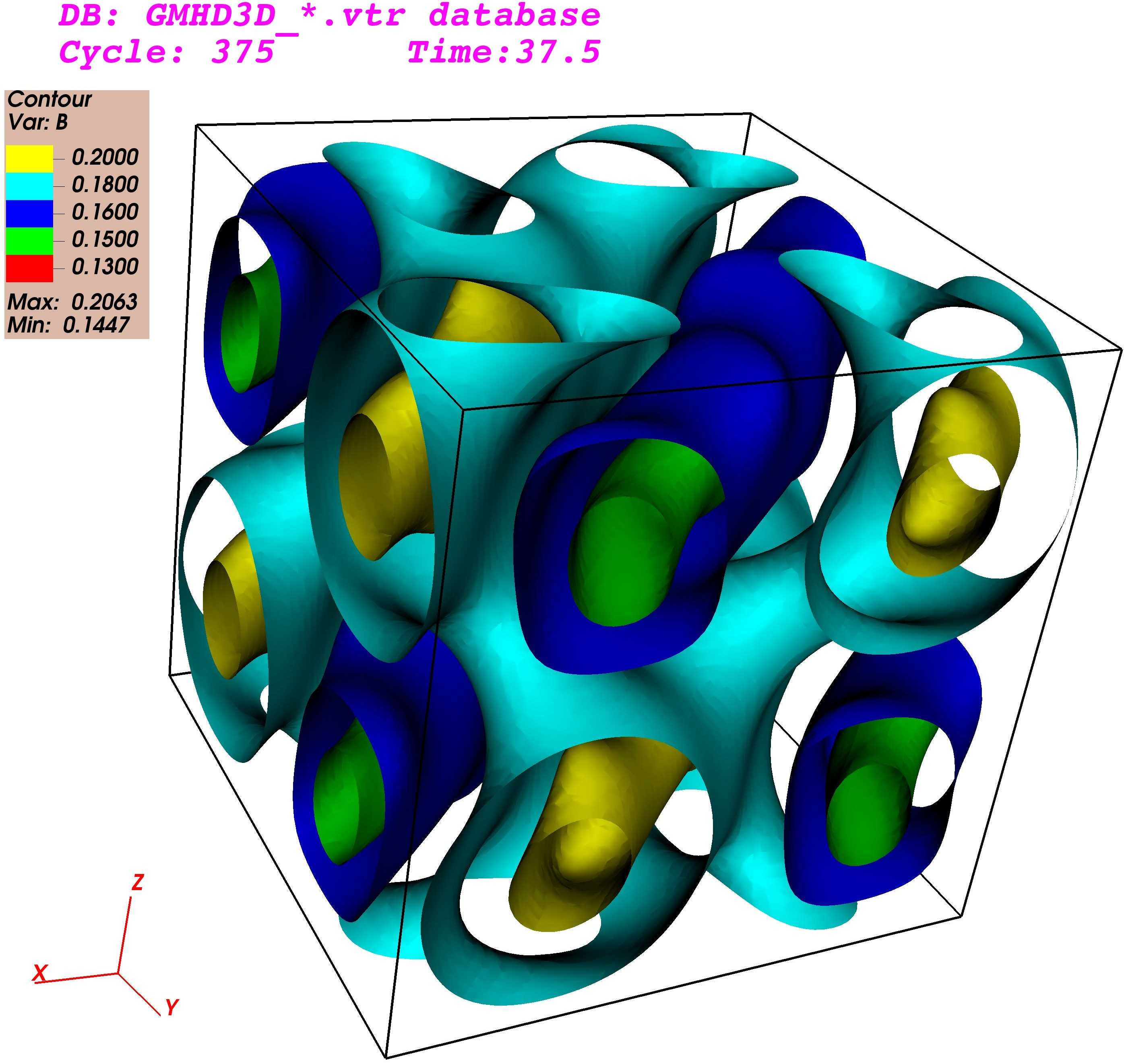}
		\caption{Time = 37.5}
	\end{subfigure}
	\begin{subfigure}{0.22\textwidth}
		\centering
		\includegraphics[scale=0.044]{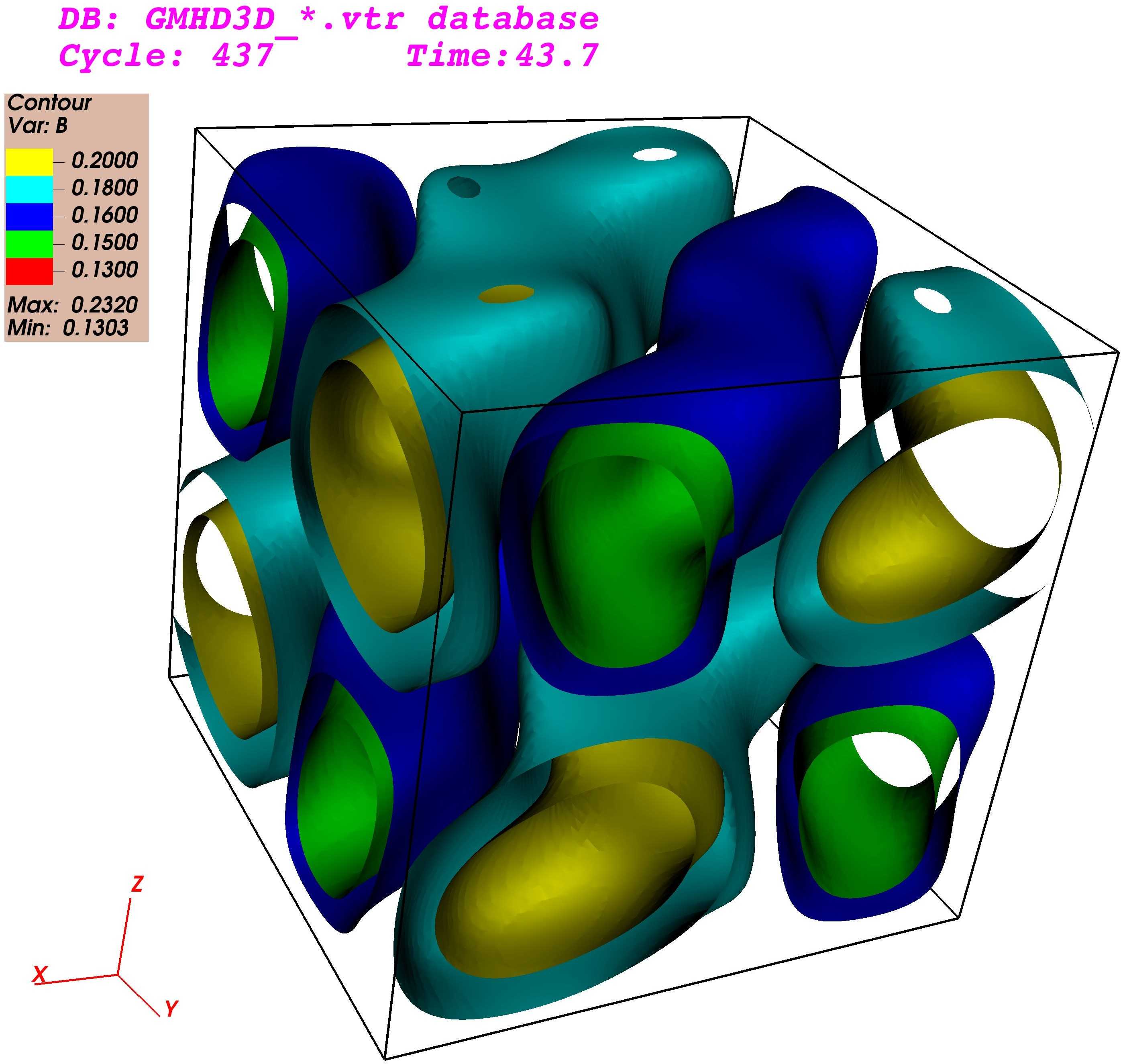}
		\caption{Time = 43.7}
	\end{subfigure}
	\begin{subfigure}{0.22\textwidth}
		\centering
		\includegraphics[scale=0.044]{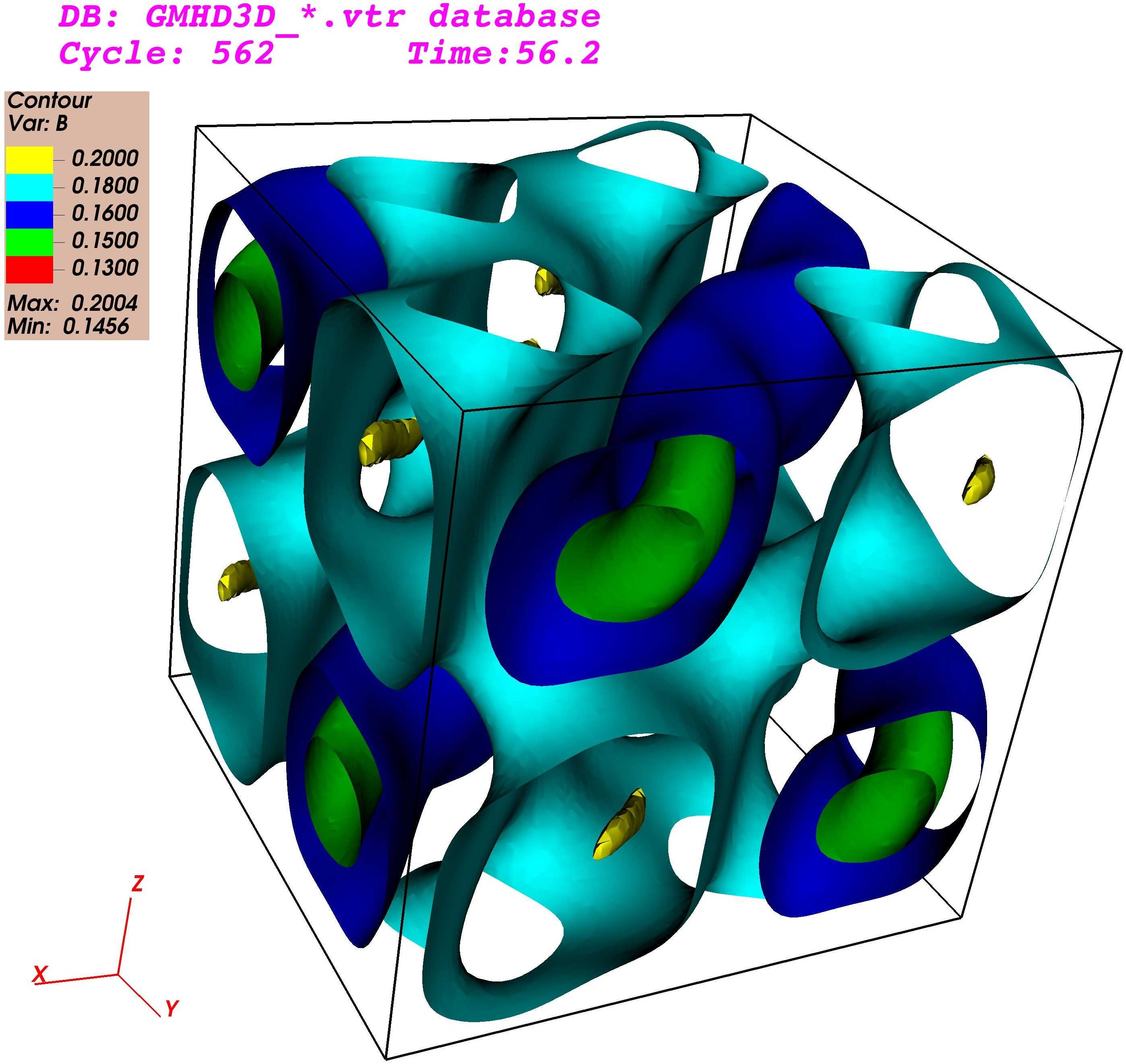}
		\caption{Time = 56.2}
	\end{subfigure}\\
	\centering
	\begin{turn}{90} 
		\LARGE{\textbf{\textcolor{blue}{\hspace{-4.0cm} $\longleftarrow$ PLUTO4.4 $\longrightarrow$}}}
	\end{turn}
	\begin{subfigure}{0.23\textwidth}
		\centering
		\includegraphics[scale=0.044]{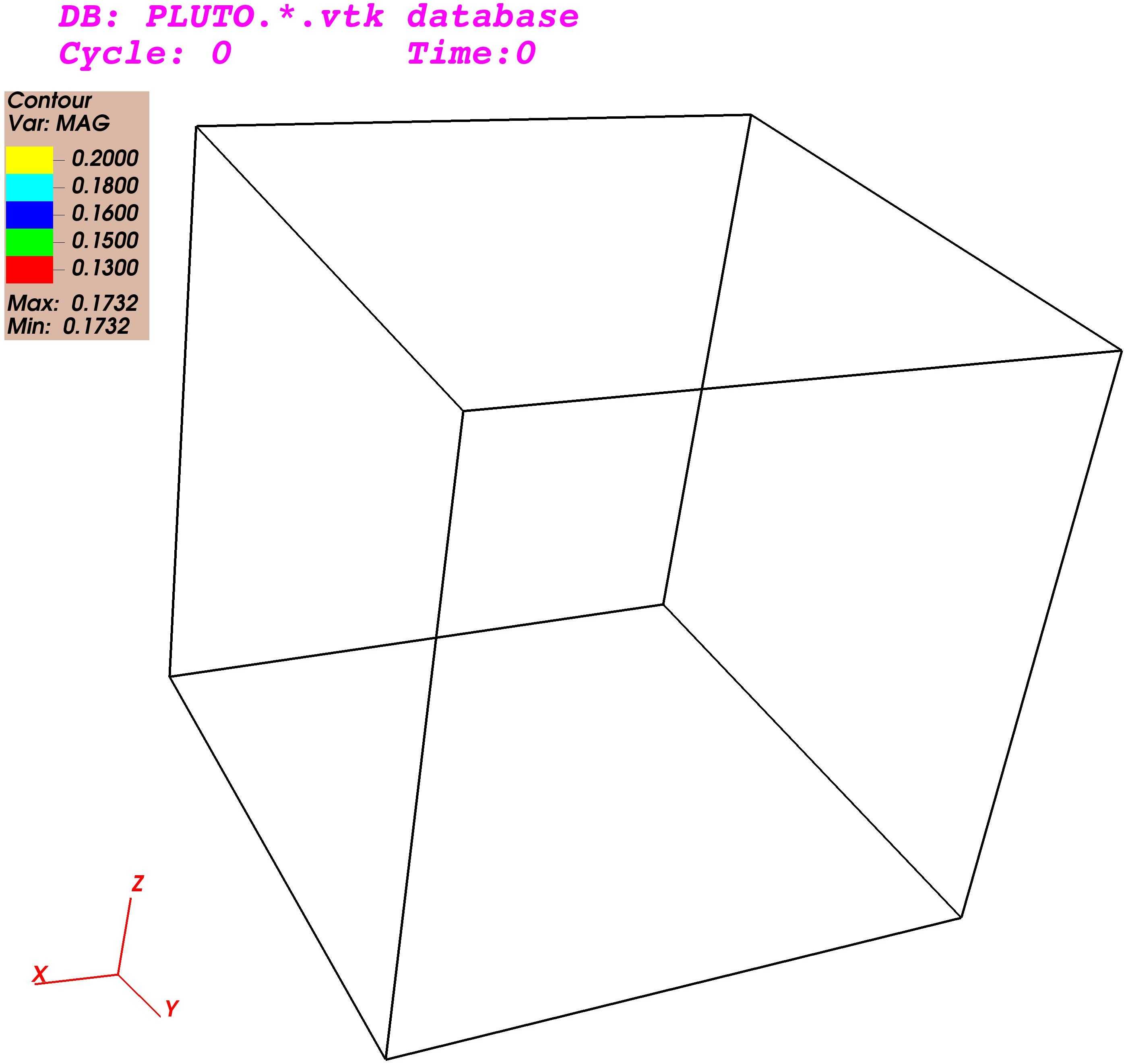}
		\caption{Time = 0.0}
	\end{subfigure}
	\begin{subfigure}{0.23\textwidth}
		\centering
		\includegraphics[scale=0.044]{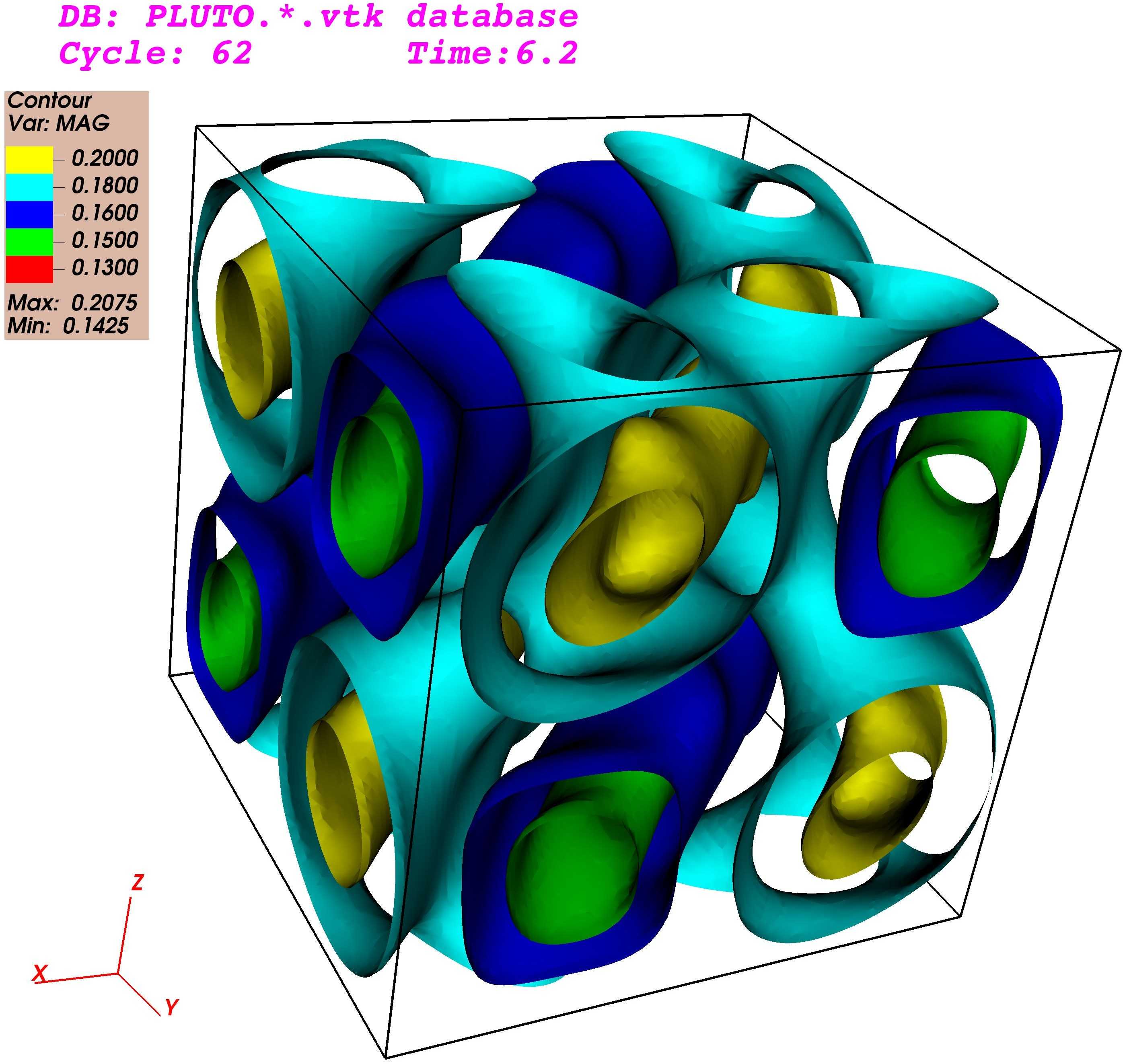}
		\caption{Time = 6.2}
	\end{subfigure}
	\begin{subfigure}{0.23\textwidth}
		\centering
		\includegraphics[scale=0.044]{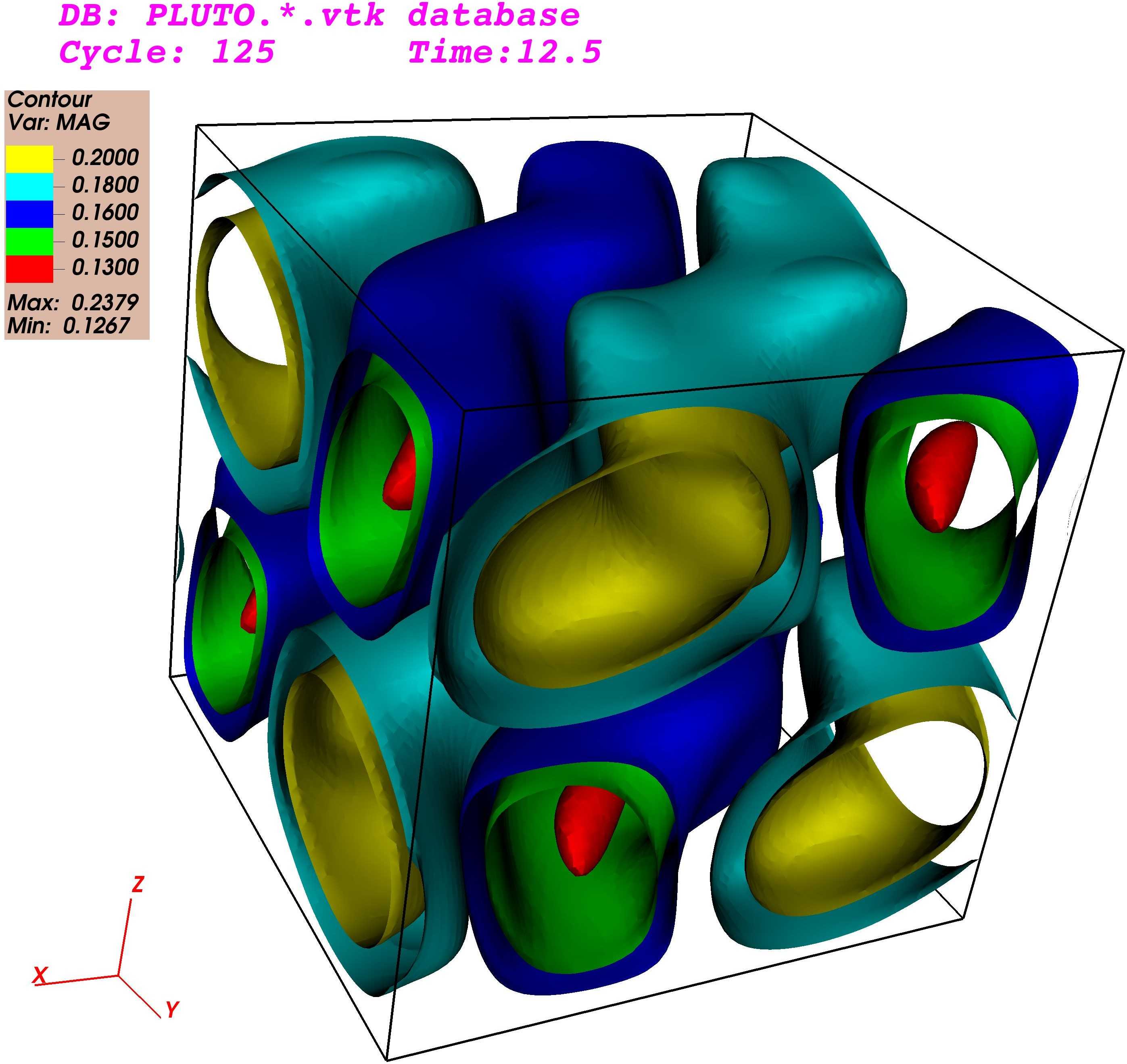}
		\caption{Time = 12.5}
	\end{subfigure}
	\begin{subfigure}{0.23\textwidth}
		\centering
		\includegraphics[scale=0.044]{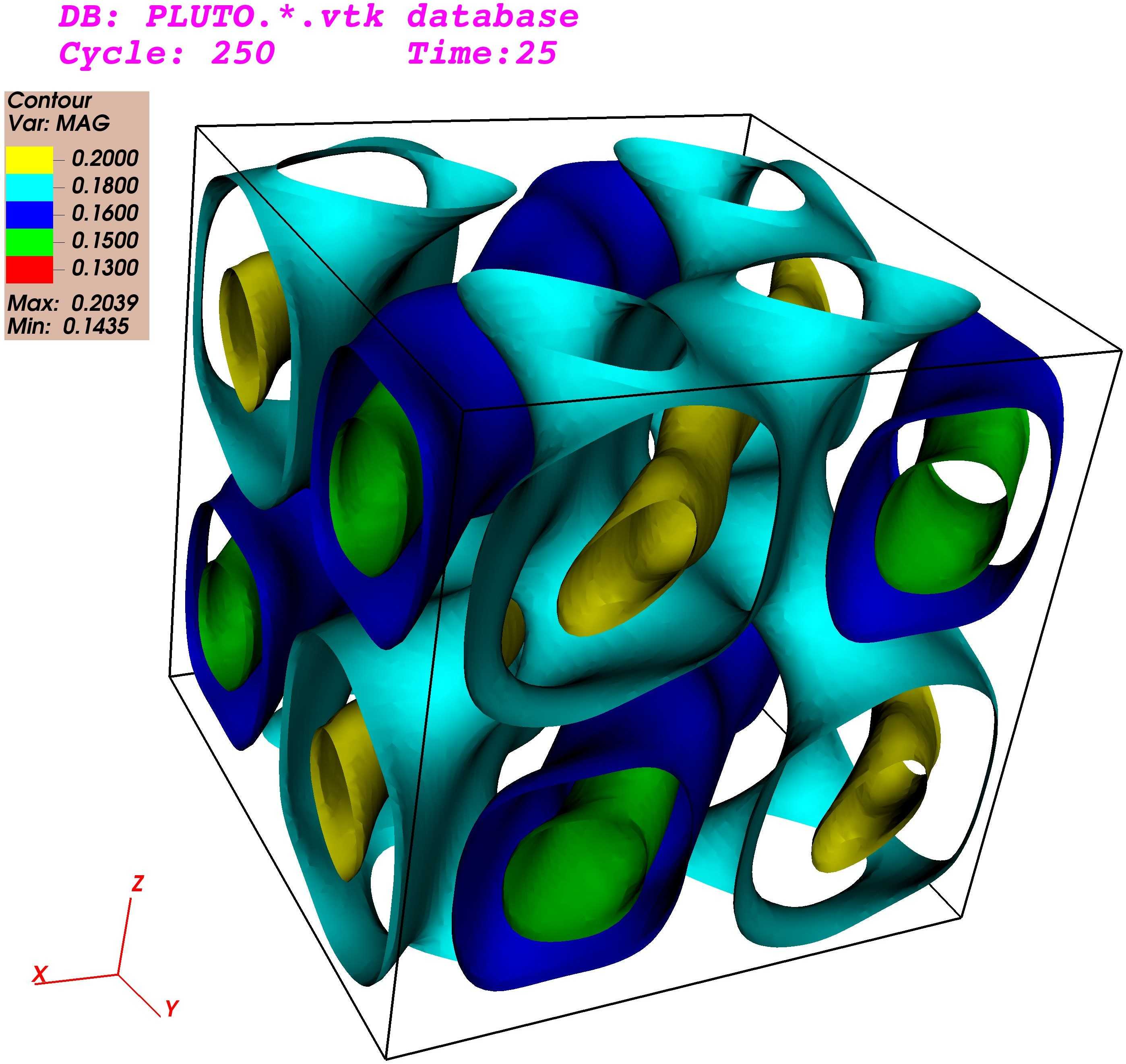}
		\caption{Time = 25.0}
	\end{subfigure}
	\begin{subfigure}{0.23\textwidth}
		\centering
		\includegraphics[scale=0.044]{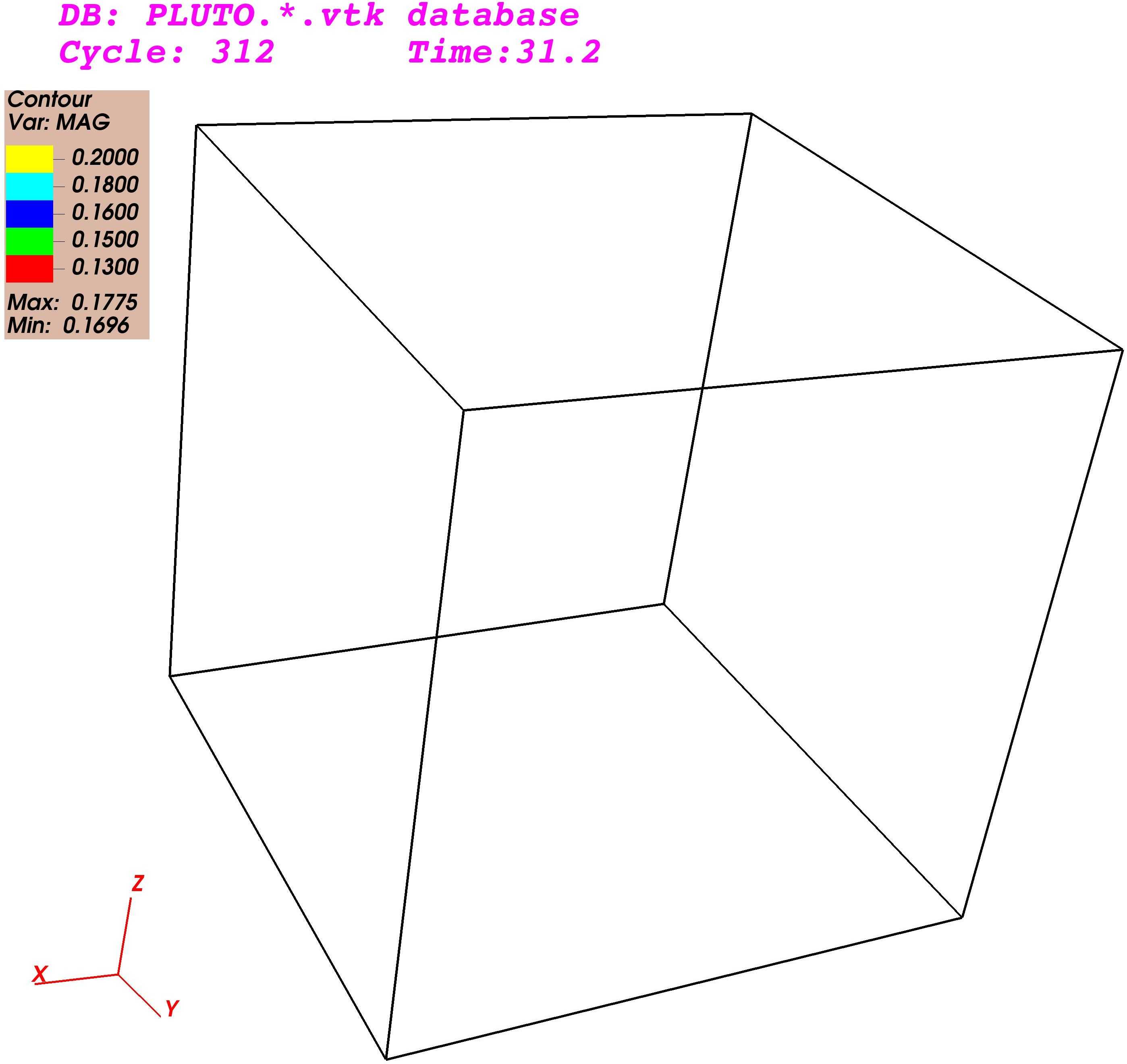}
		\caption{Time = 31.2}
	\end{subfigure}
	\begin{subfigure}{0.23\textwidth}
		\centering
		\includegraphics[scale=0.044]{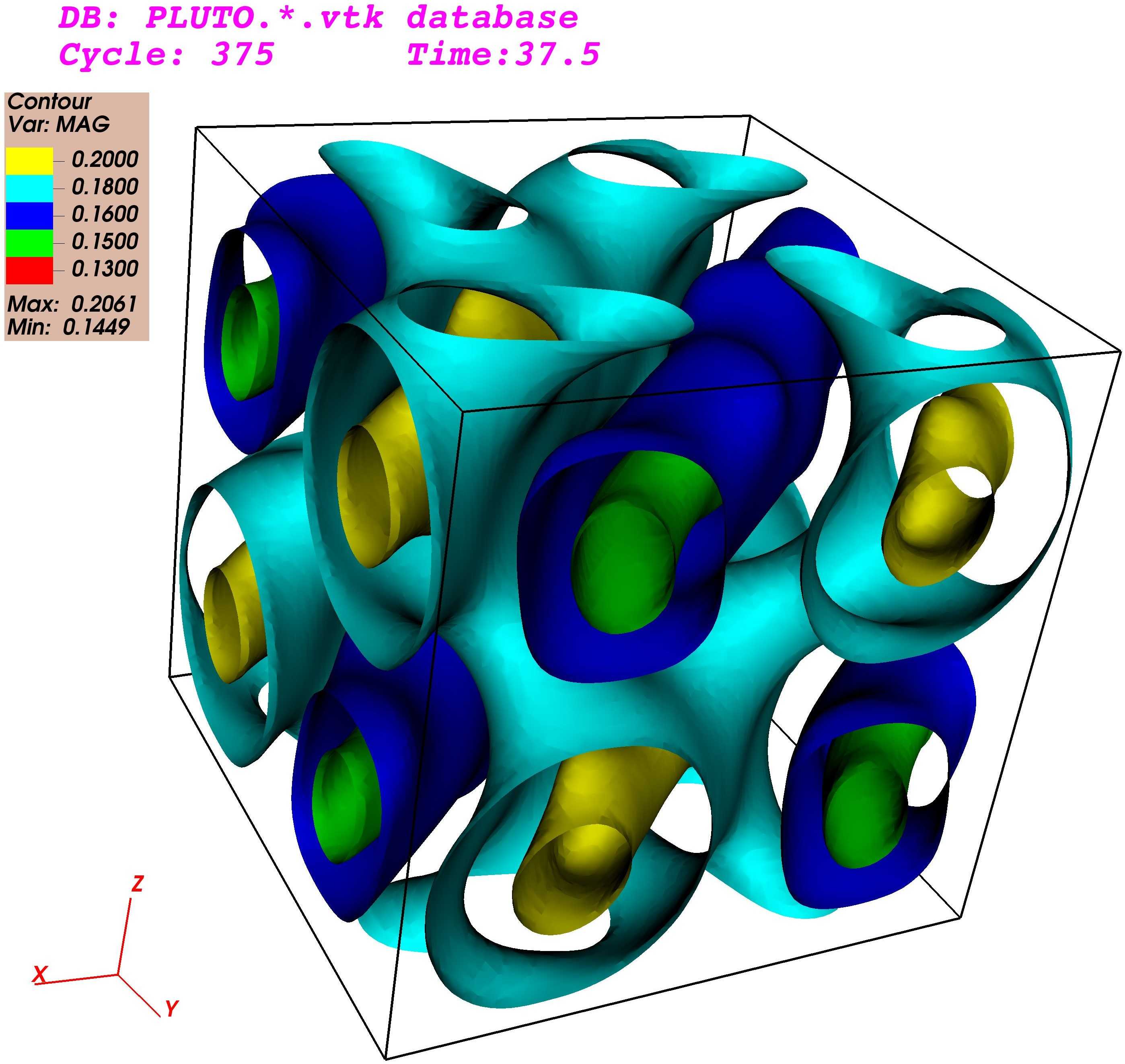}
		\caption{Time = 37.5}
	\end{subfigure}
	\begin{subfigure}{0.22\textwidth}
		\centering
		\includegraphics[scale=0.044]{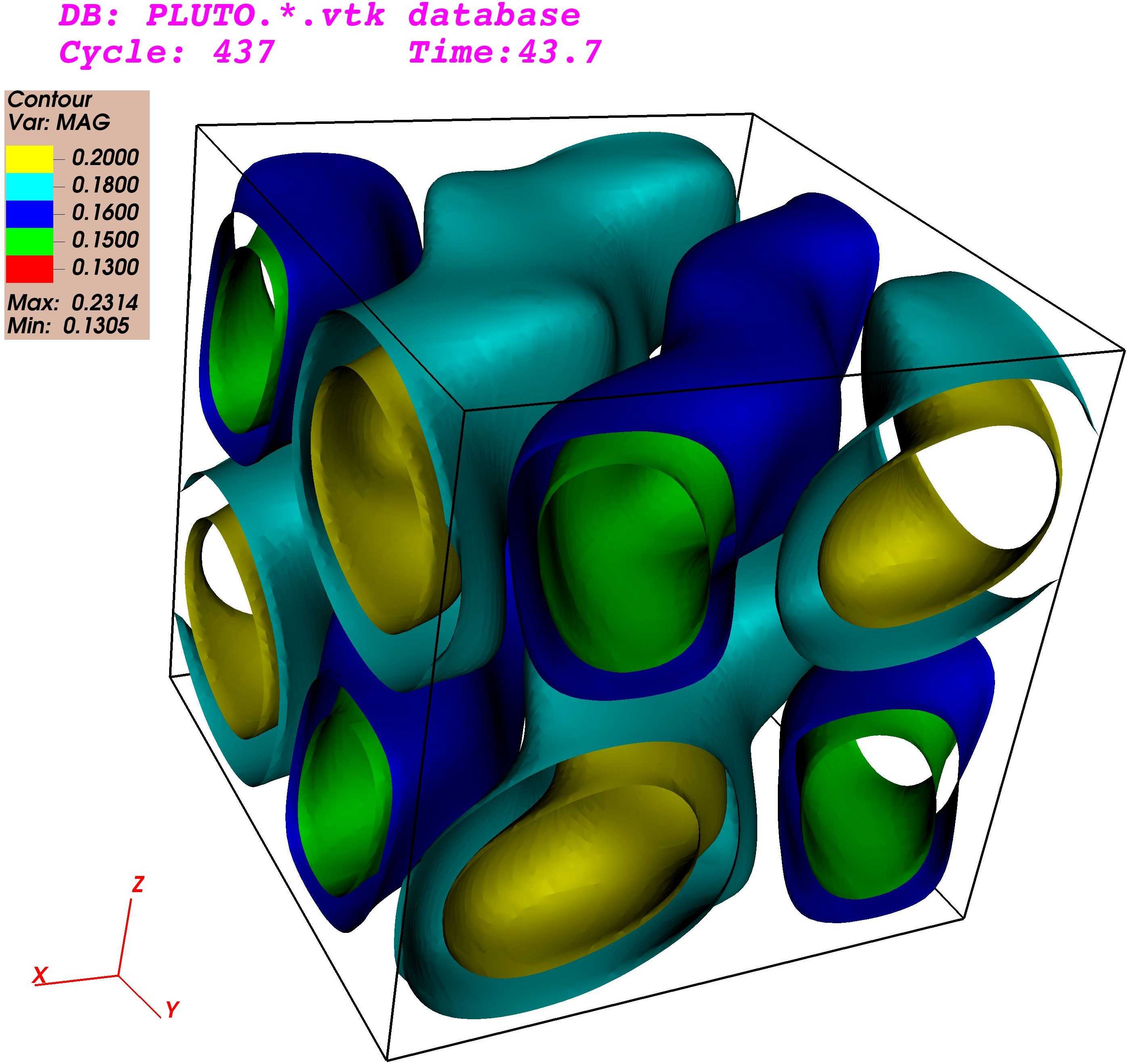}
		\caption{Time = 43.7}
	\end{subfigure}
	\begin{subfigure}{0.22\textwidth}
		\centering
		\includegraphics[scale=0.044]{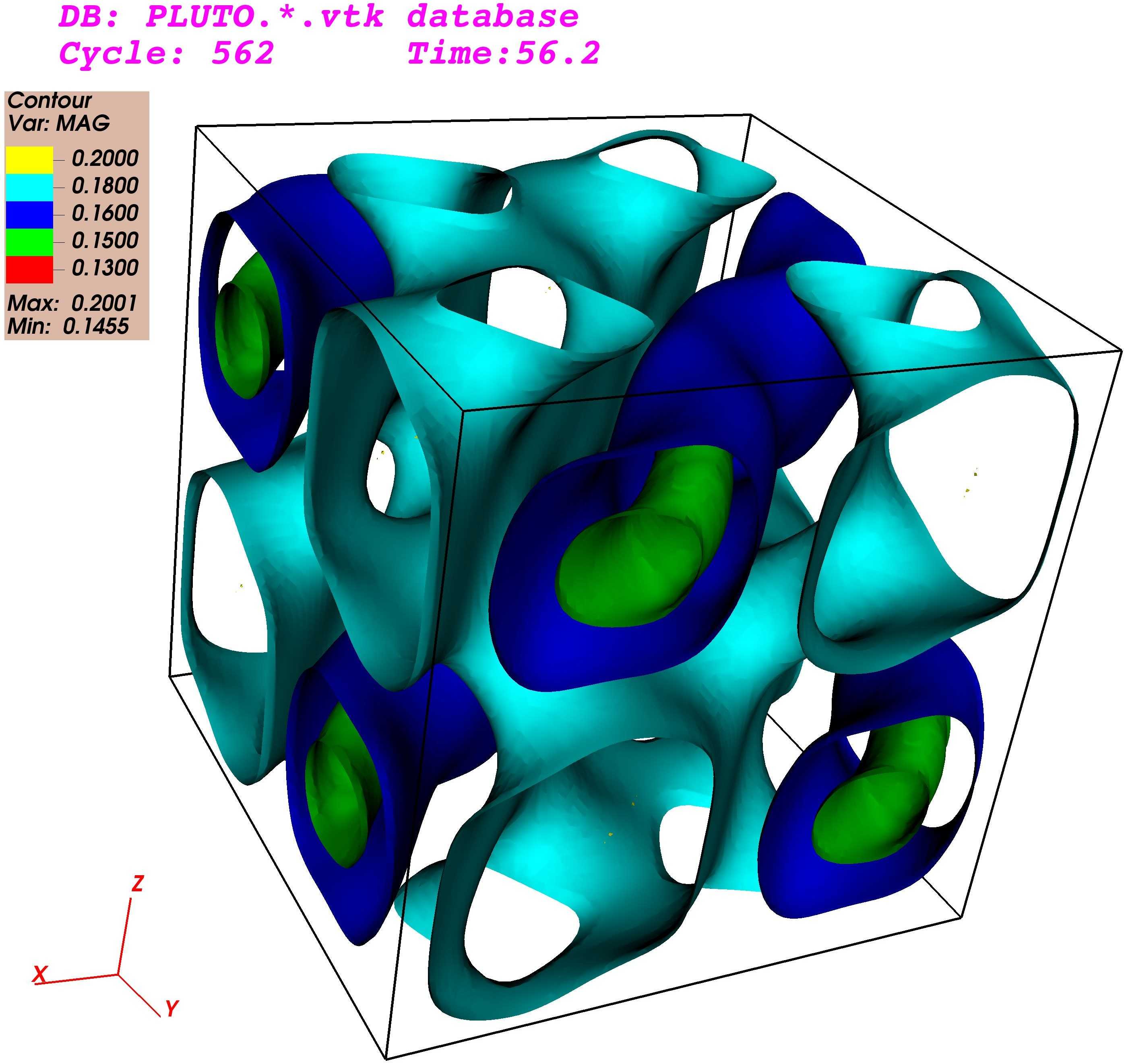}
		\caption{Time = 56.2}
	\end{subfigure}
	\caption{Recurring 3D  Taylor-Green Flow Magnetic Field iso-surface from GMHD3D code [Top two rows (a--h)] and PLUTO4.4 code [Bottom two rows (i--p)]. Values of iso-surface: \textbf{0.13 (Green), 0.15 (Yellow), 0.16 (Blue), 0.2 (Red)}. Simulation Details: Reynolds number $R_e = R_m = 1000$, Grid resolution $N = 128^3$, Time stepping $dt = 10^{-4}$, initial fluid velocity $u_0 = 1.0$, Alfven Mach number $M_A = 1.0$.}
	\label{3D TG Iso B}
\end{figure*}

Finally, we consider the Taylor-Green [TG] flow in three dimensions as the initial velocity profile. The flow profile looks like,
	\begin{equation}\label{Recurrence TG}
	\begin{aligned}
	u_x &=  u_0 [A \cos(k_0x) \sin(k_0y) \cos(k_0z)]\\
	u_y &=  - u_0 [A \sin(k_0x) \cos(k_0y) \cos(k_0z)] \\
	u_y &= 0
	\end{aligned}
	\end{equation} 
with $A = 1$ and $k_0 = 1$. Iso-surfaces of velocity and magnetic field for this flow are shown in Figs.  \ref{3D TG Iso V} \& \ref{3D TG Iso B}. Figure  \ref{3D TG Iso V} shows an iso-velocity surface with values of 0.001 (red), 0.01 (green), 0.02 (blue), 0.04 (cyan), and 0.05 (yellow).

The values of the magnetic field iso-surface are also 0.13 (Red), 0.15 (Green), 0.16 (Blue), 0.18 (Cyan), and 0.20 (Yellow) [See Fig. \ref{3D TG Iso B}].

Both the GMHD3D and PLUTO4.4 codes show that the velocity isosurface and the magnetic field isosurface continue to recur back.  This 3D Taylor-Green [TG] flow is recognized as a recurrent flow since both the iso-surfaces (velocity and magnetic field) are recurring \cite{RM_Recurrence:2019}.

%

\subsubsection{A plausible explanation for Recurrence}

To gain a thorough knowledge of the recurrence and non-recurrence phenomena, we use Thyagaraja's \cite{Thyagaraja:1979} mathematical description in terms of Rayleigh Quotient (Q(t)). Recently, Mukherjee et al. \cite{RM_Recurrence:2019} employed a modified form of the Rayleigh Quotient for MHD systems as,

\begin{equation}
Q(t) = \frac{\int_{V}[(\vec{\nabla} \times \vec{u})^2 + \frac{1}{2} (\vec{\nabla} \times \vec{B})^2]dV}{\int_{V}[\vec{|u|}^2 + \frac{1}{2} \vec{|B|}^2]dV}
\end{equation}


Physically, Q(t) is a measure of the number of active degrees of freedom possible in the system. It is already known that for typical hydrodynamic flows, the Rayleigh Quotient [Q(t)] is found to be bound in nature to demonstrate a recurrence phenomenon \cite{Thyagaraja:1979}. Recent study has shown that the Rayleigh Quotient [Q(t)] is bounded with time for Taylor-Green [TG] flow, but is unbounded for 3D Arnold-Beltrami-Childress [ABC] flow \cite{RM_Recurrence:2019} in the presence of homogeneous ambient magnetic field.

\begin{figure*}
	\includegraphics[scale=0.55]{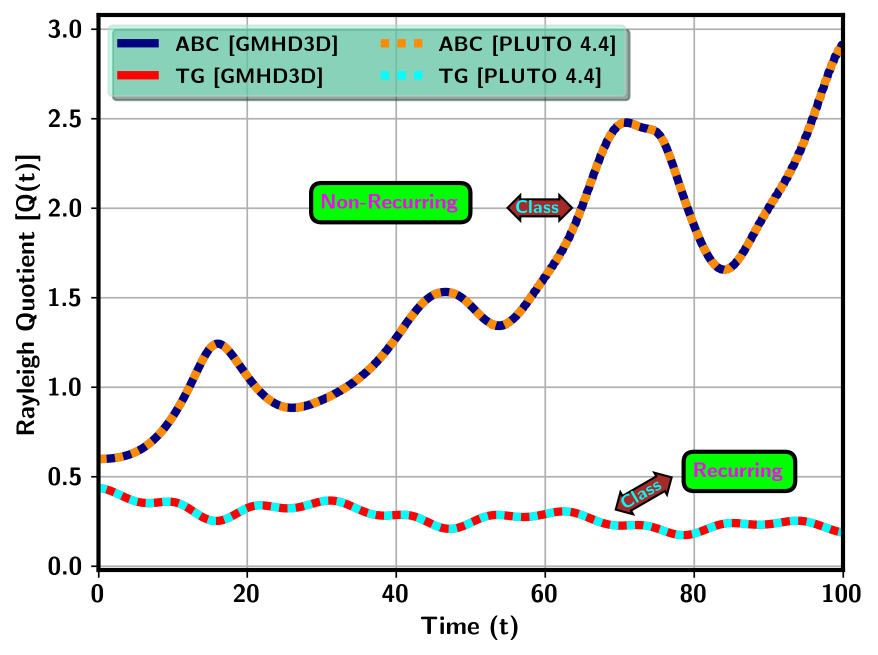}
	\caption{Rayleigh Quotient calculation from GMHD3D code and PLUTO4.4 code. \textcolor{black}{Simulation Details: Grid resolution $N = 128^3$, Time stepping $dt = 10^{-4}$}.}
	\label{RQ}
\end{figure*}

Similar features, such as the time-dependent unbounded Rayleigh Quotient for 3D Arnold-Beltrami-Childress (ABC) flow and the time-dependent bounded Rayleigh Quotient for 3D Taylor-Green (TG) flow, are also observed in this present work from both codes (GMHD3D and PLUTO4.4) [See Fig. \ref{RQ}].

\section{Summary and Conclusion}

In this study, we have examined the \textcolor{black}{performance} of two MHD codes, GMHD3D \cite{GTC} (which is developed in-house at the Institute for Plasma Research, India) and PLUTO4.4 \cite{PLUTO_Mignone:2007} (which is freely available) on a number of different physics problems. For solving coupled partial differential equations, GMHHD3D employs a pseudo-spectral technique, while PLUTO4.4 employs a finite volume method.\\


The points to take home are: \\

$\bullet$ Using the GMHD3D and PLUTO4.4 codes, we have investigated the hydrodynamic 2D Kelvin-Helmholtz (KH) instability for two broken jets traveling in opposing directions at the compressible limit. We obtain the identical growth rates for the KH instability at grid resolution $512^2$ from both codes at different sonic Mach numbers and different mode numbers, which is the same as what Keppens et al. \cite{keppens_KH:1999} has reported.\\


$\bullet$ Taylor-Green vortex evolution in 3D is investigated in the pure hydrodynamic limit utilizing both codes. The time evolution of mean square vorticity is estimated from both codes, and for a given value of Reynolds number, the PLUTO4.4 code requires a grid resolution of at least $256^3$ to replicate the GMHD3D code's $64^3$ grid resolution data. This numerical observation indicates that the spectral solver is more \textcolor{black}{superior} than a grid-based solver.\\

$\bullet$ The period doubling bifurcation of the TG vortex is reproduced using both codes as part of this investigation. The results of our numerical observation are found to agree with those of some prior studies \cite{Sharma_Sengupta:2019}.\\


$\bullet$ We explore the problem of coherent nonlinear oscillation using some well-known two-dimensional flows in the presence of a homogeneous and ambient magnetic field, and find that both codes produce almost identical results. The spatial distributions of kinetic and magnetic energy produced by both codes are similar.\\


$\bullet$ In 2D-Orszag-Tang flow, the findings from both codes are identical, but in 2D-Cat's Eye [CE] flow, the predicted oscillations of energy are found to be significantly dampened when utilizing the PLUTO4.4 solver. One possible explanation is that the PLUTO4.4 solver has a greater numerical viscosity.\\


$\bullet$ To get rid of the observed damping, we use different electric field averaging techniques that are available in PLUTO4.4. It has been observed that the PLUTO4.4 solver is most efficient with the UCT\_HLLD and CT\_CONTACT EMF averaging techniques. However, with these two techniques present, obtaining the desired oscillations in PLUTO4.4 requires at least $512^2$ grid resolution, whereas GMHD2D resolves that at $128^2$ grid resolution. This essentially demonstrates the superiority of the spectral solver over the grid based solver.\\

$\bullet$ For a variety of well-known 3D astrophysical processes, the GMHD3D and PLUTO4.4 codes produce identical oscillations of kinetic and magnetic energy in the form of Alfven waves.\\


$\bullet$ We also look closely at the different parameter regimes for a certain flow, namely the 3D Arnold–Beltrami–Childress [ABC] flow. The findings from both codes are found to be identical for a range of Alfven speeds, from sub to super-Alfvenic. When the wave number is larger ($k_0 = 8, 16$), however, the spectral solver (GMHD3D) is observed to be more \textcolor{black}{superior} than the grid based solver (PLUTO4.4), as the latter requires a grid resolution of $128^3 \& 256^3$ in order to recreate the results obtained by GMHD3D at a grid resolution of $64^3$.\\

$\bullet$ It is found that both codes yield the same outcomes for externally driven flows in both 2-dimensions and 3-dimensions.\\


$\bullet$ We finally reproduce the Recurrence phenomenon \cite{RM_Recurrence:2019} in 3-dimensional MHD plasma using both the codes. The analytical description of the same seems to support it perfectly from both the solver.\\


\textcolor{black}{The finite volume version of the PLUTO4.4 code maintains global second-order accuracy in time, whereas GMHD3D solver maintains up to 6'th order accuracy in space and 4'th order accuracy in time. Hence, the pseudo-spectral technique offers superior accuracy compared to finite volume and finite difference schemes. On the other hand, the PLUTO4.4 can handle non-periodic boundary conditions, Hall-MHD dynamics, can capture plasma flow with shock wave and so on. Also, the current version of the GMHD3D solver
does not have plasma transport terms such as thermal diffusion (for example,  Braginskii fluid equations \cite{Braginskii:2019}) as GMHD3D solves single fluid equations. Though GMHD3D can solve both dynamics energy equation and equation of state, in the current study, we have invoked equation of state (Eq.(\ref{EOS}) as closure). \textcolor{black}{A cost metric study for both the solvers (GMHD3D \& PLUTO4.4) have also been carried out on GPUs \& CPUs. It has been identified that as the number of grid points increases, the computational cost for both CPUs and GPUs also increases. It has been observed that the GPU solver is more cost-effective than the CPU solver for a substantial amount of computational load.}}\\


To conclude, we have attempted a systematic comparison of two MHD codes in both the hydrodynamic and magnetohydrodynamic limit. Our numerical analysis shows that while both algorithms produce comparable answers in most circumstances, the spectral solver surpasses the grid-based solver in periodic domain for a subset of physics-related challenges. We also believe this work highlights the advantages of a spectral solver over a grid-based solver. To the best of our knowledge, this is the first work ever attempted which makes a thorough comparision of a pseudo-spectral code with the freely available grid based MHD solver PLUTO4.4. \textcolor{black}{We intend to extend this comparative study in the near future to include advanced grid-based solvers such as FLASH, HYDRA, MIRANDA, PENCIL, etc.}


\section{ACKNOWLEDGMENTS}
The simulations and visualizations presented here are performed on GPU nodes and visualization nodes of the ANTYA cluster at the Institute for Plasma Research (IPR), India. \textcolor{black}{We would like to express our gratitude to the anonymous referees for their insightful feedback, which significantly enhanced the manuscript's quality.} We are grateful to Dr. Dipanjan Mukherjee at IUCAA and Dr. Bhargav Vaidya at IITI for their insightful discussions regarding the PLUTO code. One of the author S.B is thankful to Dr. Rupak Mukherjee at Central University of Sikkim (CUS), Gangtok, Sikkim, India for providing an initial version of GMHD3D code. S.B. is thankful to N. Vydyanathan, Bengaluru, and B. K. Sharma at NVIDIA, Bengaluru, India, for extending their help with basic GPU methods. S.B. is grateful to Dr. Soumen De Karmakar at IPR for many helpful discussions regarding GPUs, and the HPC support team of IPR for extending their help related to the ANTYA cluster.
\section{DATA AVAILABILITY}
The data underlying this article will be shared on reasonable request to the corresponding author.\\

\section{Conflict of Interest}
The authors have no conflicts to disclose.

\bibliography{biblio}

\onecolumngrid
\appendix
\section{\textbf{Coherent Nonlinear oscillations}}\label{Appen A}
\subsubsection{3D Taylor-Green [TG] Flow}

\begin{figure*}[h]
	\centering
	\begin{subfigure}{0.32\textwidth}
		\centering
		\includegraphics[scale=0.4]{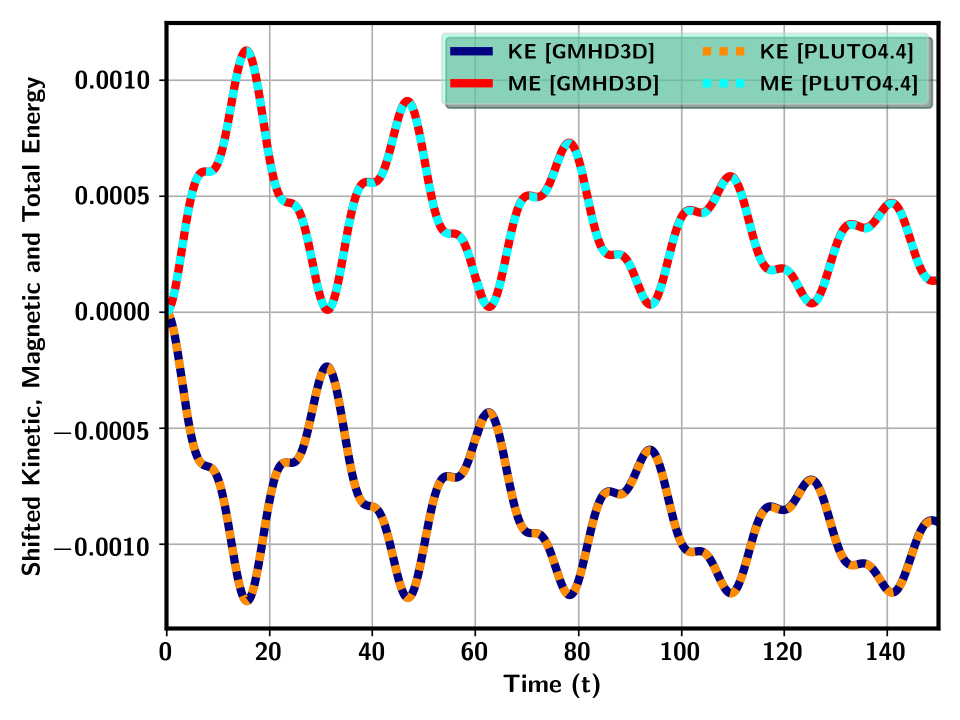}
		\caption{}
		\label{3D TG flow Energy}
	\end{subfigure}
	\begin{subfigure}{0.33\textwidth}
		\centering
		\includegraphics[scale=0.047]{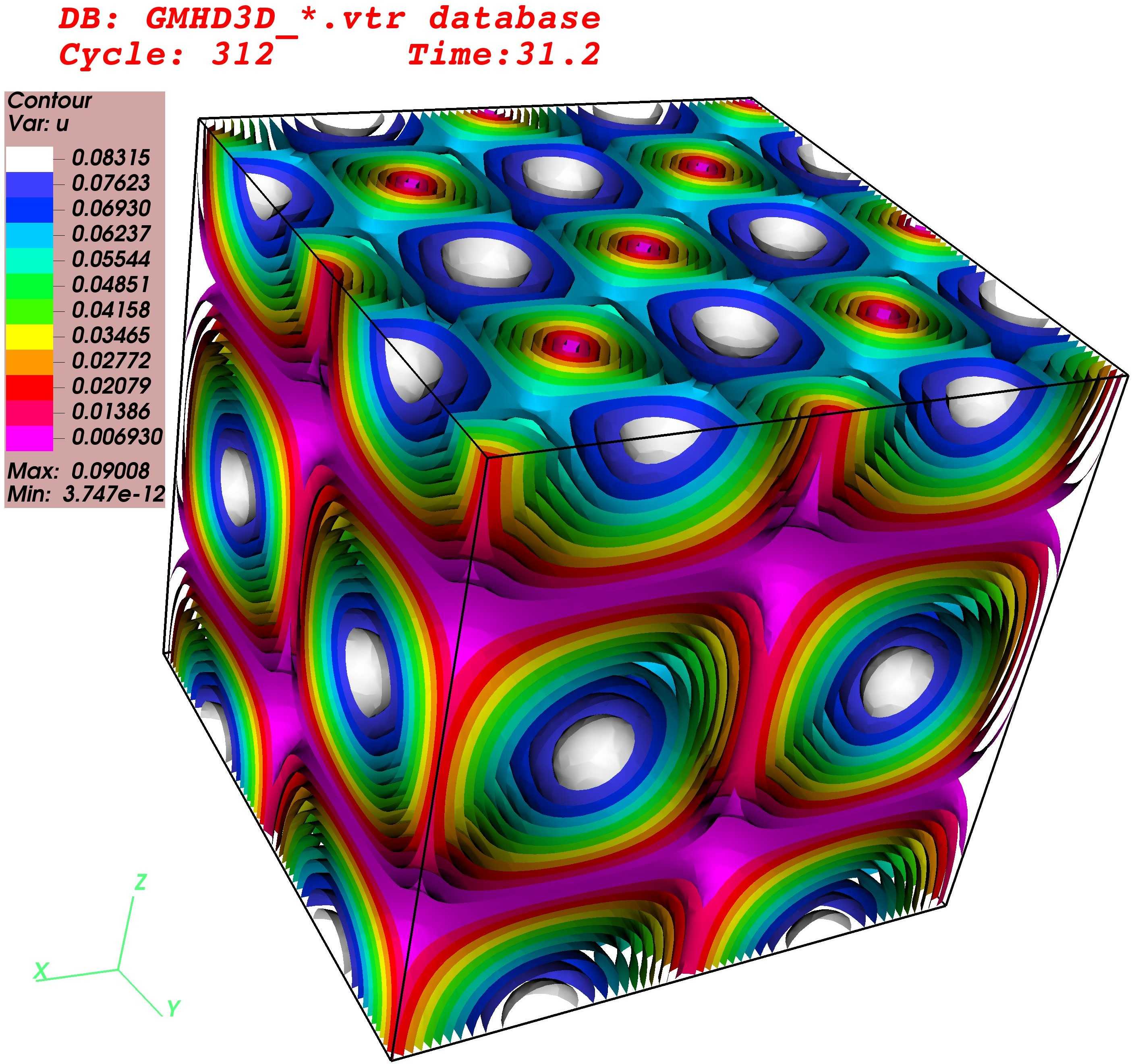}
		\caption{}
		\label{3D TG Iso V GMHD3D}
	\end{subfigure}
	\begin{subfigure}{0.33\textwidth}
		\centering
		\includegraphics[scale=0.047]{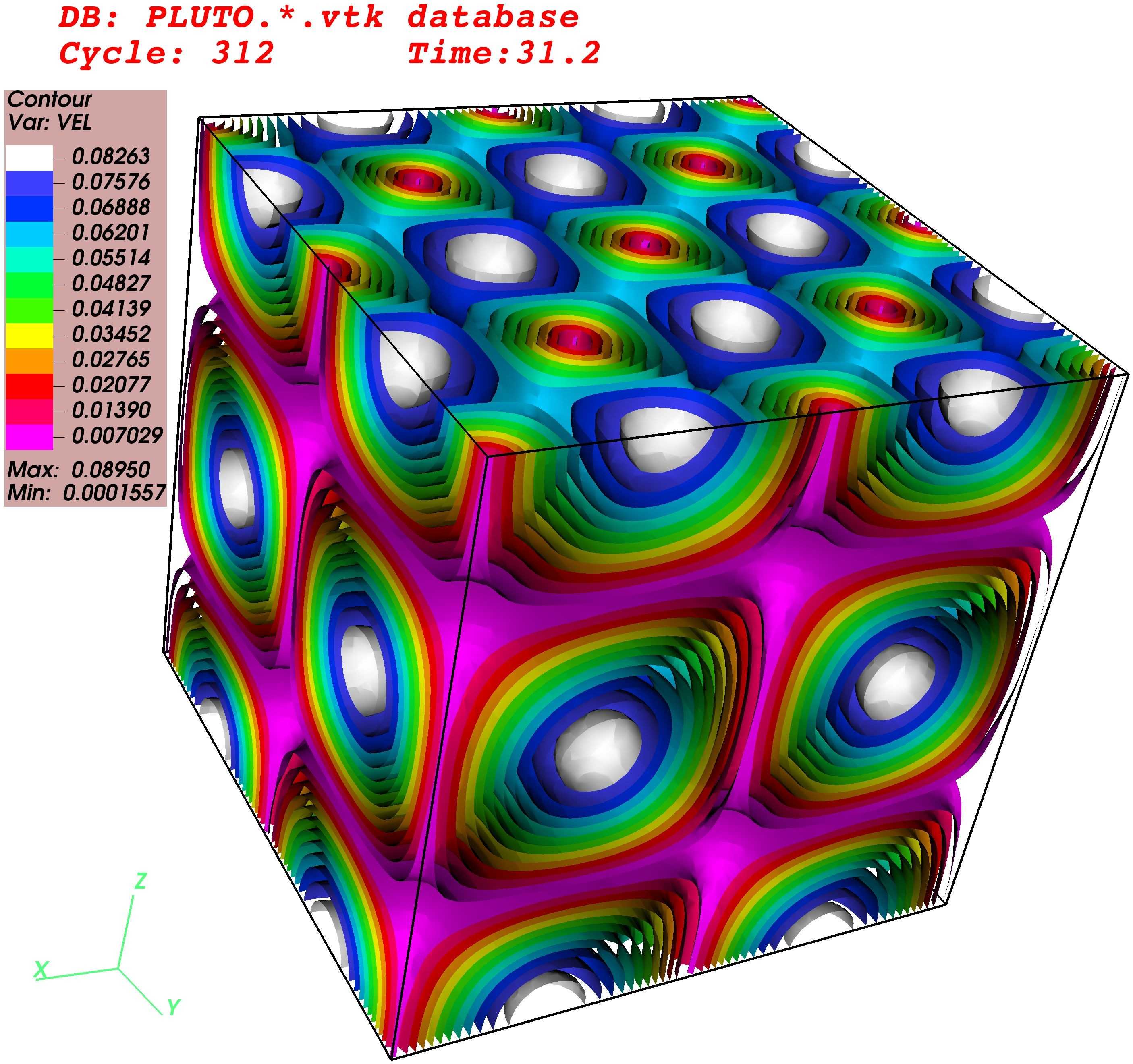}
		\caption{}
		\label{3D TG Iso V PLUTO}
	\end{subfigure}
	\caption{(a) The shifted kinetic and magnetic energies for 3D Taylor-Green Flow (TG) from GMHD3D and PLUTO4.4 code. The visualization of velocity iso-surface (Iso-V) for 3D Taylor-Green (TG) Flow at any arbitrary time from (b) GMHD3D code  and (c) PLUTO4.4 code. Simulation Details: Reynolds number $R_e = R_m = 1000$, Grid resolution $N = 128^3$, Time stepping $dt = 10^{-4}$, initial fluid velocity $u_0 = 1.0$, Alfven Mach number $M_A = 1.0$.}
\end{figure*}

One of the most well-known flows in astrophysical plasma is the Taylor-Green (TG) flow. This is a divergence free flow. The flow profile looks like, 
\begin{equation}\label{3D TG Flow}
\begin{aligned}
u_x &=  u_0 [A \cos(k_0x) \sin(k_0y) \cos(k_0z)]\\
u_y &=  - u_0 [A \sin(k_0x) \cos(k_0y) \cos(k_0z)] \\
u_y &= 0
\end{aligned}
\end{equation} 
where $A = 1.0$ and $k_0 = 1$ (the mode number). We also use the values $M_A = 1.0$ for the Alfven Mach number, $u_0 = 0.1$ for the initial fluid speed, and $M_s = 0.1$ for the sonic Mach number. In the presence of a uniform and ambient beginning magnetic field, we observe an oscillation in the kinetic and magnetic energies due to the constant conversion and exchange of energy between the two modes [See Fig. \ref{3D TG flow Energy}]. From Fig. \ref{3D TG flow Energy} it is observed that the outputs of the GMHD3D and PLUTO4.4 codes are identical.

We also visualize the velocity iso-surface (Iso-V surface) at any arbitrary time using GMHD3D data and PLUTO4.4 data, and discover that the iso-surfaces are indistinguishable from one another [See Fig. \ref{3D TG Iso V GMHD3D} \& \ref{3D TG Iso V PLUTO}].

\subsubsection{3D Archontis [No cosine] Flow}

The 3D Archontis [No cosine] flow is another interesting flow in the astrophysical plasmas. For this flow, we consider a three-dimensional velocity profile of the from,
\begin{equation}\label{3D Arc Flow}
\begin{aligned}
u_x &=  u_0 [A \sin(k_0z)]\\
u_y &=  u_0 [B \sin(k_0x)] \\
u_y &=  u_0 [C \sin(k_0y)]
\end{aligned}
\end{equation}
where $A$, $B$, and $C$ are constants with unity value, and $k_0 = 1$ is the mode number. All other parameters remain the same as in the preceding cases.

\begin{figure*}[h]
	\centering
	\begin{subfigure}{0.32\textwidth}
		\centering
		\includegraphics[scale=0.4]{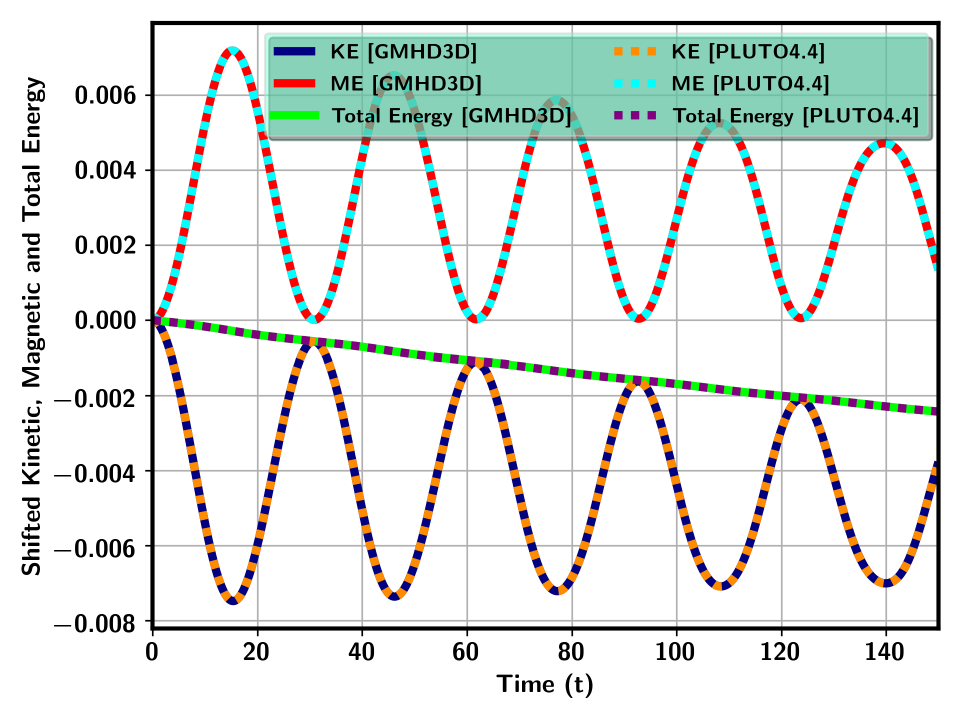}
		\caption{}
		\label{3D Archonties energy}
	\end{subfigure}
	\begin{subfigure}{0.32\textwidth}
		\centering
		\includegraphics[scale=0.047]{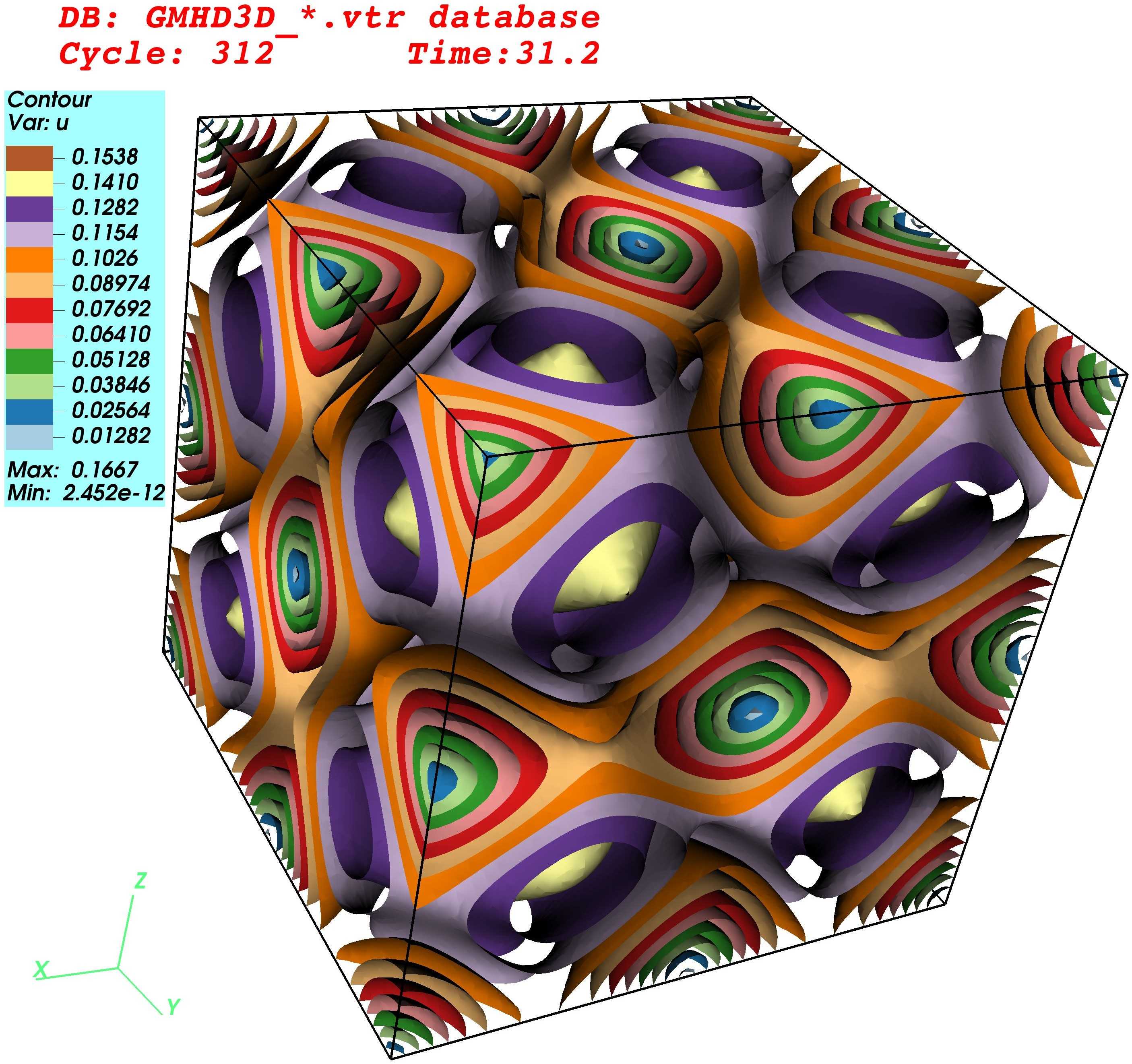}
		\caption{}
		\label{3D Archonties Iso GMHD3D}
	\end{subfigure}
	\begin{subfigure}{0.32\textwidth}
		\centering
		\includegraphics[scale=0.047]{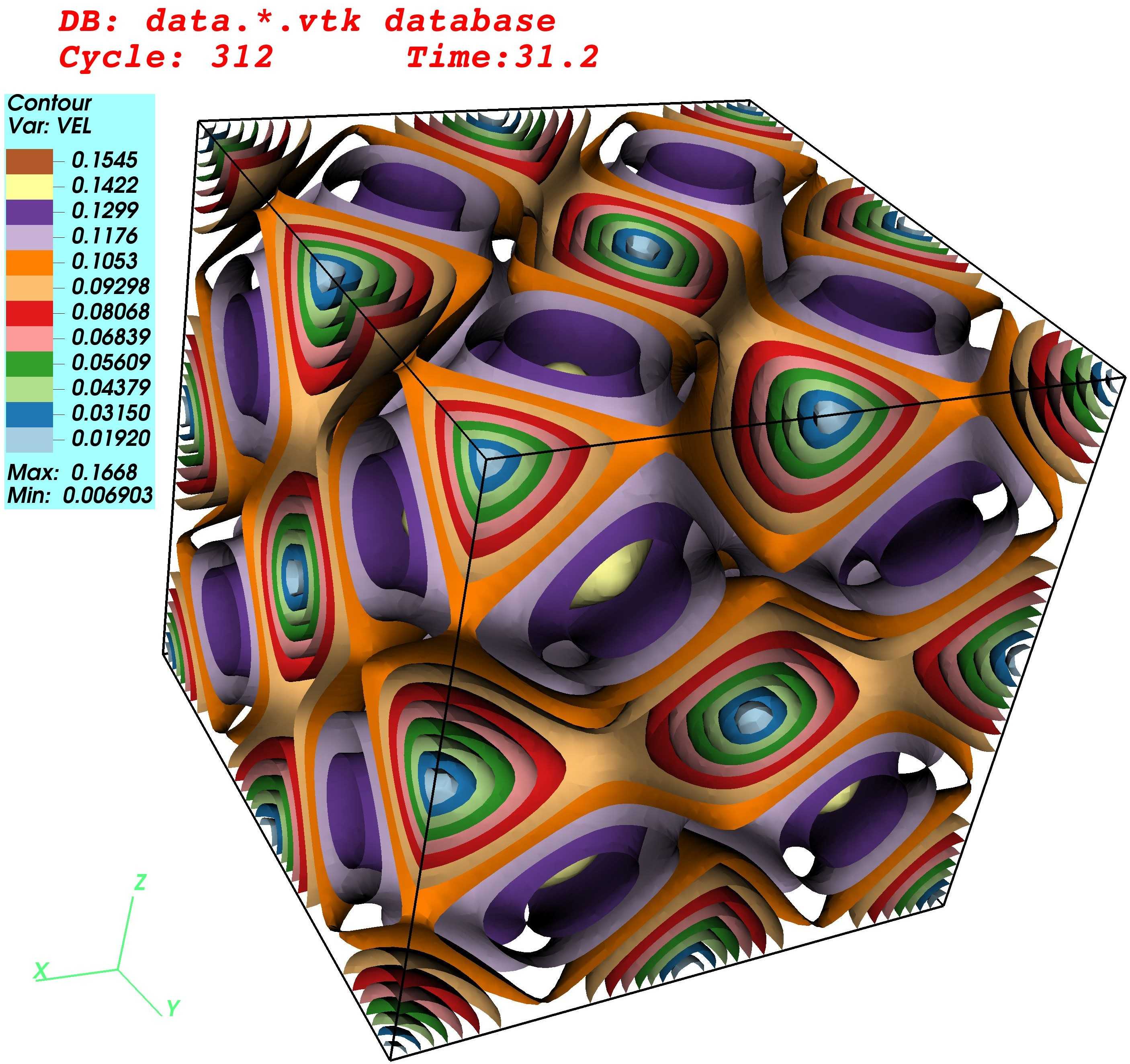}
		\caption{}
		\label{3D Archonties Iso PLUTO}
	\end{subfigure}
	\caption{(a) The shifted kinetic and magnetic energies for 3D Archontis [No cosine] flow from GMHD3D and PLUTO4.4 code. The visualization of velocity iso-surface (Iso-V) for 3D Archontis [No cosine] flow at any arbitrary time from (b) GMHD3D code  and (c) PLUTO4.4 code. Simulation Details: Reynolds number $R_e = R_m = 1000$, Grid resolution $N = 128^3$, Time stepping $dt = 10^{-4}$, initial fluid velocity $u_0 = 1.0$, Alfven Mach number $M_A = 1.0$.}
	\label{3D RO Iso V}
\end{figure*}

Like in the earlier cases, we can see the periodic exchange of energy between the kinetic and magnetic regimes [See Fig. \ref{3D Archonties energy}]. Furthermore, we find that the periods of oscillation in both codes are identical, coming in at $T = 30.680$ [See Fig. \ref{3D Archonties energy}].

Figures \ref{3D Archonties Iso GMHD3D} \& \ref{3D Archonties Iso PLUTO} illustrate the velocity iso-surface (Iso-V) of 3D Archontis flow as calculated using GMHD3D and PLUTO4.4 data, respectively. The iso-surface (Iso-V) representation reveals that the results from both codes are identical.

\subsubsection{3D Cats Eye [CE] Flow}

\begin{figure*}[h]
	\centering
	\begin{subfigure}{0.32\textwidth}
		\centering
		\includegraphics[scale=0.4]{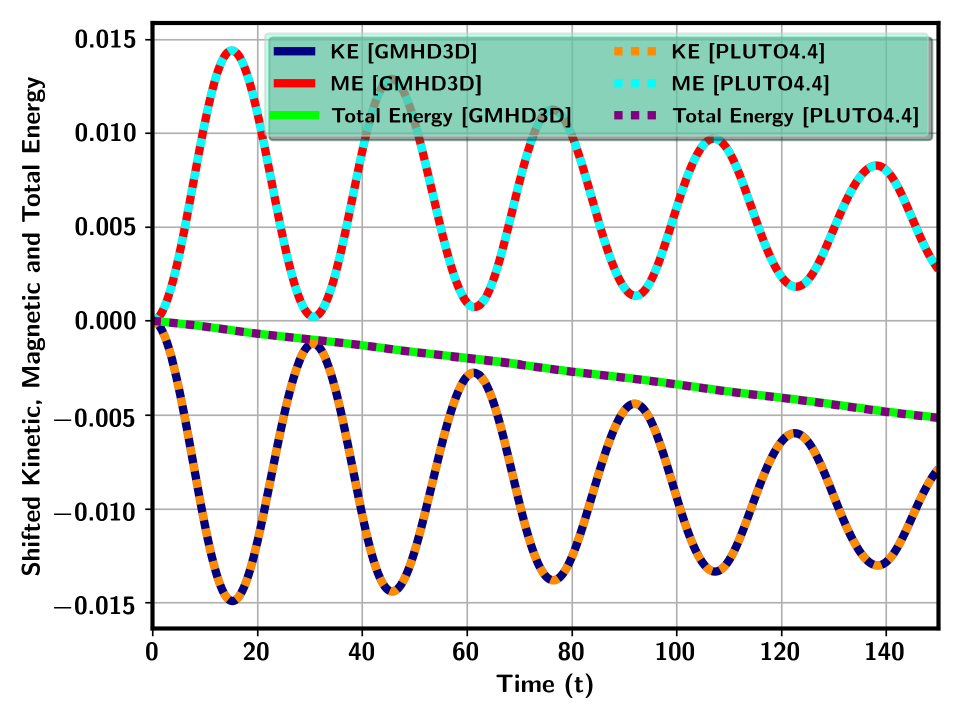}
		\caption{}
		\label{3D Cats Eye energy}
	\end{subfigure}
	\begin{subfigure}{0.32\textwidth}
		\centering
		\includegraphics[scale=0.047]{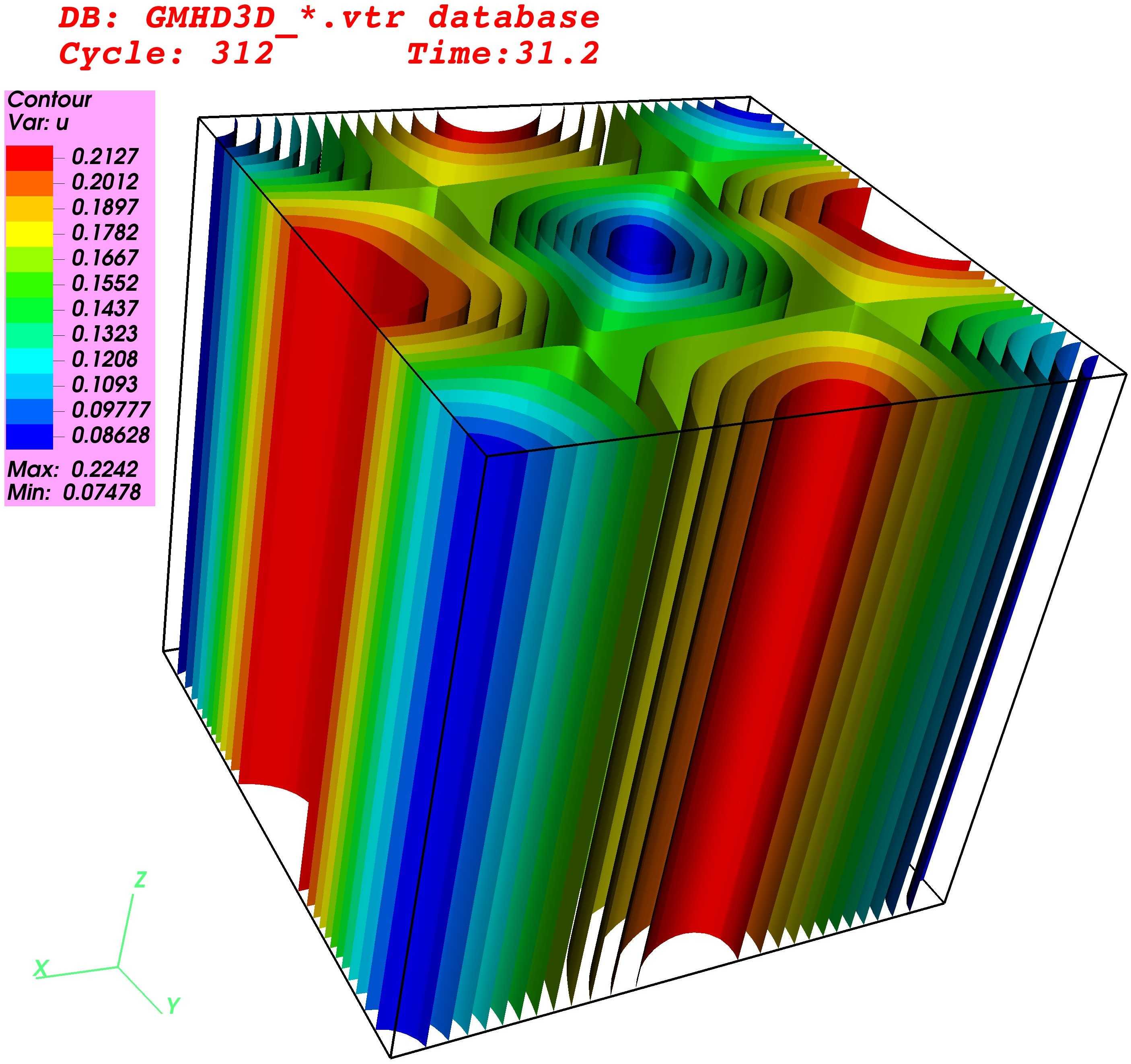}
		\caption{}
		\label{3D Cats Eye Iso V GMHD3D}
	\end{subfigure}
	\begin{subfigure}{0.32\textwidth}
		\centering
		\includegraphics[scale=0.047]{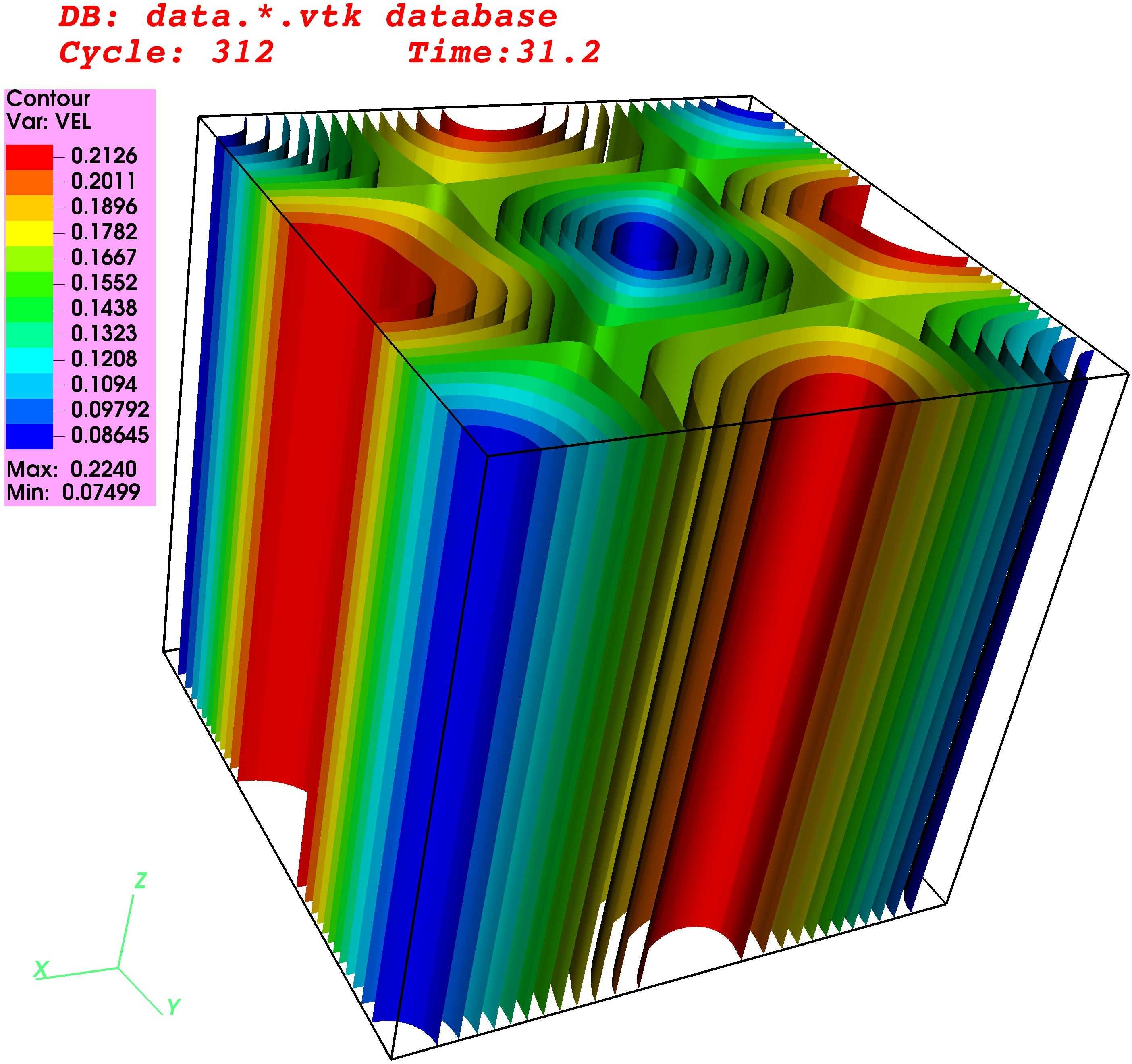}
		\caption{}
		\label{3D Cats Eye Iso V PLUTO}
	\end{subfigure}
	\caption{(a) The shifted kinetic and magnetic energies for 3D Cats Eye [CE] flow from GMHD3D and PLUTO4.4 code. The visualization of velocity iso-surface (Iso-V) for 3D Cats Eye [CE] flow at any arbitrary time from (b) GMHD3D code  and (c) PLUTO4.4 code. Simulation Details: Reynolds number $R_e = R_m = 1000$, Grid resolution $N = 128^3$, Time stepping $dt = 10^{-4}$, initial fluid velocity $u_0 = 1.0$, Alfven Mach number $M_A = 1.0$.}
	\label{3D CE Iso V}
\end{figure*}

Cats Eye [CE] flow is another well-known flow profile that we choose for the sake of completeness. The equation describing the flow profile of 3D Cats Eye [CE] flow is as follows,
\begin{equation}\label{3D CE Flow}
\begin{aligned}
u_x &=  u_0 [B \sin(k_0y)]\\
u_y &=  u_0 [A \sin(k_0x)] \\
u_y &=  u_0 [A \cos(k_0x) - B \cos(k_0y)]
\end{aligned}
\end{equation} 
where $A = \sqrt{\frac{3}{5}}$ and $B = 2A$ are real constants. The rest of the parameters are kept the same throughout the simulation, just like in the previous case. From Fig. \ref{3D Cats Eye energy} it is identified that for Cats Eye [CE] flow, the GMHD3D code and the PLUTO4.4 code reveal a periodic energy exchange between the kinetic and magnetic modes.

For additional cross-checking, we visualize the velocity iso-surface (Iso-V) using data extracted from both codes and determine that both iso-surfaces (Iso-V) are identical [See Fig. \ref{3D Cats Eye Iso V GMHD3D} \& \ref{3D Cats Eye Iso V PLUTO}].

\section{\textbf{\textcolor{black}{Coherent Nonlinear oscillations for driven flows}}}\label{Appen B}
\subsection{Test B1 [Magnetohydrodynamics]: Dynamics of a externally driven 2D flow}
Until now, we have discussed the flow dynamics without any driving force in two dimensions. Here in this subsection, we discuss about externally forced flows in 2-dimensions. Here, we start with a 2D Orszag-Tang [OT] flow profile that looks like this:
\begin{equation}\label{2D forced OT Flow}
\begin{aligned}
u_x &= - u_0 [ A \sin(k_0y)]\\
u_y &=   u_0 [ A \sin(k_0x)] \\
\end{aligned}
\end{equation}
Moreover, we force the flow profile to be equal to itself, i.e
\begin{align} 
\vec{f} &= \alpha \begin{bmatrix} 
- \sin(k_0y) \\
\sin(k_0x)\\
\end{bmatrix}
\end{align}

\begin{figure}[h]
	\includegraphics[scale=0.47]{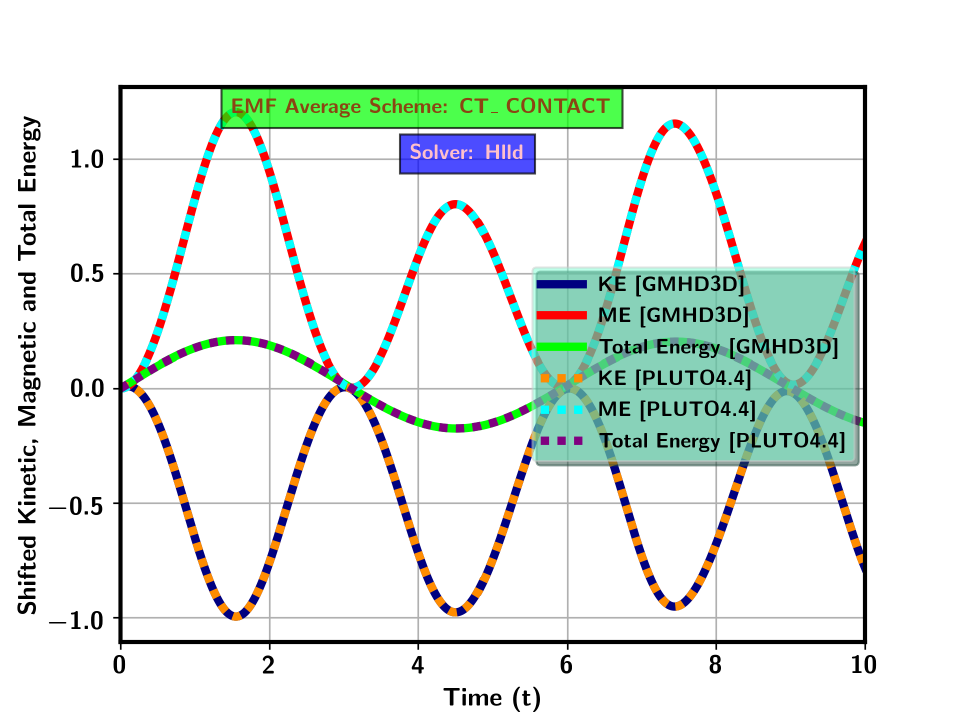}
	\caption{The shifted kinetic and magnetic energies for 2D forced Orszag-Tang Flow from GMHD2D and PLUTO4.4 (with CT\_CONTACT scheme) code at grid resolution $128^2$. Coherent non-linear oscillation with a fluctuating peak magnitude with time is observed from both the code. \textcolor{black}{Simulation Details: Time stepping $dt = 10^{-4}$}}
	\label{2D forced OT EMF}
\end{figure}

The forcing amplitude is $\alpha = 0.1$, and the drive mode number is $k_0 = 1.0$. Similar to the previous unforced Orszag-Tang [OT] scenario, we find that the kinetic and magnetic energy oscillate in the form of a coherent non-linear oscillation, but this time the peak magnitude varies with time from both the code [See Fig. \ref{2D forced OT EMF}].

\subsection{Test B2 [Magnetohydrodynamics]: Dynamics of a externally driven 3D flow (Forced ABC Flow)}
In the previous section, we have talked about Arnold-Beltrami-Childress [ABC] Flow and how it relates to coherent non-linear oscillation. Now we talk about the case where we force the 3D ABC flow with the 3D ABC flow itself. The profile of forcing can be written as,
\begin{align} 
\vec{f} &= \alpha \begin{bmatrix} 
A \sin(k_0z) + C \cos(k_0y) \\
B \sin(k_0x) + A \cos(k_0z)\\
C \sin(k_0y) + B \cos(k_0x)
\end{bmatrix}
\end{align}
where $\alpha = 0.1$, $A = B = C = 1.0$, and $k_0 = 1.0$.

\begin{figure}[h]
	\centering
	\begin{subfigure}{0.49\textwidth}
		\centering
		\includegraphics[scale=0.470]{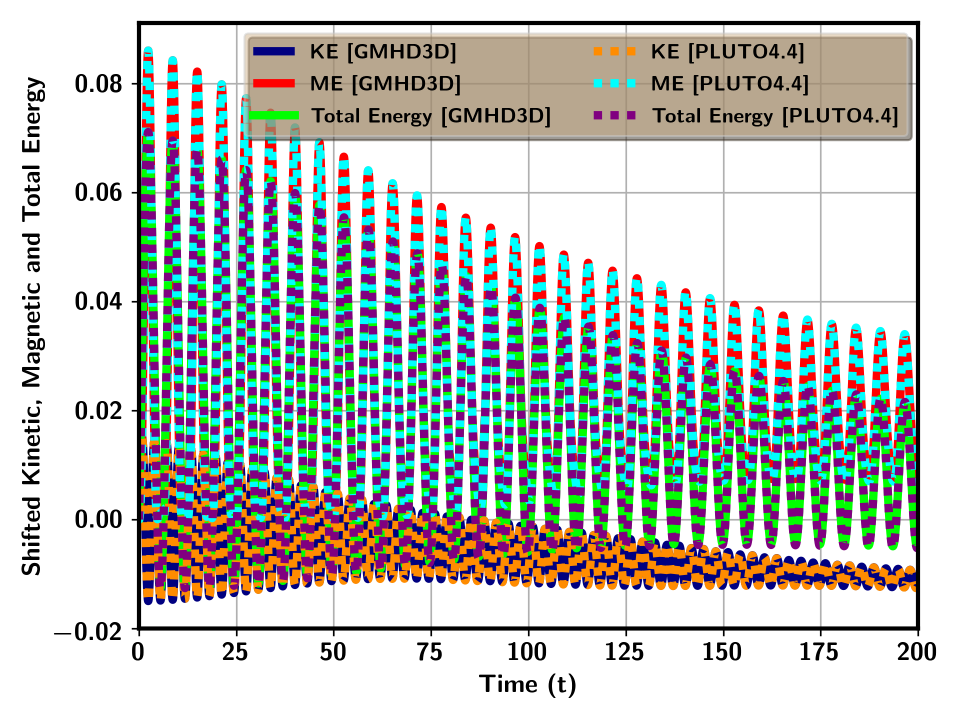}
		\caption{}
		\label{3D forced ABC}
	\end{subfigure}
	\begin{subfigure}{0.49\textwidth}
		\centering
		\includegraphics[scale=0.470]{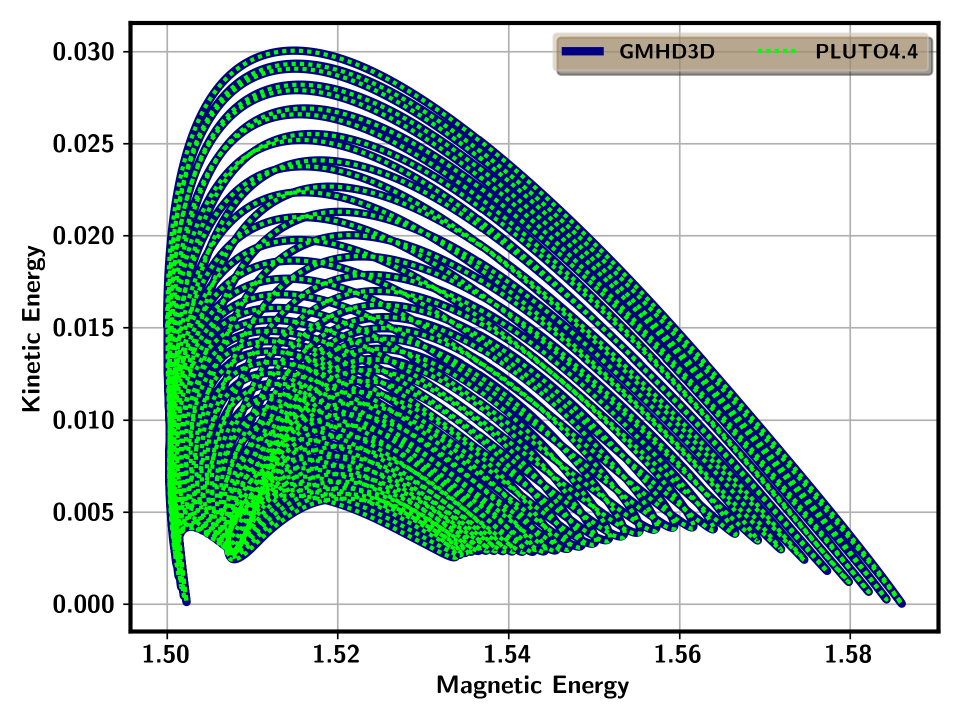}
		\caption{}
		\label{3D forced ABC Lissajous curve}
	\end{subfigure}
	\caption{(a) The shifted kinetic and magnetic energies or 3D forced Arnold–Beltrami–Childress [ABC] Flow from GMHD3D code and PLUTO4.4 code. System effectively acts as a forced relaxed system in spite of presence of driver. (b) The Lissajous curve for kinetic and magnetic energy from GMHD3D and PLUTO4.4 code. \textcolor{black}{Simulation Details: Grid resolution $N = 64^3$, Time stepping $dt = 10^{-4}$}.}
\end{figure}

Coherent non-linear oscillations of kinetic and magnetic energy are seen in both the GMHD3D and PLUTO4.4 codes, much like in forced 2D Orszag-Tang [OT] flow [See Fig. \ref{3D forced ABC}]. The system appears to operate as a forced relaxed system, despite the presence of forcing.


Both codes show an identical consistency in the dynamics of the Lissajous curve for kinetic and magnetic energy [See Fig. \ref{3D forced ABC Lissajous curve}].

%
%

\end{document}